\documentclass[a4paper,10pt]{article}
\usepackage{amsmath}
\usepackage{amssymb}
\usepackage{lmodern}
\usepackage{textgreek}
\usepackage[export]{adjustbox}
\usepackage[usenames,dvipsnames]{color}
\usepackage{sansmath}
\usepackage[margin=1.25in]{geometry}
\usepackage{times}
\usepackage{newtxtext,newtxmath}
\usepackage{framed}
\usepackage{relsize}
\usepackage{mathtools}
\usepackage{multicol}
\usepackage{xfrac}
\usepackage{graphicx}
\usepackage{tikz}
\usepackage[Symbol]{upgreek}
\usepackage{array}
\usepackage{accents}
\usepackage[T1]{fontenc}
\usepackage{esint} 
\usepackage{lipsum}
\usepackage{float}
\usepackage{bbm}
\usepackage{relsize}
\usepackage[numbers]{natbib}
\usepackage{fancyhdr}
\usepackage[font=small,labelfont=bf]{caption}
\usepackage[hang,flushmargin,bottom]{footmisc}

\makeatletter
\let\reftagform@=\tagform@
\def\tagform@#1{\maketag@@@{(\ignorespaces\textcolor{blue}{#1}\unskip\@@italiccorr)}}
\renewcommand{\eqref}[1]{\textup{\reftagform@{\ref{#1}}}}
\makeatother

\usepackage[colorlinks=true,
linkcolor=blue,
urlcolor=blue,
citecolor=blue]{hyperref}

\allowdisplaybreaks

\usetikzlibrary{decorations.pathreplacing}
\makeatletter
\def\underbrace#1{\@ifnextchar_{\tikz@@underbrace{#1}}{\tikz@@underbrace{#1}_{}}}
\def\tikz@@underbrace#1_#2{\tikz[baseline=(a.east)] {\node (a) {\(#1\)}; \draw[thick,line cap=rect,decorate,decoration={brace,amplitude=0pt}] (a.south east) -- node[below,inner sep=7pt] {\(\scriptstyle #2\)} (a.south west);}}
\makeatother

\makeatletter
\DeclareSymbolFont{largesymbolsB}{U}{esint}{m}{n}
\re@DeclareMathSymbol{\intop}{\mathop}{largesymbolsB}{'001}
        \def\int{\intop\nolimits}
\re@DeclareMathSymbol{\iintop}{\mathop}{largesymbolsB}{'003}
        
\re@DeclareMathSymbol{\iiintop}{\mathop}{largesymbolsB}{'005}
        
\re@DeclareMathSymbol{\iiiintop}{\mathop}{largesymbolsB}{'007}
        
\makeatother

\newcommand*{\dt}[1]{
  \accentset{\mbox{\small\bfseries .}}{#1}}
\newcommand*{\ddt}[1]{
  \accentset{\mbox{\small\bfseries .\hspace{0.00ex}.}}{#1}}

\newcommand\blfootnote[1]{%
  \begingroup
  \renewcommand\thefootnote{}\footnote{#1}%
  \endgroup
}

\makeatletter
\renewcommand{\@makefnmark}{\hbox{\textsuperscript{\scriptsize{\@thefnmark}}}}
\makeatother

\def\infinity{\rotatebox{90}{8}}
\newcommand{\tit}[1]{{\fontfamily{ppl}\selectfont \textit{#1}}}

\usepackage{tikz}
\usetikzlibrary{arrows}

\newcommand{\myrule} [3] []{
        \begin{tikzpicture}
            \draw[#2-#3, ultra thick, #1] (0,0) to (0.5\linewidth,0);
        \end{tikzpicture}
}

\begin{document}

\topskip120pt
\begin{titlepage}
\begin{figure}[H]
\includegraphics[width=30mm,left]{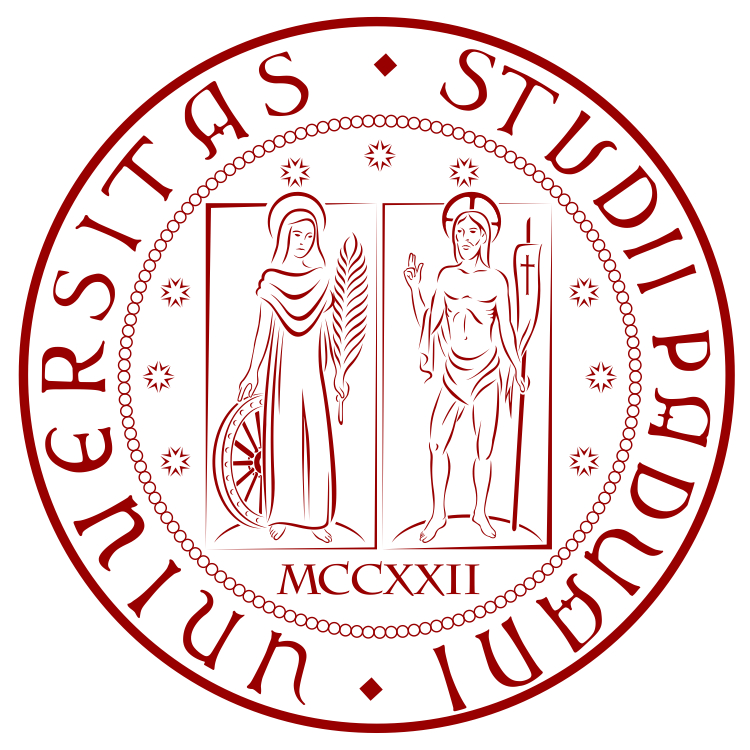}\vspace{-1in}
\includegraphics[width=40mm,right]{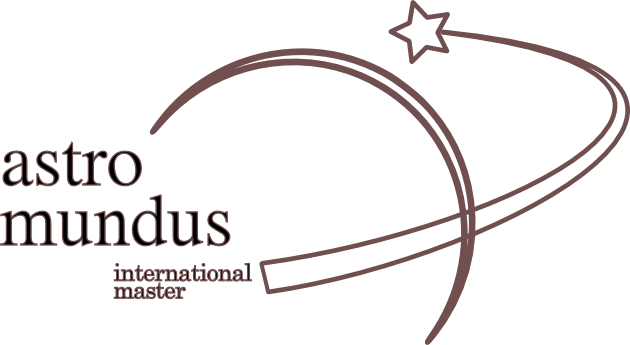}
\end{figure}

\begin{center}
\vspace{50pt}
$\;$
\linebreak 
\linebreak\linebreak\linebreak\linebreak 
\linebreak 
\linebreak 
{\Huge \textrm{Primordial non-Gaussianity in Horndeski-type model for Inflation}}
\linebreak 
\linebreak\linebreak\linebreak\linebreak 
{\Large \textrm{Avneet Singh}$^{1,2,\dagger,\ddagger}$}\blfootnote{$^\dagger$\href{avneet.singh@ligo.org}{avneet.singh@ligo.org}, $^\ddagger$\href{avneet.singh@aei.mpg.de}{avneet.singh@aei.mpg.de}}
\linebreak 
\linebreak 
{\normalsize \tit{under the advisement of}}
\linebreak 
\linebreak 
{\large \textrm{Sabino Matarrese (\tit{Supervisor})$^1$\linebreak 
Nicola Bartolo (\tit{Co-supervisor})$^1$\linebreak
Frederico Arroja (\tit{Advisor})$^1$
}}
\linebreak 
\linebreak 
\linebreak
{\normalsize $^1$\tit{Department of Physics and Astronomy, University of Padova, Via Marzolo 8, 35131 Padova, Italy}}\linebreak
{\normalsize $^2$\tit{Institute for Astro- and Particle Physics, University of Innsbruck, Technikerstra$\beta$e 25/8, A-6020 Innsbruck, Austria}} 
\linebreak 
\linebreak
\linebreak
\linebreak
{\large \textrm{{\textbf{30 September 2014}}}}\\
{\small \textrm{{\tit{edited for publication on} {25 July 2017}}}}
\linebreak 
\linebreak
\linebreak 
\linebreak
{\small \textcolor{blue}{\tit{A thesis submitted in fulfilment of the requirements for the degree of Master of Science}}}

\end{center}
\end{titlepage}

\newpage
\topskip80pt


\begin{center}
{\LARGE \textbf{Preface}}\vspace{0.4in}
\end{center}
\begin{large}
This work is a study of the concepts in Inflationary theory for the Big Bang model of the universe, Classical and Quantum field theory associated with the Inflationary framework, Horndeski single-field scalar-tensor theory, Feynman diagrams in Quantum field theory, and geometrical formulation of the ADM decomposition etc. Some new techniques and calculations are employed in the frame of this work. This work was done in requirement of the \tit{Master Thesis} for the \tit{Erasmus Mundus Program in Astronomy and Astrophysics} 2012--2014. The document draft herein contains some straightforward analysis and some advanced analytical calculations developed in the process of this review, occasionally with less-than-adequate emphasis on elaboration of preexisting concepts (appropriately referenced assuming that the reader is already familiar with the prerequisites) and on the formatting; the author apologises for the said lack of detail, wherever existent. 

\end{large}

\newpage
\topskip100pt
\begin{center}
{{\huge {\textbf{Abstract}} }}\vspace{0.3in}\\
\end{center} 
\begin{large}
We consider a member of the family of scalar-tensor theories of Inflation -- the so-called Horndeski model for Inflation, for which we calculate the non-Gaussianity (NG) and the power spectrum of curvature perturbations. We concentrate on the determination of the power spectrum, and the bi-spectrum along with the associated {$f_\mathrm{NL}$} parameter and the Spectral Index. We follow the methodology outlined within the Quantum field theory framework (\textsf{S}-matrix approach and Weinberg's `in-in' formalism) and compare -- whenever necessary -- our findings with those obtained via the traditional $\delta$N formalism. The work contained herein is an extensive review of the existing literature on this subject and the `Inflationary theory' in general, along with some newly developed techniques and calculations.
\end{large}
\newpage
\topskip120pt
\begin{center}
{{\huge {\textbf{Acknowledgments}} }}\vspace{0.3in}\\
\end{center} 
\begin{large}
\tit{
It will difficult to mention each who contributed to the process, and yet, it is very important. I would like to thank, first of all, dear Ana for her unnerving support and for the passion for Physics that I now share with her. I must also thank my mother; without her support, it would have been very much harder to get here. Moreover, my gratitude to Matteo -- with whom I share the passion for science and knowledge, my dear friends and well-wishers -- Sophie, Stella, Sandeep, Martha, William, Sebastian, Eliceth, Marlene and Leidy, for they have all affected the process of this work at different points in several different and unmodeled ways. In the end, my gratitude also goes to Prof. Sabino Matarrese, Prof. Nicola Bartolo, Dr. Frederico Arroja, and my advisor from good ol' days - Prof. Arunansu Sil. I conclude my thesis in the presence of Ef\'san S\"okmen, who must be thanked for reasons beyond comprehension.}
\end{large}
\newpage
\topskip0pt
{\hypersetup{linkbordercolor=blue}
\tableofcontents
}
\setcounter{footnote}{0}
\newpage
\topskip0pt
\section{Introduction}
\label{introduction}
Inflation has remained more or less a near-perfect theory in explaining the intricacies of the standard physics of the universe at its very birth. It has successfully resolved issues such as the horizon problem, the flatness problem, the magnetic monopole overproduction problem and the fine-tuning problem \citep{r1}, which have been found in the all-sky observations of the Cosmic Microwave Background (CMB) by Planck, WMAP etc \citep{r2,r15}. However, in order to further meet the requirements set by the specific parameters observed through CMB, such as the power spectrum ($\mathrm{P}_k$, $\mathrm{P}_\mathrm{R}$), bi-spectrum ($f\mathrm{_{NL}}$) and so on, fine-tuning of the models of Inflation is required to arrive at a concrete set of Inflationary rules which corroborate the story of the universe at later time-scales. This effort to arrive at a perfectly applicable model has led to a sea of ideas that continue to expand and evolve with passing time. Some of these models employ multiple fields and/or exotic forms of yet-to-be-confirmed physics such as Quantum Gravity, Super-symmetry etc \citep{r3}. We concentrate our study on a specific subset of such models.\\ \\
We will begin our review by considering a minimal model of Inflation and do the background study for calculating the power spectrum and the non-Gaussianity in form of bi-spectrum in the quantum field theory framework. In course of this, we will consider the primary principle behind Weinberg's `in-in formalism' for calculating the bi-spectrum \citep{r4}. Then, we will proceed to consider a specific set of Inflationary models i.e. the \tit{Horndeski} models, and perform the calculations for the power spectrum and Bi-spectrum while staying within the quantum field theory regime, i.e. employing the `in-in formalism'. We eventually compare our analytical results against the previous studies on the Horndeski models, especially in the form of expressions of the power spectrum and the bi-spectrum \citep{r7, r8}, and possibly discuss the extension of our analysis to the Tri-spectrum, which has not yet been explored for Horndeski's models -- the most general single-field scalar-tensor theory \citep{r5}.

\section{General results for the background for a minimal model}
\label{general}
We begin by considering the action I for a minimally-coupled scalar field without involving gravity, which is given by:\\
\begin{equation}
\label{eq:1}
\text{I}=\int{\mathrm{d}^4x\sqrt{-g}\bigg[\frac{1}{2}g^{\mu\nu}\partial _\mu\phi \partial _\nu\phi\ - V(\phi)\bigg]},
\end{equation}\\
where, $V(\phi)$ is the associated potential with the scalar field $\phi$. The Lagrangian ${\L}$ for our system is then given by:

\begin{equation}
\label{eq:2}
{\L}=\sqrt{-g}\bigg[\frac{1}{2}g^{\mu\nu}\partial _\mu\phi \partial _\nu\phi\ - V(\phi)\bigg]
\end{equation}\\
In such a case of a single scalar field driving the Inflation, the equations of motion are defined simply by the Euler-Lagrange equation:

\begin{equation}
\label{eq:3}
\partial _\mu\bigg(\frac{\partial {\L}}{\partial (\partial _\alpha\phi)}\bigg) - \frac{\partial {\L}}{\partial \phi} = 0.
\end{equation}\\
Using \eqref{eq:2} and \eqref{eq:3} together, we get:

\begin{equation}
\label{eq:4}
\partial _\mu\bigg[\sqrt{-g}\frac{1}{2}g^{\mu\nu}(\partial _\mu\phi)\delta^\alpha_\nu + \sqrt{-g}\frac{1}{2}g^{\mu\nu}(\partial _\nu\phi)\delta^\alpha_\mu\bigg] - \sqrt{-g}\frac{\partial V}{\partial \phi} = 0.
\end{equation}\\
Using the contraction relations, $g^{\gamma\upbeta}\delta^\alpha_\upbeta = g^{\gamma\alpha}$ and replacing the independent indices, we arrive at the following identity:

\begin{equation}
\label{eq:5}
\frac{1}{\sqrt{-g}}\partial _\nu\bigg(\sqrt{-g}g^{\mu\nu}\partial _\mu\phi\bigg) - \frac{\partial V}{\partial \phi} = 0,
\end{equation}
which is famously known as the \tit{Klein-Gordon equation}. It is often represented in the following form using the D'Alembert operator $\Box$:

\begin{equation*}
\Box = \frac{1}{\sqrt{-g}}\partial _\nu\bigg(\sqrt{-g}g^{\mu\nu}\partial _\mu\bigg),
\end{equation*}\\
such that,
\begin{equation}
\label{eq:6}
\Box\phi = \frac{\partial V}{\partial \phi}.
\end{equation}\\
In case of a Friedmann-Lema\'itre-Robertson-Walker metric (FLRW) in spherical coordinates $(r,\uptheta,\phi)$ on a flat Einstein de Sitter space-time $(k = 0)$, the metric takes the following form:

\[ g^{\mu\nu} = \left\{ \begin{array}{cccc}
-1 & 0 & 0 & 0 \\
0 & a(t)^2 & 0 & 0 \\
0 & 0 & a(t)^2 & 0 \\
0 & 0 & 0 & a(t)^2 \end{array} \right\},\] \\
where, we have re-scaled the speed of light in vacuum to unity. Thus, considering the FLRW metric and the standard relation between the scale factor $a(t)$ and the Hubble constant $\mathrm{H}$, i.e. $\dt{a}(t)/a(t) = \mathrm{H}(t)$, the Klein-Gordon equation is reduced to

\begin{equation}
\label{eq:7}
\ddt{\phi} + 3\mathrm{H}\dt{\phi} - \frac{\nabla^2\phi}{a(t)^2} + V'(\phi) = 0,
\end{equation}\\
where, in the most general scenario of a homogeneous universe, the spatial dependence of $\phi$ vanishes $(\text{i.e.}\nabla^2\phi = 0)$, yielding another reduced form of the Klein-Gordon equation,

\begin{equation}
\label{eq:8}
\ddt{\phi} + 3\mathrm{H}\dt{\phi} + V'(\phi) = 0.
\end{equation}\\
We conclude our discussion here without discussing the well-known concepts of the slow-roll parameters, particularly in this case on a frictionless and homogeneous background for the scalar field \citep{r1}. We now consider the quantum formulation of the fluctuations on a homogeneous background for the scalar field.

\section{Quantum fluctuations of a scalar field}
\label{quant}
Consider the scalar field $\phi(\tau,x)$ as:
\begin{equation}
\label{eq:9}
\phi(\tau,x) = \phi(\tau) + \delta\hspace{-0.01in}\phi(\tau,x),
\end{equation}
where, $\tau$ is the conformal time, given by $\mathrm{d}(t)/a(t) = \mathrm{d}\tau$. We note that the scalar field is essentially homogeneous in nature with only the perturbation in its background magnitude depending on spatial dimensions. Let us redefine the field for later convenience as:

\begin{equation}
\label{eq:10}
\bar{\delta}\hspace{-0.01in}\phi = a\,\delta\hspace{-0.01in}\phi.
\end{equation}
In addition, let us write down the Klein-Gordon equation for the unperturbed scalar field in conformal time:
\begin{equation}
\label{eq:11}
\Box\phi = \frac{1}{\sqrt{-g}}\partial _\nu\bigg(\sqrt{-g}g^{\mu\nu}\partial _\mu\phi\bigg) = \frac{\partial V}{\partial \phi}.
\end{equation}
We note that all the spatial derivatives must vanish and only the temporal derivatives should remain in the Klein-Gordon equation owing to the condition of homogeneity of the background field. Moreover, 
\begin{equation}
\label{eq:12}
\frac{\partial a(t)}{\partial \tau}\equiv a' = \frac{\partial a(t)}{\partial t}\frac{\partial t}{\partial \tau} = \dt{a}(t)a(t) = \mathrm{H}(t)\,a(t)^2.
\end{equation}
Using \eqref{eq:11} and \eqref{eq:12}, we get:

\begin{equation*}
\frac{1}{a^3(t)}\partial _t\bigg(-a^3(t)\partial _t\phi\bigg) - \frac{\partial V}{\partial \phi} = 0,
\end{equation*}
\begin{equation*}
\frac{1}{a^3(t)}\partial _\tau\bigg(-a^3(t)\partial _\tau\phi\frac{\partial \tau}{\partial t}\bigg)\frac{\partial \tau}{\partial t} - \frac{\partial V}{\partial \phi} = 0,
\end{equation*}
\begin{equation*}
\frac{1}{a^3(t)}\partial _\tau\bigg(-a(t)^2\partial _\tau\phi\bigg)\frac{1}{a(t)} - \frac{\partial V}{\partial \phi} = 0,
\end{equation*}
\begin{equation}
\label{eq:13}
\phi'' + 2\mathrm{H}\phi' + a(t)^2V'(\phi) = 0,
\end{equation}
where, the derivatives are with respect to the conformal time, and $\mathrm{H} = a'(t)/a(t)$ is the Hubble parameter in conformal time. 
\subsection{Second-quantization}
\label{secondquant}
Let us now recall the formulation of second-quantization. The two-dimensional (simplified case with one spatial and one temporal variable) Klein-Gordon equation for a generic scalar field $\phi$ is rewritten again as
\begin{equation*}
\Box\phi = \frac{1}{\sqrt{-g}}\partial _\nu\bigg(\sqrt{-g}g^{\mu\nu}\partial _\mu\phi\bigg) = \frac{\partial V}{\partial \phi}\indent\text{for} \indent\mu = \tau,x.
\end{equation*}
For a massive unperturbed scalar field with mass $m_\phi$ and linearized potential $V = \displaystyle{\frac{1}{2}}m^2_\phi\phi^2$ (i.e. $\partial V/\partial \phi = m^2_\phi\phi$), the Klein-Gordon equation becomes
\begin{equation*}
\Box\phi = m^2_\phi\phi,
\end{equation*}
which reduces in a 2-dimensional flat de sitter space to the simple form of
\begin{equation*}
\Box\phi = [\partial^2_\tau - \partial^2_x]\phi = m^2_\phi\phi.
\end{equation*}
The solution to this equation in time domain $\tau$ is a simple plane-wave propagating in both parallel and anti-parallel direction given by the general form of:
\begin{equation*}
\phi(\mathbf{x},\tau) = A_k(\tau)e^{i(\mathbf{k}\cdot\mathbf{x})} + B_k(\tau)e^{-i(\mathbf{k}\cdot\mathbf{x})}.
\end{equation*}
We note that the wavenumber of a mode is an induced parameter in the solution, and it is an independent quantity. Hence, we must sum over the quantity $k$ for a general solution which is simply given by
\begin{equation*}
\phi(\mathbf{x},\tau) = \sum_k[A_k(\tau)e^{i\mathbf{k}\cdot\mathbf{x}} + B_k(\tau)e^{-i\mathbf{k}\cdot\mathbf{x}}] = \frac{1}{{(2\uppi)}^{3/2}}\int{\mathrm{d}^3k[A_k(\tau)e^{i\mathbf{k}\cdot\mathbf{x}} + B_k(\tau)e^{-i\mathbf{k}\cdot\mathbf{x}}]}.
\end{equation*}
Note that since the scalar field is real, it must satisfy the following equality $$\phi(\mathbf{x},\tau) = \phi^*(\mathbf{x},\tau).$$
Under this condition, we find that $$B_k(\tau) = A_k^*(\tau).$$
Therefore, the final form of the solution reads:
\begin{equation}
\label{eq:14}
\phi(\mathbf{x},\tau) = \frac{1}{{(2\uppi)}^{3/2}}\int{\mathrm{d}^3k[A_k(\tau)e^{i\mathbf{k}\cdot\mathbf{x}} + A_k^*(\tau)e^{-i\mathbf{k}\cdot\mathbf{x}}]}.
\end{equation}
We now ``upgrade'' our classical scalar field to an operator, such that \eqref{eq:14} takes the following form:
\begin{equation}
\label{eq:15}
\bar{\phi}(x,\tau) = \frac{1}{{(2\uppi)}^{3/2}}\int{\mathrm{d}^3k[\bar{A}_k(\tau)e^{i\mathbf{k}\cdot\mathbf{x}} + \bar{A}_k^*(\tau)e^{-i\mathbf{k}\cdot\mathbf{x}}]},
\end{equation}
where, the amplitude functions $\bar{A}_k(\tau)$ and $\bar{A}_k^*(\tau)$ have been quantized and upgraded to operators. We can distinguish the time dependence of the operators in the Heisenberg picture and make the modification $\bar{A}_k(\tau) = u_k(\tau)a_k$ and $\bar{A}_k^*(\tau) = u_k^*(\tau)a_k^*$, such that
\begin{equation}
\label{eq:16}
\bar{\phi}(x,\tau) = \frac{1}{{(2\uppi)}^{3/2}}\int{\mathrm{d}^3k[u_k(\tau)\bar{a}_ke^{i\mathbf{k}\cdot\mathbf{x}} + u_k^*(\tau)\bar{a}_k^*e^{-i\mathbf{k}\cdot\mathbf{x}}]}.
\end{equation}
Now, consider the conjugate momentum for the scalar field $\bar{\phi}$, which is given by $\bar{\uppi} = \bar{\phi}'$. The following commutation relation must hold between the scalar field and its conjugate momentum:$$[\bar{\phi}, \bar{\uppi}] = i.$$
Note that this is similar to the case of a harmonic oscillator where the position and momentum operators are related by $[x,p] = i$. Moreover, the renormalization $\hbar = c = 1$ has been employed. Considering that the operators $\bar{\phi}$ and $\bar{\uppi}$ must satisfy this commutation relation, the operators could be algebraically rearranged (\tit{again} similar to the case of a quantum harmonic oscillator) such that:
\begin{equation}
\label{eq:17}
\bar{a}_k = \sqrt{\frac{m_\phi}{2}}\bar{\phi} + i\sqrt{\frac{1}{2 m_\phi}}\bar{\uppi}\indent\text{and}\indent\bar{a}_k^* = \sqrt{\frac{m_\phi}{2}}\bar{\phi} - i\sqrt{\frac{1}{2 m_\phi}}\bar{\uppi}.
\end{equation}
We can \tit{again} easily recognize these relations from the case of a harmonic oscillator and conclude that these operators are nothing but the creation and annihilation operators i.e. $\bar{a}_k \equiv a_\mathbf{k}$ and $\bar{a}_k^* \equiv a^\dagger_\mathbf{k}$. Now, the expression in \eqref{eq:16} takes the form:
\begin{equation}
\label{eq:18}
\bar{\phi}(x,\tau) = \frac{1}{{(2\uppi)}^{3/2}}\int{\mathrm{d}^3k[u_k(\tau)a_\mathbf{k}e^{i\mathbf{k}\cdot\mathbf{x}} + u_k^*(\tau)a^\dagger_\mathbf{k}e^{-i\mathbf{k}\cdot\mathbf{x}}]}.
\end{equation}
It is useful to note that in the simplest of cases (i.e. when the Klein-Gordon equation takes the form of a wave equation $\Box\phi = [\partial^2_t - \partial^2_x] = m^2_\phi\phi$), the amplitude functions are nothing but $u_k(\tau) \propto \displaystyle{e^{\pm i\Omega\tau}}$, i.e. time-dependent amplitudes of general plane-wave solutions with $\Omega = |(k^2 + m^2_\phi)^{1/2}|$. Here we conclude our discussion on second-quantization of a scalar field.

\section{Back to Cosmology}
\label{cosmo}
Let us again reconsider \eqref{eq:9} in parallel with the second-quantization principle given by \eqref{eq:18}. We can simply write down the expression for the field perturbations (i.e. the perturbation term only) by keeping the background field purely classical, i.e.  
\begin{equation}
\label{eq:19}
\bar{\delta}\hspace{-0.01in}\phi(\mathbf{x},\tau) = \frac{1}{{(2\uppi)}^{3/2}}\int{\mathrm{d}^3k[u_k(\tau)a_\mathbf{k}e^{i\mathbf{k}\cdot\mathbf{x}} + u_k^*(\tau)a^\dagger_\mathbf{k}e^{-i\mathbf{k}\cdot\mathbf{x}}]}.
\end{equation}
We now have a few cases to consider. The first and the foremost of them is when the wavelength of a perturbation remains within the horizon, i.e. on the so-called \tit{sub-horizon scales}. In such a case, the density perturbations depend on both the conformal time $\tau$ and the spatial coordinates. In that case, \eqref{eq:13} must be modified to include the spatial derivatives. The modified form of \eqref{eq:13} after including the spatial dependence is then given by:
\begin{equation}
\label{eq:20}
\phi'' + 2\mathrm{H}\phi' - \frac{\partial_x^2\phi}{a(t)^2} + V'(\phi) = 0.
\end{equation}
The above expression, when combined with \eqref{eq:9} and \eqref{eq:19}, yields
\begin{equation*}
\frac{\partial^2[\phi(\tau) + {\delta}\hspace{-0.01in}\phi(\mathbf{x},\tau)]}{\partial\tau^2} + 2\mathrm{H}\frac{\partial[\phi(\tau) + {\delta}\hspace{-0.01in}\phi(\mathbf{x},\tau)]}{\partial\tau} - \frac{\partial_x^2[\phi(\tau) + {\delta}\hspace{-0.01in}\phi(\mathbf{x},\tau)]}{a(t)^2} + a(t)^2V'[\phi(\tau) + {\delta}\hspace{-0.01in}\phi(\mathbf{x},\tau)] = 0,
\end{equation*}
which upon incorporation with the field redefinition introduced in \eqref{eq:10}, converts to the following form:

\begin{equation}
\label{eq:21}
\phi'' + \delta\hspace{-0.01in}\phi'' + 2\mathrm{H}\phi' +  2\mathrm{H}\delta\hspace{-0.01in}\phi' - \frac{\partial_x^2(\delta\hspace{-0.01in}\phi)}{a(t)^2} + a(t)^2V'[\phi(\tau) + {\delta}\hspace{-0.01in}\phi(\mathbf{x},\tau)] = 0.
\end{equation}
Now, we may expand the potential term as follows:
\begin{equation}
\label{eq:22}
V'[\phi(\tau) + {\delta}\hspace{-0.01in}\phi(\mathbf{x},\tau)] = \frac{\partial\bigg[V[\phi(\tau)] + \frac{\partial V}{\partial\phi}{\delta}\hspace{-0.01in}\phi(\mathbf{x},\tau)\bigg]}{\partial\phi} = V'(\phi) + \frac{\partial^2V}{\partial\phi^2}{\delta}\hspace{-0.01in}\phi(\mathbf{x},\tau).
\end{equation}
Then \eqref{eq:21} combined with \eqref{eq:13} and \eqref{eq:22} is written as
\begin{equation*}
\delta\hspace{-0.01in}\phi'' + 2\mathrm{H}\,\delta\hspace{-0.01in}\phi' + \frac{k^2}{a(t)^2}\delta\hspace{-0.01in}\phi(\mathbf{x},\tau) + a(t)^2m^2_\phi\delta\hspace{-0.01in}\phi(\mathbf{x},\tau) = 0,
\end{equation*}
which can be further reduced using \eqref{eq:12} and \eqref{eq:19} (considering only the positive energy mode) to
\begin{equation}
\label{eq:23}
u_k(\tau)'' + \bigg[k^2 - \frac{a''}{a} + a^2m^2_\phi\bigg] u_k(\tau) = 0.
\end{equation}
On sub-horizon scales, from \eqref{eq:12}, we can deduce that:
\begin{equation}
\label{eq:24}
\frac{a''(t)}{a(t)} = 2\mathrm{H}^2a(t)^2,
\end{equation}
assuming that the Hubble parameter is nearly constant in time such that $\ddt{a}(t)/a(t) = \mathrm{H}^2 + \dt{\mathrm{H}} \sim \mathrm{H}^2$.
We must note that the wavenumber(s) $k$ is transformed along with the field re-definition introduced in \eqref{eq:10}. This tranformation is simply given by $k\rightarrow k/a(t)$, and it can be easily interpreted from the Klein-Gordon equation for the original scalar field ($[\partial^2_\tau - \partial^2_x]\phi = m^2_\phi\phi$). Now, consider again the case of sub-horizon scales when $a(t)\uplambda<\mathrm{H}^{-1}$; this inequality can be re-written as:$$k \gg a\mathrm{H},$$
$$k^2 \gg a^2\mathrm{H}^2 \sim \frac{1}{2}\frac{a''(t)}{a(t)}.$$
Under these conditions, along with the assumption of negligible mass $m_\phi$ of the scalar field, \eqref{eq:23} reduces to 
\begin{equation}
\label{eq:25}
u_k(\tau)'' + k^2 u_k(\tau) = 0,
\end{equation}
whose solution is a plane wave given by
\begin{equation}
\label{eq:26}
u_k(\tau) = \frac{1}{\sqrt{2k}}e^{\pm ik\tau}.
\end{equation}
Similarly, for the case of scales in the \tit{super-horizon limit}, i.e. when $$k^2 \ll a^2\mathrm{H}^2 \sim \frac{1}{2}\frac{a''(t)}{a(t)},$$
the amplitudes $u_k$ satisfy the following differential equation in $\tau$:
\begin{equation}
\label{eq:27}
u_k(\tau)'' - \bigg[\frac{a''}{a} - a^2m^2_\phi\bigg] u_k(\tau) = 0.
\end{equation}
Before proceeding further, let us take a look at the definition of conformal time and the Hubble parameter, $$\frac{\dt{a}(t)}{a(t)} = \mathrm{H}(t) \sim \text{constant}.$$The solution to this differential equation yields:
\begin{equation}
\label{eq:28}
a(t) \propto e^{\mathrm{H}t}.
\end{equation}
This result, when combined with the definition of conformal time ($\mathrm{d}t/a(t) = \mathrm{d}\tau$), gives
\begin{equation*}
\int{\frac{\mathrm{d}t}{e^{\mathrm{H}t}}} = \int{\mathrm{d}\tau} = \tau \;\;\text{and},
\end{equation*}
\begin{equation}
\label{eq:29}
\tau = \frac{-1}{\mathrm{H}a(t)}.
\end{equation}
Furthermore, we can deduce that
\begin{equation}
\label{eq:30}
a'(t) = \frac{1}{\mathrm{H}\tau^2} = \mathrm{H}a^2, \indent\indent a''(t) = \frac{-2}{\mathrm{H}\tau^3} = 2\mathrm{H}^2a^3.
\end{equation}
Now reconsider \eqref{eq:27} for a scalar field with negligible mass ($m_\phi \sim 0$),
\begin{equation}
\label{eq:31}
u_k(\tau)'' - \bigg[\frac{a''}{a}\bigg] u_k(\tau) = 0.
\end{equation}
We can further reform $u_k''$ as: $$u_k' = \frac{\mathrm{d}u_k}{\mathrm{d}a}\frac{\mathrm{d}a}{\mathrm{d}\tau}$$ $$u_k'' = \frac{\mathrm{d}}{\mathrm{d}\tau}\bigg(\frac{\mathrm{d}u_k}{\mathrm{d}a}\bigg)\frac{\mathrm{d}a}{\mathrm{d}\tau} + \frac{\mathrm{d}u_k}{\mathrm{d}a}\frac{\mathrm{d}^2a}{\mathrm{d}\tau^2} = \frac{\mathrm{d}}{\mathrm{d}a}\bigg(\frac{\mathrm{d}u_k}{\mathrm{d}a}\bigg)\bigg(\frac{\mathrm{d}a}{\mathrm{d}\tau}\bigg)^2 + \frac{\mathrm{d}u_k}{\mathrm{d}a}\frac{\mathrm{d}^2a}{\mathrm{d}\tau^2}$$
\begin{equation}
\label{eq:32}
u_k'' = \frac{\mathrm{d}^2u_k}{\mathrm{d}a^2}\bigg(\frac{\mathrm{d}a}{\mathrm{d}\tau}\bigg)^2 + \frac{\mathrm{d}u_k}{\mathrm{d}a}\frac{d^2a}{\mathrm{d}\tau^2} = \frac{\mathrm{d}^2u_k}{\mathrm{d}a^2}a'^2 + \frac{\mathrm{d}u_k}{\mathrm{d}a}a''
\end{equation}
Combining \eqref{eq:29}, \eqref{eq:30}, \eqref{eq:31} and \eqref{eq:32}, the amplitudes of modes in the super-horizon limit satisfy
\begin{equation}
\label{eq:33}
\frac{a^2}{2}\frac{\mathrm{d}^2u_k}{\mathrm{d}a^2} - a\frac{\mathrm{d}u_k}{\mathrm{d}a} - u_k = 0,
\end{equation}
where, obviously $u_k(\tau) \rightarrow u_k(a)$. The solution to this second-order homogeneous differential equation is given by:
\begin{equation}
\label{eq:34}
u_k(a) = C_{k_+} a + C_{k_-} a^{-2}.
\end{equation}\\
\begin{flushleft}\myrule[line width = 0.5mm]{fast cap reversed}{fast cap reversed}\end{flushleft}
\tit{\textbf{Sidenote}}:\\
Alternatively, we can also find the solution for  differential equation in \eqref{eq:31} as follows:\\ \\
$\bullet$ From the structure of \eqref{eq:31}, it is clear that $u_k = a$ is one of the trivial solutions. We only need to find the second solution to this equation.\\ \\
$\bullet$ The trick to find the second solution, if first solution (say, $u^{(1)}_k$) is known, is to use the method of `integrating factor', i.e.$$a^2\frac{\mathrm{d}}{\mathrm{d}a}\bigg(\frac{u^{(2)}}{u^{(1)}_k}\bigg) =  \int e^{-\mathlarger{\frac{2}{a}}}\mathrm{d}a,$$ which yields the second solution as $u^{(2)}_k = a^{-2}$. Ultimately, the true solution is the linear combination of the two.
\begin{flushright}\myrule[line width = 0.5mm]{fast cap reversed}{fast cap reversed}\end{flushright}$\;$\\
For a fast expanding universe, we can neglect the decaying mode of the solution ($\propto a^{-2}$). Finally, $$|\delta\hspace{-0.01in}\phi_k| = \frac{|u_k|}{a}.$$
At horizon exit, we can define the boundary conditions for the seperate solutions for the sub-horizon and super-horizon cases such that
\begin{equation*}
u_{k:\textsf{sub-horizon}}(\tau=\tau_\textsf{exit}) = u_{k:\textsf{super-horizon}}(\tau=\tau_\textsf{exit}), 
\end{equation*}
which yields,
\begin{equation}
\label{eq:35}
C_{k_+} = \frac{1}{a(t=t_\textsf{exit})\sqrt{2k}} = \frac{\mathrm{H}}{\sqrt{2k^3}}.
\end{equation}
\subsection{General formulation for the solution}
\label{gensol}
Let us now solve for \eqref{eq:23} without any approximations. Using the relations \eqref{eq:29} and \eqref{eq:30}, we arrive at the following equality:
\begin{equation}
\label{eq:36}
u_k(\tau)'' + \bigg[k^2 - \frac{v^2_\phi -\frac{1}{4}}{\tau^2}\bigg] u_k(\tau) =  0,\indent\text{where,}\indent v^2_\phi = \frac{9}{4} - \frac{m^2_\phi}{\mathrm{H}^2}. 
\end{equation}
The general solution for this kind of equation is a function of linear combination of the \tit{Bessel functions} of the first and the second kind. Without going into much trivial discussion, we conclude that on super-horizon scales, the density perturbations take the form of \citep{r9}
\begin{equation*}
u_k(\tau) = e^{i\big(\scriptstyle v_\phi+\frac{\scriptstyle 1}{\scriptstyle 2}\big){\mathlarger{{\frac{\scriptstyle \uppi}{2}}}}}2^{v_\phi - \frac{\scriptstyle 3}{\scriptstyle 2}} \frac{\Gamma(v_\phi)}{\Gamma(3/2)} \frac{1}{\sqrt{2k}}(-k\tau)^{(\frac{1}{2}-v_\phi)},
\end{equation*}
such that,
\begin{equation}
\label{eq:37}
|\delta\hspace{-0.01in}\phi_k| = 2^{v_\phi - \frac{\scriptstyle 3}{\scriptstyle 2}} \frac{\Gamma(v_\phi)}{\Gamma(3/2)} \frac{\mathrm{H}}{\sqrt{2k^3}} {\bigg(\frac{k}{a\mathrm{H}}\bigg)}^{3/2 - v_\phi}.
\end{equation}
Needless to say, on sub-horizon scales, the standard plane wave-solution is recovered. Taking a careful look at \eqref{eq:37} suggests that $$|\delta\hspace{-0.01in}\phi_k| \geq 0,\indent\text{which implies}\indent \Gamma(v_\phi) \geq 0.$$
Note that the Gamma function $\Gamma$ is positive for positive domain values. Therefore, 
\begin{equation}
\label{eq:38}
\Gamma(v_\phi) \geq 0\indent \text{requires}\indent v_\phi \geq 0 \indent \text{or,}\indent m_\phi \leq \frac{3}{2}\mathrm{H}.
\end{equation}
For the case of a very light scalar field, i.e. $v_\phi \sim 3/2 $, \eqref{eq:37} reduces to the following simple form of
\begin{equation}
\label{eq:39}
|\delta\hspace{-0.01in}\phi_k| = \frac{\mathrm{H}}{\sqrt{2k^3}} {\bigg(\frac{k}{a\mathrm{H}}\bigg)}^{\upeta_\phi},\indent\text{such that}\indent \upeta_\phi = 3/2 - v_\phi.
\end{equation}
Note that\footnote{It is useful to relate $\upeta_\phi$ to the standard slow-roll parameter which actually equals $m^2_\phi/3\mathrm{H}^2$ for a massive field.} $\upeta_\phi = 3/2 - v_\phi = m^2_\phi/3\mathrm{H}^2$ when $m^2_\phi/3\mathrm{H}^2 \ll 1$.
\section{Power spectrum for the minimal model}
\label{powerspectrum}
In this section, we discuss the mathematics detailing the statistics of the distribution of a scalar field.
\subsection{Defining power spectrum}
\label{defineps}
Let us start by defining a \tit{two-point correlation function} (also known as the \tit{auto-correlation function} when it correlates the values of the same function). Consider a complex function $f$ in some arbitrary domain, say $\vec t$. Now, the two-point correlation (or, auto-correlation) function defined under the transformation $\vec t \rightarrow \vec t + \vec \tau$ is given by\footnote{We will skip the $1/{2\uppi}$ factor in the Fourier transforms (or, the inverse Fourier transforms) for simplicity.}
\begin{equation}
\label{eq:40}
\,\langle f(\vec t)\,f^*(\vec t + \vec \tau)\rangle = \int^{\displaystyle\infinity}_{-{\displaystyle\infinity}}f(\vec t)\,f^*(\vec t + \vec \tau)\mathrm{d}\vec t.
\end{equation}
Considering the Fourier transform of the function $f$:
\begin{equation*}
\,\langle f(\vec t)\,f^*(\vec t + \vec \tau)\rangle =\,\bigg\langle \int^{\displaystyle\infinity}_{-{\displaystyle\infinity}}\int^{\displaystyle\infinity}_{-{\displaystyle\infinity}}\mathrm{d}k_1 \mathrm{d}k_2 \,f(k_1)\,f^*(k_2)
e^{i\vec k_1\cdot\vec t}
e^{-i\vec k_2\cdot(\vec t + \vec\tau)}\bigg\rangle,
\end{equation*}
\begin{equation}
\label{eq:41}
\,\langle f(\vec t)\,f^*(\vec t + \vec \tau)\rangle = \int^{\displaystyle\infinity}_{-{\displaystyle\infinity}}\int^{\displaystyle\infinity}_{-{\displaystyle\infinity}}\mathrm{d}k_1 \mathrm{d}k_2 \,\,\langle f(k_1)\,f^*(k_2)\rangle
e^{i\vec k_1\cdot\vec t}
e^{-i\vec k_2\cdot(\vec t + \vec\tau)}.
\end{equation}\pagebreak
\begin{flushleft}\myrule[line width = 0.5mm]{fast cap reversed}{fast cap reversed}\end{flushleft}\vspace{-0.1in}
\tit{\textbf{Sidenote}}:\\
In order to prove the last identity \eqref{eq:41}, consider the following scenario. In general, two-point correlation function for two statistically discrete functions $f$ and $g$ (which in turn are functions of a discrete measurable variable $x$) is given by:
\begin{equation}
\label{eq:42}
\,\langle f(x)\,g(x)\rangle = \sum_{i = -\displaystyle\,\infinity}^{\displaystyle\infinity} \sum_{\mathrm{j} = -\displaystyle\,\infinity}^{\displaystyle\infinity}f(x_i)\,g(x_\mathrm{j})\,\mathrm{P}[f(x_i)\,g(x_\mathrm{j})],
\end{equation}
where, $\mathrm{P}$ is the \tit{joint probability distribution function} of simultaneous occurrence of $f$ and $g$. If $f$ and $g$ are independent functions, the relation reduces to the form (since $\mathrm{P}[f(x_i)\,g(x_\mathrm{j})] = \mathrm{P}[f(x_i)]\,\mathrm{P}[g(x_\mathrm{j})]$),
\begin{equation*}
\,\langle f(x)\,g(x)\rangle = \sum_{i = -\displaystyle\,\infinity}^{\displaystyle\infinity}f(x_i)\mathrm{P}[f(x_i)]\, \sum_{\mathrm{j} = -\displaystyle\,\infinity}^{\displaystyle\infinity}g(x_\mathrm{j})\,\mathrm{P}[g(x_\mathrm{j})] = \,\langle f(x)\rangle \,\langle g(x)\rangle.
\end{equation*}
In addition, the 2-point correlation function in a continuous range of variables is given by:
\begin{equation}
\label{eq:43}
\,\langle f(x_1)\,g(x_2)\rangle =\int^{\displaystyle\infinity}_{-{\displaystyle\infinity}} \int^{\displaystyle\infinity}_{-{\displaystyle\infinity}}f(x_1)\,g(x_2)\,\mathrm{P}[f(x_1)\,g(x_2)]\mathrm{d}[f(x_1)]\mathrm{d}[f(x_2)].
\end{equation}
Now, for simplicity, we take the case of one discrete random function $f(x)$. Consider a scenario where we make repeated measurements of $f$ over the specified domain at some point $x_i$. The ensemble average for such a set of measurements would be given by
\begin{equation}
\label{eq:44}
\,\langle f(x)\rangle = \lim_{i \rightarrow \displaystyle\infinity}\sum f(x_i)\mathrm{P}[f(x_i)] = \int^{\displaystyle\infinity}_{-{\displaystyle\infinity}}f(x)\mathrm{P}[f(x)]\mathrm{d}[f(x)].
\end{equation}
Introducing the Fourier transform for the function $f(x)$, 
\begin{equation}
\label{eq:45}
f(x)=\int^{\displaystyle\infinity}_{-{\displaystyle\infinity}}f(k)e^{ikx}\mathrm{d}k,
\end{equation}
for repeated measurements of $f(x)$ at $x=x_i$, the variation in the Fourier space occurs only in the amplitudes of the Fourier modes. Therefore, we can write the above equation for $i$th measurement as
\begin{equation}
\label{eq:46}
f(x_i)=\int^{\displaystyle\infinity}_{-{\displaystyle\infinity}}f_{k_i}(k)e^{ikx}\mathrm{d}k.
\end{equation}
Thus, using \eqref{eq:46}, \eqref{eq:44} reduces to
\begin{equation}
\label{eq:47}
\,\langle f(x)\rangle = \lim_{i \rightarrow \displaystyle\infinity}\sum \int^{\displaystyle\infinity}_{-{\displaystyle\infinity}}f_{k_i}(k)e^{ikx}\mathrm{d}k\,\mathrm{P}[f(x_i)] = \int^{\displaystyle\infinity}_{-{\displaystyle\infinity}}\int^{\displaystyle\infinity}_{-{\displaystyle\infinity}}f(k)e^{ikx}\mathrm{d}k\mathrm{P}[f(x)]\,\mathrm{d}[f(x)].
\end{equation}
Moreover, one must intuitively recognize that the probability density function $\mathrm{P}[f(x_i)]$ in real domain and the probability density function $\mathrm{P}[f_{k_i}]$ in Fourier space are the same; this is because the probability of random occurrence of some value for the function $f$ is equivalent to the probability of variance in the corresponding mode amplitudes. Hence,
\begin{equation}
\label{eq:48}
\mathrm{P}[f(x_i)] = \mathrm{P}[f_{k_i}(k)].
\end{equation}
Accordingly, from \eqref{eq:47} and \eqref{eq:48}, we conclude that
\begin{equation}
\label{eq:49}
\,\langle f(x)\rangle =\lim_{i \rightarrow \displaystyle\infinity}\int^{\displaystyle\infinity}_{-{\displaystyle\infinity}}\mathrm{d}k \,e^{ikx}\sum f_{k_i}(k)\mathrm{P}[f_{k_i}(k)] = \int^{\displaystyle\infinity}_{-{\displaystyle\infinity}}\mathrm{d}k e^{ikx}\,\langle f_{k_i} \rangle.
\end{equation}
Hence, we have proved the identity \eqref{eq:49} used to derive the expression \eqref{eq:41}. This identity may be proven in a similar but more involved manner for the more complex case of a two-point correlation function.
\begin{flushright}\myrule[line width = 0.5mm]{fast cap reversed}{fast cap reversed}\end{flushright}
Now, for a perturbation in the scalar field $\phi$, we use the analogous form
\begin{equation}
\label{eq:50}
\,\langle \delta\hspace{-0.01in}\phi(\vec x)\,\delta\hspace{-0.01in}\phi^*(\vec x + \vec r)\rangle =  \int^{\displaystyle\infinity}_{-{\displaystyle\infinity}}\int^{\displaystyle\infinity}_{-{\displaystyle\infinity}}\mathrm{d}k_1 \mathrm{d}k_2 \,\,\langle \delta\hspace{-0.01in}\phi(k_1)\,\delta\hspace{-0.01in}\phi^*(k_2)\rangle
e^{i\vec k_1\cdot\vec x}
e^{-i\vec k_2\cdot(\vec x + \vec r)}.
\end{equation}
The quantity $\,\langle \delta\hspace{-0.01in}\phi(k_1)\,\delta\hspace{-0.01in}\phi^*(k_2)\rangle$ is defined as the \tit{power spectrum} of the density fluctuations. We can derive the explicit relation for power spectrum in the following way using the inverse Fourier transform:
\begin{equation}
\label{eq:51}
\,\langle \delta\hspace{-0.01in}\phi(k_1)\,\delta\hspace{-0.01in}\phi^*(k_2)\rangle =  \int^{\displaystyle\infinity}_{-{\displaystyle\infinity}}\int^{\displaystyle\infinity}_{-{\displaystyle\infinity}}\mathrm{d}\vec x\, \mathrm{d}(\vec x + \vec r)\,\,\langle \delta\hspace{-0.01in}\phi(\vec x)\,\delta\hspace{-0.01in}\phi^*(\vec x + \vec r)\rangle
e^{-i\vec k_1\cdot\vec x}
e^{i\vec k_2\cdot(\vec x + \vec r)},
\end{equation}
where $\,\langle \delta\hspace{-0.01in}\phi(\vec x)\,\delta\hspace{-0.01in}\phi^*(\vec x + \vec r)\rangle = \xi(\vec x, \vec r)$ is the \tit{spatial correlation function}. Therefore,
\begin{equation*}
\,\langle \delta\hspace{-0.01in}\phi(k_1)\,\delta\hspace{-0.01in}\phi^*(k_2)\rangle =  \int^{\displaystyle\infinity}_{-{\displaystyle\infinity}}\mathrm{d}\vec r\,\xi(\vec x,\vec r) e^{i\vec k_2\cdot\vec r}
\int^{\displaystyle\infinity}_{-{\displaystyle\infinity}}\mathrm{d}\vec x\, e^{-i(\vec k_1 - \vec k_2)\cdot\vec x},
\end{equation*}
where the following identities hold; $$\int^{\displaystyle\infinity}_{-{\displaystyle\infinity}}\mathrm{d}\vec x\, e^{-i(\vec k_1 - \vec k_2)\cdot\vec x} = (2\uppi)^2\,\delta (\vec k_1 - \vec k_2),$$
and\footnote{The additional factor of $1/2|\vec k|^3$ in the definition of power spectrum is chosen for convenience. It leads to Lorentz invariance of the power spectrum function.},
\begin{equation}
\label{eq:52}
\int^{\displaystyle\infinity}_{-{\displaystyle\infinity}}\mathrm{d}\vec r\,\xi(\vec x,\vec r) e^{i\vec k\cdot\vec r} \equiv \frac{1}{2|\vec k|^3}\mathrm{P}_k (\vec k).
\end{equation}
The homogenity and isotropy of the density of perturbations (translational and rotational invariance of $\xi$) requires that $\xi(\vec x, \vec r)\equiv \xi(|\vec r|)$, which when combined with \eqref{eq:51}, reduces $\langle \delta\hspace{-0.01in}\phi(k_1)\,\delta\hspace{-0.01in}\phi^*(k_2)\rangle$ to
\begin{equation*}
\,\langle \delta\hspace{-0.01in}\phi(k_1)\,\delta\hspace{-0.01in}\phi^*(k_2)\rangle = (2\uppi)^2\,\delta (\vec k_1 - \vec k_2)\int^{\displaystyle\infinity}_{-{\displaystyle\infinity}}\mathrm{d}|\vec r|\,\xi(|\vec r|) e^{i|\vec k||\vec r|} \indent\text{since}\indent \vec r \rightarrow |\vec r|\;\;\text{implies}\;\;\vec k \rightarrow |\vec k|,
\end{equation*}
Thus, the power spectrum is finally written by combining the above expression with the definition \eqref{eq:52},
\begin{equation}
\label{eq:53}
\,\langle \delta\hspace{-0.01in}\phi(k_1)\,\delta\hspace{-0.01in}\phi^*(k_2)\rangle = \frac{2 \uppi^2}{|\vec k|^3}\mathrm{P}_k (|\vec k|)\,\delta (\vec k_1 - \vec k_2) = \frac{2 \uppi^2}{k^3}\mathrm{P}_k (k)\,\delta (\vec k_1 - \vec k_2).
\end{equation}
The above result clearly shows that individual modes are essentially uncorrelated.

\subsection{Power spectrum for the quantized form of a generic scalar field in de sitter stage}
\label{psquant}
We have now completely defined the power spectrum through our previous discussion. Moving on, reconsider \eqref{eq:19},
\begin{equation*}
\bar{\delta}\hspace{-0.01in}\phi(\mathbf{x},\tau) = \frac{1}{{(2\uppi)}^{3/2}}\int{\mathrm{d}^3k[u_k(\tau)a_\mathbf{k}e^{i\mathbf{k}\cdot\mathbf{x}} + u_k^*(\tau)a^\dagger_\mathbf{k}e^{-i\mathbf{k}\cdot\mathbf{x}}]}.
\end{equation*}
Let us take the Fourier transform\footnote{We ignore the limits over the integrals. They are assumed to go from $-\displaystyle\,\infinity$ to $\displaystyle\infinity$, unless mentioned otherwise.} of the above equation,
\begin{equation*}
\bar{\delta}\hspace{-0.01in}\phi_\mathbf{k}(\tau) = \int\mathrm{d}^3x\, \bar{\delta}\hspace{-0.01in}\phi(\mathbf{x},\tau)\,e^{-i\mathbf{k'}\cdot \mathbf{x}} = \frac{1}{{(2\uppi)}^3}\int \mathrm{d}^3x\,e^{-i\mathbf{k'}\cdot\mathbf{x}}\int {\mathrm{d}^3k\,[u_k(\tau)a_\mathbf{k}e^{i\mathbf{k}\cdot\mathbf{x}} + u_k^*(\tau)a^\dagger_\mathbf{k}e^{-i\mathbf{k}\cdot\mathbf{x}}]},
\end{equation*}
\begin{equation*}
\bar{\delta}\hspace{-0.01in}\phi_\mathbf{k}(\tau) = \frac{1}{{(2\uppi)}^3}\int {\mathrm{d}^3k\int  \mathrm{d}^3x\,[u_k(\tau)a_\mathbf{k}e^{i(\mathbf{k' - k}).\mathbf{x}} + u_k^*(\tau)a^\dagger_\mathbf{k}e^{-i(\mathbf{k' + k}).\mathbf{x}}]},
\end{equation*}
\begin{equation*}
\bar{\delta}\hspace{-0.01in}\phi_\mathbf{k}(\tau) = \frac{1}{{(2\uppi)}^3}\int {\mathrm{d}^3k\,[u_k(\tau)\,a_\mathbf{k}\,\delta (\mathbf{k} - \mathbf{k'}) + u_k^*(\tau)\,a^\dagger_\mathbf{k}\,\delta (\mathbf{k'} + \mathbf{k})]}.
\end{equation*}
Note that since the amplitudes $u_k(\tau)$ are functions of wavenumber and not the wavevector, i.e. $u_{\mathbf{-k}}(\tau) = u_{\mathbf{k}}(\tau) = u_k(\tau)$, we may rewrite
\begin{equation}
\label{eq:54}
\bar{\delta}\hspace{-0.01in}\phi_\mathbf{k}(\tau) = \frac{1}{{(2\uppi)}^3}\bigg[u_k(\tau)\,a_\mathbf{k} + u_k^*(\tau)\,a^\dagger_\mathbf{-k}\bigg]\indent\text{and}\indent \bar{\delta}\hspace{-0.01in}\phi_\mathbf{k^*}(\tau) = \frac{1}{{(2\uppi)}^3}\bigg[u^*_k(\tau)\,a^\dagger_\mathbf{k} + u_k(\tau)\,a_\mathbf{-k}\bigg].
\end{equation}
Thus, once we have the expression for the amplitudes $u_k(\tau)$ for the scalar field perturbations in Fourier space, we can go ahead and calculate the power spectrum. The expectation values of the field operators are given by
\begin{equation}
\label{eq:55}
\,\langle\bar{\delta}\hspace{-0.01in}\phi_\mathbf{k_1}(\tau)\,\bar{\delta}\hspace{-0.01in}\phi_\mathbf{k_2^*}(\tau)\rangle = \,\langle \chi_\mathbf{n}|\bar{\delta}\hspace{-0.01in}\phi_\mathbf{k_1}(\tau)\,\bar{\delta}\hspace{-0.01in}\phi_\mathbf{k^*_2}(\tau)|\chi_\mathbf{n}\rangle,
\end{equation}
where, $\chi_\mathbf{n}$ is the wave-function corresponding to the Hermitian operator in some random state $\mathbf{n}$. Combining \eqref{eq:54} and \eqref{eq:55} gives:
\begin{equation*}
\,\langle\bar{\delta}\hspace{-0.01in}\phi_\mathbf{k_1}(\tau)\,\bar{\delta}\hspace{-0.01in}\phi_\mathbf{k_2^*}(\tau)\rangle = \frac{1}{(2\uppi)^6}\,\bigg\langle \chi_\mathbf{n}\bigg|(u_{k_1}u^*_{k_2})\,a_\mathbf{k_1}a^\dagger_\mathbf{k_2} + (u_{k_1}u_{k_2})\,a_\mathbf{k_1}a_\mathbf{-k_2} + (u^*_{k_1}u^*_{k_2})\,a^\dagger_\mathbf{k_1}a^\dagger_\mathbf{-k_2} + (u^*_{k_1}u_{k_2})\,a^\dagger_\mathbf{-k_1}a_\mathbf{k_2}\bigg|\chi_\mathbf{n}\bigg\rangle.
\end{equation*}
Considering the properties of creation and annihilation operators,
\begin{equation}
\label{eq:56}
\,\langle \chi_\mathbf{n}|a_\mathbf{k_1}a^\dagger_\mathbf{k_2}|\chi_\mathbf{n}\rangle = \delta(\mathbf{k_1}-\mathbf{k_2})\indent\text{and}
\indent \,\langle \chi_\mathbf{n}|a_\mathbf{k_1}a_\mathbf{k_2}|\chi_\mathbf{n}\rangle = \,\langle \chi_\mathbf{n}|a^\dagger_\mathbf{k_1}a^\dagger_\mathbf{k_2}|\chi_\mathbf{n}\rangle = 0,
\end{equation}
only the first and last term can give non-zero contribution to the expectation value. Thus, \eqref{eq:55} reduces to
\begin{equation*}
\,\langle\bar{\delta}\hspace{-0.01in}\phi_\mathbf{k_1}(\tau)\,\bar{\delta}\hspace{-0.01in}\phi_\mathbf{k_2^*}(\tau)\rangle = \frac{1}{(2\uppi)^6}\bigg[(u_{k_1}u^*_{k_2})\,\delta(\mathbf{k_1}-\mathbf{k_2}) + (u^*_{k_1}u_{k_2})\,\delta(\mathbf{-k_1}-\mathbf{k_2})\bigg].
\end{equation*}
We may assume the simple case when $\mathbf{k_1} = \mathbf{k_2}$ as an example, giving us
\begin{equation}
\label{eq:57}
\,\langle{\delta}\hspace{-0.01in}\phi_\mathbf{k_1}(\tau)\,{\delta}\hspace{-0.01in}\phi_\mathbf{k_2^*}(\tau)\rangle = \frac{1}{a(t)^2} \,\langle\bar{\delta}\hspace{-0.01in}\phi_\mathbf{k_1}(\tau)\,\bar{\delta}\hspace{-0.01in}\phi_\mathbf{k_2^*}(\tau)\rangle = \frac{1}{(2\uppi)^6}\frac{|u_k|^2}{a(t)^2}\,\delta(\mathbf{k_1}-\mathbf{k_2}) = |\delta\hspace{-0.01in}\phi_\mathbf{k}|^2.
\end{equation}
Combining the above with the definition of power spectrum in \eqref{eq:53}, we find that
\begin{equation}
\label{eq:58}
\mathrm{P}_k (|\vec k|) = \frac{k^3}{2\uppi^2}|\delta\hspace{-0.01in}\phi_\mathbf{k}|^2 \propto \frac{k^3}{2\uppi^2}\frac{|u_k|^2}{a(t)^2}.
\end{equation}
Further, from the discussion of the solution to the Klein-Gordon equation in \eqref{eq:39}, we can write the power spectrum as\footnote{We ignore the proportionality factor of $1/(2\uppi)^6$ in \eqref{eq:57}. It comes from the mere definition of Fourier transforms.}:
\begin{equation}
\label{eq:59}
\mathrm{P}_k (|\vec k|) = \bigg(\frac{\mathrm{H}}{2\uppi}\bigg)^2 {\bigg(\frac{k}{a\mathrm{H}}\bigg)}^{ 3 - 2v_\phi}.
\end{equation}
It is convenient to define $n_{\phi}$, called the \tit{spectral index} of the power spectrum, in the following way:
\begin{equation}
\label{eq:60}
n_{\phi} - 1 = \frac{\mathrm{d\,ln}[\mathrm{P}_k (|\vec k|)]}{\mathrm{d\,(ln\,}k)},
\end{equation}
which, in the case of a light scalar field, is given as
\begin{equation}
\label{eq:61}
n_{\phi} - 1 = 3 -2v_\phi = 2\upeta_\phi.
\end{equation}
Thus, the expression in \eqref{eq:60} may be rewritten as
\begin{equation}
\label{eq:62}
\mathrm{P}_k (|\vec k|) \propto k^{\displaystyle(\displaystyle n_{\phi} - 1)}.
\end{equation}
We deduce that for the case of a massless scalar field ($v_\phi = 3/2$), $n_{\phi} = 1$, which yields
\begin{equation}
\label{eq:63}
\mathrm{P}_k (|\vec k|) = \text{constant},
\end{equation}
i.e. the power spectrum is scale invariant. It can be shown that the power spectrum is in fact given by: $$\mathrm{P}_k (|\vec k|) = \bigg(\frac{\mathrm{H}}{2\uppi}\bigg)^2.$$
We may also briefly discuss the case when the scalar field is extremely massive, i.e. $m_\phi \gg \dfrac{3}{2}\mathrm{H}$. From section \ref{quant} and section \ref{cosmo}, we borrow the expression for mode amplitudes $u_k(\tau)$, which in the case of massive scalar field ($v_\phi \sim i \dfrac{m_\phi}{\mathrm{H}}$) is reduced to the following important form:
\begin{equation*}
u_k(\tau) = e^{i\big(\scriptstyle v_\phi+\frac{\scriptstyle 1}{\scriptstyle 2}\big){\mathlarger{{\frac{\scriptstyle \uppi}{2}}}}}2^{v_\phi - \frac{\scriptstyle 3}{\scriptstyle 2}} \frac{\Gamma(v_\phi)}{\Gamma(3/2)} \frac{1}{\sqrt{2k}}(-k\tau)^{\mathlarger{\big(\frac{1}{2}-v_\phi\big)}} \propto  e^{-2\big(\mathlarger{\frac{m_\phi}{\vphantom{\tilde{\mathrm{H}}}\mathrm{H}}\big)}^{\mathlarger 2}}.
\end{equation*}
We see that for a massive scalar field, the amplitude damps exponentially with $v_\phi$. The spectral index is then easily calculated using \eqref{eq:58} and \eqref{eq:60}.
\subsection{Quantum fluctuations in a quasi-de sitter expansion stage}
\label{quaside}
Up until now, we had assumed that the Hubble parameter $\mathrm{H}$ is a constant in time. However, it is not entirely true. In fact, from the second condition of slow-roll for the Inflaton field, another parameter\footnote{It is indeed true that both parameters must satisfy $|\upeta_\phi|\text{,}\,\epsilon_\phi \ll 1$ for ``successful'' Inflation.} exists besides the $\upeta_\phi$ parameter \citep{r15}, such that
\begin{equation}
\label{eq:64}
\epsilon_\phi \equiv \frac{\mathbf{\dt{\mathrm{H}}}}{\;\mathrm{H}^2},\;\;\;\upeta_\phi\equiv\frac{\dt{\epsilon}_\phi}{\mathrm{H}\epsilon_\phi}\indent\text{where}\indent\epsilon_\phi,\;\upeta_\phi \ll 1.
\end{equation}
We had assumed $\epsilon_\phi = 0$ in previous discussions; in this section, we drop this assumption. Instead, we consider the case when $\epsilon_\phi$ is roughly a constant\footnote{This assumption is indeed close to ideal since the second derivative $\mathbf{\ddt{\mathrm{H}}}$ is of the order of $\epsilon_\phi^2$, which also holds true for $\upeta_\phi$. It is worth mentioning that the slow-roll parameters by themselves follow the criteria that $\dt{\upeta}_\phi, \dt{\epsilon}_\phi \sim O(\upeta_\phi^2, \epsilon_\phi^2)$}. In that case, we can solve \eqref{eq:64} and get the solution for the Hubble parameter as
\begin{equation}
\label{eq:65}
\frac{\dt{a}}{a} = \mathrm{H} = \bigg(\epsilon_\phi t + \frac{1}{\text{ }\mathrm{H}_c}\bigg)^{-1}\indent\text{where,}\indent \mathrm{H}_c\,\text{is the constant of integration},
\end{equation}
which when solved for $a(t)$ gives
\begin{equation}
\label{eq:66}
a(t) = \bigg(\epsilon_\phi t + 
\frac{1}{\text{ }\mathrm{H}_c}\bigg)^{-1/\epsilon_\phi} = \mathrm{H}(t)^{1/\epsilon_\phi}.
\end{equation}
From the definition of the conformal time [$\mathrm{d}\tau = \mathrm{d}t/a(t)$] given in section \ref{quant}, we can write $a(t)$ in terms of the conformal time as
\begin{equation}
\label{eq:67}
a(t) = -\frac{1}{\tau}\frac{1}{\mathrm{H}(1-\epsilon_\phi)} = [\tau (\epsilon_\phi - 1)]^{1/(\epsilon_\phi -1 )}.
\end{equation}
In addition,
\begin{equation}
\label{eq:68}
\frac{a''}{a} = \frac{2}{\tau^2}\bigg[1 + \frac{3}{2}\epsilon_\phi\bigg].
\end{equation}
For the present case, the mode amplitudes satisfy the Klein-Gordon equation of motion \eqref{eq:36}, for which we have the modified expression for $v_\phi$:
\begin{equation}
\label{eq:69}
v_\phi = \frac{3}{2} + \epsilon_\phi - \upeta_\phi.
\end{equation}
Note that since the first derivatives of the slow-roll parameters are of the order $\dt{\upeta}_\phi, \dt{\epsilon}_\phi \sim O(\upeta_\phi^2, \epsilon_\phi^2)$, $v_\phi$ is roughly a constant. Thus, after accounting for this correction to the field fluctuations introduced by $\epsilon_\phi$, \eqref{eq:39} for the amplitudes of modes is modified to the following form:
\begin{equation}
\label{eq:70}
|\delta\hspace{-0.01in}\phi_k| = \frac{\mathrm{H}}{\sqrt{2k^3}} {\bigg(\frac{k}{a\mathrm{H}}\bigg)}^{\upeta_\phi - \epsilon_\phi}.
\end{equation}
We further find that
\begin{equation}
\label{eq:71}
|\delta\dt{\phi}_k| = |\delta\hspace{-0.01in}\phi_k|\bigg[(1 + \epsilon_\phi - \upeta_\phi)\frac{\mathbf{\dt{\mathrm{H}}}}{\mathrm{H}} + (\epsilon_\phi - \upeta_\phi)\frac{\dt{a}}{a}\bigg].
\end{equation}
Using \eqref{eq:64}, we solve to get
\begin{equation*}
|\delta\dt{\phi}_k| = |\delta\hspace{-0.01in}\phi_k|\bigg[-(1 + \epsilon_\phi - \upeta_\phi)\epsilon_\phi \mathrm{H} + (\epsilon_\phi - \upeta_\phi)\mathrm{H}\bigg].
\end{equation*}
Lastly, ignoring the higher order contributions from the cross-terms ($\epsilon_\phi\upeta_\phi$) and the quadratic terms ($\epsilon_\phi^2$), we arrive at the following simplified expression:
\begin{equation}
\label{eq:72}
\frac{|\delta\dt{\phi}_k|}{|\delta\hspace{-0.01in}\phi_k|} = -\upeta_\phi \mathrm{H}\ll 1.
\end{equation}
We conclude from the above expression that the scalar field perturbations are more or less unchanged (often termed as `frozen') once they cross the horizon. We also find that the power spectrum in this case takes the following form:
\begin{equation}
\label{eq:73}
\mathrm{P}_k (|\vec k|) = \bigg(\frac{\mathrm{H}}{2\uppi}\bigg)^2 {\bigg(\frac{k}{a\mathrm{H}}\bigg)}^{2(\epsilon_\phi - \upeta_\phi)}.
\end{equation}
\section{Wick's theorem and higher-order correlation functions}
\label{wick}
It is beneficial to introduce higher-order correlation functions at this point. Consider a quantity $\mathcal{L}(t,\mathbf{x})$ which is Gaussian-distributed for an infinite set of measurements. Then, for such a quantity, it is a well-known result\footnote{A Gaussian distribution is an even-function around the argument corresponding to its maxima. Therefore, expectation value of any odd function with the Gaussian profile as its probability density function must naturally vanish. On the other hand, expectation values of all even functions will survive.} that except for even-ordered (order 2, 4, 6 etc) correlation functions, all other odd higher-order correlation functions (order 1, 3, 5 etc) vanish. This translates into saying that power spectrum, i.e. two-point correlation function, is all we need to characterize such a quantity as long as it is Gaussian-distributed; this is because, according to Wick's theorem, all even non-vanishing higher-order correlation functions can be broken down into permuted summation of all possible pairs of 2-point correlation functions. For example, a 4$^\mathrm{th}$ order correlation function for a Gaussian-distributed quantity $\mathcal{L}(t,\mathbf{x})\,$ using Wick's theorem can be written as follows:
\begin{equation}
\label{eq:74}
\begin{multlined}
\,\langle \mathcal{L}(t,\mathbf{x_1})\,\mathcal{L}(t,\mathbf{x_2})\,\mathcal{L}(t,\mathbf{x_3})\,\mathcal{L}(t,\mathbf{x_4})\,\rangle = \,\langle \mathcal{L}(t,\mathbf{x_1})\,\mathcal{L}(t,\mathbf{x_3})\rangle\,\,\langle\mathcal{L}(t,\mathbf{x_2})\,\mathcal{L}(t,\mathbf{x_4})\,\rangle \\ + \,\langle \mathcal{L}(t,\mathbf{x_1})\,\mathcal{L}(t,\mathbf{x_3})\rangle\,\,\langle\mathcal{L}(t,\mathbf{x_2})\,\mathcal{L}(t,\mathbf{x_4})\,\rangle + \,\langle \mathcal{L}(t,\mathbf{x_1})\,\mathcal{L}(t,\mathbf{x_4})\rangle\,\,\langle\mathcal{L}(t,\mathbf{x_2})\,\mathcal{L}(t,\mathbf{x_3})\,\rangle.
\end{multlined}
\end{equation}
However, such relations do not hold if the distribution deviates from Gaussianity. For instance, when the non-Gaussianity is small, the Wick's theorem could be modified to include a `connected term' which is non-zero when the distribution deviates from Gaussianity\footnote{This is equivalent to saying that the higher-order even correlation functions can no longer be written solely in terms of the 2-point correlation function.}. Hence, a 4-point correlation function for a slightly non-Gaussian distribution takes the following form:
\begin{equation}
\label{eq:75}
\begin{multlined}
\,\langle \mathcal{L}(t,\mathbf{x_1})\,\mathcal{L}(t,\mathbf{x_2})\,\mathcal{L}(t,\mathbf{x_3})\,\mathcal{L}(t,\mathbf{x_4})\,\rangle = \,\langle \mathcal{L}(t,\mathbf{x_1})\,\mathcal{L}(t,\mathbf{x_3})\rangle\,\,\langle\mathcal{L}(t,\mathbf{x_2})\,\mathcal{L}(t,\mathbf{x_4})\,\rangle \\ + \,\langle \mathcal{L}(t,\mathbf{x_1})\,\mathcal{L}(t,\mathbf{x_3})\rangle\,\,\langle\mathcal{L}(t,\mathbf{x_2})\,\mathcal{L}(t,\mathbf{x_4})\,\rangle + \,\langle \mathcal{L}(t,\mathbf{x_1})\,\mathcal{L}(t,\mathbf{x_4})\rangle\,\,\langle\mathcal{L}(t,\mathbf{x_2})\,\mathcal{L}(t,\mathbf{x_3})\,\rangle \\ + \,\langle \mathcal{L}(t,\mathbf{x_1})\,\mathcal{L}(t,\mathbf{x_2})\,\mathcal{L}(t,\mathbf{x_3})\,\mathcal{L}(t,\mathbf{x_4})\,\rangle_c.
\end{multlined}
\end{equation}
In conclusion, in order to quantify small non-Gaussianities in a perturbed Gaussian distribution, the simplest tool we can conjure is to calculate the 3-point correlation function (also called bi-spectrum in Fourier domain), which would otherwise vanish for a purely Gaussian distribution. 
\subsection{The curious case of self-interacting scalar fields}
\label{self}
We have seen before in section \ref{general} that the (linearized) potential associated with a scalar field $\phi$ is given by $$V(\phi) = \frac{1}{2} m_\phi^2 \phi^2.$$
For a self-interacting scalar field, there are terms which extend beyond the quadratic term which are seen in the Taylor series' expansion:
\begin{equation}
\label{eq:76}
V(\phi+\delta\hspace{-0.01in}\phi) = \underbrace{{V(\phi)} + \dfrac{\mathrm{d}V}{\mathrm{d}\phi}\delta \phi +  \dfrac{1}{2!}\dfrac{\mathrm{d}^2V}{\mathrm{d}\phi^2}(\delta \phi)^2}_{\textsf{limited by quadratic potential}} + \underbrace{\dfrac{1}{3!}\dfrac{\mathrm{d}^3V}{\mathrm{d}\phi^3}(\delta \phi)^3 + \dfrac{1}{4!}\dfrac{\mathrm{d}^4V}{\mathrm{d}\phi^4}(\delta \phi)^4 + ...}_{\textsf{further interaction terms}}
\end{equation}
Before proceeding further, it is necessary to introduce the `interaction picture' in Quantum mechanical formulation of perturbed fields.
\subsubsection{Interaction picture}
\label{intpic}
In quantum formulation of field theory, there are two formulations that are often differentiated based on the ease of obtaining a solution for a given system that is evolving in time. Schr\"odinger picture, on one hand, is a formulation in which we assume that it is the state vectors that evolve in time and not the operators, which corresponding to the observables of the system. Heisenberg picture, on the other hand, is the formulation in which the operators are assumed to be evolving in time, while the state vectors are kept independent of time. Interaction picture is the formulation in which both the state vectors and the operators carry a part of time dependence of the observables. In quantum field theory, the idea behind interaction picture is the same as the first-order time-dependent perturbation theory in quantum mechanics where we achieve the solution in two parts: \tit{a}) a complete and well-known analytical part of the entire solution, and \tit{b}) an unknown interaction part which could be analyzed separately. In interaction picture, the state vectors and the operators (as given in the Schr\"odinger picture) are transformed by a unitary transformation. To begin with, we re-write the perturbed Hamiltonian in Schr\"odinger picture as\footnote{Note that the sub-script `NI' stands for non-interacting part of the Hamiltonian, while the sub-script `IG' stands for the interacting part of the Hamiltonian. In addition, the sub-script $\mathcal{S}$ stands for quantities in the Schr\"odinger picture.}
\begin{equation}
\label{eq:77}
\mathcal{H}_\mathcal{S} = \mathcal{H}_\mathrm{NI} + \mathcal{H}_{\mathrm{IG}}(t),
\end{equation}
where, we have separated the Hamiltonian in a way that $\mathcal{H}_\mathrm{NI}$ is well-understood and exactly solvable, while $\mathcal{H}_{\mathrm{IG}}$ is the part which is hard to realize and analyze. Usually, the explicit dependency\footnote{Note that the explicit time-dependency here refers to the `dependency' arising from time-dependent force-fields such as electric field, magnetic field etc. in the Hamiltonian, and not the inherent temporal `evolution'.} on time of the Hamiltonian are carried into the second term $\mathcal{H}_{\mathrm{IG}}$ in order to simplify the solving process. Moreover, the state vectors $\chi$ in the interaction picture are defined to evolve in time, from time $t_\mathrm{o}$ to $t$, as
\begin{equation}
\label{eq:78}
|\chi_{_\mathrm{\scriptstyle I}}(t)\rangle = e^{i\mathcal{H}_\mathrm{NI}(t-t_\mathrm{o})}\,|\chi_{_\mathcal{\scriptstyle S}}(t_\mathrm{o})\rangle.
\end{equation}
Meanwhile, the operators evolve as
\begin{equation}
\label{eq:79}
\mathcal{O}_{_\mathrm{\scriptstyle I}}(t) = e^{i\mathcal{H}_\mathrm{NI}(t-t_\mathrm{o})}\,\mathcal{O}_{_\mathcal{\scriptstyle S}}(t_\mathrm{o})e^{-i\mathcal{H}_\mathrm{NI}(t-t_\mathrm{o})},
\end{equation}
where, again $\hbar = 1$, and we consider for simplicity that the initial time $t_\mathrm{o}=0$. Remember that in the Schr\"odinger picture, the operators are generally time-independent. Therefore, in general, $\mathcal{O}_{_\mathcal{\scriptstyle S}}(t)$ can simply be replaced by $\mathcal{O}_{_\mathcal{\scriptstyle S}}$ unless there is an implicit and inherent dependence of the operator on time. When $\mathcal{O}_{_\mathcal{\scriptstyle S}} = \mathcal{H}_\mathrm{NI}$, the interaction picture coincides with the Schr\"odinger picture since
\begin{equation}
\label{eq:80}
\mathcal{O}_{_\mathrm{\scriptstyle I| NI}}(t) = e^{i\mathcal{H}_\mathrm{NI}t}\,\mathcal{H}_\mathrm{NI}\,e^{-i\mathcal{H}_\mathrm{NI}t} = \mathcal{H}_\mathrm{NI}\indent \text{for}\indent \mathcal{O}_{_\mathcal{\scriptstyle S}} = \mathcal{H}_\mathrm{NI}.
\end{equation}
Note that we have used the property that the Hamiltonian commutes with its differentiable functionals\footnote{In fact, all operators commute with their differentiable functions.}. The same is not true for $\mathcal{H}_{\mathrm{IG}}(t)$ unless the commutation $[\mathcal{H}_{\mathrm{IG}},\,\mathcal{H}_{\mathrm{NI}}] = 0$ holds, such that 
\begin{equation}
\label{eq:81}
\mathcal{O}_{_\mathrm{\scriptstyle I|IG}}(t) = e^{i\mathcal{H}_\mathrm{NI}t}\,\mathcal{H}_{\mathrm{IG}}\,e^{-i\mathcal{H}_\mathrm{NI}t} \neq \mathcal{H}_{\mathrm{IG}}\indent \text{for}\indent \mathcal{O}_{_\mathcal{\scriptstyle S}} = \mathcal{H}_\mathrm{IG}.
\end{equation} 
We have now set the basics of quantum field theory in the interaction picture. We now proceed forward with the perturbed case of our scalar field $\phi$. \\
\begin{flushleft}\myrule[line width = 0.5mm]{fast cap reversed}{fast cap reversed}\end{flushleft}
\tit{\textbf{Sidenote}}:\\
We make important remarks about switching between different pictures and recall the important commutation relations between operators. It is advisable to read the Appendix \ref{A.1} for some finer details on this section.\\ \\
$\bullet$ Firstly, for operators $\mathcal{A}$ and $\mathcal{B}$,
\begin{equation}
\label{eq:82}
e^{\mathcal{A}+\mathcal{B}} = e^\mathcal{A} e^\mathcal{B} = e^\mathcal{B}e^\mathcal{A},
\end{equation}
only if, $[\mathcal{A}\,\mathcal{B}] = 0$, i.e. if $\mathcal{A}$ and $\mathcal{B}$ commute.\\ \\
$\bullet$ Secondly, to switch between the Schr\"odinger picture and the Heisenberg picture, we follow the following relations,
\begin{equation}
\label{eq:83}
\mathcal{O}_\mathrm{H} = e^{i\mathcal{H_\mathcal{S}}t}\,\mathcal{O}_\mathcal{S}\,e^{-i\mathcal{H_\mathcal{S}}t}\indent\text{for}\indent
\mathcal{S}\rightarrow\mathrm{H},
\end{equation}
and,
\begin{equation}
\label{eq:84}
\mathcal{O}_\mathcal{S} = e^{-i\mathcal{H_\mathcal{S}}t}\,\mathcal{O}_\mathrm{H}\,e^{i\mathcal{H_\mathcal{S}}t}\indent\text{for}\indent
\mathrm{H}\rightarrow\mathcal{S},
\end{equation}
where, the operator of the form $e^{-i\mathcal{H}t}\equiv {\hat{\mathrm{U}}}$ is usually called a \tit{propagator}.
We have already discussed that,
\begin{equation}
\label{eq:85}
\mathcal{O}_{_\mathrm{\scriptstyle I}} = e^{i\mathcal{H}_\mathrm{NI}t}\,\mathcal{O}_{_\mathcal{\scriptstyle S}}e^{-i\mathcal{H}_\mathrm{NI}t}\indent\text{for}\indent \mathcal{S}\rightarrow\mathrm{I}.
\end{equation}
Moreover, the time-evolution of operators can alternatively be written from \eqref{eq:85} as 
\begin{equation}
\label{eq:86}
i\hbar\frac{\mathrm{d}(\mathcal{O}_\mathrm{I})}{\mathrm{d}t} = [\mathcal{O}_\mathrm{I},\mathcal{H}_\mathrm{NI}].
\end{equation}
\begin{flushright}\myrule[line width = 0.5mm]{fast cap reversed}{fast cap reversed}\end{flushright}$\;$\\
Furthermore, it can also be easily proved using \eqref{eq:78}, \eqref{eq:79} and \eqref{eq:80} that the following relation holds:
$$i\hbar \dfrac{\partial |\chi_{_\mathrm{\scriptstyle I}}(t)\rangle}{\partial t} = \mathcal{O}_{_\mathrm{\scriptstyle I| IG}}(t)|\chi_{_\mathrm{\scriptstyle I}}(t)\rangle.$$
which is the Schr\"odinger equation in interaction picture with the corresponding Hamiltonian $\mathcal{O}_{_\mathrm{\scriptstyle I| IG}}(t)$.

\subsection{Back to our quantized scalar field}
\label{backtoquant}
For a self-interacting scalar field, we find that the potential has terms beyond the quadratic term such that its expression takes the following form:
\begin{equation}
\label{eq:87}
V(\phi) = \frac{1}{2} m_\phi^2 \phi^2 + g(\phi)\indent\text{where,}\indent g(\phi)\text{ is of the form }\gamma\phi^p,
\end{equation}
where, obviously $p>2$. \eqref{eq:74} in that case looks like,
\begin{equation}
\label{eq:88}
V(\phi+\delta\hspace{-0.01in}\phi) = \underbrace{\dfrac{1}{2} m_\phi^2 (\phi + \delta\hspace{-0.01in}\phi)^2 + {{g(\phi)} + \dfrac{\mathrm{d}g}{\mathrm{d}\phi}\delta \phi +  \dfrac{1}{2!}\dfrac{\mathrm{d}^2g}{\mathrm{d}\phi^2}(\delta \phi)^2}}_{\textsf{up to second order in }{\delta\hspace{-0.01in}\phi}} + \underbrace{{\dfrac{1}{3!}\dfrac{\mathrm{d}^3g}{\mathrm{d}\phi^3}(\delta \phi)^3 + \dfrac{1}{4!}\dfrac{\mathrm{d}^4g}{\mathrm{d}\phi^4}(\delta \phi)^4 + ...}}_{\textsf{higher-order interactions of order 3 or higher}}.
\end{equation}
Thus, the interaction part is represented by the contributions beyond the second order by
\begin{equation}
\label{eq:89}
V_{\mathrm{int}}(\phi,\delta\hspace{-0.01in}\phi) = {\frac{1}{3!}\frac{\mathrm{d}^3g}{\mathrm{d}\phi^3}(\delta \phi)^3 + \frac{1}{4!}\frac{\mathrm{d}^4g}{\mathrm{d}\phi^4}(\delta \phi)^4 + ...}.
\end{equation}
The corresponding quantized Hamiltonian is written by upgrading the field to an operator\footnote{Remember that we have quantized only(!) the tiny perturbation $\delta\hspace{-0.01in}\phi(\mathbf{x},\tau)$ to the scalar field as per \eqref{eq:9} and \eqref{eq:19}.} such as $\bar\phi(\mathbf{x},\tau) \rightarrow \phi(\tau) + \bar\delta\hspace{-0.01in}\phi(\mathbf{x},\tau)$. This leads to
\begin{equation}
\label{eq:90}
\mathcal{H}_{\mathcal{S}}(\mathbf{x},t) = \frac{1}{2}g^{\mu\nu}\partial _\mu[\bar\phi(\mathbf{x},\tau)] \partial _\nu[\bar\phi(\mathbf{x},\tau)] + \frac{1}{2} m_\phi^2 [\bar\phi(\mathbf{x},\tau)]^2 + V_{\mathrm{int}}(\phi(\tau),\bar\delta\hspace{-0.01in}\phi) = \mathcal{H}_\mathrm{NI} + \mathcal{H}_{\mathrm{IG}}(\mathbf{x},t),
\end{equation}
such that\footnote{Take a note here that $\phi$ here represents the background field [$\phi(\tau)$] and the tiny fluctuations in its value are $\delta\hspace{-0.01in}\phi(\mathbf{x},\tau)$, such that $\phi(\tau,x) = \phi(\tau) + \delta\hspace{-0.01in}\phi(\tau,\mathbf{x})$, similar to the way we represented them in \eqref{eq:9} before.} $\mathcal{H}_{\mathrm{int}}(t) = V_{\mathrm{int}}(\phi,\bar\delta\hspace{-0.01in}\phi)$. When upgraded to a quantized operator and integrated over spatial coordinates, the Hamiltonian takes the form of
\begin{equation}
\label{eq:91}
\mathcal{H}_{\mathcal{S}}(t) = \int\mathrm{d}^3x\,\mathcal{H}_\mathrm{NI} + \int\mathrm{d}^3x\,\mathcal{H}_{\mathrm{IG}}(\mathbf{x},t) \equiv\mathcal{H}_\mathrm{O}(t) + \mathcal{H}_{\mathrm{int}}(t).
\end{equation}
In order to write the N-point correlation function for the new Fourier modes (which have been perturbed by the interaction terms in the Hamiltonian) of perturbed scalar field $\phi$ in interaction picture, we proceed by writing\footnote{Remember that the operators $\bar\delta\hspace{-0.01in}\phi_\mathbf{k_i}$ are in the Heisenberg picture and we need to transform them to Schr\"odinger picture first in order to meet the criteria of the interaction picture. Moreover, the expectation value is calculated for the operators $\bar\delta\hspace{-0.01in}\phi_\mathbf{k_i}$ at the same time $t$. When calculated at different times for different operators, the formulation is slightly different. Refer to Peskin and Schr\"oder: \tit{An introduction to Quantum Field Theory}, Chapter \tit{4}, Sec: \tit{4.2}, Page: \tit{84--86} for more details.}, 
\begin{equation}
\label{eq:92}
[\bar\delta\hspace{-0.01in}\phi_\mathbf{k_1}\,\bar\delta\hspace{-0.01in}\phi_\mathbf{k_2}\,\bar\delta\hspace{-0.01in}\phi_\mathbf{k_3}\, ...\, \bar\delta\hspace{-0.01in}\phi_\mathbf{k_N}]_{_\mathrm{\scriptstyle I}} = e^{i\mathcal{H}_\mathrm{NI}t}\,[\bar\delta\hspace{-0.01in}\phi_\mathbf{k_1}\,\bar\delta\hspace{-0.01in}\phi_\mathbf{k_2}\,\bar\delta\hspace{-0.01in}\phi_\mathbf{k_3}\, ...\, \bar\delta\hspace{-0.01in}\phi_\mathbf{k_N}]_{_\mathcal{\scriptstyle S}}\,e^{-i\mathcal{H}_\mathrm{NI}t}\indent\text{for}\indent \mathcal{S}\rightarrow\mathrm{I}.
\end{equation}\\
\begin{flushleft}\myrule[line width = 0.5mm]{fast cap reversed}{fast cap reversed}\end{flushleft}
\tit{\textbf{Sidenote}}:\\
The Hamiltonian in Schr\"odinger picture for a free non-interacting scalar field is similar to that of a harmonic oscillator and is given by,
\begin{equation}
\label{eq:93}
\begin{multlined}
\mathcal{H}_\mathrm{NI} = \sqrt{-g}\bigg[\frac{1}{2}g^{\mu\nu}\partial _\mu\bar\phi(\mathbf{x},\tau) \partial _\nu\bar\phi(\mathbf{x},\tau) + V(\bar\phi(\mathbf{x},\tau))\bigg] \\ \indent\indent\indent\indent\indent =  \sqrt{-g}\bigg[\frac{1}{2}g^{\mu\nu}\partial _\mu\bar\phi(\mathbf{x},\tau) \partial _\nu\bar\phi(\mathbf{x},\tau) + \frac{1}{2}m_\phi^2\bar\phi(\mathbf{x},\tau)^2\bigg].
\end{multlined}
\end{equation}
Furthermore, the interacting part of the Hamiltonian $\mathcal{H}_\mathrm{IG}\,(\equiv \mathcal{H}_\mathrm{int})$ is a function of $\bar\phi(\mathbf{x},\tau)$ $[\phi(\tau)+\bar\delta\hspace{-0.01in}\phi(\mathbf{x},\tau)]$, as we can see from \eqref{eq:90}. This merely translates into saying that $\mathcal{H}_\mathrm{NI}$, $\mathcal{H}_\mathrm{int}$ and $\mathcal{H}_\mathcal{S}$ all commute$^{12}$ amongst themselves in pairs since they are all differentiable functionals of $\bar\delta\hspace{-0.01in}\phi(\mathbf{x},\tau)$. Refer to Appendix \ref{A.2} for finer details when $\mathcal{H}_\mathrm{NI}$, $\mathcal{H}_\mathrm{int}$ and $\mathcal{H}_\mathcal{S}$ fail to commute. Moreover, $\bar\delta\hspace{-0.01in}\phi_\mathbf{k_i}$ and $\bar\delta\hspace{-0.01in}\phi$ do not commute since they belong to separate Hilbert spaces.
\begin{flushright}\myrule[line width = 0.5mm]{fast cap reversed}{fast cap reversed}\end{flushright}$\;$\\
Following the argument above. combined with the assertion made in \eqref{eq:82}, we can conclude that
\begin{equation}
\label{eq:94}
[\bar\delta\hspace{-0.01in}\phi_\mathbf{k_1}\,\bar\delta\hspace{-0.01in}\phi_\mathbf{k_2}\,\bar\delta\hspace{-0.01in}\phi_\mathbf{k_3}\, ...\, \bar\delta\hspace{-0.01in}\phi_\mathbf{k_N}]_{_\mathrm{\scriptstyle I}} = e^{-i\mathcal{H}_\mathrm{int}t}\,e^{i\mathcal{H}_\mathrm{S}t}\,[\bar\delta\hspace{-0.01in}\phi_\mathbf{k_1}\,\bar\delta\hspace{-0.01in}\phi_\mathbf{k_2}\,\bar\delta\hspace{-0.01in}\phi_\mathbf{k_3}\, ...\, \bar\delta\hspace{-0.01in}\phi_\mathbf{k_N}]_{_\mathcal{\scriptstyle S}}\,e^{-i\mathcal{H}_\mathcal{S}t}\,e^{i\mathcal{H}_\mathrm{int}t}.
\end{equation}
Further, using \eqref{eq:84}, we get
\begin{equation}
\label{eq:95}
[\bar\delta\hspace{-0.01in}\phi_\mathbf{k_1}\,\bar\delta\hspace{-0.01in}\phi_\mathbf{k_2}\,\bar\delta\hspace{-0.01in}\phi_\mathbf{k_3}\, ...\, \bar\delta\hspace{-0.01in}\phi_\mathbf{k_N}]_{_\mathrm{\scriptstyle I}} = e^{-i\mathcal{H}_\mathrm{int}t}\,[\bar\delta\hspace{-0.01in}\phi_\mathbf{k_1}\,\bar\delta\hspace{-0.01in}\phi_\mathbf{k_2}\,\bar\delta\hspace{-0.01in}\phi_\mathbf{k_3}\, ...\, \bar\delta\hspace{-0.01in}\phi_\mathbf{k_N}]_{_\mathrm{\scriptstyle H}}\,e^{i\mathcal{H}_\mathrm{int}t},
\end{equation}
or, alternatively,
\begin{equation}
\label{eq:96}
[\bar\delta\hspace{-0.01in}\phi_\mathbf{k_1}\,\bar\delta\hspace{-0.01in}\phi_\mathbf{k_2}\,\bar\delta\hspace{-0.01in}\phi_\mathbf{k_3}\, ...\, \bar\delta\hspace{-0.01in}\phi_\mathbf{k_N}]_{_\mathrm{\scriptstyle H}} = e^{i\mathcal{H}_\mathrm{int}t}\,[\bar\delta\hspace{-0.01in}\phi_\mathbf{k_1}\,\bar\delta\hspace{-0.01in}\phi_\mathbf{k_2}\,\bar\delta\hspace{-0.01in}\phi_\mathbf{k_3}\, ...\, \bar\delta\hspace{-0.01in}\phi_\mathbf{k_N}]_{_\mathrm{\scriptstyle I}}\,e^{-i\mathcal{H}_\mathrm{int}t}.
\end{equation}
Since we now have the perturbed formulation for the operators, we can write the expectation value\footnote{Refer to Appendix \ref{A.3} for detailed calculation of this expressions assisted by the identities \eqref{eq:98} and \eqref{eq:99}.} for the N-point correlation function as,
\begin{equation}
\label{eq:97}
\begin{multlined}
\,\langle\bar\delta\hspace{-0.01in}\phi_\mathbf{k_1}\,\bar\delta\hspace{-0.01in}\phi_\mathbf{k_2}\,\bar\delta\hspace{-0.01in}\phi_\mathbf{k_3}\, ...\, \bar\delta\hspace{-0.01in}\phi_\mathbf{k_N}\rangle_{_\mathrm{\scriptstyle H}} = \,\langle \mathfrak{\Omega}|[\bar\delta\hspace{-0.01in}\phi_\mathbf{k_1}\,\bar\delta\hspace{-0.01in}\phi_\mathbf{k_2}\,\bar\delta\hspace{-0.01in}\phi_\mathbf{k_3}\, ...\, \bar\delta\hspace{-0.01in}\phi_\mathbf{k_N}]_{_\mathrm{\scriptstyle H}}|\mathfrak{\Omega}\rangle \\ = \,\langle \mathfrak{\Omega}| e^{i\mathcal{H}_\mathrm{int}t}\,[\bar\delta\hspace{-0.01in}\phi_\mathbf{k_1}\,\bar\delta\hspace{-0.01in}\phi_\mathbf{k_2}\,\bar\delta\hspace{-0.01in}\phi_\mathbf{k_3}\, ...\, \bar\delta\hspace{-0.01in}\phi_\mathbf{k_N}]_{_\mathrm{\scriptstyle I}}\,e^{-i\mathcal{H}_\mathrm{int}t}|\mathfrak{\Omega}\rangle = \,\langle 0| e^{i\mathcal{H}_\mathrm{int}t}\,[\bar\delta\hspace{-0.01in}\phi_\mathbf{k_1}\,\bar\delta\hspace{-0.01in}\phi_\mathbf{k_2}\,\bar\delta\hspace{-0.01in}\phi_\mathbf{k_3}\, ...\, \bar\delta\hspace{-0.01in}\phi_\mathbf{k_N}]_{_\mathrm{\scriptstyle I}}\,e^{-i\mathcal{H}_\mathrm{int}t}| 0 \rangle, 
\end{multlined}
\end{equation}
where, $|\mathfrak{\Omega\rangle}$ is the new interacting vacuum ground state. We shall take note that $\mathcal{H}_\mathrm{int}$ itself is a variant in time (or, conformal time), which leads to the following development,
\begin{equation}
\label{eq:98}
\,\langle\bar\delta\hspace{-0.01in}\phi_\mathbf{k_1}\,\bar\delta\hspace{-0.01in}\phi_\mathbf{k_2}\,\bar\delta\hspace{-0.01in}\phi_\mathbf{k_3}\, ...\, \bar\delta\hspace{-0.01in}\phi_\mathbf{k_N}\rangle_{_\mathrm{\scriptstyle H}} = \,\bigg\langle 0\,\bigg| e^{{i\displaystyle\int_{\tau_\mathrm{o}}^{\tau}}\mathcal{H}_\mathrm{int}(\tau')\,\mathrm{d}\tau'}\,\bigg[\bar\delta\hspace{-0.01in}\phi_\mathbf{k_1}\,\bar\delta\hspace{-0.01in}\phi_\mathbf{k_2}\,\bar\delta\hspace{-0.01in}\phi_\mathbf{k_3}\, ...\, \bar\delta\hspace{-0.01in}\phi_\mathbf{k_N}\bigg]_{\mathrm{\scriptstyle I}}\,e^{{-i\displaystyle\int_{\tau_\mathrm{o}}^{\tau}}\mathcal{H}_\mathrm{int}(\tau')\,\mathrm{d}\tau'}\bigg |\,0 \bigg\rangle,
\end{equation}
where, we have also revoked the previously simplifying assumption about the initial time ($t_\mathrm{o}=0$) such that now $t_\mathrm{o}, \tau_\mathrm{o}\neq 0$, and in fact, $t_\mathrm{o},\tau_\mathrm{o}\rightarrow -\,\infinity$. One important remark to be made here is that we have used the interaction picture so that we could relate the N-point correlation function to the perturbations in the Hamiltonian contained in the interaction terms of the potential. Now, expanding the propagator to first-order approximation in $\mathcal{H}_\mathrm{int}$ as follows,
\begin{equation}
\label{eq:99}
\hat{\mathrm{U}}^{+/-}(\tau,\tau_\mathrm{o}) = e^{{\pm i\displaystyle\int_{\tau_\mathrm{o}}^{\tau}}\mathcal{H}_\mathrm{int}(\tau')\,\mathrm{d}\tau'} = \mathcal{I} \pm i\int_{\tau_\mathrm{o}}^{\tau}\mathcal{H}_\mathrm{int}(\tau')\,\mathrm{d}\tau',
\end{equation} 
we get,
\begin{equation*}
\begin{multlined}
\,\langle\bar\delta\hspace{-0.01in}\phi_\mathbf{k_1}\,\bar\delta\hspace{-0.01in}\phi_\mathbf{k_2}\,\bar\delta\hspace{-0.01in}\phi_\mathbf{k_3}\, ...\, \bar\delta\hspace{-0.01in}\phi_\mathbf{k_N}\rangle_{_\mathrm{\scriptstyle H}} = \\ \,\bigg\langle 0\,\bigg| \Bigg\{\mathcal{I} + i\int_{\tau_\mathrm{o}}^{\tau}\mathcal{H}_\mathrm{int}(\tau')\,\mathrm{d}\tau'\Bigg\}\,\Bigg\{\bar\delta\hspace{-0.01in}\phi_\mathbf{k_1}\,\bar\delta\hspace{-0.01in}\phi_\mathbf{k_2}\,\bar\delta\hspace{-0.01in}\phi_\mathbf{k_3}\, ...\, \bar\delta\hspace{-0.01in}\phi_\mathbf{k_N}\Bigg\}_{\mathrm{\scriptstyle I}}\,\Bigg\{\mathcal{I} - i\int_{\tau_\mathrm{o}}^{\tau}\mathcal{H}_\mathrm{int}(\tau')\,\mathrm{d}\tau'\Bigg\}\bigg |\,0 \bigg\rangle,
\end{multlined}
\end{equation*}
\\
\begin{equation*}
\begin{multlined}
\,\langle\bar\delta\hspace{-0.01in}\phi_\mathbf{k_1}\,\bar\delta\hspace{-0.01in}\phi_\mathbf{k_2}\,\bar\delta\hspace{-0.01in}\phi_\mathbf{k_3}\, ...\, \bar\delta\hspace{-0.01in}\phi_\mathbf{k_N}\rangle_{_\mathrm{\scriptstyle H}} = \\ \,\bigg\langle 0\,\bigg|\Bigg\{\bar\delta\hspace{-0.01in}\phi_\mathbf{k_1}\,\bar\delta\hspace{-0.01in}\phi_\mathbf{k_2}\,\bar\delta\hspace{-0.01in}\phi_\mathbf{k_3}\, ...\, \bar\delta\hspace{-0.01in}\phi_\mathbf{k_N}\Bigg\}_{\mathrm{\scriptstyle I}} + \Bigg[\Bigg\{i\int_{\tau_\mathrm{o}}^{\tau}\mathcal{H}_\mathrm{int}(\tau')\,\mathrm{d}\tau'\Bigg\}, \Bigg\{\bar\delta\hspace{-0.01in}\phi_\mathbf{k_1}\,\bar\delta\hspace{-0.01in}\phi_\mathbf{k_2}\,\bar\delta\hspace{-0.01in}\phi_\mathbf{k_3}\, ...\, \bar\delta\hspace{-0.01in}\phi_\mathbf{k_N}\Bigg\}_{\mathrm{\scriptstyle I}}\Bigg ] \\ - \Bigg\{i\int_{\tau_\mathrm{o}}^{\tau}\mathcal{H}_\mathrm{int}(\tau')\,\mathrm{d}\tau'\Bigg\}\Bigg\{\bar\delta\hspace{-0.01in}\phi_\mathbf{k_1}\,\bar\delta\hspace{-0.01in}\phi_\mathbf{k_2}\,\bar\delta\hspace{-0.01in}\phi_\mathbf{k_3}\, ...\, \bar\delta\hspace{-0.01in}\phi_\mathbf{k_N}\Bigg\}_{\mathrm{\scriptstyle I}}\Bigg\{ i\int_{\tau_\mathrm{o}}^{\tau}\mathcal{H}_\mathrm{int}(\tau')\,\mathrm{d}\tau'\Bigg\}\bigg |\,0 \bigg\rangle,
\end{multlined}
\end{equation*}
\\
\begin{equation}
\label{eq:100}
\begin{multlined}
\,\langle\bar\delta\hspace{-0.01in}\phi_\mathbf{k_1}\,\bar\delta\hspace{-0.01in}\phi_\mathbf{k_2}\,\bar\delta\hspace{-0.01in}\phi_\mathbf{k_3}\, ...\, \bar\delta\hspace{-0.01in}\phi_\mathbf{k_N}\rangle_{_\mathrm{\scriptstyle H}} = \\ \,\bigg\langle 0\,\bigg|\Bigg\{\bar\delta\hspace{-0.01in}\phi_\mathbf{k_1}\,\bar\delta\hspace{-0.01in}\phi_\mathbf{k_2}\,\bar\delta\hspace{-0.01in}\phi_\mathbf{k_3}\, ...\, \bar\delta\hspace{-0.01in}\phi_\mathbf{k_N}\Bigg\}_{\mathrm{\scriptstyle I}}\bigg |\,0 \bigg\rangle  + \,\bigg\langle 0\,\bigg| \Bigg[\Bigg\{i\int_{\tau_\mathrm{o}}^{\tau}\mathcal{H}_\mathrm{int}(\tau')\,\mathrm{d}\tau'\Bigg\}, \Bigg\{\bar\delta\hspace{-0.01in}\phi_\mathbf{k_1}\,\bar\delta\hspace{-0.01in}\phi_\mathbf{k_2}\,\bar\delta\hspace{-0.01in}\phi_\mathbf{k_3}\, ...\, \bar\delta\hspace{-0.01in}\phi_\mathbf{k_N}\Bigg\}_{\mathrm{\scriptstyle I}}\Bigg ]\bigg |\,0 \bigg\rangle \\ - \,\bigg\langle 0\,\bigg|\Bigg\{i\int_{\tau_\mathrm{o}}^{\tau}\mathcal{H}_\mathrm{int}(\tau')\,\mathrm{d}\tau'\Bigg\}\Bigg\{\bar\delta\hspace{-0.01in}\phi_\mathbf{k_1}\,\bar\delta\hspace{-0.01in}\phi_\mathbf{k_2}\,\bar\delta\hspace{-0.01in}\phi_\mathbf{k_3}\, ...\, \bar\delta\hspace{-0.01in}\phi_\mathbf{k_N}\Bigg\}_{\mathrm{\scriptstyle I}}\Bigg\{ i\int_{\tau_\mathrm{o}}^{\tau}\mathcal{H}_\mathrm{int}(\tau')\,\mathrm{d}\tau'\Bigg\}\bigg |\,0 \bigg\rangle.
\end{multlined}
\end{equation}
Here, $|0 \rangle$ is the free-field vacuum state, i.e. the eigenstate devoid of perturbations in the Hamiltonian (eigenstate corresponding to $\mathcal{H}_\mathrm{NI}$). Moreover, remember that the eigenstate $|0 \rangle$ is a function of the final time $\tau$ and not the integrable time $\tau'$. Therefore, when acted upon by the interaction Hamiltonian, the $|0 \rangle$ eigenstate remains unaffected. The last term in \eqref{eq:100} could then be written as
\begin{equation}
\label{eq:101}
\begin{multlined}
\Bigg\{i\int_{\tau_\mathrm{o}}^{\tau}\,\bigg\langle 0\,\big|\mathcal{H}_\mathrm{int}(\tau')\,\mathrm{d}\tau'\Bigg\}\Bigg\{\bar\delta\hspace{-0.01in}\phi_\mathbf{k_1}\,\bar\delta\hspace{-0.01in}\phi_\mathbf{k_2}\,\bar\delta\hspace{-0.01in}\phi_\mathbf{k_3}\, ...\, \bar\delta\hspace{-0.01in}\phi_\mathbf{k_N}\Bigg\}_{\mathrm{\scriptstyle I}}\Bigg\{ i\int_{\tau_\mathrm{o}}^{\tau}\mathcal{H}_\mathrm{int}(\tau')\,\big |\,0 \bigg\rangle \mathrm{d}\tau'\Bigg\} = \\ \,\bigg\langle 0\,\bigg|\Bigg\{\bar\delta\hspace{-0.01in}\phi_\mathbf{k_1}\,\bar\delta\hspace{-0.01in}\phi_\mathbf{k_2}\,\bar\delta\hspace{-0.01in}\phi_\mathbf{k_3}\, ...\, \bar\delta\hspace{-0.01in}\phi_\mathbf{k_N}\Bigg\}_{\mathrm{\scriptstyle I}}\bigg |\,0 \bigg\rangle,\indent\text{since}\indent \mathcal{H}_\mathrm{int}(\tau')\,\big |\,0 \bigg\rangle = \big |\,0 \bigg\rangle,
\end{multlined}
\end{equation}
and according to which, \eqref{eq:100} now reduces to the simpler form of
\begin{equation}
\label{eq:102}
\begin{multlined}
\,\langle\bar\delta\hspace{-0.01in}\phi_\mathbf{k_1}\,\bar\delta\hspace{-0.01in}\phi_\mathbf{k_2}\,\bar\delta\hspace{-0.01in}\phi_\mathbf{k_3}\, ...\, \bar\delta\hspace{-0.01in}\phi_\mathbf{k_N}\rangle_{_\mathrm{\scriptstyle H}} = \,\bigg\langle 0\,\Bigg| \Bigg[\Bigg\{i\int_{\tau_\mathrm{o}}^{\tau}\mathcal{H}_\mathrm{int}(\tau')\,\mathrm{d}\tau'\Bigg\}, \Bigg\{\bar\delta\hspace{-0.01in}\phi_\mathbf{k_1}\,\bar\delta\hspace{-0.01in}\phi_\mathbf{k_2}\,\bar\delta\hspace{-0.01in}\phi_\mathbf{k_3}\, ...\, \bar\delta\hspace{-0.01in}\phi_\mathbf{k_N}\Bigg\}_{\mathrm{\scriptstyle I}}\Bigg ]\Bigg |\,0 \Bigg\rangle.
\end{multlined}
\end{equation}
The operators $\bar\delta\hspace{-0.01in}\phi_\mathbf{k_i}$ are independent of the integrable time $\tau'$ and functions of only the final time $\tau$. Hence, they can be pulled into the integral to yield the following expression:
\begin{equation}
\label{eq:103}
\begin{multlined}
\,\langle\bar\delta\hspace{-0.01in}\phi_\mathbf{k_1}\,\bar\delta\hspace{-0.01in}\phi_\mathbf{k_2}\,\bar\delta\hspace{-0.01in}\phi_\mathbf{k_3}\, ...\, \bar\delta\hspace{-0.01in}\phi_\mathbf{k_N}\rangle_{_\mathrm{\scriptstyle H}} = i\int_{\tau_\mathrm{o}}^{\tau}\mathrm{d}\tau'\,\bigg\langle 0\,\Bigg| \Bigg[\mathcal{H}_\mathrm{int}(\tau'), \Bigg\{\bar\delta\hspace{-0.01in}\phi_\mathbf{k_1}\,\bar\delta\hspace{-0.01in}\phi_\mathbf{k_2}\,\bar\delta\hspace{-0.01in}\phi_\mathbf{k_3}\, ...\, \bar\delta\hspace{-0.01in}\phi_\mathbf{k_N}(\tau)\Bigg\}_{\mathrm{\scriptstyle I}}\Bigg ]\Bigg |\,0 \Bigg\rangle.
\end{multlined}
\end{equation}
It is worth noting that the operators within the commutator are acting on the vacuum state at different times $\tau$ and $\tau'$. Furthermore, in order to solve the above equation, we must be able to calculate the following quantity at different arguments of time $\tau_1,\,\tau_2,\,...\,\tau_\mathrm{N}$ such that
$$\,\langle0|\,\{\bar\delta\hspace{-0.01in}\phi_\mathbf{k_1}(\tau_1)\,\bar\delta\hspace{-0.01in}\phi_\mathbf{k_2}(\tau_2)\,\bar\delta\hspace{-0.01in}\phi_\mathbf{k_3}(\tau_3)\, ...\, \bar\delta\hspace{-0.01in}\phi_\mathbf{k_N}(\tau_\mathrm{N})\}_{\mathrm{\scriptstyle I}}|0\rangle\indent\text{such that}\indent\tau_1,\,\tau_2,\,...\,\tau_\mathrm{N}\in\{\tau',\tau\},$$
which in turn can be calculated via Wick's theorem as long as we have the 2-point correlation functions for all possible permutations in $\{\bar\delta\hspace{-0.01in}\phi_\mathbf{k_1}(\tau_1)\,\bar\delta\hspace{-0.01in}\phi_\mathbf{k_2}(\tau_2)\,\bar\delta\hspace{-0.01in}\phi_\mathbf{k_3}(\tau_3)\, ...\, \bar\delta\hspace{-0.01in}\phi_\mathbf{k_N}(\tau_\mathrm{N})\}_{\mathrm{\scriptstyle I}}$. Let us now try to calculate the expression in \eqref{eq:103} in terms of the 2-point correlation functions (also known alternatively as 'the Feynman propagators') using \eqref{eq:54} and \eqref{eq:56}, which can be expressed as simply as,
\begin{equation}
\label{eq:104}
\begin{multlined}
\,\langle0|\,\{\bar\delta\hspace{-0.01in}\phi_\mathbf{k_i}(\tau_1)\,\bar\delta\hspace{-0.01in}\phi_\mathbf{k_j}(\tau_2)\}_{\mathrm{\scriptstyle I}}|0\rangle = \delta(\mathbf{k_i}+\mathbf{k_j})\,D(\mathbf{k_i},\mathbf{k_j},\tau_1,\tau_2) = \delta(\mathbf{k_i}+\mathbf{k_j})\,u_{k}(\tau)\,u_{k}^*(\tau') \\ \indent\text{for}\indent\tau_1,\,\tau_2\in\{\tau',\tau\};\; \tau'<\tau,\indent\text{and,}\indent
|\mathbf{k_i}|=|\mathbf{k_j}|=k.
\end{multlined}
\end{equation}
Remember that the order of the time arguments in the above expression of the 2-point correlation function (which extends to the syntax of \{$D(\mathbf{k_i},\mathbf{k_j},\tau_1,\tau_2)$\} is very important, i.e. $D(\mathbf{k_i},\mathbf{k_j},\tau_1,\tau_2)\neq D(\mathbf{k_i},\mathbf{k_j},\tau_2,\tau_1)$ in general.
\subsection{Feynman diagrams}
\label{feyndiag}
In order to simplify the expression in \eqref{eq:103}, we introduce the Feynman diagrams which simplify the process of evaluating the Wick's theorem, i.e. expressing the 3-point correlation function as an explicit function of the 2-point correlations (i.e. the Feynman propagators). Feynman diagrams are an easy way to express the Wick's theorem graphically. Let us now introduce the interacting part of the Hamiltonian $\mathcal{H}_\mathrm{int}(\tau') = \displaystyle\int\mathrm{d}^3x\dfrac{\gamma}{p!}[\bar\delta\hspace{-0.01in}\phi(\mathbf{x},\tau')]^p$, where $p>2$.\blfootnote{$^{\dagger}$Refer to Appendix \ref{A.2} for discussion on path-ordering and Dyson expansion.}\blfootnote{$^{\ddagger}$In addition, each internal point must connect to at least one other internal point (if present) such that the total number of lines originating and concluding at an internal point is equal to the power-law index $p$. A line connecting two internal points is known as an \tit{internal line} or the \tit{interaction line}.}\\
\begin{flushleft}\myrule[line width = 0.5mm]{fast cap reversed}{fast cap reversed}\end{flushleft}
\tit{\textbf{Sidenote}}:\\
We follow the standard prescribed recipe (simplified for our case) in order to `draw' and `solve' the Feynman diagrams in momentum space. It is a canonical `geometrical' formulation of Wick's theorem that is useful in solving for N-point correlation function given in \eqref{eq:103}, i.e.
$$\,\langle\bar\delta\hspace{-0.01in}\phi_\mathbf{k_1}\,\bar\delta\hspace{-0.01in}\phi_\mathbf{k_2}\,\bar\delta\hspace{-0.01in}\phi_\mathbf{k_3}\, ...\, \bar\delta\hspace{-0.01in}\phi_\mathbf{k_N}\rangle_{_\mathrm{\scriptstyle H}} = i\int_{\tau_\mathrm{o}}^{\tau}\mathrm{d}\tau'\,\bigg\langle 0\,\Bigg| \Bigg[\mathcal{H}_\mathrm{int}(\tau'), \Bigg\{\bar\delta\hspace{-0.01in}\phi_\mathbf{k_1}\,\bar\delta\hspace{-0.01in}\phi_\mathbf{k_2}\,\bar\delta\hspace{-0.01in}\phi_\mathbf{k_3}\, ...\, \bar\delta\hspace{-0.01in}\phi_\mathbf{k_N}(\tau)\Bigg\}_{\mathrm{\scriptstyle I}}\Bigg ]\Bigg |\,0 \Bigg\rangle.$$
$\bullet$ Draw dots corresponding to each $\bar\delta\hspace{-0.01in}\phi_\mathbf{k_i}$ for every wave-vector $k_1,k_2\,...\,k_\mathrm{N}$. We call them the \tit{external points}.\\ \\
$\bullet$ Draw dots corresponding to each of the interaction terms in the expression above. In our case, there is only one such term of $\mathcal{H}_\mathrm{int}(\tau')$. Intuitively, the number of interaction terms depend on the number of terms that feature in the Dyson expansion in the interaction picture$^\dagger$. However, we chose to expand the propagator in \eqref{eq:99} only up to the first order, and therefore, we have only one interaction term in our case. The number of these interaction terms represent the \tit{internal points}.\\ \\
$\bullet$ Now, we connect the dots in a way that each internal point connects to exactly one of the external points. A line connecting an external point to an internal point is known as an \tit{external line}$^\ddagger$. All dots are typically called \tit{vertices}. In the end, all possible permutations and combinations of diagrams are considered.\\ \\
$\bullet$ Further, each external line or internal line is then assigned a Feynman propagator corresponding to the two points that it connects. Each vertex is assigned a delta function $\delta(\mathbf{k_1}+\mathbf{k_2}+\, ... \,+\mathbf{k_N})$ as a form of conservation of momentum, and an interaction factor $-i\gamma$ at each vertex. Note that an interaction term of the form $\mathcal{H}_\mathrm{int}(\tau') = \displaystyle\int\mathrm{d}^3x\dfrac{\gamma}{p!}[\bar\delta\hspace{-0.01in}\phi(\mathbf{x},\tau')]^p$ contributes the interaction factor $-i\gamma$ to the diagram, which is essentially the `weight' of the interaction.\\ \\
$\bullet$ The final expression for a N-ordered correlation function is the multiplication of all the terms within one diagram, and then eventually performing a summation over all possible diagrams.
\begin{flushright}\myrule[line width = 0.5mm]{fast cap reversed}{fast cap reversed}\end{flushright}
\subsection{The 3-point correlation function for a cubic potential}
\label{threepoint}
We follow the rules prescribed in the previous section to calculate the 3-point correlation function $\,\langle\bar\delta\hspace{-0.01in}\phi_\mathbf{k_1}\,\bar\delta\hspace{-0.01in}\phi_\mathbf{k_2}\,\bar\delta\hspace{-0.01in}\phi_\mathbf{k_3}\rangle_{_\mathrm{\scriptstyle H}}$ for a cubic potential in momentum space, i.e. $\mathrm{N}=3$ and $p=3$. Thus, we evaluate the quantity
\begin{equation}
\label{eq:105}
\begin{multlined}
\,\langle\bar\delta\hspace{-0.01in}\phi_\mathbf{k_1}\,\bar\delta\hspace{-0.01in}\phi_\mathbf{k_2}\,\bar\delta\hspace{-0.01in}\phi_\mathbf{k_3}\rangle_{_\mathrm{\scriptstyle H}} = i\int_{\tau_\mathrm{o}}^{\tau}\mathrm{d}\tau'\,\bigg\langle 0\,\Bigg| \Bigg[\mathcal{H}_\mathrm{int}(\tau'), \Bigg\{\bar\delta\hspace{-0.01in}\phi_\mathbf{k_1}\,\bar\delta\hspace{-0.01in}\phi_\mathbf{k_2}\,\bar\delta\hspace{-0.01in}\phi_\mathbf{k_3}(\tau)\Bigg\}_{\mathrm{\scriptstyle I}}\Bigg ]\Bigg |\,0 \Bigg\rangle =\\ - \, i\int_{\tau_\mathrm{o}}^{\tau}\mathrm{d}\tau'\,\bigg\langle 0\,\Bigg|  \Bigg\{\bar\delta\hspace{-0.01in}\phi_\mathbf{k_1}\,\bar\delta\hspace{-0.01in}\phi_\mathbf{k_2}\,\bar\delta\hspace{-0.01in}\phi_\mathbf{k_3}(\tau)\Bigg\}_{\mathrm{\scriptstyle I}}\mathcal{H}_\mathrm{int}(\tau')\Bigg |\,0 \Bigg\rangle \,-\, (-i)\int_{\tau_\mathrm{o}}^{\tau}\mathrm{d}\tau'\,\bigg\langle 0\,\Bigg| \mathcal{H}_\mathrm{int}(\tau') \Bigg\{\bar\delta\hspace{-0.01in}\phi_\mathbf{k_1}\,\bar\delta\hspace{-0.01in}\phi_\mathbf{k_2}\,\bar\delta\hspace{-0.01in}\phi_\mathbf{k_3}(\tau)\Bigg\}_{\mathrm{\scriptstyle I}}\Bigg |\,0 \Bigg\rangle \, \\ \indent\text{where}\indent\mathcal{H}_\mathrm{int}(\tau') = \displaystyle\int\mathrm{d}^3x\dfrac{\gamma}{3!}[\bar\delta\hspace{-0.01in}\phi(\mathbf{x},\tau')]^3.
\end{multlined}
\end{equation}
 \\
Following the rules prescribed in the previous section, figure \ref{fig1} below shows the Feynman representation for the first term in \eqref{eq:105}. In a similar fashion, the Feynman diagram for the second term in \eqref{eq:105} is also shown below in figure \ref{fig2}. Note that the change in order of time arguments in both cases correspond to the featuring of time arguments $\tau$ and $\tau'$ in \eqref{eq:105}. Finally, we write the evaluated form of \eqref{eq:105} using the Feynman diagrams shown below:
\begin{equation}
\label{eq:106}
\begin{multlined}
\,\langle\bar\delta\hspace{-0.01in}\phi_\mathbf{k_1}\,\bar\delta\hspace{-0.01in}\phi_\mathbf{k_2}\,\bar\delta\hspace{-0.01in}\phi_\mathbf{k_3}\rangle_{_\mathrm{\scriptstyle H}} = i\int_{\tau_\mathrm{o}}^{\tau}\mathrm{d}\tau'\,\bigg\langle 0\,\Bigg| \Bigg[\mathcal{H}_\mathrm{int}(\tau'), \Bigg\{\bar\delta\hspace{-0.01in}\phi_\mathbf{k_1}\,\bar\delta\hspace{-0.01in}\phi_\mathbf{k_2}\,\bar\delta\hspace{-0.01in}\phi_\mathbf{k_3}(\tau)\Bigg\}_{\mathrm{\scriptstyle I}}\Bigg ]\Bigg |\,0 \Bigg\rangle = \\ - \, i\int_{\tau_\mathrm{o}}^{\tau}\mathrm{d}\tau'\,\bigg\langle 0\,\Bigg|  \Bigg\{\bar\delta\hspace{-0.01in}\phi_\mathbf{k_1}\,\bar\delta\hspace{-0.01in}\phi_\mathbf{k_2}\,\bar\delta\hspace{-0.01in}\phi_\mathbf{k_3}(\tau)\Bigg\}_{\mathrm{\scriptstyle I}}\mathcal{H}_\mathrm{int}(\tau')\Bigg |\,0 \Bigg\rangle \,-\, (-i)\int_{\tau_\mathrm{o}}^{\tau}\mathrm{d}\tau'\,\bigg\langle 0\,\Bigg| \mathcal{H}_\mathrm{int}(\tau') \Bigg\{\bar\delta\hspace{-0.01in}\phi_\mathbf{k_1}\,\bar\delta\hspace{-0.01in}\phi_\mathbf{k_2}\,\bar\delta\hspace{-0.01in}\phi_\mathbf{k_3}(\tau)\Bigg\}_{\mathrm{\scriptstyle I}}\Bigg |\,0 \Bigg\rangle \, \\ = -i\gamma\,\delta(\mathbf{k_1}+\mathbf{k_2}+\mathbf{k_3})\int_{\tau_\mathrm{o}}^{\tau}\mathrm{d}\tau' [D(\mathbf{k_1},\mathbf{k_2},\tau,\tau')\,D(\mathbf{k_2},\mathbf{k_3},\tau,\tau')\,D(\mathbf{k_1},\mathbf{k_3},\tau,\tau') - \\ D(\mathbf{k_1},\mathbf{k_2},\tau',\tau)\,D(\mathbf{k_2},\mathbf{k_3},\tau',\tau)\,D(\mathbf{k_1},\mathbf{k_3},\tau',\tau)] \\ \indent\text{where}\indent\mathcal{H}_\mathrm{int}(\tau') = \displaystyle\int\mathrm{d}^3x\dfrac{\gamma}{3!}[\bar\delta\hspace{-0.01in}\phi(\mathbf{x},\tau')]^3.
\end{multlined}
\end{equation}\vspace{-0.2in}
\begin{multicols}{2}
\begin{figure}[H]
\centering
\includegraphics[width=65mm]{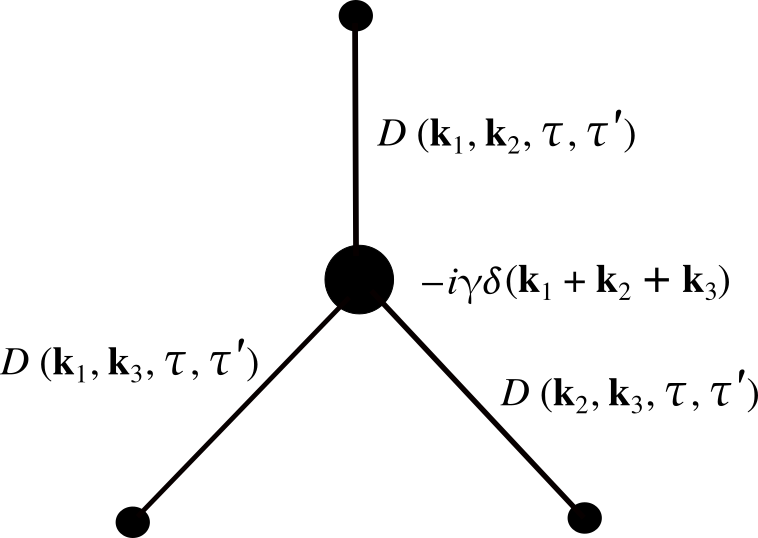}
\caption{Feynman diagram for `leading term' for $\mathrm{N}=3$ and $p=3$}
\label{fig1}
\end{figure}
\begin{figure}[H]
\centering
\includegraphics[width=65mm]{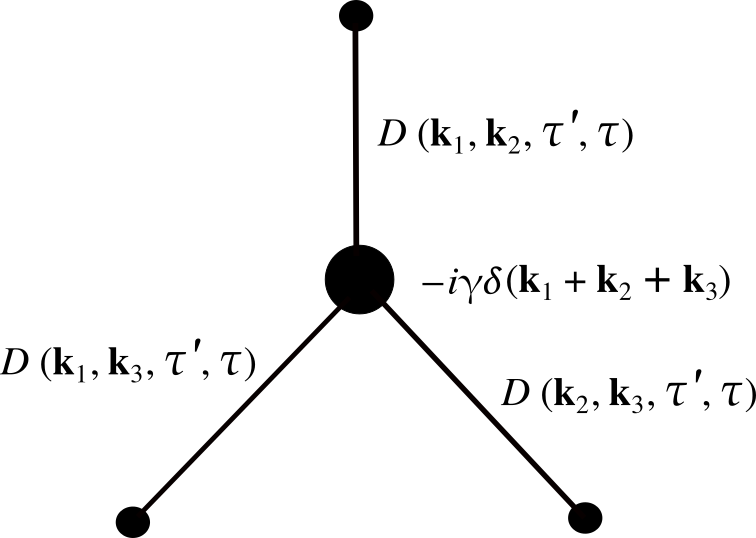}
\caption{Feynman diagram for `lagging term' for $\mathrm{N}=3$ and $p=3$}
\label{fig2}
\end{figure} 
\end{multicols}
{\noindent}Note the ordering of the time arguments $\tau$ and $\tau'$ in the Feynman propagators appearing the expression above, and we set $\tau_\mathrm{o}\rightarrow -\,\infinity$. \\ \\
Now, we follow the definition of the Feynman propagator provided in \eqref{eq:104} in terms of $u_k(\tau)$ and $u^*_k(\tau')$, and rewrite the full result for a 3-point correlation function such that,
\begin{equation}
\label{eq:107}
\begin{multlined}
\,\langle\bar\delta\hspace{-0.01in}\phi_\mathbf{k_1}\,\bar\delta\hspace{-0.01in}\phi_\mathbf{k_2}\,\bar\delta\hspace{-0.01in}\phi_\mathbf{k_3}\rangle_{_\mathrm{\scriptstyle H}} = -i\gamma\,\delta(\mathbf{k_1}+\mathbf{k_2}+\mathbf{k_3})\int_{-\displaystyle\,\infinity}^{\tau}\mathrm{d}\tau' [u_{k_1}(\tau)\,u_{k_1}^*(\tau')\,u_{k_2}(\tau)\,u_{k_2}^*(\tau')\,u_{k_3}(\tau)\,u_{k_3}^*(\tau') \,- \\ u_{k_1}(\tau')\,u_{k_1}^*(\tau)\,u_{k_2}(\tau')\,u_{k_2}^*(\tau)\,u_{k_3}(\tau')\,u_{k_3}^*(\tau)] \\ \indent\text{where,}\indent\mathcal{H}_\mathrm{int}(\tau') = \displaystyle\int\mathrm{d}^3x\dfrac{\gamma}{3!}[\bar\delta\hspace{-0.01in}\phi(\mathbf{x},\tau')]^3.
\end{multlined}
\end{equation}
This result can further be simplified if $u_k(\tau)$ is a known solution to equation \eqref{eq:36}. For example, for a massless field ($m_\phi = 0$),
\begin{equation}
\label{eq:108}
\begin{multlined}
u_k(\tau) = \Bigg(1-\dfrac{i}{k\tau}\Bigg)\dfrac{e^{-ik\tau}}{\sqrt{2k}}.
\end{multlined}
\end{equation}
The solution to \eqref{eq:107} is quite cumbersome to calculate. We simply performed a numerical computation for the solution rather than writing an analytic expansion. The result for $\tau\rightarrow 0$, i.e. right after horizon exit, follows
\begin{equation}
\label{eq:109}
\begin{multlined}
\,\langle\bar\delta\hspace{-0.01in}\phi_\mathbf{k_1}\,\bar\delta\hspace{-0.01in}\phi_\mathbf{k_2}\,\bar\delta\hspace{-0.01in}\phi_\mathbf{k_3}\rangle_{_\mathrm{H}} \xrightarrow{\tau\rightarrow 0} -\dfrac{\gamma \mathrm{H}^2}{12}\dfrac{\delta(\mathbf{k_1}+\mathbf{k_2}+\mathbf{k_3})}{({k_1}{k_2}{k_3})^3}\Bigg[-\sum_{i=1} ^{3}\Bigg\{k_i^3\Bigg(c_e+\zeta_3(k_i)+\mathrm{log\Bigg[-\tau\sum_{i=1} ^{3}k_i \Bigg]} \Bigg)\Bigg\}\Bigg].
\end{multlined}
\end{equation}
where, $c_e$ is the Euler's constant with an approximate value of 0.577, and $\zeta_3(k_i)$ is given by
\begin{equation}
\label{eq:110}
\begin{multlined}
\zeta_3(k_i) = \dfrac{-\Bigg[\displaystyle\sum_{i=1}^3 k_i\Bigg]^4 +2 \Bigg[\displaystyle\sum_{i=1}^3 k_i\Bigg]^2 \Bigg[\displaystyle\sum_{i,j=1;\,i<j}^3 k_i k_j\Bigg]+\Bigg[\displaystyle\sum_{i=1}^3 k_i\Bigg] \Bigg[\displaystyle\sum_{i,j,l=1;\,i<j<l}^3 k_i k_j k_l\Bigg]}{\Bigg[\displaystyle\sum_{i=1}^3 k_i\Bigg]^4 -3 \Bigg[\displaystyle\sum_{i=1}^3 k_i\Bigg]^2 \Bigg[\displaystyle\sum_{i,j=1;\,i<j}^3 k_i k_j\Bigg]+3\Bigg[\displaystyle\sum_{i=1}^3 k_i\Bigg] \Bigg[\displaystyle\sum_{i,j,l=1;\,i<j<l}^3 k_i k_j k_l\Bigg]}.
\end{multlined}
\end{equation}\pagebreak
\section{Calculating the bi-spectrum in Horndeski models}
\label{hornbi}
In writing the Lagrangian of an arbitrary system, it has been proven that unless the associated equations of motion are less than or equal to 2 in order, they experience the so-called \tit{Ostrogradski's instability} \citep{r12}. The most general Lagrangian for such systems avoiding Ostrogradski's instability was first derived by Horndeski in 1974 \citep{r16}. His derivation encompassed all dimensions ranging from 1 up to infinity. Hence, the Horndeski's Lagrangian is the most general single-field scalar-tensor theory in existence. However, the equivalent expression of the Lagrangian for gravitational theories in a 4-dimensional system of curved space-time was derived by \tit{\citeauthor{r10}} \citep{r10}:

\begin{equation}
\label{eq:111}
\text{I}=\int{\mathrm{d}^4x\sqrt{-g}\bigg[\frac{\mathrm{M}_\mathsf{pl}^2}{2}\mathrm{R} +  \mathrm{P}[\phi,f(\phi)] + \mathrm{G}_3[\phi,f(\phi)]\Box\phi +\mathcal{L}_4 + \mathcal{L}_5\bigg]},
\end{equation}\\
where, $\mathrm{M}_\mathsf{pl}$ is the reduced Planck mass. Moreover,

\begin{equation}
\label{eq:112}
\mathcal{L}_4=\mathrm{G}_4[\phi,f(\phi)]\mathrm{R}+
\mathrm{G}_4^{f}[(\Box\phi)^2-(\nabla_\mu\nabla_\nu\phi)(\nabla^\mu\nabla^\nu\phi)], 
\end{equation}
\begin{equation}
\label{eq:113}
\mathcal{L}_5=\mathrm{G}_5[\phi,f(\phi)]\mathrm{G}_{\mu\nu}(\nabla^\mu\nabla^\nu\phi)-\frac{1}{6}
\mathrm{G}_5^{f}[(\Box\phi)^3-3(\Box\phi)(\nabla_\mu\nabla_\nu\phi)(\nabla^\mu\nabla^\nu\phi)+2(\nabla^\mu\nabla_\alpha\phi)(\nabla^\alpha\nabla_\upbeta\phi)(\nabla^\upbeta\nabla_\mu\phi)],
\end{equation}
such that, 
$$\mathrm{G}_5^{f}\equiv\frac{\partial\mathrm{G}_5[\phi,f(\phi)]}{\partial[f(\phi)]}.$$
In the expression for action given in \eqref{eq:111}, there are 2 independent entities, i.e. the scalar field $\phi$ and the metric $g_{\mu\nu}$. Therefore, in order to solve for the equations of motion, we take variation of the action with respect to $\phi$ and $g_{\mu\nu}$. For the case of a flat-FLRW universe, the metric will take the form:

\begin{equation}
\label{eq:114}
\left\{ \begin{array}{cccc}
-1 & 0 & 0 & 0\\
0 & a(t)^2 & 0 & 0\\
0 & 0 & a(t)^2 & 0\\
0 & 0 & 0 & a(t)^2\end{array} \right\}.
\end{equation}
Therefore, it remains to be calculated the variation of action with respect to $g_{00}$, $g_{tt}$ and $\phi$ only in the limit of a flat-FLRW metric.
\subsection{The equations of motion and constraints}
\label{eqmcon}
\subsubsection{Constraint via $\mathbf{g_{\textbf{00}}}$}
\label{g00}
Following the discussion in the previous section, we write the metric in the limit of a flat-FLRW metric for the case of $g_{00}$. It is equivalent to solving the Einstein's field equations for the density parameter $\uprho$. 
\begin{equation}
\label{eq:115}
g_{\mu\nu}=\left\{ \begin{array}{cccc}
g_{00} & 0 & 0 & 0\\
0 & a(t)^2 & 0 & 0\\
0 & 0 & a(t)^2 & 0\\
0 & 0 & 0 & a(t)^2\end{array} \right\}_{\displaystyle{g_{00}\rightarrow -1.}}
\end{equation}
Remember that the equations of motion for a Lagrangian of the form $\L(x,\dt{x},t)$ are given by the Euler-Lagrange equation,
\begin{equation}
\label{eq:116}
\partial _{\mu}\partial_{\nu}\bigg(\frac{\partial {\L}}{\partial (\partial _{\mu}\partial_{\nu} x)}\bigg) + \partial _\alpha\bigg(\frac{\partial {\L}}{\partial (\partial _\alpha x)}\bigg) - \frac{\partial {\L}}{\partial x} = 0.
\end{equation}
However, if we assume homogeneous and isotropic background of all parameters, the equation takes the simpler form of 
\begin{equation}
\label{eq:117}
\frac{\partial^2}{\partial t^2}\bigg(\frac{\partial {\L}}{\partial\ddt{x}}\bigg) + \frac{\partial}{\partial t}\bigg(\frac{\partial {\L}}{\partial \dt{x}}\bigg) - \frac{\partial {\L}}{\partial x} = 0.
\end{equation}
Let us now consider each term in the Lagrangian separately and solve for \eqref{eq:117}.\\ \\
\textbf{\tit{a}) $\mathcal{R}$-\tit{term} $\boldmath\displaystyle{\sqrt{-g}\bigg[\frac{\mathrm{M}_\mathsf{pl}^2}{2}\mathrm{R}\bigg]}$:} For the $\mathcal{R}$-term, we expand the Ricci Scalar R into its degenerate form to get\footnote{$\;$Refer to \eqref{eq:24} for the expression for R$_{\mu\nu}$.}:
\begin{equation}
\label{eq:118}
\mathcal{R}-\mathrm{term}: \sqrt{-g}\bigg[\frac{\mathrm{M}_\mathsf{pl}^2}{2}\mathrm{R}\bigg]=a^3 \frac{\mathrm{M}_\mathsf{pl}^2}{2}\sqrt{-g_{00}}\,\frac{3a\dt{a}\dt{g_{00}} - 6a\ddt{a}g_{00} - 6\dt{a}^2 g_{00}}{a^2 g_{00}^2}.
\end{equation}
The corresponding equation of motion for the $\mathcal{R}$-term in the limit of our assumed flat-FLRW metric ($g_{00}=-1$, $\dt{g_{00}}=0$) is
\begin{equation}
\label{eq:119}
\frac{\partial}{\partial t}\bigg(\frac{\partial {\L}}{\partial \dt{g_{00}}}\bigg) - \frac{\partial {\L}}{\partial g_{00}}=\,-a^3 \frac{\mathrm{M}_\mathsf{pl}^2}{2}\frac{3}{2}\frac{\dt{a}}{a^2}(3a\dt{g_{00}}-2\dt{a})\xrightarrow{ \dt{g_{00}}\rightarrow 0} \frac{1}{2}\,a^3 [3\mathrm{M}_\mathsf{pl}^2 \mathrm{H}^2].
\end{equation}
\textbf{\tit{b}) $\mathcal{P}$-\tit{term}} {\boldmath$\sqrt{-g}\,\mathbf{P}[\phi,f(\phi)]$}: For the $\mathcal{P}$-term, we need to take care of the factor $f(\phi)$. Note that
\begin{equation}
\label{eq:120}
f(\phi)=-\frac{\partial^\mu\phi\,\partial_\mu\phi}{2}=-g^{\mu\nu}\frac{\partial_\nu\phi\,\partial_\mu\phi}{2}=-g_{\mu\nu}\frac{\partial^\nu\phi\,\partial^\mu\phi}{2},
\end{equation}
which reduces in case of a homogeneous and isotropic distribution of the scalar field $\phi$ on a flat-FLRW background to
\begin{equation}
\label{eq:121}
f(\phi)=-\frac{1}{g_{00}}\frac{\dt{\phi}^2}{2}.
\end{equation}
The equation of motion for $\mathcal{P}$-term is then written as
\begin{equation}
\label{eq:122}
\frac{\partial {\L}}{\partial g_{00}}=a^3\frac{\partial(\sqrt{-g_{00}})}{\partial g_{00}}\mathrm{P}[\phi,f(\phi)] + a^3\sqrt{(-g_{00})}\frac{\partial \mathrm{P}}{\partial f}\frac{\partial f}{\partial g_{00}}\xrightarrow{g_{00}\rightarrow -1}\frac{1}{2}a^3[\mathrm{P}-(\partial^f\!\mathrm{P})\,{\dt{\phi}^2}].
\end{equation}\\
\begin{flushleft}\myrule[line width = 0.5mm]{fast cap reversed}{fast cap reversed}\end{flushleft}
\tit{\textbf{Sidenote}}:\\
In this sidenote, we calculate the expressions for the Christoffel Symbols, the d'Alembert operator $\Box$, the double covariant derivative $\nabla_\mu\nabla_\nu$, and the couble contravariant derivative $\nabla^\mu\nabla^\nu$ to be used later in our calculations.\\ \\
$\bullet$ $\nabla_\mu\nabla_\nu$: The double covariant derivative for a scalar field is given by
\begin{equation}
\label{eq:123}
\nabla_\mu\nabla_\nu = \partial_\mu\partial_\nu - \Gamma^{\gamma}_{\mu\nu}\partial_\gamma.
\end{equation}
$\bullet$ $\nabla^\mu\nabla^\nu$: The double contravariant derivative in terms of the contravariant partial derivatives for a scalar field can be calculated by contracting \eqref{eq:12} such that
\begin{equation}
\label{eq:124}
\nabla^\mu\nabla^\nu =g^{\mu\alpha}
g^{\nu\gamma}\nabla_{\alpha}\nabla_{\gamma}\;\;\;\;\text{where,}\;\;\;g^{\mu\nu}=\frac{1}{g_{\mu\nu}}.
\end{equation}
$\bullet$ $\Box$: The d'Alembert operator is written as
\begin{equation}
\label{eq:125}
\Box\equiv \nabla_\mu\nabla^\mu = \nabla^\mu\nabla_\mu = g_{\mu\nu}\nabla^\mu\nabla^\nu=g^{\mu\nu}
\nabla_\mu\nabla_\nu\;\;\;\text{since,}\;\;\;\nabla_\mu g_{\alpha\gamma}=0.
\end{equation}
$\bullet$ $\Gamma^{\gamma}_{\mu\nu}$: Meanwhile, for the given metric in \eqref{eq:115}, the non-zero Christoffel Symbols are given by:
\begin{equation}
\label{eq:126}
\Gamma^0
_{00}=\frac{\dt{g_{00}}}{2g_{00}},\;\;\Gamma^0_{11}=\Gamma^0_{22}=\Gamma^0_{33}=-\frac{a\dt{a}}{g_{00}},\;\;\Gamma^1_{01}=\Gamma^1_{10}=\Gamma^2_{02}=\Gamma^2_{20}=
\Gamma^3_{03}=\Gamma^3_{30}=\frac{\dt{a}}{a}.
\end{equation}
\begin{flushright}\myrule[line width = 0.5mm]{fast cap reversed}{fast cap reversed}\end{flushright}$\;$\\
\textbf{\tit{c}) $\mathcal{G}$-\tit{term}} {\boldmath$\sqrt{-g}\,\mathrm{G}_3[\phi,f(\phi)]\Box\phi$}: The exact expression of the $\mathcal{G}$-term is evaluated to be,
\begin{equation}
\label{eq:127}
\text{$\mathcal{G}$-term}: \sqrt{-g}\,\mathrm{G}_3[\phi,f(\phi)]\Box\phi=a^3 \sqrt{-g_{00}}\,\mathrm{G}_3[\phi,f(\phi)]\,g^{\mu\nu}\nabla_\mu\nabla_\nu{\phi},
\end{equation}
\begin{equation}
\label{eq:128}
\text{$\mathcal{G}$-term}: \sqrt{-g}\,\mathrm{G}_3[\phi,f(\phi)]\Box\phi=a^3 \sqrt{-g_{00}}\,\mathrm{G}_3[\phi,f(\phi)]\,\Bigg[\frac{\ddt{\phi}}{g_{00}}-\frac{\dt{g_{00}}\dt{\phi}}{2g_{00}^2}+3\dt{\phi}\frac{\dt{a}}{ag_{00}}\Bigg].
\end{equation}
The associated term in the equation of motion is thus given by
\begin{equation}
\label{eq:129}
\frac{\partial}{\partial t}\bigg(\frac{\partial {\L}}{\partial \dt{g_{00}}}\bigg) - \frac{\partial {\L}}{\partial g_{00}}=\frac{1}{2}a^3\,[-3\mathrm{H}\,(\partial^{f\!}\mathrm{G}_3)\dt{\phi}^3 + (\partial^\phi\mathrm{G}_3)\dt{\phi}^2].
\end{equation}
\textbf{\tit{d}) $\mathcal{L}_4$-\tit{term}}: The G$_4$R term in $\mathcal{L}_4$ yields
\begin{equation}
\label{eq:130}
\frac{\partial}{\partial t}\bigg(\frac{\partial {\L}}{\partial \dt{g_{00}}}\bigg) - \frac{\partial {\L}}{\partial g_{00}}=\frac{1}{2}a^3\,[6\mathrm{H}^2\mathrm{G_4}+6\mathrm{H}(\partial^{\phi\!}\mathrm{G}_4)\dt{\phi} -6\dt{\phi}^2(\frac{\ddt{a}}{a}+\mathrm{H}^2)(\partial^{f\!}\mathrm{G_4})
+6\mathrm{H}(\partial^{f\!}\mathrm{G_4})\dt{\phi}\ddt{\phi}].
\end{equation}
The second term of $\partial^{f\!}\mathrm{G_4}$ in $\mathcal{L}_4$ gives the contribution
\begin{equation}
\begin{multlined}
\label{eq:131}
\{\partial^{f\!}\mathrm{G_4}|\mathcal{L}_4\}\mathrm{-term:} \sqrt{-g}\,\mathrm{G}^f_4[\phi,f(\phi)][(\Box\phi)^2 - (\nabla_\mu\nabla_\nu\phi)(\nabla^\mu\nabla^\nu\phi).
\end{multlined}
\end{equation}
\begin{flushleft}\myrule[line width = 0.5mm]{fast cap reversed}{fast cap reversed}\end{flushleft}
\tit{\textbf{Sidenote}}:\\
Using \eqref{eq:123}, it can be shown that
\begin{equation}
\label{eq:132}
\left\{ \begin{array}{cccc}
\nabla_0\nabla_0 & \nabla_0\nabla_1 & \nabla_0\nabla_2 & \nabla_0\nabla_3\\
\nabla_1\nabla_0 & \nabla_1\nabla_1 & \nabla_1\nabla_2 & \nabla_1\nabla_3\\
\nabla_2\nabla_0 & \nabla_2\nabla_1 & \nabla_2\nabla_2 & \nabla_2\nabla_3\\
\nabla_3\nabla_0 & \nabla_3\nabla_1 & \nabla_3\nabla_2 & \nabla_3\nabla_3\end{array} \right\}\phi=\frac{1}{g_{00}}\left\{ \begin{array}{cccc}
\Big\{g_{00}\ddt{\phi}-\frac{1}{2}\dt{g_{00}}\dt{\phi}\Big\} & 0 & 0 & 0\\ 
0 & \{\dt{\phi}\dt{a}a\} & 0 & 0\\ 
0 & 0 & \{\dt{\phi}\dt{a}a\} & 0\\ 
0 & 0 & 0 & \{\dt{\phi}\dt{a}a\}\end{array}\right\}.
\end{equation}
\begin{flushright}\myrule[line width = 0.5mm]{fast cap reversed}{fast cap reversed}\end{flushright}$\;$\\
Using \eqref{eq:123}, \eqref{eq:124} and \eqref{eq:132}, we get
\begin{equation}
\begin{multlined}
\label{eq:133}
\{\partial^{f\!}\mathrm{G_4}|\mathcal{L}_4\}\mathrm{-term:} \sqrt{-g}\,\mathrm{G}^f_4[\phi,f(\phi)][(\Box\phi)^2 - (\nabla_\mu\nabla_\nu\phi)(\nabla^\mu\nabla^\nu\phi)]\\=a^3 \sqrt{-g_{00}}\,\mathrm{G}_4^f[\phi,f(\phi)]\,
\Bigg[\frac{6}{a^4}(\nabla_1\nabla_1\phi)^2 + \frac{6}{a^2g_{00}}\{(\nabla_0\nabla_0\phi)(\nabla_1\nabla_1
\phi)\}\Bigg]\\
=a^3\sqrt{-g_{00}}\,\mathrm{G}_4^f[\phi,f(\phi)]\,\Bigg\{\frac{6}{a^2}\Bigg\}\Bigg\{\frac{\dt{\phi}^2\dt{a}^2}{g_{00}^2} + \frac{\dt{\phi}\dt{a}a}{g_{00}^2}\Bigg(\ddt{\phi}-\frac{1}{2}\frac{\dt{g_{00}}}{g_{00}}\dt{\phi}\Bigg)\Bigg\}.
\end{multlined}
\end{equation}
Lastly, the equation of motion for $\{\partial^{f\!}\mathrm{G_4}|\mathcal{L}_4\}\mathrm{-term}$ is calculated to give
\begin{equation}
\label{eq:134}
\begin{multlined}
\frac{\partial}{\partial t}\bigg(\frac{\partial {\L}}{\partial \dt{g_{00}}}\bigg) - \frac{\partial {\L}}{\partial g_{00}}=\frac{1}{2}a^3\,\Bigg[6\dt{\phi}^2(\frac{\ddt{a}}{a}-\mathrm{H}^2)(\partial^{f\!}\mathrm{G_4})
-6\mathrm{H}(\partial^{f\!}\mathrm{G_4})\dt{\phi}\ddt{\phi}
+6\mathrm{H}(\partial^{(f\!,\,\phi)}\mathrm{G_4})\dt{\phi}^3\\-6(\partial^{(f\!,\,f)}\mathrm{G_4}
)\mathrm{H}^2\dt{\phi}^4\Bigg].
\end{multlined}
\end{equation}
\textbf{\tit{e}) $\mathcal{L}_5$-\tit{term}:} The G$_5$ term in $\mathcal{L}_5$ requires $\mathrm{G}_{\mu\nu}$, which is given by
\begin{equation}
\label{eq:135}
\begin{multlined}
\mathrm{G}_{\mu\nu}=\mathrm{R}_{\mu\nu}-\frac{R}{2}g_{\mu\nu},
\end{multlined}
\end{equation}
\begin{equation}
\label{eq:136}
\begin{multlined}
\mathrm{G}_{\mu\nu}\equiv\{\mathrm{G}_{00},\mathrm{G}_{kk}\}=\Bigg\{3\frac{\dt{a}^2}{a^2},\frac{-1}{g_{00}^2}[a\dt{a}\dt{g}-(2a\ddt{a}+\dt{a}^2)g_{00}]\Bigg\},
\end{multlined}
\end{equation}
where,
\begin{equation}
\label{eq:137}
\mathrm{R}_{\mu\nu}\equiv\{\mathrm{R}_{00},\mathrm{R}_{kk}\}=\Bigg\{\frac{3}{2}\frac{\dt{a}}{a}\frac{\dt{g_{00}}}{g_{00}}-3\frac{\ddt{a}}{a},\frac{1}{2g_{00}^2}[a\dt{a}\dt{g}-2g_{00}a\ddt{a}+4g_{00}\dt{a}^2]\Bigg\}.
\end{equation}
The contribution from the G$_5$ term in $\mathcal{L}_5$ turns out to be
\begin{equation}
\label{eq:138}
\begin{multlined}
\{\mathrm{G_5}|\mathcal{L}_5\}\mathrm{-term:} \sqrt{-g_{00}}\,a^3\mathrm{G}_5[\phi,f(\phi)]\mathrm{G}_{\mu\nu}\nabla^\mu
\nabla^\nu\phi\\=\sqrt{-g_{00}}\,a^3 \mathrm{G}_5[\phi,f(\phi)]\Bigg[3\frac{\dt{a}^2}{a^2}\frac{1}{g_{00}^2}\Bigg(\ddt{\phi}-\frac{1}{2}\frac{\dt{g_{00}}}{g_{00}}             \dt{\phi}\Bigg)-\frac{3}{a^4}\frac{\dt{\phi}\dt{a}a}{g_{00}^3}(a\dt{a}\dt{g_{00}}-2\ddt{a}ag_{00}-\dt{a}^2g_{00})\Bigg].
\end{multlined}
\end{equation}
The associated equation of motion is given by
\begin{equation}
\label{eq:139}
\begin{multlined}
\frac{\partial}{\partial t}\bigg(\frac{\partial {\L}}{\partial \dt{g_{00}}}\bigg) - \frac{\partial {\L}}{\partial g_{00}}=\frac{1}{2}a^3\Bigg[9(\partial^\phi\mathrm{G}_{5})\dt{\phi}^2\mathrm{H}^2+
(\partial^f\mathrm{G}_{5})\Bigg\{6\ddt{\phi}\dt{\phi}^2\mathrm{H}^2+
-6\dt{\phi}^3\frac{\ddt{a}\dt{a}}{a^2}-3\dt{\phi}^3\frac{\dt{a}^3}{a^3}\Bigg\}\Bigg].
\end{multlined}
\end{equation}
In addition, the contribution from the G$^f_5$ term in $\mathcal{L}_5$ is written as
\begin{equation}
\label{eq:140}
\begin{multlined}
\{\mathrm{G}^f_5|\mathcal{L}_5\}\mathrm{-term:} -\frac{1}{6}\sqrt{-g_{00}}\,a^3
\mathrm{G}_5^{f}[(\Box\phi)^3-3(\Box\phi)(\nabla_\mu\nabla_\nu\phi)(\nabla^\mu\nabla^\nu\phi)+2(\nabla^\mu\nabla_\alpha\phi)(\nabla^\alpha\nabla_\upbeta\phi)(\nabla^\upbeta\nabla_\mu\phi)]\\=
-\sqrt{-g_{00}}\,a^3
\mathrm{G}_5^{f}\Bigg[\frac{(\nabla_1\nabla_1\phi)^3}{a^6}+3\frac{(\nabla_0\nabla_0\phi)}{g_{00}}\frac{(\nabla_1\nabla_1\phi)^2}{a^4}\Bigg],
\end{multlined}
\end{equation}

\begin{equation}
\label{eq:141}
\begin{multlined}
\{\mathrm{G^f_5}|\mathcal{L}_5\}\mathrm{-term:} =
\sqrt{-g_{00}}\,
\mathrm{G}_5^{f}\Bigg[\frac{\dt{a}^3\dt{\phi}^3}{g_{00}^3}+3\frac{\ddt{\phi}\dt{\phi}^2\dt{a}^2a}{g_{00}^3}-\frac{3\dt{g_{00}}\dt{\phi}^3\dt{a}^2a}{2g_{00}^4}\Bigg].
\end{multlined}
\end{equation}
Thus, the contribution from the G$^f_5$ term in $\mathcal{L}_5$ to the equation of motion is equal to
\begin{equation}
\label{eq:142}
\begin{multlined}
\frac{\partial}{\partial t}\bigg(\frac{\partial {\L}}{\partial \dt{g_{00}}}\bigg) - \frac{\partial {\L}}{\partial g_{00}}=\frac{1}{2}a^3\Bigg[-(\partial^{(f\!,\,f)}\mathrm{G_5})\dt{\phi}^5\mathrm{H}^3+(\partial^{f}\mathrm{G_5})\Bigg\{
-6\ddt{\phi}\dt{\phi}^2\mathrm{H}^2-2\dt{\phi}^3\mathrm{H}^3+
6\dt{\phi}^3\frac{\ddt{a}\dt{a}}{a^2}\Bigg\}\\+3(\partial^{(f\!,\,\phi)}\mathrm{G_5})\dt{\phi}^4\mathrm{H}^2\Bigg].
\end{multlined}
\end{equation}
In the end, the combined expression expression for the equation of motion from the variation of $g_{00}$ up to first order is written by summing the contributions from individually calculated terms such that
\begin{equation}
\label{eq:143}
\begin{multlined}
\mathfrak{C}_1=\frac{\partial}{\partial t}\bigg(\frac{\partial {\L}}{\partial \dt{g_{00}}}\bigg) - \frac{\partial {\L}}{\partial g_{00}}=3\mathrm{M}_\mathsf{pl}^2\mathrm{H}^2\Bigg(1+\frac{2\mathrm{G}_4}{\mathrm{M}_\mathsf{pl}^2}\Bigg)+\mathrm{P}+6\mathrm{H}(\partial^\phi\mathrm{G_4})\dt{\phi}+[(\partial^\phi\mathrm{G_3})-12\mathrm{H}^2(\partial^f\mathrm{G_4})\\
+9\mathrm{H}^2(\partial^\phi\mathrm{G_5})-(\partial^f\mathrm{P})]\dt{\phi}^2+[6\partial^{(f\!,\,\phi)}\mathrm{G_4})-3(\partial^{f}\mathrm{G_3})-5\mathrm{H}^2(\partial^{f}\mathrm{G_5})]\mathrm{H}\dt{\phi}^3+3[(\partial^{(f\!,\,\phi)}\mathrm{G_5})-\\
2(\partial^{(f\!,\,f)}\mathrm{G_4})]\mathrm{H}^2\dt{\phi}^4-(\partial^{(f\!,\,f)}\mathrm{G_5})\mathrm{H}^3\dt{\phi}^5=0.
\end{multlined}
\end{equation}
\subsubsection{Constraints via $\mathbf{g}_{\textbf{\textit{kk}}}$}
\label{gkk}
In the case of $g_{k\!k}$, we can rewrite the metric in the limit of a flat-FLRW form such that:
\begin{equation}
\label{eq:144}
g_{\mu\nu}=\left\{ \begin{array}{cccc}
-1 & 0 & 0 & 0\\
0 & g_{k\!k} & 0 & 0\\
0 & 0 & g_{k\!k} & 0\\
0 & 0 & 0 & g_{k\!k}\end{array} \right\}_{\displaystyle{g_{k\!k}\rightarrow a(t)^2.}}
\end{equation}\\
\begin{flushleft}\myrule[line width = 0.5mm]{fast cap reversed}{fast cap reversed}\end{flushleft}
\tit{\textbf{Sidenote}}:\\
Corresponding to the given metric in \eqref{eq:144}, the Christoffel symbols, the Einstein tensor, the Ricci tensor and the Ricci scalar are given by: 
\begin{equation}
\label{eq:145}
\Gamma^0
_{00}=0,\;\;\Gamma^0_{11}=\Gamma^0_{22}=\Gamma^0_{33}=\frac{\dt{g}_{k\!k}}{2},\;\;\Gamma^1_{01}=\Gamma^1_{10}=\Gamma^2_{02}=\Gamma^2_{20}=
\Gamma^3_{03}=\Gamma^3_{30}=\frac{\dt{g}_{k\!k}}{2g_{k\!k}},
\end{equation}
\begin{equation}
\label{eq:146}
\mathrm{R}_{\mu\nu}=\frac{3\ddt{g}_{k\!k}}{g_{k\!k}}\;,\;\;\;f(\phi)=\frac{\dt{\phi}^2}{2},
\end{equation}
\begin{equation}
\label{eq:147}
\mathrm{G}_{\mu\nu}\equiv\{\mathrm{G}_{00},\mathrm{G}_{kk}\}=\Bigg\{3\frac{\dt{g}_{k\!k}^2}{4{g_{k\!k}}^2},\frac{-1}{4g_{k\!k}}[4{g_{k\!k}}\ddt{g}_{k\!k}-\dt{{g_{k\!k}}}^2]\Bigg\},\quad\text{and}
\end{equation}
\begin{equation}
\label{eq:148}
\mathrm{R}_{\mu\nu}\equiv\{\mathrm{R}_{00},\mathrm{R}_{kk}\}=\Bigg\{3\frac{\dt{g}_{k\!k}^2}{4{g_{k\!k}}^2}-\frac{3\ddt{g}_{k\!k}}{2{g_{k\!k}}},\frac{\ddt{g}_{k\!k}}{2}+\frac{\dt{{g_{k\!k}}}^2}{4{g_{k\!k}}}\Bigg\}.
\end{equation}
Moreover,
\begin{equation}
\label{eq:149}
\left\{ \begin{array}{cccc}
\nabla_0\nabla_0 & \nabla_0\nabla_1 & \nabla_0\nabla_2 & \nabla_0\nabla_3\\
\nabla_1\nabla_0 & \nabla_1\nabla_1 & \nabla_1\nabla_2 & \nabla_1\nabla_3\\
\nabla_2\nabla_0 & \nabla_2\nabla_1 & \nabla_2\nabla_2 & \nabla_2\nabla_3\\
\nabla_3\nabla_0 & \nabla_3\nabla_1 & \nabla_3\nabla_2 & \nabla_3\nabla_3\end{array} \right\}\phi=\frac{1}{2}\left\{ \begin{array}{cccc}
2\ddt{\phi} & 0 & 0 & 0\\ 
0 & -\dt{\phi}\dt{g}_{k\!k} & 0 & 0\\ 
0 & 0 & -\dt{\phi}\dt{g}_{k\!k} & 0\\ 
0 & 0 & 0 & -\dt{\phi}\dt{g}_{k\!k}\end{array}\right\}.
\end{equation}
\begin{flushright}\myrule[line width = 0.5mm]{fast cap reversed}{fast cap reversed}\end{flushright}$\;$\\
We again use the variational principle for the $g_{k\!k}$ terms, similar to how it is implemented in the previous section. This is again equivalent to solving the Einstein's field equation for the pressure parameter $p$. The equivalent total expression (for all the $\mathcal{R}$, $\mathcal{P}$, G$_3$, $\mathcal{L}_4$ and $\mathcal{L}_5$ terms) is evaluated to be
\begin{equation}
\label{eq:150}
\begin{split}
\mathfrak{C}_2=\frac{\partial}{\partial t}\bigg(\frac{\partial {\L}}{\partial \dt{g}_{k\!k}}\bigg) - \frac{\partial {\L}}{\partial {g}_{k\!k}}\xrightarrow{{g}_{k\!k}\rightarrow a(t)^2} 3\mathrm{M}_\mathsf{pl}^2\mathrm{H}^2\Bigg(1+\frac{2\mathrm{G}_4}{\mathrm{M}_\mathsf{pl}^2}\Bigg)+\mathrm{P}+4\mathrm{H}(\partial^\phi\mathrm{G_4})\dt{\phi}+2\Bigg[(\partial^\phi
\mathrm{G_5})\dt{\phi}^2-2(\partial^f\mathrm{G_4})\dt{\phi}^2\\
-\mathrm{H}(\partial^f\mathrm{G_5})\dt{\phi}^3+\mathrm{M}_\mathsf{pl}^2
\Bigg(1+\frac{2\mathrm{G}_4}{\mathrm{M}_\mathsf{pl}^2}\Bigg)\Bigg]
\dt{\mathrm{H}}+\Big[2(\partial^\phi\mathrm{G_4})+4\mathrm{H}
\Big\{(\partial^\phi\mathrm{G_5})-(\partial^f\mathrm{G_4})
\Big\}\dt{\phi}+\Big\{2(\partial^{(f\!,\,\phi)}\mathrm{G_4})-\\(\partial^f\mathrm{G_3})-3\mathrm{H}^2
(\partial^f\mathrm{G_5})\Big\}\dt{\phi}^2 + 2\mathrm{H}
\Big\{(\partial^{(f\!,\,\phi)}\mathrm{G_5})-2(\partial^{(f\!,\,f)}\mathrm{G_4})\Big\}\dt{\phi}^3-\mathrm{H}^2
(\partial^{(f\!,\,f)}\mathrm{G_5})\dt{\phi}^4\Big]\ddt{\phi}
+\Big[2(\partial^{(\phi\!,\,\phi)}\mathrm{G_4})\\+3\mathrm{H}^2(\partial^\phi\mathrm{G_5})
-(\partial^\phi\mathrm{G_3})-6\mathrm{H}^2(\partial^f
\mathrm{G_4})\Big]\dt{\phi}^2 +2\mathrm{H}\Big[(\partial^{(\phi\!,\,\phi)}\mathrm{G_5})
-2(\partial^{(f\!,\,\phi)}\mathrm{G_4})-\mathrm{H}^2(\partial^f
\mathrm{G_5})\Big]\dt{\phi}^3\\-\mathrm{H}^2(\partial^{(f\!,\,\phi)}\mathrm{G_5})\dt{\phi}^4=0.\;\;\;\;\;\;
\end{split}
\end{equation}
\subsubsection{Equation of motion for $\phi$:}
\label{eqmphi}
Furthermore, one may also write the equation of motion for the scalar field $\phi$. The metric in this case is
simply the flat-FLRW metric in its explicit form,
\begin{equation}
\label{eq:151}
g_{\mu\nu}=\left\{ \begin{array}{cccc}
-1 & 0 & 0 & 0\\
0 & a(t)^2 & 0 & 0\\
0 & 0 & a(t)^2 & 0\\
0 & 0 & 0 & a(t)^2\end{array} \right\}.
\end{equation}
We summarize the expressions for the several terms that we use below.\\
\begin{flushleft}\myrule[line width = 0.5mm]{fast cap reversed}{fast cap reversed}\end{flushleft}
\tit{\textbf{Sidenote}}:\\
Corresponding to the given metric in \eqref{eq:151}, the Christoffel symbols, the Einstein tensor, the Ricci tensor and the Ricci scalar are given by
\begin{equation}
\label{eq:152}
\Gamma^0
_{00}=0,\;\;\Gamma^0_{11}=\Gamma^0_{22}=\Gamma^0_{33}=\dt{a}a,\;\;\Gamma^1_{01}=\Gamma^1_{10}=\Gamma^2_{02}=\Gamma^2_{20}=
\Gamma^3_{03}=\Gamma^3_{30}=\frac{\dt{a}}{a},
\end{equation}
\begin{equation}
\label{eq:153}
\mathrm{R}_{\mu\nu}=\frac{6\ddt{a}a+6\dt{a}^2}{a^2}\;,\;\;\;f(\phi)=\frac{\dt{\phi}^2}{2},
\end{equation}
\begin{equation}
\label{eq:154}
\mathrm{G}_{\mu\nu}\equiv\{\mathrm{G}_{00},\mathrm{G}_{kk}\}=\Bigg\{3\frac{\dt{a}^2}{a^2},-(2\ddt{a}a+\dt{a}^2)\},\quad\text{and}
\end{equation}
\begin{equation}
\label{eq:155}
\mathrm{R}_{\mu\nu}\equiv\{\mathrm{R}_{00},\mathrm{R}_{kk}\}=\Bigg\{-3\frac{\dt{a}}{a},\frac{1}{2}\Big[-2\ddt{a}a+4\dt{a}^2\Big]\Bigg\}.
\end{equation}
Moreover,
\begin{equation}
\label{eq:156}
\left\{ \begin{array}{cccc}
\nabla_0\nabla_0 & \nabla_0\nabla_1 & \nabla_0\nabla_2 & \nabla_0\nabla_3\\
\nabla_1\nabla_0 & \nabla_1\nabla_1 & \nabla_1\nabla_2 & \nabla_1\nabla_3\\
\nabla_2\nabla_0 & \nabla_2\nabla_1 & \nabla_2\nabla_2 & \nabla_2\nabla_3\\
\nabla_3\nabla_0 & \nabla_3\nabla_1 & \nabla_3\nabla_2 & \nabla_3\nabla_3\end{array} \right\}\phi=\frac{1}{2}\left\{ \begin{array}{cccc}
2\ddt{\phi} & 0 & 0 & 0\\ 
0 & -2\dt{\phi}\dt{a}a & 0 & 0\\ 
0 & 0 & -2\dt{\phi}\dt{a}a & 0\\ 
0 & 0 & 0 & -2\dt{\phi}\dt{a}a\end{array}\right\}.
\end{equation}
\begin{flushright}\myrule[line width = 0.5mm]{fast cap reversed}{fast cap reversed}\end{flushright}$\;$\\
Finally, the equation of motion for the scalar field is given by
{\allowdisplaybreaks
\begin{align*}
\mathfrak{C}_3=\frac{\partial}{\partial t}\bigg(\frac{\partial {\L}}{\partial \dt{\phi}}\bigg) - \frac{\partial {\L}}{\partial \phi}=\Big[6(\partial^\phi\mathrm{G_4})+12\Big\{(\partial^\phi
\mathrm{G_5})-(\partial^f\mathrm{G_4})\Big\}\mathrm{H}\dt{\phi}
+3\Big\{2(\partial^{(f\!,\,\phi)}\mathrm{G_4}) -(\partial^f\mathrm{G_3})-\\3\mathrm{H}^2
(\partial^f\mathrm{G_5})\Big\}\dt{\phi}^2+6\Big\{
(\partial^{(f\!,\,\phi)}\mathrm{G_5})-2(\partial^{(f\!,\,f)}\mathrm{G_4})\Big\}\mathrm{H}\dt{\phi}^3-3\mathrm{H}^2
(\partial^{(f\!,\,f)}\mathrm{G_5})\dt{\phi}^4\Big]\dt{\mathrm{H}}+\Big[
3\Big\{(\partial^{(f\!,\,f\!,\,\phi)}\mathrm{G_5})-\\2(\partial^{(f\!,\,f\!,\,f)}\mathrm{G_4})\Big\}\mathrm{H}^2\dt{\phi}^4-\mathrm{H}^3
(\partial^{(f\!,\,f\!,\,f)}\mathrm{G_5})\dt{\phi}^5+\Big\{
6(\partial^{(f\!,\,f\!,\,\phi)}\mathrm{G_4})-3
(\partial^{(f\!,\,f)}\mathrm{G_3})-7\mathrm{H}^2(\partial^{(f\!,\,f)}\mathrm{G_5})\Big\}\mathrm{H}\dt{\phi}^3+\\2(
\partial^{\phi}\mathrm{G_4})-(
\partial^{f}\mathrm{P})+\Big\{15(\partial^{(f\!,\,\phi)}\mathrm{G_5})\mathrm{H}^2-24(\partial^{(f\!,\,f)}\mathrm{G_4})\mathrm{H}^2+(\partial^{(f\!,\,\phi)}\mathrm{G_3})-(\partial^{(f\!,\,\phi)}\mathrm{P})\Big\}\dt{\phi}^2+\Big\{18(\partial^{(f\!,\,\phi)}\mathrm{G_4})\\-6\mathrm{H}^2(\partial^{f}\mathrm{G_5})-
6(\partial^{f}\mathrm{G_3})\Big\}\mathrm{H}\dt{\phi}+6\Big\{
(\partial^{\phi}\mathrm{G_5})-(\partial^{f}\mathrm{G_4})\Big\}
\mathrm{H}^2\Big]\ddt{\phi}+\Big[3\Big\{(\partial^{(f\!,\,f\!,\,\phi)}\mathrm{G_5})-\mathrm{H}^2(\partial^{(f\!,\,f)}\mathrm{G_5})\\-2(\partial^{(f\!,\,f\!,\,\phi)}\mathrm{G_4})\Big\}\Big]\mathrm{H}^2\dt{\phi}^4+\Big[
\Big\{7(\partial^{(f\!,\,\phi)}\mathrm{G_5})-18
(\partial^{(f\!,\,f)}\mathrm{G_4})\Big\}\mathrm{H}^2
+3\Big\{2(\partial^{(f\!,\,f\!,\,\phi)}\mathrm{G_4})-(\partial^{(f\!,\,\phi)}\mathrm{G_3})\Big\}\Big]\mathrm{H}\dt{\phi}^3\\+\Big[
3\Big\{(\partial^{(\phi\!,\,\phi)}\mathrm{G_5})+
6(\partial^{(f\!,\,\phi)}\mathrm{G_4})-3(\partial^{f}\mathrm{G_3})\Big\}
\mathrm{H}^2-9\mathrm{H}^4(\partial^{f}\mathrm{G_5})-
(\partial^{(f\!,\,\phi)}\mathrm{P})+(\partial^{(\phi\!,\,\phi)}\mathrm{G_3})\Big]\dt{\phi}+\\3\Big[6
(\partial^{\phi}\mathrm{G_5})\mathrm{H}^2 -6\mathrm{H}^2
(\partial^{f}\mathrm{G_4})-(\partial^{f}\mathrm{P})+
2(\partial^{\phi}\mathrm{G_3})\Big]\mathrm{H}\dt{\phi}-
\mathrm{H}^3(\partial^{(f\!,\,f\!,\,\phi)}\mathrm{G_5})\dt{\phi}^5+12\mathrm{H}^2(\partial^{\phi}
\mathrm{G_4})+(\partial^{\phi}\mathrm{P})=0.\tag{157}
\label{eq:157}
\end{align*}}\!
Note that one could also arrive at the results given in \eqref{eq:143} and \eqref{eq:150} by simply solving
the Einstein's field equation ($\mathrm{G}_{\mu\nu}=8\uppi\mathrm{G}\,\mathrm{T}_{\mu\nu}$), where $\mathrm{T}_{\mu\nu}\equiv\{\mathrm{T}_{00},\mathrm{T}_{kk}\} = \{\uprho,p\}$. The equations \eqref{eq:143} and \eqref{eq:150} are merely the constraint equations equivalent to the solutions for total energy density $\uprho$ and pressure $p$ obtained by solving Einstein's field equation, while \eqref{eq:157} is equivalent to the Klein-Gordon equation. In \eqref{eq:143}, we can recognise the term for $\uprho$ from Friedmann's equation,
\begin{equation}
\tag{158}\label{eq:158}
\begin{multlined}
-3\mathrm{M}_\mathsf{pl}^2\mathrm{H}^2=\uprho=6\mathrm{M}_\mathsf{pl}^2
\mathrm{H}^2\mathrm{G}_4+\mathrm{P}+6\mathrm{H}(\partial^\phi\mathrm{G_4})\dt{\phi}+[(\partial^\phi\mathrm{G_3})-12\mathrm{H}^2(\partial^f\mathrm{G_4})\\
+9\mathrm{H}^2(\partial^\phi\mathrm{G_5})-(\partial^f\mathrm{P})]\dt{\phi}^2+[6\partial^{(f\!,\,\phi)}\mathrm{G_4})-3(\partial^{f}\mathrm{G_3})-5\mathrm{H}^2(\partial^{f}\mathrm{G_5})]\mathrm{H}\dt{\phi}^3+3[(\partial^{(f\!,\,\phi)}\mathrm{G_5})-\\
2(\partial^{(f\!,\,f)}\mathrm{G_4})]\mathrm{H}^2\dt{\phi}^4-(\partial^{(f\!,\,f)}\mathrm{G_5})\mathrm{H}^3\dt{\phi}^5=\mathfrak{C}_1 -\uprho.
\end{multlined}
\end{equation}
Now, the Bianchi identity states that
\begin{equation}
\tag{159}\label{eq:159}
\begin{multlined}
\nabla_\mu\mathrm{G}^{\mu\nu}=\nabla_\mu\mathrm{T}^{\mu\nu}=0,
\end{multlined}
\end{equation}
which yields the well-known corollary to Friedmann's equation:
\begin{equation}
\tag{160}\label{eq:160}
\begin{multlined}
\dt{\uprho}+3\mathrm{H}(\uprho+p)=0.
\end{multlined}
\end{equation}
Moreover, we know that
\begin{equation}
\tag{161}\label{eq:161}
\begin{multlined}
\uprho=3\mathrm{M}_\mathsf{pl}^2\mathrm{H}^2.
\end{multlined}
\end{equation}
Using \eqref{eq:160} and \eqref{eq:161}, we can identify $p$ in \eqref{eq:150}:
\begin{equation}
\tag{162}\label{eq:162}
\begin{multlined}
p=-\frac{\dt{\uprho}}{3\mathrm{H}}-\uprho=-3\mathrm{M}_\mathsf{pl}^2\mathrm{H}^2-2
\mathrm{M}_\mathsf{pl}^2\dt{\mathrm{H}}=6\mathrm{M}_\mathsf{pl}^2
\mathrm{H}^2\mathrm{G}_4++\mathrm{P}+4\mathrm{H}(\partial^\phi\mathrm{G_4})\dt{\phi}+2\Bigg[(\partial^\phi
\mathrm{G_5})\dt{\phi}^2-2(\partial^f\mathrm{G_4})\dt{\phi}^2\\
-\mathrm{H}(\partial^f\mathrm{G_5})\dt{\phi}^3+\mathrm{M}_\mathsf{pl}^2
\Bigg(1+\frac{2\mathrm{G}_4}{\mathrm{M}_\mathsf{pl}^2}\Bigg)\Bigg]
\dt{\mathrm{H}}+\Big[2(\partial^\phi\mathrm{G_4})+4\mathrm{H}
\Big\{(\partial^\phi\mathrm{G_5})-(\partial^f\mathrm{G_4})
\Big\}\dt{\phi}+\Big\{2(\partial^{(f\!,\,\phi)}\mathrm{G_4})-\\(\partial^f\mathrm{G_3})-3\mathrm{H}^2
(\partial^f\mathrm{G_5})\Big\}\dt{\phi}^2 + 2\mathrm{H}
\Big\{(\partial^{(f\!,\,\phi)}\mathrm{G_5})-2(\partial^{(f\!,\,f)}\mathrm{G_4})\Big\}\dt{\phi}^3-\mathrm{H}^2
(\partial^{(f\!,\,f)}\mathrm{G_5})\dt{\phi}^4\Big]\ddt{\phi}
+\Big[2(\partial^{(\phi\!,\,\phi)}\mathrm{G_4})\\+3\mathrm{H}^2(\partial^\phi\mathrm{G_5})
-(\partial^\phi\mathrm{G_3})-6\mathrm{H}^2(\partial^f
\mathrm{G_4})\Big]\dt{\phi}^2 +2\mathrm{H}\Big[(\partial^{(\phi\!,\,\phi)}\mathrm{G_5})
-2(\partial^{(f\!,\,\phi)}\mathrm{G_4})-\mathrm{H}^2(\partial^f
\mathrm{G_5})\Big]\dt{\phi}^3\\-\mathrm{H}^2(\partial^{(f\!,\,\phi)}\mathrm{G_5})\dt{\phi}^4=\mathfrak{C}_2-\uprho.\;\;\;\;
\end{multlined}
\end{equation}
Following \eqref{eq:158} and \eqref{eq:162}, the Bianchi identity can alternatively be written as:
\begin{equation}
\tag{163}\label{eq:163}
\begin{multlined}
\dt{\uprho}+3\mathrm{H}(\uprho+p)=-\dt{\mathfrak{C}}_1-3\mathrm{H}
(\mathfrak{C}_1-\mathfrak{C}_2)+\{\partial_t(\mathfrak{C}_1-
\uprho)-3\mathrm{H}(\mathfrak{C}_2-\mathfrak{C}_1+\uprho+p)\}=0,
\end{multlined}
\end{equation}
where again, it can proven that
\begin{equation}
\tag{164}\label{eq:164}
\begin{multlined}
\{\partial_t(\mathfrak{C}_1-
\uprho)-3\mathrm{H}(\mathfrak{C}_2-\mathfrak{C}_1+\uprho+p)\}=
\dt{\phi}\mathfrak{C}_3,
\end{multlined}
\end{equation}
such that \eqref{eq:163} reduces to
\begin{equation}
\tag{165}\label{eq:165}
\begin{multlined}
\dt{\uprho}+3\mathrm{H}(\uprho+p)=-\dt{\mathfrak{C}}_1-3\mathrm{H}
(\mathfrak{C}_1-\mathfrak{C}_2)+\dt{\phi}\mathfrak{C}_3=0.
\end{multlined}
\end{equation}
\subsubsection{The slow-roll approximation}
\label{slowroll}
We can now try to calculate the slow-roll parameter $\epsilon$ for our system. However, we need to define the following parameters beforehand.\\ \\
\textbf{\tit{a}) \tit{Primary slow-roll terms}:}
The primary first-order slow-roll terms for H, $\phi$ and P$[\phi,f(\phi)]$ are defined as
\begin{equation}
\tag{166}\label{eq:166}
\epsilon=-\frac{\dt{\mathrm{H}}}{\mathrm{H}^2},\;\;\;\;\;\;\;\aleph_\phi=\frac{\ddt{\phi}}{\mathrm{H}\dt{\phi}},\;\;\;\;\;\;\;
\aleph_{\mathrm{P}^f}=\frac{f(\phi)}{\mathrm{M}^2_\mathsf{pl}\mathrm{F}\,\mathrm{H}^2}(\partial^f\mathrm{P}).
\end{equation}
\textbf{\tit{b}) \tit{Second-order slow-roll terms}:}
The second order slow-roll terms for G$_3$, G$_4$ and G$_5$ are defined as
\begin{equation}
\tag{167}\label{eq:167}
\aleph_{\mathrm{G_3}^f}=\frac{\dt{\phi}f(\phi)}{\mathrm{M}^2_\mathsf{pl}\mathrm{F}\,\mathrm{H}}(\partial^f\mathrm{G_3}),\;\;\;\;\;\;\;
\aleph_{\mathrm{G_4}^f}=\frac{{f(\phi)}}{\mathrm{M}^2_\mathsf{pl}\mathrm{F}}(\partial^f\mathrm{G_4}),\;\;\;\;\;\;\;
\aleph_{\mathrm{G_5}^f}=\frac{\dt{\phi}\mathrm{H}f(\phi)}{\mathrm{M}^2_\mathsf{pl}\mathrm{F}}(\partial^f\mathrm{G_5}),
\end{equation}
\begin{equation}
\tag{168}\label{eq:168}
\aleph_{\mathrm{G_3}^\phi}=\frac{f(\phi)}{\mathrm{M}^2_\mathsf{pl}\mathrm{F}\,\mathrm{H}}(\partial^\phi\mathrm{G_3}),\;\;\;\;\;\;\;
\aleph_{\mathrm{G_4}^\phi}=\frac{\dt{\phi}}{\mathrm{M}^2_\mathsf{pl}\mathrm{F}\,\mathrm{H}}(\partial^\phi\mathrm{G_4}),\;\;\;\;\;\;\;
\aleph_{\mathrm{G_5}^\phi}=\frac{f(\phi)}{\mathrm{M}^2_\mathsf{pl}\mathrm{F}}(\partial^\phi\mathrm{G_5}).
\end{equation}
\textbf{\tit{c}) \tit{Third-order slow-roll terms}:}
The third order slow-roll terms for G$_4$ and G$_5$ are defined as
\begin{equation}
\tag{169}\label{eq:169}
\aleph_{\mathrm{G_4}^{f\!f}}=\frac{f(\phi)^2}{\mathrm{M}^2_\mathsf{pl}\mathrm{F}}(\partial^{(f\!,\,f)}\mathrm{G_4}),\;\;\;\;\;\;\;
\aleph_{\mathrm{G_4}^{f\!\phi}}=\frac{\dt{\phi}{f(\phi)}}{\mathrm{M}^2_\mathsf{pl}\mathrm{F}\,\mathrm{H}}(\partial^{(f\!,\,\phi)}\mathrm{G_4}),\;\;\;\;\;\;\;
\aleph_{\mathrm{G_4}^{\phi\phi}}=\frac{f(\phi)}{\mathrm{M}^2_\mathsf{pl}\mathrm{F}\,\mathrm{H}^2}(\partial^{(\phi\phi)}\mathrm{G_4}),
\end{equation}
\begin{equation}
\tag{170}\label{eq:170}
\aleph_{\mathrm{G_5}^{f\!f}}=\frac{\dt{\phi}\mathrm{H}f(\phi)}{\mathrm{M}^2_\mathsf{pl}\mathrm{F}}(\partial^{(f\!,\,f)}\mathrm{G_5}),\;\;\;\;\;\;\;
\aleph_{\mathrm{G_5}^{f\!\phi}}=\frac{{f(\phi)^2}}{\mathrm{M}^2_\mathsf{pl}\mathrm{F}}(\partial^{(f\!,\,\phi)}\mathrm{G_5}),\;\;\;\;\;\;\;
\aleph_{\mathrm{G_5}^{\phi\phi}}=\frac{\dt{\phi}f(\phi)}{\mathrm{M}^2_\mathsf{pl}\mathrm{F}\,\mathrm{H}}(\partial^{(\phi\phi)}\mathrm{G_5}).
\end{equation}
where, we have redefined F as:
\begin{equation}
\tag{171}\label{eq:171}
\mathrm{F}=1+\frac{2\mathrm{G}_4}{\mathrm{M}^2_\mathsf{pl}}.
\end{equation}
We eliminate the term for P from \eqref{eq:143} and \eqref{eq:150}, and arrive at the following expression:
\begin{equation}
\tag{172}\label{eq:172}
\begin{multlined}
[1-4\aleph_{\mathrm{G_4}^{f}}-2\aleph_{\mathrm{G_5}^{f}}+2
\aleph_{\mathrm{G_5}^{\phi}}]\epsilon=\aleph_{\mathrm{P}^{f}}
+3\aleph_{\mathrm{G_3}^{f}}-2\aleph_{\mathrm{G_3}^{\phi}}+6
\aleph_{\mathrm{G_4}^{f}}-\aleph_{\mathrm{G_4}^{\phi}}-6
\aleph_{\mathrm{G_5}^{\phi}}+3\aleph_{\mathrm{G_5}^{f}}\\
+12\aleph_{\mathrm{G_4}^{f\!f}}+2\aleph_{\mathrm{G_5}^{f\!f}}
-10\aleph_{\mathrm{G_4}^{f\!\phi}}+2\aleph_{\mathrm{G_4}^{\phi\phi}}
-8\aleph_{\mathrm{G_5}^{f\!\phi}}+2\aleph_{\mathrm{G_5}^{\phi
\phi}}-\aleph_{\phi}[\aleph_{\mathrm{G_3}^{f}}+4
\aleph_{\mathrm{G_4}^{f}}-\aleph_{\mathrm{G_4}^{\phi}}+
8\aleph_{\mathrm{G_4}^{f\!f}}\\+3\aleph_{\mathrm{G_5}^{f}}-
4\aleph_{\mathrm{G_5}^{\phi}}+2\aleph_{\mathrm{G_5}^{f\!f}}-
2\aleph_{\mathrm{G_4}^{f\!\phi}}-4\aleph_{\mathrm{G_5}^{f\!\phi}}].
\end{multlined}
\end{equation}
Now, since $\epsilon\ll 1$, all terms of $\aleph$ in \eqref{eq:172} must follow $|\aleph_i|\ll 1$, such that
\begin{equation}
\tag{173}\label{eq:173}
\begin{multlined}
\epsilon=\aleph_{\mathrm{P}^{f}}
+3\aleph_{\mathrm{G_3}^{f}}-2\aleph_{\mathrm{G_3}^{\phi}}+6
\aleph_{\mathrm{G_4}^{f}}\\-\aleph_{\mathrm{G_4}^{\phi}}-6
\aleph_{\mathrm{G_5}^{\phi}}+3\aleph_{\mathrm{G_5}^{f}}
+12\aleph_{\mathrm{G_4}^{f\!f}}+2\aleph_{\mathrm{G_5}^{f\!f}}
-10\aleph_{\mathrm{G_4}^{f\!\phi}}+2\aleph_{\mathrm{G_4}^{\phi\phi}}
-8\aleph_{\mathrm{G_5}^{f\!\phi}}+2\aleph_{\mathrm{G_5}^{\phi
\phi}}.
\end{multlined}
\end{equation}
Moreover, we can rewrite down the expression for the second-order slow-roll parameter $\upeta$ as
\begin{equation}
\tag{174}\label{eq:174}
\begin{multlined}
\upeta=-\frac{1}{2}\frac{\ddt{\mathrm{H}}}{\dt{\mathrm{H}}\,\mathrm{H}}=-\frac{1}{2}\frac{1}{\epsilon\mathrm{H}}\frac{\ddt{\mathrm{H}}}{\mathrm{H}^2}=
-\frac{1}{2}\frac{1}{\epsilon\mathrm{H}}(-\dt{\epsilon}
+2\mathrm{H}\epsilon^2)\xrightarrow{\epsilon\ll 1}\frac{1}{2}
\frac{\dt{\epsilon}}{\epsilon\mathrm{H}},
\end{multlined}
\end{equation}
such that,
\begin{equation}
\tag{175}\label{eq:175}
\begin{multlined}
|\upeta|=\Bigg|\frac{1}{2}\frac{\dt{\epsilon}}{\epsilon\mathrm{H}}\Bigg|=\sum_\nu\Bigg|\frac{1}{2}\frac{\dt{\aleph_v}}{\epsilon\mathrm{H}}\Bigg|\ll 1,
\end{multlined}
\end{equation}
where, $\aleph_\nu$ represents $\aleph$ terms appearing in \eqref{eq:173}. Note that \eqref{eq:175} is valid as long as for each $\nu$, we have
\begin{equation}
\tag{176}\label{eq:176}
\begin{multlined}
\Bigg|\frac{1}{2}\frac{\dt{\aleph_v}}{\epsilon\mathrm{H}}\Bigg|\ll 1.
\end{multlined}
\end{equation}
Let us take, for example, the case of $\aleph_{\mathrm{G_4}^{f}}$:
\begin{equation}
\tag{177}\label{eq:177}
\begin{multlined}
\Bigg|\frac{1}{2}\frac{\dt{\aleph_v}}{\epsilon\mathrm{H}}\Bigg|=\Bigg|\frac{1}{2}\frac{\dt{\aleph}_{{\mathrm{G_4}^{f}}}}{\epsilon\mathrm{H}}\Bigg|=\frac{1}{2}\Bigg|\frac{{\aleph_{{\mathrm{G_4}^{f\!\phi}}}}}{\epsilon\mathrm{H}}+2\aleph_\phi\frac{{\aleph_{
{\mathrm{G_4}^{f\!f}}}}}{\epsilon}+\frac{{\aleph_{{\mathrm{G_4}^{f}}}}}{\epsilon}[2\aleph_\phi-\aleph_\mathrm{F}]\Bigg|\ll 1\;\;\;
\text{where,}\;\;\;\aleph_\mathrm{F}=\frac{\dt{\mathrm{F}}}{\mathrm{F}\,\mathrm{H}}.
\end{multlined}
\end{equation}
Clearly, it can be deducted from \eqref{eq:177} that ${\aleph_{{\mathrm{G_4}^{f\!\phi}}}}
=\mathcal{O}(\epsilon^2)$. Similarly, it can be proven that
\begin{equation}
\tag{178}\label{eq:178}
\begin{multlined}
[{\aleph_{{\mathrm{G_4}^{f\!\phi}}}},
{\aleph_{{\mathrm{G_4}^{\phi\phi}}}},
{\aleph_{{\mathrm{G_5}^{f\!\phi}}}},
{\aleph_{{\mathrm{G_5}^{\phi\phi}}}}]=\mathcal{O}(\epsilon^2).
\end{multlined}
\end{equation}
Thus, \eqref{eq:173} reduces to
\begin{equation}
\tag{179}\label{eq:179}
\begin{multlined}
\epsilon={\aleph_{{\mathrm{P}^{f}}}}+
3{\aleph_{{\mathrm{G_3}^{f}}}}-
2{\aleph_{{\mathrm{G_3}^{\phi}}}}+
6{\aleph_{{\mathrm{G_4}^{f}}}}
-{\aleph_{{\mathrm{G_4}^{\phi}}}}-6
{\aleph_{{\mathrm{G_5}^{\phi}}}}+3
{\aleph_{{\mathrm{G_5}^{f}}}}+12
{\aleph_{{\mathrm{G_4}^{f\!f}}}}+2
{\aleph_{{\mathrm{G_5}^{f\!f}}}}.
\end{multlined}
\end{equation}

\subsection{The power spectrum}
\label{powerspect}
In order to calculate the spectrum of the initial perturbations in the scalar field, we introduce perturbative terms in the scalar field $\phi$ encoding the fluctuations on top of a homogeneous and isotropic background. The power spectrum, for instance, is calculated by introducing perturbations up to the first order in the scalar field. It is beneficial to use the ADM formalism in differential geometry for the calculation of the power spectrum. \\ \\
The ADM metric in its general form is given by\footnote{Refer to Appendix \ref{A.4} for details on ADM formulation in Numerical Relativity.}:
\begin{equation}
\tag{180}
\label{eq:180}
\begin{multlined}
g_{\mu\nu}\mathrm{d}x^\mu\mathrm{d}x^\nu=\mathrm{N}_\tau^2
\mathrm{d}t^2+h_{\mu\nu}^{(3)}(\mathrm{d}x^\mu+\mathrm{N}^\mu\mathrm{d}t)(\mathrm{d}x^\nu+
\mathrm{N}^\nu\mathrm{d}t).
\end{multlined}
\end{equation}
In its perturbed form, the metric in the ADM formalism may be written as
\begin{equation}
\tag{181}\label{eq:181}
g_{\mu\nu}=\left\{ \begin{array}{cccc}
-[(1+\alpha)^2-a(t)^{-2}e^{-2\mathcal{R}}(\partial\Theta)^2] 
& 2\partial_x\Theta & 2\partial_y\Theta & 2\partial_z\Theta\\
2\partial_x\Theta & a(t)^2e^{2\mathcal{R}} & 0 & 0\\
2\partial_y\Theta & 0 & a(t)^2e^{2\mathcal{R}} & 0\\
2\partial_z\Theta & 0 & 0 & a(t)^2e^{2\mathcal{R}}\end{array} \right\},
\end{equation}
where, $\alpha$, $\mathcal{R}$ and $\Theta$ are scalar perturbations. In order to arrive at this form of the metric, we have used the comoving gauge \citep{r14}, with perturbations parametrised with $\alpha$, $\mathcal{R}$ and $\Theta$ as follows:
\begin{equation}
\tag{182}\label{eq:182}
\mathrm{d}\phi(\vec{x},t=t_\mathrm{o})=0,\;\;\;\;\mathrm{N}_\tau=-(1+\alpha)\;\;\;\;
h_{\mu\nu}^{(3)}=e^{2\mathcal{R}}a(t)^2\;\;\;\;\mathrm{N}^\mu=e^{-2\mathcal{R}}a(t)^{-2}\partial_\mu\Theta.
\end{equation}
In doing this, we have simply used an alternative parametrisation of the perturbations in terms of $\alpha$, $\mathcal{R}$ and $\Theta$ instead of writing the action explicitly in terms of $\mathrm{N}_\tau$ and $\mathrm{N}^\mu$, as it is usually done in numerical relativity. In this case, the scalar field $\phi$ (when $\mathrm{d}\phi=0$) and $\mathcal{R}$ (instead of $h_{\mu\nu}^{(3)}$) become the independent degrees of freedom, while the new Lagrange multipliers are now $\partial_\mu\Theta$ and $\alpha$, instead of $\mathrm{N}^\mu$ and $\mathrm{N}_\tau$ respectively. Note that by employing this gauge transformation, we have separately fixed the time and space re-parametrisations. We now include the redefined ADM metric \eqref{eq:181} into the action given in \eqref{eq:111}, the constraints $\mathfrak{C}_i$ from \eqref{eq:143}, \eqref{eq:150} and \eqref{eq:157}, and perturb it to second order\footnote{We occasionally used the \textsf{MAXIMA} and \textsf{XAct} (\textsf{XPert}) package for \textsf{Mathematica} 9.0 to perturb the action up to the required order.}:
{\allowdisplaybreaks
\begin{align*}
\mathrm{I}=\int\mathrm{d}t\,\mathrm{d}^3\!x\,a(t)^3\Bigg[
\Bigg\{-3\dt{\mathcal{R}}^2+\frac{2}{a(t)^2}
(\partial^2\Theta)\dt{\mathcal{R}}-\frac{2}{a(t)^2}(\partial^2\mathcal{R})\alpha\Bigg\}\Big\{
\mathrm{M}^2_{\mathsf{pl}}\mathrm{F}
-4f(\phi)(\partial^{f}\mathrm{G_4})-2\mathrm{H}\dt{\phi}
(\partial^{f}\mathrm{G_5})\\+2f(\phi)(\partial^{\phi}\mathrm{G_5})\Big\}+\Bigg\{-\frac{1}{a(t)^2}\alpha(\partial^2\Theta)+3\alpha\dt{\mathcal{R}}\Bigg\}
\Big\{2
\mathrm{M}^2_{\mathsf{pl}}\mathrm{H\,F}
-2f(\phi)\dt{\phi}(\partial^{f}\mathrm{G_3})-16\mathrm{H}
[f(\phi)(\partial^{f}\mathrm{G_4})\\+f(\phi)^2(\partial^{(f\!,\,f)}\mathrm{G_4})]+2\dt{\phi}[(\partial^{\phi}\mathrm{G_4})+2f(\phi)(\partial^{(f\!,\,\phi)}\mathrm{G_4})]-2\mathrm{H}^2\dt{\phi}[5f(\phi)
(\partial^{f}\mathrm{G_5})+2f(\phi)^2(\partial^{(f\!,\,f)}\mathrm{G_5})]\\+4\mathrm{H}f(\phi)[3(\partial^{\phi}\mathrm{G_5})+2f(\phi)(\partial^{(f\!,\,\phi)}\mathrm{G_5})]
\Big\}+\frac{1}{3}\alpha^2\Big\{
-9\mathrm{M}^2_{\mathsf{pl}}\mathrm{H}^2\mathrm{F}+3
[f(\phi)(\partial^{f}\mathrm{P})+2f(\phi)^2(\partial^{(f\!,\,f)}\mathrm{P})]\\+18\dt{\phi}\mathrm{H}[2f(\phi)(\partial^{f}\mathrm{G_3})+f(\phi)^2(\partial^{(f\!,\,f)}\mathrm{G_5})]-6f(\phi)[(\partial^{\phi}\mathrm{G_3})+f(\phi)(\partial^{(f\!,\,\phi)}\mathrm{G_3})]+18\hspace{0.01in}\mathrm{H}^2[7f(\phi)(\partial^{f}\mathrm{G_4})\\+16f(\phi)^2(\partial^{(f\!,\,f)}\mathrm{G_4})+4f(\phi)^3(\partial^{(f\!,\,f\!,\,f)}\mathrm{G_4})]-18\dt{\phi}\mathrm{H}[(\partial^{\phi}\mathrm{G_4})+5f(\phi)(\partial^{(f\!,\,\phi)}\mathrm{G_4})+2f(\phi)^2(\partial^{(f\!,\,f\!,\,\phi)}\mathrm{G_4})]\\+6\mathrm{H}^3\dt{\phi}[15f(\phi)(\partial^{f}\mathrm{G_5})+13f(\phi)^2(\partial^{(f\!,\,f)}\mathrm{G_5})+2f(\phi)^3(\partial^{(f\!,\,f\!,\,f)}\mathrm{G_5})]-18\hspace{0.01in}\mathrm{H}^2f(\phi)[6(\partial^{\phi}\mathrm{G_5})+\\9f(\phi)(\partial^{(f\!,\,\phi)}\mathrm{G_5})+2f(\phi)^2(\partial^{(f\!,\,f\!,\,\phi)}\mathrm{G_5})]\Big\}+\frac{1}{a(t)^2}
(\partial\mathcal{R})^2\Big\{\mathrm{M}^2_{\mathsf{pl}}
\mathrm{F}-2f(\phi)(\partial^{\phi}\mathrm{G_5})-2
f(\phi)(\partial^{f}\mathrm{G_5})\ddt{\phi}\,\Big\}\Bigg].
\tag{183}\label{eq:183}
\end{align*}}
where, we can abbreviate:
\begin{equation}
\tag{184}\label{eq:184}
\begin{multlined}
\textbf{{${\Gamma}$}}_1=\mathrm{M}^2_{\mathsf{pl}}\mathrm{F}
-4f(\phi)(\partial^{f}\mathrm{G_4})-2\mathrm{H}\dt{\phi}
f(\phi)(\partial^{f}\mathrm{G_5})+2f(\phi)(\partial^{\phi}\mathrm{G_5}),
\end{multlined}
\end{equation}\vspace{0.01in}
\begin{equation}
\tag{185}\label{eq:185}
\begin{multlined}
\textbf{{${\Gamma}$}}_2=2\mathrm{M}^2_{\mathsf{pl}}\mathrm{H\,F}
-2f(\phi)\dt{\phi}(\partial^{f}\mathrm{G_3})-16\mathrm{H}
[f(\phi)(\partial^{f}\mathrm{G_4})+f(\phi)^2(\partial^{(f\!,\,f)}\mathrm{G_4})]+2\dt{\phi}[(\partial^{\phi}\mathrm{G_4})+ 2f(\phi)\\(\partial^{(f\!,\,\phi)}\mathrm{G_4})]-2\mathrm{H}^2\dt{\phi}[5f(\phi)
(\partial^{f}\mathrm{G_5})+2f(\phi)^2(\partial^{(f\!,\,f)}\mathrm{G_5})]+4\mathrm{H}f(\phi)[3(\partial^{\phi}\mathrm{G_5})+2f(\phi)(\partial^{(f\!,\,\phi)}\mathrm{G_5})],
\end{multlined}
\end{equation}\vspace{0.01in}
\begin{equation}
\tag{186}\label{eq:186}
\begin{multlined}
\textbf{{${\Gamma}$}}_3=-9\mathrm{M}^2_{\mathsf{pl}}\mathrm{H}^2\mathrm{F}+3
[f(\phi)(\partial^{f}\mathrm{P})+2f(\phi)^2(\partial^{(f\!,\,f)}\mathrm{P})]+18\dt{\phi}\mathrm{H}[2f(\phi)(\partial^{f}\mathrm{G_3})+f(\phi)^2(\partial^{(f\!,\,f)}\mathrm{G_3})]-\\6f(\phi)[(\partial^{\phi}\mathrm{G_3})+f(\phi)(\partial^{(f\!,\,\phi)}\mathrm{G_3})]+18\hspace{0.01in}\mathrm{H}^2[7f(\phi)(\partial^{f}\mathrm{G_4})+16f(\phi)^2(\partial^{(f\!,\,f)}\mathrm{G_4})+4f(\phi)^3(\partial^{(f\!,\,f\!,\,f)}\mathrm{G_4})]-\\18\dt{\phi}\mathrm{H}[(\partial^{\phi}\mathrm{G_4})+5f(\phi)(\partial^{(f\!,\,\phi)}\mathrm{G_4})+2f(\phi)^2(\partial^{(f\!,\,f\!,\,\phi)}\mathrm{G_4})]+6\mathrm{H}^3\dt{\phi}[15f(\phi)(\partial^{f}\mathrm{G_5})+13f(\phi)^2(\partial^{(f\!,\,f)}\mathrm{G_5})\\+2f(\phi)^3(\partial^{(f\!,\,f\!,\,f)}\mathrm{G_5})]-18\hspace{0.01in}\mathrm{H}^2f(\phi)[6(\partial^{\phi}\mathrm{G_5})+9f(\phi)(\partial^{(f\!,\,\phi)}\mathrm{G_5})+2f(\phi)^2(\partial^{(f\!,\,f\!,\,\phi)}\mathrm{G_5})],
\end{multlined}
\end{equation}\vspace{0.01in}
\begin{equation}
\tag{187}\label{eq:187}
\begin{multlined}
\textbf{{${\Gamma}$}}_4=\mathrm{M}^2_{\mathsf{pl}}
\mathrm{F}-2f(\phi)(\partial^{\phi}\mathrm{G_5})-2
f(\phi)(\partial^{f}\mathrm{G_5})\ddt{\phi},
\end{multlined}
\end{equation}
yielding the following condensed form:
\begin{equation}
\tag{188}\label{eq:188}
\begin{multlined}
\mathrm{I}=\int\mathrm{d}t\,\mathrm{d}^{3}\!x\,a(t)^3\Bigg[
\Bigg\{-3\dt{\mathcal{R}}^2+\frac{2}{a(t)^2}
(\partial^2\Theta)\dt{\mathcal{R}}-\frac{2}{a(t)^2}(\partial^2\mathcal{R})\alpha\Bigg\}\textbf{{${\Gamma}$}}_1+\Bigg\{-\frac{1}{a(t)^2}\alpha(\partial^2\Theta)+3\alpha\dt{\mathcal{R}}\Bigg\}
\textbf{{${\Gamma}$}}_2+\\ \frac{1}{3}\alpha^2\textbf{{${\Gamma}$}}_3+\frac{1}{a(t)^2}
(\partial\mathcal{R})^2\textbf{{${\Gamma}$}}_4\Bigg].
\end{multlined}
\end{equation}
Note that we need not expand $\mathrm{N}_\tau$ and $\mathrm{N}^\mu$ up to the second order, but only the $\mathcal{R}$ terms. This is because  in the effective action\footnote{Refer to \eqref{eq:A.4.7} in Appendix.}, the Lagrange multipliers are in a multiplicative relation with the equations of motion of the system, and hence they vanish.  \\ \\
Now, we can easily compute the modified Hamiltonian and momentum constraints by re-arranging the action for the Lagrange multipliers $\alpha$ and $\partial_\mu\Theta$ ($\partial^2\Theta=\sum_\mu(\partial_\mu\Theta)^2$) in \eqref{eq:188},
\begin{equation}
\tag{189}\label{eq:189}
\begin{multlined}
\alpha=2\frac{\textbf{{${\Gamma}$}}_1}{\textbf{{${\Gamma}$}}_2}\dt{\mathcal{R}},
\end{multlined}
\end{equation}
\begin{equation}
\tag{190}\label{eq:190}
\begin{multlined}
(\partial^2\Theta)=a(t)^2\frac{2}{3}\frac{\textbf{{${\Gamma}$}}_3}{\textbf{{${\Gamma}$}}_2}\alpha-2\frac{\textbf{{${\Gamma}$}}_1}{\textbf{{${\Gamma}$}}_2}(\partial^2\mathcal{R})+3a(t)^2\dt{\mathcal{R}}.
\end{multlined}
\end{equation}\\
We substitute the Hamiltonian and momentum constraints into the action in \eqref{eq:188}, and use integration by parts on each \textbf{{${\Gamma}$}} term separately, and arrive at the following compressed form:
\begin{equation}
\tag{191}\label{eq:191}
\begin{multlined}
\mathrm{I}_1=\int\mathrm{d}t\,\mathrm{d}^{3}\!x\,a(t)^3 \mathrm{Q}\Bigg[\dt{\mathcal{R}}^2-\frac{v^2_\mathrm{c}}{a(t)^2}(\partial\mathcal{R})^2\Bigg],
\end{multlined}
\end{equation}
where,
\begin{equation}
\tag{192}\label{eq:192}
\begin{multlined}
v^2_\mathrm{c}=\frac{3(2\textbf{{${\Gamma}$}}^2_1\textbf{{${\Gamma}$}}_2\mathrm{H}-\textbf{{${\Gamma}$}}^2_2\textbf{{${\Gamma}$}}_4+4
\textbf{{${\Gamma}$}}_1\dt{\textbf{{${\Gamma}$}}}_1\textbf{{${\Gamma}$}}_2-2
\textbf{{${\Gamma}$}}^2_1\dt{\textbf{{${\Gamma}$}}}_2)}{\textbf{{${\Gamma}$}}_1(4\textbf{{${\Gamma}$}}_1\textbf{{${\Gamma}$}}_3+9\textbf{{${\Gamma}$}}^2_2)},
\end{multlined}
\end{equation}
\begin{equation}
\tag{193}\label{eq:193}
\begin{multlined}
\mathrm{Q}=\frac{\textbf{{${\Gamma}$}}_1
(4\textbf{{${\Gamma}$}}_3\textbf{{$
{\Gamma}$}}_1+9\textbf{{${\Gamma}$}}^2_2)}{3\textbf{{${\Gamma}$}}^2_2}.
\end{multlined}
\end{equation}
In order to simplify, we treat the action $\mathrm{I}_1$ in \eqref{eq:191} as the effective action, with $\mathcal{R}$ being the effective scalar field representing the scalar perturbations in the spatial components of the metric in \eqref{eq:182}. We thus quantise $\mathcal{R}$ instead of the scalar field $\phi$ explicitly using the formalism of second-quantization discussed in section \ref{secondquant},
\begin{equation}
\tag{194}\label{eq:194}
\begin{multlined}
\mathcal{R}(\tau,x)=\frac{1}{(2\uppi)^{3/2}}\int\mathrm{d}^3k\,[A_k(\tau)e^{i\mathbf{k}\cdot\mathbf{x}}+A^*_k(\tau)e^{-i\mathbf{k}\cdot\mathbf{x}}],\\
\end{multlined}
\end{equation}
where, we have reintroduced the conformal time $\tau$ such that $\mathrm{d}(t)/a(t) = \mathrm{d}\tau$. In operator form, according to second-quantization,
\begin{equation}
\tag{195}\label{eq:195}
\begin{multlined}
\bar{\mathcal{R}}(\tau,x)=\frac{1}{(2\uppi)^{3/2}}\int\mathrm{d}^3k\,[\bar{A}_k(\tau)e^{i\mathbf{k}\cdot\mathbf{x}}+\bar{A}^*_k(\tau)e^{-i\mathbf{k}\cdot\mathbf{x}}],\\\bar{\mathcal{R}}(\tau,x)=\frac{1}{(2\uppi)^{3/2}}\int\mathrm{d}^3k\,[u_k(\tau)\bar{a}_k e^{i\mathbf{k}\cdot\mathbf{x}}+u^*_k(\tau)
\bar{a}^*_k e^{-i\mathbf{k}\cdot\mathbf{x}}].
\end{multlined}
\end{equation}
In a way similar to that used in section \ref{general} and section \ref{quant}, we can solve for the mode amplitudes $u_k(\tau)$ by substituting \eqref{eq:195} in the effective equation of motion for $\mathcal{R}$ derived from the effective action \eqref{eq:191}. We assume for the time being that the functions $\mathrm{Q}$, $\mathrm{H}$ and $v_\mathrm{c}$ are constant in time (termed as the \tit{de sitter expansion}). The expression for the mode function is then written as
\begin{equation}
\tag{196}\label{eq:196}
\begin{multlined}
u_k(\tau)=\frac{1}{2\mathrm{Q}^{\frac{1}{2}}\,v_\mathrm{c}}\frac{\mathrm{H}}{({v_\mathrm{c}})^{1/2}}\frac{1}{k^{3/2}}(1+ikv_\mathrm{c}\tau)e^{-ikv_\mathrm{c}\tau}.
\end{multlined}
\end{equation}
Moreover, recall from \eqref{eq:58} that the power spectrum is given by
\begin{equation}
\tag{197}\label{eq:197}
\begin{multlined}
\mathrm{P}_\mathrm{R} (|\vec k|) \equiv \mathrm{P}_k (k)= \frac{k^3}{2\uppi^2}\frac{|u_k|^2}{a(t)^2}=\frac{k^3}{2\uppi^2}\frac{|u_k(\tau)u^*_k(\tau)|}{a(t)^2}=\frac{\mathrm{H}^2}{8\uppi^2\mathrm{Q}\,v^3_\mathrm{c}}(1+k^2 v^2_\mathrm{c}\tau^2).
\end{multlined}
\end{equation}
Now, for modes much larger than the Hubble horizon ($k\propto\uplambda^{-1}\gg 1/v_\mathrm{c}\tau$), the power spectrum reduces to:
\begin{equation}
\tag{198}\label{eq:198}
\begin{multlined}
\mathrm{P}_\mathrm{R} (|\vec k|)=\frac{\mathrm{H}^2}{8\uppi^2\mathrm{Q}\,v^3_\mathrm{c}}(1+k^2 v^2_\mathrm{c}\tau^2)\equiv\frac{\mathrm{H}^2}{8\uppi^2\mathrm{Q}\,v^3_\mathrm{c}}\;\;\;\;\text{for,}\;\;\;\uplambda\gg v_\mathrm{c}\tau.
\end{multlined}
\end{equation}
Take note that the choice of the comoving gauge \eqref{eq:182} has enabled us to treat terms $\mathrm{P}$, $\mathrm{G}_3$, $\mathrm{G}_4$ and $\mathrm{G}_5$, and their higher-order derivatives, as mere coefficients, thereby making our calculation of the power spectrum extremely simple. Furthermore, we can introduce a new parameter $\epsilon_\mathrm{c}$ such that,
\begin{equation}
\tag{199}\label{eq:199}
\begin{multlined}
\mathrm{P}_\mathrm{R} (|\vec k|)=\frac{\mathrm{H}^2}{8\uppi^2\mathrm{M}^2_\mathsf{pl}\,\mathrm{F}\,\epsilon_\mathrm{c} v_\mathrm{c}}\;\;\;\;\text{for,}\;\;\;\epsilon_\mathrm{c}=\frac{\mathrm{Q}\,v^2_\mathrm{c}}{\mathrm{M}^2_\mathsf{pl}\mathrm{F}}.
\end{multlined}
\end{equation}
In terms of the slow-roll parameters introduced in section \ref{slowroll}, $\epsilon_\mathrm{c}$ can further be written using \eqref{eq:179} as
\begin{equation}
\tag{200}\label{eq:200}
\begin{multlined}
\epsilon_\mathrm{c}=\epsilon+\aleph_{\mathrm{G_3}^{f}}+
\aleph_{\mathrm{G_4}^{\phi}}+8\aleph_{\mathrm{G_4}^{f\!f}}+
\aleph_{\mathrm{G_5}^{f}}+2\aleph_{\mathrm{G_5}^{f\!f}}+\mathcal{O}
(\epsilon^2)=\\ \aleph_{\mathrm{P}^{f}}+4\aleph_{\mathrm{G_3}^{f}}-
2\aleph_{\mathrm{G_3}^{\phi}}+6\aleph_{\mathrm{G_4}^{f}}+20
\aleph_{\mathrm{G_4}^{f\!f}}+4\aleph_{\mathrm{G_5}^{f}}+4
\aleph_{\mathrm{G_5}^{f\!f}}-6\aleph_{\mathrm{G_5}^{\phi}}+
\mathcal{O}(\epsilon^2).
\end{multlined}
\end{equation}
In order to calculate the tilt in the spectral index ($n_\mathcal{R}$), we adopt the assumption that $\mathrm{Q}$, $\mathrm{H}$ and $v_\mathrm{c}$ are slowly changing in time (termed as the \tit{quasi de sitter expansion}). Ideally, one would solve again for the mode amplitudes, but in our case, one can simply draw an analogy from our discussion in sections \ref{general}, \ref{quant}, \ref{cosmo}, and draw the terms that contribute to the spectral index via their slow variation. These terms in our case are contained within the leading coefficient in \eqref{eq:199},
\begin{equation*}
\begin{multlined}
\frac{\mathrm{H}^2}{8\uppi^2\mathrm{M}^2_\mathsf{pl}\,\mathrm{F}\,\epsilon_\mathrm{c} v_\mathrm{c}}.
\end{multlined}
\end{equation*}
These terms are $\mathrm{H}^2$, $\mathrm{Q}^{-1}$, $\epsilon^{-1}_\mathrm{c}$ and $v^{-1}_\mathrm{c}$, and they contribute via their slow variation as follows:
\begin{equation}
\tag{201}\label{eq:201}
\begin{multlined}
\mathrm{H}^2\rightarrow\frac{1}{\mathrm{H}}\frac{\partial_\tau(\mathrm{H}^2)}{\mathrm{H}^2}=-2\frac{\dt{\mathrm{H}}}{\mathrm{H}^2}=-2\epsilon,
\end{multlined}
\end{equation}
\begin{equation}
\tag{202}\label{eq:202}
\begin{multlined}
\mathrm{Q}^{-1}\rightarrow\frac{1}{\mathrm{H}}\frac{\partial_\tau(\mathrm{Q}^{-1})}{\mathrm{Q}^{-1}}=-\frac{\dt{\mathrm{Q}}}{\mathrm{H\,Q}}=-\aleph_{\epsilon_\mathrm{c}},
\end{multlined}
\end{equation}
\begin{equation}
\tag{203}\label{eq:203}
\begin{multlined}
\mathrm{F}^{-1}\rightarrow\frac{1}{\mathrm{H}}\frac{\partial_\tau(\mathrm{F}^{-1})}{\mathrm{F}^{-1}}=-\frac{\dt{\mathrm{F}}}{\mathrm{H\,F}}=-\aleph_{\mathrm{F}},
\end{multlined}
\end{equation}

\begin{equation}
\tag{204}\label{eq:204}
\begin{multlined}
v^{-1}_\mathrm{c}\rightarrow\frac{1}{\mathrm{H}}\frac{\partial_\tau(v^{-1}_\mathrm{c})}{v^{-1}_\mathrm{c}}=-\frac{\dt{v}_\mathrm{c}}{\mathrm{H}\,v_\mathrm{c}}=-\aleph_{v_\mathrm{c}},
\end{multlined}
\end{equation}
such that the spectral index $n_\mathcal{R}$ is given by
\begin{equation}
\tag{205}\label{eq:205}
\begin{multlined}
n_{\mathcal{R}} - 1 = \frac{\mathrm{d\,ln}[\mathrm{P}_\mathrm{R} (|\vec k|)]}{\mathrm{d\,(ln\,}k)}\xrightarrow{v_\mathrm{c}k=a(t)\mathrm{H}}-2\epsilon-\aleph_{\epsilon_\mathrm{c}}-
\aleph_{\mathrm{F}}-\aleph_{v_\mathrm{c}},
\end{multlined}
\end{equation}
where, the relation $v_\mathrm{c}k=a(t)\mathrm{H}\neq{-\tau^{-1}}$ holds at the horizon crossing. \\ \\
In order to calculate the power spectrum of the tensor perturbations (in form of gravitational waves), we rewrite the metric including tensor perturbations,
\begin{equation}
\tag{206}\label{eq:206}
\begin{multlined}
g_{\mu\nu}=\left\{ \begin{array}{cccc}
-1 & 0 & 0 & 0\\
0 & a(t)^2(1+\vvmathbb{h}^{\mathrm{tt}}_{11}) & a(t)^2\vvmathbb{h}^{\mathrm{tt}}_{12} & a(t)^2\vvmathbb{h}^{\mathrm{tt}}_{13}\vspace{0.05in}\\
0 & a(t)^2\vvmathbb{h}^{\mathrm{tt}}_{21} & a(t)^2(1+\vvmathbb{h}^{\mathrm{tt}}_{22}) & a(t)^2\vvmathbb{h}^{\mathrm{tt}}_{23}\vspace{0.05in}\\
0 & a(t)^2\vvmathbb{h}^{\mathrm{tt}}_{31} & a(t)^2\vvmathbb{h}^{\mathrm{tt}}_{32} & a(t)^2(1+\vvmathbb{h}^{\mathrm{tt}}_{33})\end{array} \right\},
\end{multlined}
\end{equation}
where, $\vvmathbb{h}^{\mathrm{tt}}_{\mu\nu}=\vvmathbb{h}_{+}e^+_{\mu\nu} + \vvmathbb{h}_{\times}e^\times_{\mu\nu}$, $\vvmathbb{h}_{+}$, $\vvmathbb{h}_{\times}$ are the two polarisations, and $e^+_{\mu\nu}$, $e^\times_{\mu\nu}$ are the polarization `unit' tensors. The action in this case reduces to
\begin{equation}
\tag{207}\label{eq:207}
\begin{multlined}
\mathrm{I}_t=\int\mathrm{d}t\,\mathrm{d}^{3}\!x\,a(t)^3 \mathrm{Q}_\mathrm{t}\Bigg[\dt{\vvmathbb{h}}^2_{(+,\times)}-\frac{v^2_\mathrm{t}}{a(t)^2}\{\partial \vvmathbb{h}_{(+,\times)}\}^2\Bigg],
\end{multlined}
\end{equation}
such that,
\begin{equation}
\tag{208}\label{eq:208}
\begin{multlined}
\mathrm{Q}_\mathrm{t}\equiv\frac{{{\Gamma}_1}}{4},\;\;\;\;\;v^2_\mathrm{t}=\frac{{{\Gamma}_4}}{{{\Gamma}_1}}.
\end{multlined}
\end{equation}
The power spectrum and the spectral index are then calculated for this form of tensor perturbations, and are given by:
\begin{equation}
\tag{209}\label{eq:209}
\begin{multlined}
\mathrm{P}_\mathrm{T} (|\vec k|)=\frac{\mathrm{H}^2}{2\uppi^2\mathrm{Q_t}\,v^3_\mathrm{t}}\sim\frac{2\mathrm{H}^2}{\uppi^2\mathrm{M^2_{pl}F}},
\end{multlined}
\end{equation}
\begin{equation}
\tag{210}\label{eq:210}
\begin{multlined}
n_{\mathrm{T}}=-2\epsilon-\aleph_{\mathrm{F}}.
\end{multlined}
\end{equation}
Moreover, the standard tensor-to-scalar ratio $r$ can be calculated from \eqref{eq:199}, \eqref{eq:200} and \eqref{eq:209} as,
\begin{equation}
\tag{211}\label{eq:211}
\begin{multlined}
r=\frac{\mathrm{P}_{\mathrm{T}}}{\mathrm{P}_{\mathrm{R}}}\sim 16\epsilon_\mathrm{c}v_\mathrm{c}
\end{multlined}
\end{equation}
\subsection{Bi-spectrum}
\label{bispect}
The calculation of bi-spectrum is similar but analytically complex upon the inclusion of third-order perturbations in $\alpha$, $\mathcal{R}$ and $\Theta$. The resulting action is given by
\begin{align*}
\mathrm{I}_2=\int\mathrm{d}t\,\mathrm{d}^3\!x\,a(t)^3\Bigg[
\beth_1\alpha^3+\alpha^2\Bigg\{\beth_2\mathcal{R}+\beth_3\dt
{\mathcal{R}}+\beth_4\frac{\partial^2\mathcal{R}}{a(t)^2}+\beth_5\frac{\partial^2\Theta}{a(t)^2}\Bigg\}
+\alpha\Bigg\{\beth_6(\partial_\mu\mathcal{R})\frac{\partial
_\mu\Theta}{a(t)^2}+\beth_7\dt{\mathcal{R}}\mathcal{R}+\\\beth_8
\dt{\mathcal{R}}\frac{\partial^2\mathcal{R}}{a(t)^2}+\beth_9\frac{\{(\partial_\mu\partial_\nu\Theta)(\partial_\mu\partial_\nu\Theta)-(\partial^2\Theta)^2\}}{a(t)^4}\Bigg\}+\beth_{10}\frac{\{(\partial_\mu\partial_\nu
\Theta)(\partial_\mu\partial_\nu\mathcal{R})-(\partial^2\Theta)(\partial^2\mathcal{R})\}}{a(t)^4}+\beth_{11}\mathcal{R}
\frac{\partial^2\Theta}{a(t)^2}\\+\beth_{12}\dt{\mathcal{R}}\frac{\partial^2\Theta}{a(t)^2}+\beth_{13}\mathcal{R}\frac{\partial^2\mathcal{R}}{a(t)^2}+\beth_{14}\frac{(\partial\mathcal{R})^2}{a(t)^2}+\Big\{\beth_{15}+\beth_{16}
\dt{\mathcal{R}}\Big\}\dt{\mathcal{R}}^2+\beth_{17}\mathcal{R}\frac{(\partial\mathcal{R})^2}{a(t)^2}+\beth_{18}\dt{\mathcal{R}}^2\mathcal{R}+
\beth_{19}\dt{\mathcal{R}}\frac{(\partial_\mu\mathcal{R})(\partial_\Theta)}{a(t)^2}\\+\Bigg\{\beth_{20}\dt{\mathcal{R}}
+\beth_{21}
\mathcal{R}\Bigg\}\dt{\mathcal{R}}\frac{\partial^2\Theta}{a(t)^2}+\Big\{\beth_{22}\dt{\mathcal{R}}+\beth_{23}{\mathcal{R}}\Big\}\frac{\{(\partial_\mu\partial_\nu\Theta)(\partial_\mu\partial_\nu\Theta)-(\partial^2\Theta)^2\}}{a(t)^4}+\beth_{24}(\partial_\mu\mathcal{R})(\partial_\mu{\Theta})\frac{\partial^2\Theta}{a(t)^4}\Bigg].
\tag{212}\label{eq:212}
\end{align*}
where, the $\beth_i$ terms are abbreviated as\footnote{The terms in red are additional higher-order coefficients induced by the third-order perturbation terms, when compared to the second-order action in \eqref{eq:183}.}
\begin{equation}
\tag{213}\label{eq:213}
\begin{multlined}
\beth_1=3\mathrm{M}^2_{\mathsf{pl}}\mathrm{H}^2\mathrm{F}-
\Bigg[f(\phi)(\partial^{f}\mathrm{P})+4f(\phi)^2(\partial^{(f\!,\,f)}\mathrm{P})+{{\color{red}\frac{4}{3}f(\phi)^3(\partial^{(f\!,\,f\!,\,f)}\mathrm{P})}}\Bigg]+2\dt{\phi}\mathrm{H}\Bigg[10f(\phi)(\partial^{f}\mathrm{G_3})+\\11f(\phi)^2(\partial^{(f\!,\,f)}\mathrm{G_3})+{{\color{red}2f(\phi)^3(\partial^{(f\!,\,f\!,\,f)}\mathrm{G_3})}}\Bigg]+2f(\phi)\Bigg[(\partial^{\phi}\mathrm{G_3})+\frac{7}{3}f(\phi)(\partial^{(f\!,\,\phi)}\mathrm{G_3})+{{\color{red}\frac{2}{3}f(\phi)^2(\partial^{(f\!,\,\phi ,f)}\mathrm{G_3})}}\Bigg]\\-\hspace{0.01in}2\mathrm{H}^2\Big[33f(\phi)(\partial^{f}\mathrm{G_4})+126f(\phi)^2(\partial^{(f\!,\,f)}\mathrm{G_4})+68f(\phi)^3(\partial^{(f\!,\,f\!,\,f)}\mathrm{G_4})+{{\color{red}8f(\phi)^4(\partial^{(f\!,\,f\!,\,f\!,\,f)}\mathrm{G_4})}}\Big]+\\2\dt{\phi}\mathrm{H}\Big[3(\partial^{\phi}\mathrm{G_4})+27f(\phi)(\partial^{(f\!,\,\phi)}\mathrm{G_4})+24f(\phi)^2(\partial^{(f\!,\,f\!,\,\phi)}\mathrm{G_4})+{{\color{red}4f(\phi)^3(\partial^{(f\!,\,f\!,\,f\!,\,\phi)}\mathrm{G_4})}}\Big]-\\ \mathrm{H}^3\dt{\phi}\Bigg[70f(\phi)(\partial^{f}\mathrm{G_5})+98f(\phi)^2(\partial^{(f\!,\,f)}\mathrm{G_5})+32f(\phi)^3(\partial^{(f\!,\,f\!,\,f)}\mathrm{G_5})+{{\color{red}\frac{8}{3}f(\phi)^4(\partial^{(f\!,\,f\!,\,f\!,\,f)}\mathrm{G_5})}}\Bigg]\\+2\hspace{0.01in}\mathrm{H}^2f(\phi)
\Big[30(\partial^{\phi}\mathrm{G_5})+75f(\phi)(\partial^{(f\!,\,\phi)}\mathrm{G_5})+36f(\phi)^2(\partial^{(f\!,\,f\!,\,\phi)}\mathrm{G_5})+{{\color{red}4f(\phi)^3(\partial^{(f\!,\,f\!,\,f\!,\,\phi)}\mathrm{G_5})}}\Big]{{\color{red}:\Leftrightarrow{{\Gamma_3}}}},
\end{multlined}
\end{equation}
\begin{equation}
\tag{214}\label{eq:214}
\begin{multlined}
\beth_2={{\Gamma_3}}=-9\mathrm{M}^2_{\mathsf{pl}}\mathrm{H}^2\mathrm{F}+3
[f(\phi)(\partial^{f}\mathrm{P})+2f(\phi)^2(\partial^{(f\!,\,f)}\mathrm{P})]+18\dt{\phi}\mathrm{H}[2f(\phi)(\partial^{f}\mathrm{G_3})+f(\phi)^2(\partial^{(f\!,\,f)}\mathrm{G_3})]-\\6f(\phi)[(\partial^{\phi}\mathrm{G_3})+f(\phi)(\partial^{(f\!,\,\phi)}\mathrm{G_3})]+18\hspace{0.01in}\mathrm{H}^2[7f(\phi)(\partial^{f}\mathrm{G_4})+16f(\phi)^2(\partial^{(f\!,\,f)}\mathrm{G_4})+4f(\phi)^3(\partial^{(f\!,\,f\!,\,f)}\mathrm{G_4})]-\\18\dt{\phi}\mathrm{H}[(\partial^{\phi}\mathrm{G_4})+5f(\phi)(\partial^{(f\!,\,\phi)}\mathrm{G_4})+2f(\phi)^2(\partial^{(f\!,\,f\!,\,\phi)}\mathrm{G_4})]+6\mathrm{H}^3\dt{\phi}[15f(\phi)(\partial^{f}\mathrm{G_5})+13f(\phi)^2(\partial^{(f\!,\,f)}\mathrm{G_5})\\+2f(\phi)^3(\partial^{(f\!,\,f\!,\,f)}\mathrm{G_5})]-18\hspace{0.01in}\mathrm{H}^2f(\phi)[6(\partial^{\phi}\mathrm{G_5})+9f(\phi)(\partial^{(f\!,\,\phi)}\mathrm{G_5})+2f(\phi)^2(\partial^{(f\!,\,f\!,\,\phi)}\mathrm{G_5})],
\end{multlined}
\end{equation}
\begin{equation}
\tag{215}\label{eq:215}
\begin{multlined}
\beth_3=-3\,\beth_5=-3\Bigg\{2\mathrm{M}^2_{\mathsf{pl}}\mathrm{H\,F}
-2\dt{\phi}[f(\phi)(\partial^{f}\mathrm{G_3})+{{\color{red}f(\phi)^2
(\partial^{f\!,\,f}\mathrm{G_3})}}]-4\mathrm{H}
[7f(\phi)(\partial^{f}\mathrm{G_4})+\\16f(\phi)^2(\partial^{(f\!,\,f)}\mathrm{G_4})+{{\color{red}4f(\phi)^3(\partial^{(f\!,\,f\!,\,f)}\mathrm{G_4})}}]+2\dt{\phi}[(\partial^{\phi}\mathrm{G_4})+ 5f(\phi)(\partial^{(f\!,\,\phi)}\mathrm{G_4})+ {{\color{red}2f(\phi)^2(\partial^{(f\!,\,\phi,f)}\mathrm{G_4})}}]\\-2\mathrm{H}^2\dt{\phi}[15f(\phi)
(\partial^{f}\mathrm{G_5})+13f(\phi)^2(\partial^{(f\!,\,f)}\mathrm{G_5})+{{\color{red}2f(\phi)^3(\partial^{(f\!,\,f\!,\,f)}\mathrm{G_5})}}]+4\mathrm{H}f(\phi)[6(\partial^{\phi}\mathrm{G_5})+\\9f(\phi)(\partial^{(f\!,\,\phi)}\mathrm{G_5})+{{\color{red}2f(\phi)^2(\partial^{(f\!,\,f\!,\,\phi)}\mathrm{G_5})}}]\Bigg\}{{\color{red}:\Leftrightarrow{{\Gamma_2}}}},
\end{multlined}
\end{equation}
\begin{equation}
\tag{216}\label{eq:216}
\begin{multlined}
\beth_4=-4[f(\phi)(\partial^{f}\mathrm{G_4})+{{\color{red}2f(\phi)^2(\partial^{f\!,\,f}\mathrm{G_4})}}]-8\mathrm{H}\dt{\phi}
[f(\phi)(\partial^{f}\mathrm{G_5})+{{\color{red}f(\phi)^2(\partial^{f\!,\,f}
\mathrm{G_5})}}]
+\\4f(\phi)[(\partial^{\phi}\mathrm{G_5})+{{\color{red}2f(\phi)(\partial^{\phi}\mathrm{G_5})}}]{{\color{red}:\Leftrightarrow{{\Gamma_1}}}},
\end{multlined}
\end{equation}
\begin{equation}
\tag{217}\label{eq:217}
\begin{multlined}
\beth_6=-\frac{1}{9}\beth_7=\beth_{11}=-{{\Gamma_2}}=-\Bigg[2\mathrm{M}^2_{\mathsf{pl}}\mathrm{H\,F}
-2f(\phi)\dt{\phi}(\partial^{f}\mathrm{G_3})-16\mathrm{H}
[f(\phi)(\partial^{f}\mathrm{G_4})+f(\phi)^2(\partial^{(f\!,\,f)}\mathrm{G_4})]+\\2\dt{\phi}[(\partial^{\phi}\mathrm{G_4})+ 2f(\phi)(\partial^{(f\!,\,\phi)}\mathrm{G_4})]-2\mathrm{H}^2\dt{\phi}[5f(\phi)
(\partial^{f}\mathrm{G_5})+2f(\phi)^2(\partial^{(f\!,\,f)}\mathrm{G_5})]+4\mathrm{H}f(\phi)[3(\partial^{\phi}\mathrm{G_5})+\\2f(\phi)(\partial^{(f\!,\,\phi)}\mathrm{G_5})]\Bigg],
\end{multlined}
\end{equation}
\begin{equation}\\
\tag{218}\label{eq:218}
\begin{multlined}
\beth_8=2\,\beth_{10}=2\,\beth_{16}=-2\,\beth_{20}=-4\,\beth_{22}=
4\dt{\phi}f(\phi)(\partial^{(f\!,\,\phi)}\mathrm{G_5}),
\end{multlined}
\end{equation}
\begin{equation}
\tag{219}\label{eq:219}
\begin{multlined}
\beth_9=\frac{1}{4}\beth_{12}=-\frac{1}{6}\beth_{15}=
-\frac{1}{2}\mathrm{M}^2_{\mathsf{pl}}\mathrm{F}
+4[f(\phi)(\partial^{f}\mathrm{G_4})+{{\color{red}
f(\phi)^2(\partial^{f\!,\,f}\mathrm{G_4})}}]+\mathrm{H}\dt{\phi}
[5f(\phi)(\partial^{f}\mathrm{G_5})+\\{{\color{red}
2f(\phi)^2(\partial^{f\!,\,f}
\mathrm{G_5})}}]
-f(\phi)[3(\partial^{\phi}\mathrm{G_5})+{{\color{red}
2f(\phi)(\partial^{\phi}\mathrm{G_5})}}]{{\color{red}:\Leftrightarrow{{\Gamma_1}}}},
\end{multlined}
\end{equation}
\begin{equation}
\tag{220}\label{eq:220}
\begin{multlined}
\beth_9=2\,\beth_{14}=\frac{2}{9}\beth_{18}=-\beth_{19}=-\beth_{21}=-\frac{4}{3}\beth_{23}=
\beth_{24}=-2{{\Gamma_1}}=-2\Bigg[\mathrm{M}^2_{\mathsf{pl}}\mathrm{F}
-4f(\phi)(\partial^{f}\mathrm{G_4})-\\2\mathrm{H}\dt{\phi}
f(\phi)(\partial^{f}\mathrm{G_5})+2f(\phi)(\partial^{\phi}\mathrm{G_5})\Bigg],
\end{multlined}
\end{equation}
\begin{equation}
\tag{221}\label{eq:221}
\begin{multlined}
\beth_{17}={{\Gamma_4}}=\mathrm{M}^2_{\mathsf{pl}}
\mathrm{F}-2f(\phi)(\partial^{\phi}\mathrm{G_5})-2
f(\phi)(\partial^{f}\mathrm{G_5})\ddt{\phi}.
\end{multlined}
\end{equation}
Let us recall the Hamiltonian and Lagrangian constraints:
\begin{equation*}
\begin{multlined}
\alpha=2\frac{\textbf{{${\Gamma}$}}_1}{\textbf{{${\Gamma}$}}_2}\dt{\mathcal{R}},
\;\;\;\;\;
(\partial^2\Theta)=a(t)^2\frac{2}{3}\frac{\textbf{{${\Gamma}$}}_3}{\textbf{{${\Gamma}$}}_2}\alpha-2\frac{\textbf{{${\Gamma}$}}_1}{\textbf{{${\Gamma}$}}_2}(\partial^2\mathcal{R}),
\end{multlined}
\end{equation*}
which we can use to eliminate $\alpha$ from the action in \eqref{eq:212}, and reduce it to
\begin{equation}
\tag{222}\label{eq:222}
\begin{multlined}
\mathrm{I}_2=\int\mathrm{d}t\,\mathrm{d}^3\!x\,a(t)^3\Bigg[
\Bigg\{\beth_{16}+\beth_{15}\Bigg[2\frac{{\Gamma_1}}{{\Gamma_2}}\Bigg]+\beth_{3}\Bigg[2\frac{{\Gamma_1}}{{\Gamma_2}}\Bigg]^2+\beth_{1}\Bigg[2\frac{{\Gamma_1}}{{\Gamma_2}}\Bigg]^3\Bigg\}\dt{\mathcal{R}}^3
+\Bigg[2\frac{{\Gamma_1}}{{\Gamma_2}}\Bigg]\Bigg\{\beth_{8}+\beth_{4}\Bigg[2\frac{{\Gamma_1}}{{\Gamma_2}}\Bigg]\Bigg\}\dt{\mathcal{R}}^2
\frac{\partial^2\mathcal{R}}{a(t)^2}\\
+\Bigg\{\beth_{20}+\beth_{12}\Bigg[2\frac{{\Gamma_1}}{{\Gamma_2}}\Bigg]+\beth_{5}\Bigg[2\frac{{\Gamma_1}}{{\Gamma_2}}\Bigg]^2\Bigg\}\dt{\mathcal{R}}^2
\frac{\partial^2\Theta}{a(t)^2}+
\Bigg\{\beth_{18}+\beth_{7}\Bigg[2\frac{{\Gamma_1}}{{\Gamma_2}}\Bigg]+\beth_{2}\Bigg[2\frac{{\Gamma_1}}{{\Gamma_2}}\Bigg]^2\Bigg\}\mathcal{R}\dt{\mathcal{R}}^2
+ \Bigg[\Bigg\{\beth_{22}+\beth_{9}\Bigg[2\frac{{\Gamma_1}}{{\Gamma_2}}\Bigg]\\\Bigg\}\dt{\mathcal{R}}+
\beth_{23}\mathcal{R}\Bigg]\Bigg\{\frac{\{(\partial_\mu\partial_\nu\Theta)(\partial_\mu\partial_\nu\Theta)-(\partial^2\Theta)^2\}}{a(t)^4}\Bigg\}+\Bigg\{\beth_{10}\Bigg[2\frac{{\Gamma_1}}{{\Gamma_2}}\Bigg]\Bigg\} \Bigg\{\frac{\{(\partial_\mu\partial_\nu\Theta)(\partial_\mu\partial_\nu\mathcal{R})-(\partial^2\Theta)(\partial^2\mathcal{R})\}}{a(t)^4}\dt{\mathcal{R}}\Bigg\}\\+\Bigg\{\beth_{17}+\beth_{13}
\Bigg[\partial_\tau({{\Gamma_1}}/ {{\Gamma_2}})\Bigg]+\Bigg[\frac{{\Gamma_1}}{{\Gamma_2}}\Bigg]\Bigg[\dt{\beth}_{13}+\mathrm{H}
\beth_{13}\Bigg]\Bigg\}\mathcal{R}\frac{(\partial
\mathcal{R})^2}{a(t)^2}+ \beth_{24}\Bigg\{\frac{(\partial_\mu\mathcal{R})(\partial_\nu\Theta)(\partial^2\Theta)}{a(t)^4}\Bigg\}
\Bigg].\;\;\;\;\;\;\;\;\;
\end{multlined}
\end{equation}
In order to eliminate $\Theta$, we first introduce an auxiliary field $\upchi$ as an independent degree of freedom \citep{r11} such that
\begin{equation}
\tag{223}\label{eq:223}
\begin{multlined}
\Theta=-2\frac{{\Gamma_1}}{{\Gamma_2}}\mathcal{R}+a(t)^2\frac{\upchi}{{{\Gamma_1}}}.
\end{multlined}
\end{equation}
According to this re-definition, \eqref{eq:190} reduces to the following:
\begin{equation}
\tag{224}\label{eq:224}
\begin{multlined}
\partial^2\upchi=\frac{1}{3}\Bigg\{\beth_{18}+\beth_{7}\Bigg[2\frac{{\Gamma_1}}{{\Gamma_2}}\Bigg]+\beth_{2}\Bigg[2\frac{{\Gamma_1}}{{\Gamma_2}}\Bigg]^2\Bigg\}\dt{\mathcal{R}}.
\end{multlined}
\end{equation}\\
\begin{flushleft}\myrule[line width = 0.5mm]{fast cap reversed}{fast cap reversed}\end{flushleft}
\tit{\textbf{Sidenote}}:\\
In this note, we illustrate the calculation to \eqref{eq:223} beginning at \eqref{eq:190}. To begin with,
\begin{equation}
\tag{225}\label{eq:225}
\begin{multlined}
(\partial^2\Theta)=a(t)^2\frac{2}{3}\frac{\textbf{{${\Gamma}$}}_3}{\textbf{{${\Gamma}$}}_2}\alpha-2\frac{\textbf{{${\Gamma}$}}_1}{\textbf{{${\Gamma}$}}_2}(\partial^2\mathcal{R}).
\end{multlined}
\end{equation}
Now, substituting the expression for $\Theta$ from \eqref{eq:220} and recalling that $\displaystyle{\alpha=\Bigg[2\frac{{\Gamma_1}}{{\Gamma_2}}\Bigg]\dt{\mathcal{R}}}$, we get
\begin{equation}
\tag{226}\label{eq:226}
\begin{multlined}
(\partial^2\Theta)= \partial^2\Bigg[-2\frac{{\Gamma_1}}{{\Gamma_2}}\mathcal{R}+a(t)^2\frac{\upchi}{{{\Gamma_1}}}\Bigg] =a(t)^2\frac{2}{3}\frac{\textbf{{${\Gamma}$}}_3}{\textbf{{${\Gamma}$}}_2}\Bigg[2\frac{{\Gamma_1}}{{\Gamma_2}}\Bigg]\dt{\mathcal{R}}-2\frac{\textbf{{${\Gamma}$}}_1}{\textbf{{${\Gamma}$}}_2}(\partial^2\mathcal{R}),
\end{multlined}
\end{equation}
\begin{equation*}
\begin{multlined}
\partial^2\Bigg[-2\frac{{\Gamma_1}}{{\Gamma_2}}\mathcal{R}+a(t)^2\frac{\upchi}{{{\Gamma_1}}}\Bigg]=a(t)^2\frac{4}{3}\frac{\textbf{{${\Gamma}$}}_3{{\Gamma_1}}}{\textbf{{${\Gamma}$}}^2_2}\dt{\mathcal{R}}-2\frac{\textbf{{${\Gamma}$}}_1}{\textbf{{${\Gamma}$}}_2}(\partial^2\mathcal{R})+3a(t)^2\dt{\mathcal{R}},
\end{multlined}
\end{equation*}
\begin{equation*}
\begin{multlined}
\partial^2\Bigg[-2\frac{{\Gamma_1}}{{\Gamma_2}}\mathcal{R}+a(t)^2\frac{\upchi}{{{\Gamma_1}}}\Bigg]=a(t)^2\frac{4}{3}\frac{\textbf{{${\Gamma}$}}_3{{\Gamma_1}}}{\textbf{{${\Gamma}$}}^2_2}\dt{\mathcal{R}}-2\frac{\textbf{{${\Gamma}$}}_1}{\textbf{{${\Gamma}$}}_2}(\partial^2\mathcal{R})+3a(t)^2\dt{\mathcal{R}},
\end{multlined}
\end{equation*}
\begin{equation*}
\begin{multlined}
-2\frac{{\Gamma_1}}{{\Gamma_2}}(\partial^2\mathcal{R})+a(t)^2\frac{(\partial^2\upchi)}{{{\Gamma_1}}}=a(t)^2\frac{4}{3}\frac{\textbf{{${\Gamma}$}}_3{{\Gamma_1}}}{\textbf{{${\Gamma}$}}^2_2}\dt{\mathcal{R}}-2\frac{\textbf{{${\Gamma}$}}_1}{\textbf{{${\Gamma}$}}_2}(\partial^2\mathcal{R})+3a(t)^2\dt{\mathcal{R}},
\end{multlined}
\end{equation*}
\begin{equation*}
\begin{multlined}
{\partial^2\upchi}=\frac{4}{3}\frac{\textbf{{${\Gamma}$}}_3{
{\Gamma^2_1}}}{\textbf{{${\Gamma}$}}^2_2}\dt{\mathcal{R}}+3\dt{\mathcal{R}}{
{\Gamma^2_1}}=\frac{1}{3}\Bigg\{-9{{\Gamma_2}}+9{{\Gamma_2}}\Bigg[2\frac{{\Gamma_1}}{{\Gamma_2}}\Bigg]+{{\Gamma_3}}\Bigg[2\frac{{\Gamma_1}}{{\Gamma_2}}\Bigg]^2\Bigg\}\dt{\mathcal{R}},
\end{multlined}
\end{equation*}
\begin{equation}
\tag{227}\label{eq:227}
\begin{multlined}
{\partial^2\upchi}=\frac{1}{3}\Bigg\{-9{{\Gamma_2}}+9{{\Gamma_2}}\Bigg[2\frac{{\Gamma_1}}{{\Gamma_2}}\Bigg]+{{\Gamma_3}}\Bigg[2\frac{{\Gamma_1}}{{\Gamma_2}}\Bigg]^2\Bigg\}\dt{\mathcal{R}}=\\ \frac{1}{3}\Bigg\{\beth_{18}+\beth_{7}\Bigg[2\frac{{\Gamma_1}}{{\Gamma_2}}\Bigg]+\beth_{2}\Bigg[2\frac{{\Gamma_1}}{{\Gamma_2}}\Bigg]^2\Bigg\}\dt{\mathcal{R}}.
\end{multlined}
\end{equation}
\begin{flushright}\myrule[line width = 0.5mm]{fast cap reversed}{fast cap reversed}\end{flushright}$\;$\\
We can now plug \eqref{eq:223} in \eqref{eq:222} and arrive at the following expression for the action,
\begin{align*}
\mathrm{I}_2=\int\mathrm{d}t\,\mathrm{d}^3\!x\,a(t)^3\Bigg[
a(t)^{-1}\Bigg(\Bigg[\Bigg\{\beth_{16}+\beth_{15}\Bigg[2\frac{{\Gamma_1}}{{\Gamma_2}}\Bigg]+\beth_{3}\Bigg[2\frac{{\Gamma_1}}{{\Gamma_2}}\Bigg]^2+\beth_{1}\Bigg[2\frac{{\Gamma_1}}{{\Gamma_2}}\Bigg]^3\Bigg\}+\Bigg\{\beth_{20}+\beth_{12}\Bigg[2\frac{{\Gamma_1}}{{\Gamma_2}}\Bigg]+\\ \beth_{5}\Bigg[2\frac{{\Gamma_1}}{{\Gamma_2}}\Bigg]^2\Bigg\}\frac{1}{3{{\Gamma_1}}}\Bigg\{\beth_{18}+\beth_{7}\Bigg[2\frac{{\Gamma_1}}{{\Gamma_2}}\Bigg]+\beth_{2}\Bigg[2\frac{{\Gamma_1}}{{\Gamma_2}}\Bigg]^2\Bigg\}-\Bigg\{\beth_{22}+\beth_{9}\Bigg[2\frac{{\Gamma_1}}{{\Gamma_2}}\Bigg]\Bigg\}
\frac{1}{9{{\Gamma^2_1}}}\Bigg\{\beth_{18}+\beth_{7}\Bigg[2\frac{{\Gamma_1}}{{\Gamma_2}}\Bigg]+\\ \beth_{2}\Bigg[2\frac{{\Gamma_1}}{{\Gamma_2}}\Bigg]^2\Bigg\}^2\Bigg]
\dt{\mathcal{R}}^3 \Bigg[\Bigg\{\beth_{18}+\beth_{7}\Bigg[2\frac{{\Gamma_1}}{{\Gamma_2}}\Bigg]+\beth_{2}\Bigg[2\frac{{\Gamma_1}}{{\Gamma_2}}\Bigg]^2\Bigg\}-\beth_{23}
\frac{1}{9{{\Gamma^2_1}}}\Bigg\{\beth_{18}+\beth_{7}\Bigg[2\frac{{\Gamma_1}}{{\Gamma_2}}\Bigg]+ \beth_{2}\Bigg[2\frac{{\Gamma_1}}{{\Gamma_2}}\Bigg]^2\Bigg\}^2\Bigg]\mathcal{R}
\dt{\mathcal{R}}^2 \\ 
+\Bigg[\beth_{24}\frac{1}{3{{\Gamma^2_1}}}\Bigg\{\beth_{18}+\beth_{7}\Bigg[2\frac{{\Gamma_1}}{{\Gamma_2}}\Bigg]+ \beth_{2}\Bigg[2\frac{{\Gamma_1}}{{\Gamma_2}}\Bigg]^2\Bigg\}\Bigg]\dt{\mathcal{R}}
(\partial_\mu\mathcal{R})(\partial_\mu\upchi)+\frac{1}{{{\Gamma^2_1}}}\Bigg[\Bigg\{\beth_{22}+\beth_{9}\Bigg[2\frac{{\Gamma_1}}{{\Gamma_2}}\Bigg]\Bigg\}\dt{\mathcal{R}}+\\
\beth_{23}\mathcal{R}\Bigg](\partial_\mu\partial_\nu\upchi)^2
\Bigg)+a(t)\Bigg(\Bigg[2\frac{{\Gamma_1}}{{\Gamma_2}}\Bigg\{\beth_{8}+\beth_{4}\Bigg[2\frac{{\Gamma_1}}{{\Gamma_2}}\Bigg]\Bigg\}-2\frac{{\Gamma_1}}{{\Gamma_2}}\Bigg\{\beth_{20}+\beth_{12}\Bigg[2\frac{{\Gamma_1}}{{\Gamma_2}}\Bigg]+\beth_{5}\Bigg[2\frac{{\Gamma_1}}{{\Gamma_2}}\Bigg]^2\Bigg\}+\\ \frac{4}{3{\Gamma_2}}\Bigg\{\beth_{18}+\beth_{7}\Bigg[2\frac{{\Gamma_1}}{{\Gamma_2}}\Bigg]+\beth_{2}\Bigg[2\frac{{\Gamma_1}}{{\Gamma_2}}\Bigg]^2\Bigg\}
\Bigg\{\beth_{22}+\beth_{9}\Bigg[2\frac{{\Gamma_1}}{{\Gamma_2}}\Bigg]\Bigg\}-\frac{2}{{\Gamma_2}}\Bigg\{\beth_{18}+\beth_{7}\Bigg[2\frac{{\Gamma_1}}{{\Gamma_2}}\Bigg]+\\ \beth_{2}\Bigg[2\frac{{\Gamma_1}}{{\Gamma_2}}\Bigg]^2\Bigg\}\beth_{10}\Bigg]
\dt{\mathcal{R}}^2(\partial^2\mathcal{R})+\Bigg[\frac{4}{3{\Gamma_2}}\Bigg\{\beth_{18}+\beth_{7}\Bigg[2\frac{{\Gamma_1}}{{\Gamma_2}}\Bigg]+\beth_{2}\Bigg[2\frac{{\Gamma_1}}{{\Gamma_2}}\Bigg]^2\Bigg\}\beth_{23}\Bigg]
\dt{\mathcal{R}}\mathcal{R}(\partial^2\mathcal{R})
+\Bigg[\Bigg\{\beth_{17}+\\ \beth_{13}
\Bigg[\partial_\tau({{\Gamma_1}}/ {{\Gamma_2}})\Bigg]+\Bigg[\frac{{\Gamma_1}}{{\Gamma_2}}\Bigg]\Bigg[\dt{\beth}_{13}+\mathrm{H}
\beth_{13}\Bigg]\Bigg\}\Bigg]\mathcal{R}(\partial^2\mathcal{R})
+\Bigg[\frac{2}{3{\Gamma_2}}\Bigg\{\beth_{18}+\beth_{7}\Bigg[2\frac{{\Gamma_1}}{{\Gamma_2}}\Bigg]+\beth_{2}\Bigg[2\frac{{\Gamma_1}}{{\Gamma_2}}\Bigg]^2\Bigg\}\beth_{24}\Bigg]
\dt{\mathcal{R}}(\partial^2\mathcal{R})+\\
\frac{2}{{\Gamma_2}}\Bigg[\beth_{10}-2\Bigg\{\beth_{22}+\beth_{9}\Bigg[2\frac{{\Gamma_1}}{{\Gamma_2}}\Bigg]\Bigg\}
\Bigg]\dt{\mathcal{R}}(\partial_\mu\partial_\nu\upchi)(\partial_\mu\partial_\nu\mathcal{R})-\Bigg[\frac{4}{{\Gamma_2}}\beth_{23}\Bigg]{\mathcal{R}}(\partial_\mu\partial_\nu\upchi)(\partial_\mu\partial_\nu\mathcal{R})
+\\\Bigg[\frac{2}{{\Gamma_2}}\beth_{24}\Bigg](\partial^2\mathcal{R})(\partial_\mu\upchi)(\partial_\mu\mathcal{R})
\Bigg)+a(t)^{-1}\Bigg(\Bigg\{2\frac{{\Gamma_1}}{{\Gamma_2}}\Bigg\}^2\Bigg[\Bigg\{\beth_{22}+\beth_{9}\Bigg[2\frac{{\Gamma_1}}{{\Gamma_2}}\Bigg]\Bigg\}-\beth_{10}
\Bigg]\dt{\mathcal{R}}\{(\partial_\mu\partial_\nu\mathcal{R})
^2-(\partial^2\mathcal{R})^2\}+\\ \Bigg[\beth_{23}\Bigg\{2\frac{{\Gamma_1}}{{\Gamma_2}}\Bigg\}^2\Bigg]{\mathcal{R}}\{(\partial_\mu\partial_\nu\mathcal{R})
^2-(\partial^2\mathcal{R})^2\}+\Bigg[\beth_{24}\Bigg\{2\frac{{\Gamma_1}}{{\Gamma_2}}\Bigg\}^2\Bigg](\partial\mathcal{R})^2(\partial^2\mathcal{R})
\Bigg)\Bigg],
\tag{228}\label{eq:228}
\end{align*}
which we can abbreviate by introducing ${\mathrm{Q}_3}$:
\begin{equation}
\tag{229}\label{eq:229}
\begin{multlined}
{\mathrm{Q}_3}=\Bigg\{\beth_{18}+\beth_{7}\Bigg[2\frac{{\Gamma_1}}{{\Gamma_2}}\Bigg]+\beth_{2}\Bigg[2\frac{{\Gamma_1}}{{\Gamma_2}}\Bigg]^2\Bigg\}=3{\mathrm{Q}},
\end{multlined}
\end{equation}
such that
\begin{align*}
\mathrm{I}_2=\int\mathrm{d}t\,\mathrm{d}^3\!x\,a(t)^3\Bigg[
a(t)^{-1}\Bigg(\Bigg[\Bigg\{\beth_{16}+\beth_{15}\Bigg[2\frac{{\Gamma_1}}{{\Gamma_2}}\Bigg]+\beth_{3}\Bigg[2\frac{{\Gamma_1}}{{\Gamma_2}}\Bigg]^2+\beth_{1}\Bigg[2\frac{{\Gamma_1}}{{\Gamma_2}}\Bigg]^3\Bigg\}+\Bigg\{\beth_{20}+\beth_{12}\Bigg[2\frac{{\Gamma_1}}{{\Gamma_2}}\Bigg]+\\ \beth_{5}\Bigg[2\frac{{\Gamma_1}}{{\Gamma_2}}\Bigg]^2\Bigg\}\frac{1}{3{{\Gamma_1}}}{\mathrm{Q}_3}-\Bigg\{\beth_{22}+\beth_{9}\Bigg[2\frac{{\Gamma_1}}{{\Gamma_2}}\Bigg]\Bigg\}
\frac{1}{9{{\Gamma^2_1}}}{\mathrm{Q}^2_3}\Bigg]
\dt{\mathcal{R}}^3 \Bigg[{\mathrm{Q}_3}-\beth_{23}
\frac{1}{9{{\Gamma^2_1}}}{\mathrm{Q}^2_3}\Bigg]\mathcal{R}
\dt{\mathcal{R}}^2 
+\Bigg[\beth_{24}\frac{1}{3{{\Gamma^2_1}}}{\mathrm{Q}_3}\Bigg]\\ \dt{\mathcal{R}}
(\partial_\mu\mathcal{R})(\partial_\mu\upchi)+\frac{1}{{{\Gamma^2_1}}}\Bigg[\Bigg\{\beth_{22}+\beth_{9}\Bigg[2\frac{{\Gamma_1}}{{\Gamma_2}}\Bigg]\Bigg\}\dt{\mathcal{R}}+
\beth_{23}\mathcal{R}\Bigg](\partial_\mu\partial_\nu\upchi)^2
\Bigg)+a(t)\Bigg(\Bigg[2\frac{{\Gamma_1}}{{\Gamma_2}}\Bigg\{\beth_{8}+\beth_{4}\Bigg[2\frac{{\Gamma_1}}{{\Gamma_2}}\Bigg]\Bigg\}\\-2\frac{{\Gamma_1}}{{\Gamma_2}}\Bigg\{\beth_{20}+ \beth_{12}\Bigg[2\frac{{\Gamma_1}}{{\Gamma_2}}\Bigg]+\beth_{5}\Bigg[2\frac{{\Gamma_1}}{{\Gamma_2}}\Bigg]^2\Bigg\}+\frac{4}{3{\Gamma_2}}{\mathrm{Q}_3}
\Bigg\{\beth_{22}+\beth_{9}\Bigg[2\frac{{\Gamma_1}}{{\Gamma_2}}\Bigg]\Bigg\}-\frac{2}{{\Gamma_2}}{\mathrm{Q}_3}\beth_{10}\Bigg]
\dt{\mathcal{R}}^2(\partial^2\mathcal{R})+\\ \Bigg[\frac{4}{3{\Gamma_2}}{\mathrm{Q}_3}\beth_{23}\Bigg]
\dt{\mathcal{R}}\mathcal{R}(\partial^2\mathcal{R})
+\Bigg[\Bigg\{\beth_{17}+ \beth_{13}
\Bigg[\partial_\tau({{\Gamma_1}}/ {{\Gamma_2}})\Bigg]+\Bigg[\frac{{\Gamma_1}}{{\Gamma_2}}\Bigg]\Bigg[\dt{\beth}_{13}+\mathrm{H}
\beth_{13}\Bigg]\Bigg\}\Bigg]\mathcal{R}(\partial^2\mathcal{R})
+\\ \Bigg[\frac{2}{3{\Gamma_2}}{\mathrm{Q}_3}\beth_{24}\Bigg]
\dt{\mathcal{R}}(\partial^2\mathcal{R})+
\frac{2}{{\Gamma_2}}\Bigg[\beth_{10}-2\Bigg\{\beth_{22}+\beth_{9}\Bigg[2\frac{{\Gamma_1}}{{\Gamma_2}}\Bigg]\Bigg\}
\Bigg]\dt{\mathcal{R}}(\partial_\mu\partial_\nu\upchi)(\partial_\mu\partial_\nu\mathcal{R})-\\\Bigg[\frac{4}{{\Gamma_2}}\beth_{23}\Bigg]{\mathcal{R}}(\partial_\mu\partial_\nu\upchi)(\partial_\mu\partial_\nu\mathcal{R})
+\Bigg[\frac{2}{{\Gamma_2}}\beth_{24}\Bigg](\partial^2\mathcal{R})(\partial_\mu\upchi)(\partial_\mu\mathcal{R})
\Bigg)+a(t)^{-1}\Bigg(\Bigg\{2\frac{{\Gamma_1}}{{\Gamma_2}}\Bigg\}^2\Bigg[\Bigg\{\beth_{22}+\beth_{9}\Bigg[2\frac{{\Gamma_1}}{{\Gamma_2}}\Bigg]\Bigg\}-\beth_{10}
\Bigg]\\\dt{\mathcal{R}}\{(\partial_\mu\partial_\nu\mathcal{R})
^2-(\partial^2\mathcal{R})^2\}+ \Bigg[\beth_{23}\Bigg\{2\frac{{\Gamma_1}}{{\Gamma_2}}\Bigg\}^2\Bigg]{\mathcal{R}}\{(\partial_\mu\partial_\nu\mathcal{R})
^2-(\partial^2\mathcal{R})^2\}+\Bigg[\beth_{24}\Bigg\{2\frac{{\Gamma_1}}{{\Gamma_2}}\Bigg\}^2\Bigg](\partial\mathcal{R})^2(\partial^2\mathcal{R})
\Bigg)\Bigg].
\tag{230}\label{eq:230}
\end{align*}
Now, in a way similar to the case of power spectrum, we can use integration by parts \citep{r11} in order to reduce the action to the following expression:
\begin{align*}
\mathrm{I}_2=\int\mathrm{d}t\,\mathrm{d}^3\!x\,\Bigg[
a(t)^3\text{{\boldmath $\vvmathbb{R}_1$}}\mathcal{R}\dt{\mathcal{R}}^2
+a(t)\text{{\boldmath $\vvmathbb{R}_2$}}\mathcal{R}(\partial\mathcal{R})^2+a(t)^3\text{{\boldmath $\vvmathbb{R}_3$}}\dt{\mathcal{R}}^3+a(t)^3\text{{\boldmath $\vvmathbb{R}_4$}}\dt{\mathcal{R}}(\partial_\mu\upchi)(\partial_\mu\mathcal{R})+\\a(t)^3\text{{\boldmath $\vvmathbb{R}_5$}}(\partial\upchi)^2(\partial^2\mathcal{R})
+a(t)\text{{\boldmath $\vvmathbb{R}_6$}}\dt{\mathcal{R}}^2(\partial^2\mathcal{R})+
a(t)^{-1}\text{{\boldmath $\vvmathbb{R}_7$}}[(\partial\mathcal{R})^2(\partial^2\mathcal{R})-\mathcal{R}
\partial_\mu\partial_\nu\{\partial_\mu\mathcal{R}\}(\partial_\nu\mathcal{R})]+\\a(t)\text{{\boldmath $\vvmathbb{R}_8$}}[(\partial_\mu\mathcal{R})(\partial_\mu\upchi)(\partial^2\mathcal{R})-\mathcal{R}
\partial_\mu\partial_\nu\{\partial_\mu\mathcal{R}\}(\partial_\nu{\upchi})]+\text{{\boldmath $\vvmathbb{R}_9$}}\Bigg],
\tag{231}\label{eq:231}
\end{align*}
where, we have introduced parameters {\boldmath $\vvmathbb{R}_i$} given by
\begin{equation}
\tag{232}\label{eq:232}
\begin{multlined}
\text{{\boldmath $\vvmathbb{R}_1$}}=\Bigg\{3\mathrm{Q}_3-2\frac{
{\Gamma_1}}{v^2_{\mathrm{c}}{\Gamma_2}}(\dt{\mathrm{Q}}_3+3\mathrm{H}\hspace{0.01in}\mathrm{Q_3})-\mathrm{Q_3}{\partial_t\Bigg(2\frac{{\Gamma_1}}{{\Gamma_2}}\Bigg)}\Bigg\},
\end{multlined}
\end{equation}
\begin{equation}
\tag{233}\label{eq:233}
\begin{multlined}
\text{{\boldmath $\vvmathbb{R}_2$}}=\Bigg\{\Bigg(\beth_{17}+ \beth_{13}
\Bigg[\partial_\tau({{\Gamma_1}}/ {{\Gamma_2}})\Bigg]+\Bigg[\frac{{\Gamma_1}}{{\Gamma_2}}\Bigg]\Bigg[\dt{\beth}_{13}+\mathrm{H}
\beth_{13}\Bigg]\Bigg)+\frac{1}{a(t)}\partial_t\Bigg(2\frac{{\Gamma_1}}{{\Gamma_2}}a(t)\mathrm{Q_3}\Bigg)\Bigg\},
\end{multlined}
\end{equation}
\begin{equation}
\tag{234}\label{eq:234}
\begin{multlined}
\text{{\boldmath $\vvmathbb{R}_3$}}=\Bigg\{\Bigg(\beth_{16}+\beth_{15}\Bigg[2\frac{{\Gamma_1}}{{\Gamma_2}}\Bigg]+\beth_{3}\Bigg[2\frac{{\Gamma_1}}{{\Gamma_2}}\Bigg]^2+ \beth_{1}\Bigg[2\frac{{\Gamma_1}}{{\Gamma_2}}\Bigg]^3\Bigg)+\frac{\mathrm{Q}_3}{{\Gamma_1}}\Bigg(\beth_{20}+ \beth_{12}\Bigg[2\frac{{\Gamma_1}}{{\Gamma_2}}\Bigg]+\beth_{5}\Bigg[2\frac{{\Gamma_1}}{{\Gamma_2}}\Bigg]^2\Bigg)+2\frac{
{\Gamma_1}}{v^2_{\mathrm{c}}{\Gamma_2}}\mathrm{Q_3}\Bigg\},
\end{multlined}
\end{equation}
\begin{equation}
\tag{235}\label{eq:235}
\begin{multlined}
\text{{\boldmath $\vvmathbb{R}_4$}}=\frac{\mathrm{Q_3}}{{\Gamma_1}}\Bigg[\frac{\beth_{23}+\beth_{24}}{{\Gamma_1}}-{{\Gamma_1}}\partial_t
\Bigg(\frac{1}{{\Gamma^2_1}}\Bigg\{\beth_{22}+\beth_{9}\Bigg[2\frac{{\Gamma_1}}{{\Gamma_2}}\Bigg]\Bigg\}\Bigg)+3\frac{\mathrm{H}}{{\Gamma_1}}\Bigg\{\beth_{22}+\beth_{9}\Bigg[2\frac{{\Gamma_1}}{{\Gamma_2}}\Bigg]\Bigg\}
\Bigg],
\end{multlined}
\end{equation}
\begin{equation}
\tag{236}\label{eq:236}
\begin{multlined}
\text{{\boldmath $\vvmathbb{R}_5$}}=\frac{1}{2}\Bigg[\frac{\beth_{23}}{{\Gamma^2_1}}-\partial_t
\Bigg(\frac{1}{{\Gamma^2_1}}\Bigg\{\beth_{22}+\beth_{9}\Bigg[2\frac{{\Gamma_1}}{{\Gamma_2}}\Bigg]\Bigg\}\Bigg)+3\frac{\mathrm{H}}{{\Gamma^2_1}}\Bigg\{\beth_{22}+\beth_{9}\Bigg[2\frac{{\Gamma_1}}{{\Gamma_2}}\Bigg]\Bigg\}
\Bigg],
\end{multlined}
\end{equation}
\begin{equation}
\tag{237}\label{eq:237}
\begin{multlined}
\text{{\boldmath $\vvmathbb{R}_6$}}=\Bigg\{\beth_{8}+\beth_{4}\Bigg[2\frac{{\Gamma_1}}{{\Gamma_2}}\Bigg]\Bigg\}-2\frac{
{\Gamma_1}}{{\Gamma_2}}\Bigg\{\beth_{20}+\beth_{12}\Bigg[2\frac{{\Gamma_1}}{{\Gamma_2}}\Bigg]+ \beth_{5}\Bigg[2\frac{{\Gamma_1}}{{\Gamma_2}}\Bigg]^2\Bigg\},
\end{multlined}
\end{equation}
\begin{equation}
\tag{238}\label{eq:238}
\begin{multlined}
\text{{\boldmath $\vvmathbb{R}_7$}}=4\beth_{23}\frac{{\Gamma^2_1}}{{\Gamma^2_2}}-\frac{a(t)}{3}\partial_t\Bigg(
4\frac{{\Gamma^2_1}}{a(t){\Gamma^2_2}}\Bigg[
-\beth_{10}+\Bigg\{\beth_{22}+\beth_{9}\Bigg[2\frac{{\Gamma_1}}{{\Gamma_2}}\Bigg]\Bigg\}\Bigg]\Bigg)+\frac{8}{3}\frac{{\Gamma^2_1}}{{\Gamma^2_2}}{\beth_{24}}-\\ \frac{\mathrm{Q_3}v^2_\mathrm{c}}{{{\Gamma_2}}}\Bigg[
\beth_{10}-2\Bigg\{\beth_{22}+\beth_{9}\Bigg[2\frac{{\Gamma_1}}{{\Gamma_2}}\Bigg]\Bigg\}\Bigg],
\end{multlined}
\end{equation}
\begin{equation}
\tag{238}\label{eq:239}
\begin{multlined}
\text{{\boldmath $\vvmathbb{R}_8$}}=-\frac{4}{{\Gamma_2}}\beth_{23}-\frac{1}{2} a(t)^2\partial_t\Bigg(
\frac{2}{a(t)^2{{\Gamma_2}}}\Bigg[
\beth_{10}-2\Bigg\{\beth_{22}+\beth_{9}\Bigg[2\frac{{\Gamma_1}}{{\Gamma_2}}\Bigg]\Bigg\}\Bigg]\Bigg)+\frac{2}{{\Gamma_2}}{\beth_{24}}-\\ \frac{2\mathrm{Q_3}v^2_\mathrm{c}}{{{\Gamma^2_1}}}\Bigg\{\beth_{22}+\beth_{9}\Bigg[2\frac{{\Gamma_1}}{{\Gamma_2}}\Bigg]\Bigg\},
\end{multlined}
\end{equation}
\begin{equation}
\tag{240}\label{eq:240}
\begin{multlined}
\text{{\boldmath $\vvmathbb{R}_9$}}=-2\Bigg[\frac{1}{{\Gamma^2_1}}\Bigg\{\beth_{22}+\beth_{9}\Bigg[2\frac{{\Gamma_1}}{{\Gamma_2}}\Bigg]\Bigg\}\Bigg\{(\partial_k\mathcal{R})(\partial_k\upchi)-\frac{1}{\partial^2[\partial_\mu\partial_\nu\{(\partial_\mu\mathcal{R})(\partial_\nu\upchi\}]}\Bigg\}-2\frac{
{\Gamma_1}}{v^2_{\mathrm{c}}{\Gamma_2}}\mathcal{R}\dt{\mathcal{R}}+\frac{1}{{{2\Gamma_2}}a(t)^2}\Bigg[
\beth_{10}-\\2\Bigg\{\beth_{22}+\beth_{9}\Bigg[2\frac{{\Gamma_1}}{{\Gamma_2}}\Bigg]\Bigg\}\Bigg]\Bigg\{(\partial_k\mathcal{R})^2-\frac{1}{\partial^2[\partial_\mu\partial_\nu\{(\partial_\mu\mathcal{R})(\partial_\nu\mathcal{R}\}]}\Bigg\}\Bigg]\Bigg[a(t)^3(\dt{\mathrm{Q}_3}\dt{\mathcal{R}}+\mathrm{Q_3}\ddt{\mathcal{R}})+3a(t)^2\dt{a}(t)\mathrm{Q_3}\dt{\mathcal{R}}\;\;\;\;\;\;\\ - a(t)\mathrm{Q_3}v^2_\mathrm{c}(\partial^2\mathcal{R})\Bigg].\;\;\;\;\;
\end{multlined}
\end{equation}
Now, we revert back to our discussion of the higher order correlation functions in section \ref{wick} and \ref{hornbi}. The 3-point correlation can be rewritten from the expression given in \eqref{eq:103}\footnote{It helps to realize that for a 3-point correlation, there must be interactions for each term (\text{{\boldmath $\vvmathbb{R}_1$}}, \text{{\boldmath $\vvmathbb{R}_2$}}... \text{{\boldmath $\vvmathbb{R}_9$}}) in the interaction part of the Hamiltonian given by \eqref{eq:242}, Hence, we require three terms for the propagators [\text{{\boldmath $\vvmathbb{D}$}}$(\mathbf{k_i},\mathbf{k_j},\tau',\tau)$ and, \text{{\boldmath $\vvmathbb{D}$}}$(\mathbf{k_i},\mathbf{k_j},\tau,\tau')$], and their time and spatial derivatives for full computation of contributions by each term in the interaction part of the Hamiltonian.},
\begin{equation}
\tag{241}\label{eq:241}
\begin{multlined}
\,\langle\bar\delta\mathcal{R}_\mathbf{k_1}\,\bar\delta\mathcal{R}_\mathbf{k_2}\,\bar\delta\mathcal{R}_\mathbf{k_3}\rangle_{_\mathrm{\scriptstyle H}} = i\int_{\tau_\mathrm{o}}^{\tau}\mathrm{d}\tau'\,\bigg\langle 0\,\Bigg| \Bigg[\mathcal{H}_\mathrm{int}(\tau'), \Bigg\{\bar\delta\hspace{-0.01in}\mathcal{R}_\mathbf{k_1}\,\bar\delta\hspace{-0.01in}\mathcal{R}_\mathbf{k_2}\,\bar\delta\hspace{-0.01in}\mathcal{R}_\mathbf{k_3}\}_{\mathrm{\scriptstyle I}}\Bigg ]\Bigg |\,0 \Bigg\rangle\\=i\int_{-\infinity}^{0}\mathrm{d}\tau'\,\bigg\langle 0\,\Bigg| \Bigg[\mathcal{H}_\mathrm{int}(\tau'), \Bigg\{\bar\delta\hspace{-0.01in}\mathcal{R}_\mathbf{k_1}\,\bar\delta\hspace{-0.01in}\mathcal{R}_\mathbf{k_2}\,\bar\delta\hspace{-0.01in}\mathcal{R}_\mathbf{k_3}\}_{\mathrm{\scriptstyle I}}\Bigg ]\Bigg |\,0 \Bigg\rangle\propto\\-i\delta(\mathbf{k_1}+\mathbf{k_2}+\mathbf{k_3})\int_{-\infinity}^{\tau\rightarrow 0}\mathrm{d}\tau' [\text{{\boldmath $\vvmathbb{D}$}}(\mathbf{k_1},\mathbf{k_2},\tau,\tau')\,\text{{\boldmath $\vvmathbb{D}$}}(\mathbf{k_2},\mathbf{k_3},\tau,\tau')\,\text{{\boldmath $\vvmathbb{D}$}}(\mathbf{k_1},\mathbf{k_3},\tau,\tau') - \\ \text{{\boldmath $\vvmathbb{D}$}}(\mathbf{k_1},\mathbf{k_2},\tau',\tau)\,\text{{\boldmath $\vvmathbb{D}$}}(\mathbf{k_2},\mathbf{k_3},\tau',\tau)\,\text{{\boldmath $\vvmathbb{D}$}}(\mathbf{k_1},\mathbf{k_3},\tau',\tau)],
\end{multlined}
\end{equation}
where, we have reintroduced the conformal time $\tau$ such that $\mathrm{d}(t)/a(t) = \mathrm{d}\tau$. We have also set the limits to $\tau_\mathrm{o}\rightarrow -\,\infinity$ and $\tau\rightarrow 0$, as explained in section \ref{threepoint}. Moreover, the interaction part of the Hamiltonian $\mathcal{H}_\mathrm{int}(t)$ is given by:
\begin{align*}
\mathcal{H}_\mathrm{int}(t)=\int\mathrm{d}^3\!x\,\Bigg[
a(t)^3\text{{\boldmath $\vvmathbb{R}_1$}}\mathcal{R}\dt{\mathcal{R}}^2
+a(t)\text{{\boldmath $\vvmathbb{R}_2$}}\mathcal{R}(\partial\mathcal{R})^2+a(t)^3\text{{\boldmath $\vvmathbb{R}_3$}}\dt{\mathcal{R}}^3+a(t)^3\text{{\boldmath $\vvmathbb{R}_4$}}\dt{\mathcal{R}}(\partial_\mu\upchi)(\partial_\mu\mathcal{R})+\\a(t)^3\text{{\boldmath $\vvmathbb{R}_5$}}(\partial\upchi)^2(\partial^2\mathcal{R})
+a(t)\text{{\boldmath $\vvmathbb{R}_6$}}\dt{\mathcal{R}}^2(\partial^2\mathcal{R})+
a(t)^{-1}\text{{\boldmath $\vvmathbb{R}_7$}}[(\partial\mathcal{R})^2(\partial^2\mathcal{R})-\mathcal{R}
\partial_\mu\partial_\nu\{\partial_\mu\mathcal{R}\}(\partial_\nu\mathcal{R})]+\\a(t)\text{{\boldmath $\vvmathbb{R}_8$}}[(\partial_\mu\mathcal{R})(\partial_\mu\upchi)(\partial^2\mathcal{R})-\mathcal{R}
\partial_\mu\partial_\nu\{\partial_\mu\mathcal{R}\}(\partial_\nu{\upchi})]+\text{{\boldmath $\vvmathbb{R}_9$}}\Bigg].
\tag{242}\label{eq:242}
\end{align*}
We will now calculate the contribution from each term separately in the limit $\tau\sim-[a(\tau)\mathrm{H}]^{-1}$.
\subsubsection{Interactions in bi-spectrum}
\label{intbi}
\textbf{\tit{a}) \text{{\boldmath $\vvmathbb{R}_1$}} \tit{term}:} Let us first transform our interaction term to a variable in terms of the conformal time $\tau$. This is done simply by using the definition of conformal time [$\mathrm{d}(t)/a(t) = \mathrm{d}\tau$] such that 
\begin{equation}
\tag{243}\label{eq:243}
\begin{multlined}
\mathcal{H}^{\text{{\boldmath $\vvmathbb{R}_1$}}}_\mathrm{int}(\tau')=\int\mathrm{d}^3\!x\,\Bigg[
a(\tau')\text{{\boldmath $\vvmathbb{R}_1$}}\mathcal{R}{(\partial_{\tau'}\mathcal{R}})^2\Bigg].
\end{multlined}
\end{equation}
In \text{{\boldmath $\vvmathbb{R}_1$}} term, we have three interactions - one with $\mathcal{R}$ and two with ${(\partial_\tau\mathcal{R}})$. The propagators \text{{\boldmath $\vvmathbb{D}$}}$(\mathbf{k_i},\mathbf{k_j},\tau',\tau)$ and \text{{\boldmath $\vvmathbb{D}$}}$(\mathbf{k_i},\mathbf{k_j},\tau,\tau')$ are thus given by\footnote{\text{{\boldmath $\vvmathbb{D}$}}$(\mathbf{k_i},\mathbf{k_j},\tau',\tau)$ $\neq$ \text{{\boldmath $\vvmathbb{D}$}}$(\mathbf{k_i},\mathbf{k_j},\tau,\tau')$}
\begin{equation}
\tag{244}\label{eq:244}
\begin{multlined}
\text{{\boldmath $\vvmathbb{D}$}}(\mathbf{k_1},\mathbf{k_2},\tau,\tau')=u_{k_1}(\tau)\,u_{k_1}^*(\tau'),
\end{multlined}
\end{equation}
\begin{equation}
\tag{245}\label{eq:245}
\begin{multlined}
\text{{\boldmath $\vvmathbb{D}$}}(\mathbf{k_2},\mathbf{k_3},\tau,\tau')=\partial_{\tau'}\{u_{k_2}(\tau)\,u_{k_2}^*(\tau')\},
\end{multlined}
\end{equation}
\begin{equation}
\tag{246}\label{eq:246}
\begin{multlined}
\text{{\boldmath $\vvmathbb{D}$}}(\mathbf{k_1},\mathbf{k_3},\tau,\tau')=\partial_{\tau'}\{u_{k_3}(\tau)\,u_{k_3}^*(\tau')\},
\end{multlined}
\end{equation}
\begin{equation}
\tag{247}\label{eq:247}
\begin{multlined}
\text{{\boldmath $\vvmathbb{D}$}}(\mathbf{k_1},\mathbf{k_2},\tau',\tau)=u_{k_1}(\tau')\,u_{k_1}^*(\tau),
\end{multlined}
\end{equation}
\begin{equation}
\tag{248}\label{eq:248}
\begin{multlined}
\text{{\boldmath $\vvmathbb{D}$}}(\mathbf{k_2},\mathbf{k_3},\tau',\tau)=\partial_{\tau'}\{u_{k_2}(\tau')\,u_{k_2}^*(\tau)\},
\end{multlined}
\end{equation}
\begin{equation}
\tag{249}\label{eq:249}
\begin{multlined}
\text{{\boldmath $\vvmathbb{D}$}}(\mathbf{k_2},\mathbf{k_3},\tau,\tau')=\partial_{\tau'}\{u_{k_3}(\tau')\,u_{k_3}^*(\tau)\}.
\end{multlined}
\end{equation}
Thus, the overall contribution from the \text{{\boldmath $\vvmathbb{R}_1$}} term is,
\begin{equation}
\tag{250}\label{eq:250}
\begin{multlined}
\,\langle\bar\delta\mathcal{R}_\mathbf{k_1}\,\bar\delta\mathcal{R}_\mathbf{k_2}\,\bar\delta\mathcal{R}_\mathbf{k_3}\rangle_{\text{{\boldmath $\vvmathbb{R}_1$}}} = 
-i\text{{\boldmath $\vvmathbb{R}_1$}}\delta(\mathbf{k_1}+\mathbf{k_2}+\mathbf{k_3})\int_{-\infinity}^{\tau\rightarrow 0}\mathrm{d}\tau'\, a(\tau')[\{u_{k_1}(\tau)\,u_{k_1}^*(\tau')\}(\partial_{\tau'}\{u_{k_2}(\tau)\,u_{k_2}^*(\tau')\})\times\vspace{0.1in}\\(\partial_{\tau'}\{u_{k_3}(\tau)\,u_{k_3}^*(\tau')\})-\{u_{k_1}(\tau')\,u_{k_1}^*(\tau)\}(\partial_{\tau'}\{u_{k_2}(\tau')\,u_{k_2}^*(\tau)\})(\partial_{\tau'}\{u_{k_3}(\tau')\,u_{k_3}^*(\tau)\})].
\end{multlined}
\end{equation}
However, these are not the only terms that contribute to $\langle\bar\delta\mathcal{R}_\mathbf{k_1}\,\bar\delta\mathcal{R}_\mathbf{k_2}\,\bar\delta\mathcal{R}_\mathbf{k_3}\rangle_{\text{{\boldmath $\vvmathbb{R}_1$}}}$. In fact, as shown in section \ref{feyndiag}, summation must be made over all possible Feynman diagrams, i.e. over all possible permutations of the interaction terms. While we have only represented one of the many possible diagrams, the remaining terms  are symmetric in $k_1,k_2$ and $k_3$ with the same coefficients and therefore, they need not be calculated explicitly. We represent these terms with '\textsf{sym}$(k_1,k_2,k_3)$'. Therefore, the correct expression for $\langle\bar\delta\mathcal{R}_\mathbf{k_1}\,\bar\delta\mathcal{R}_\mathbf{k_2}\,\bar\delta\mathcal{R}_\mathbf{k_3}\rangle_{\text{{\boldmath $\vvmathbb{R}_1$}}}$ is given by:
\begin{equation}
\tag{251}\label{eq:251}
\begin{multlined}
\,\langle\bar\delta\mathcal{R}_\mathbf{k_1}\,\bar\delta\mathcal{R}_\mathbf{k_2}\,\bar\delta\mathcal{R}_\mathbf{k_3}\rangle_{\text{{\boldmath $\vvmathbb{R}_1$}}} = 
-i\text{{\boldmath $\vvmathbb{R}_1$}}\delta(\mathbf{k_1}+\mathbf{k_2}+\mathbf{k_3}){\displaystyle\int_{-\infinity}^{\tau\rightarrow 0}}\mathrm{d}\tau'\, a(\tau')[\{u_{k_1}(\tau)\,u_{k_1}^*(\tau')\}(\partial_{\tau'}\{u_{k_2}(\tau)\,u_{k_2}^*(\tau')\})\vspace{0.1in}\times\\(\partial_{\tau'}\{u_{k_3}(\tau)\,u_{k_3}^*(\tau')\})-\{u_{k_1}(\tau')\,u_{k_1}^*(\tau)\}(\partial_{\tau'}\{u_{k_2}(\tau')\,u_{k_2}^*(\tau)\})(\partial_{\tau'}\{u_{k_3}(\tau')\,u_{k_3}^*(\tau)\})] + \text{\textsf{sym}}(k_1,k_2,k_3).
\end{multlined}
\end{equation}
Now, from the expression of mode amplitude $u_{k}(\tau)$ derived in \eqref{eq:196},
\begin{equation*}
\begin{multlined}
u_k(\tau)=\frac{1}{2\mathrm{Q}^{\frac{1}{2}}\,v_\mathrm{c}}\frac{\mathrm{H}}{({v_\mathrm{c}})^{1/2}}\frac{1}{k^{3/2}}(1+ikv_\mathrm{c}\tau)e^{-ikv_\mathrm{c}\tau},
\end{multlined}
\end{equation*}
we get
\begin{equation}
\tag{252}\label{eq:252}
\begin{multlined}
\,\langle\bar\delta\mathcal{R}_\mathbf{k_1}\,\bar\delta\mathcal{R}_\mathbf{k_2}\,\bar\delta\mathcal{R}_\mathbf{k_3}\rangle_{\text{{\boldmath $\vvmathbb{R}_1$}}} \xrightarrow{\tau\rightarrow 0}
-i\text{{\boldmath $\vvmathbb{R}_1$}}\frac{\mathrm{H}^6}{2^6\mathrm{Q}^3v^9_\mathrm{c}} \delta(\mathbf{k_1}+\mathbf{k_2}+\mathbf{k_3})\frac{1}{(k_1k_2k_3)^3}\Bigg[\Bigg\{e^{-i(k_1+k_2+k_3)v_\mathrm{c}\tau}{{\mathlarger{\mathlarger{\mathlarger{\mathlarger{\uppi}}}}}}_{i=1}^3(1+k_iv_\mathrm{c}\tau)\vspace{0.1in}\\\int_{-\infinity}^{\tau\rightarrow 0}\mathrm{d}\tau'\, a(\tau')\Bigg[
e^{i(k_1+k_2+k_3)v_\mathrm{c}\tau'}(1-k_1v_\mathrm{c}\tau')k^2_2k^2_3v^4_\mathrm{c}\tau'^2\Bigg]\Bigg\}-\Bigg\{e^{i(k_1+k_2+k_3)v_\mathrm{c}\tau}{{\mathlarger{\mathlarger{\mathlarger{\mathlarger{\uppi}}}}}}_{i=1}^3(1-k_iv_\mathrm{c}\tau)\vspace{0.1in}\\\int_{-\infinity}^{\tau\rightarrow 0}\mathrm{d}\tau'\, a(\tau')\Bigg[
e^{-i(k_1+k_2+k_3)v_\mathrm{c}\tau'}(1+k_1v_\mathrm{c}\tau')k^2_2k^2_3v^4_\mathrm{c}\tau'^2\Bigg]\Bigg\}\Bigg] + \text{\textsf{sym}}(k_1,k_2,k_3)=\\
-i\text{{\boldmath $\vvmathbb{R}_1$}}\frac{\mathrm{H}^6}{2^6\mathrm{Q}^3v^9_\mathrm{c}} \delta(\mathbf{k_1}+\mathbf{k_2}+\mathbf{k_3})\frac{1}{(k_1k_2k_3)^3}\Bigg[\Bigg\{\int_{-\infinity}^{\tau\rightarrow 0}\mathrm{d}\tau'\, a(\tau')\Bigg[
e^{i(k_1+k_2+k_3)v_\mathrm{c}\tau'}(1-k_1v_\mathrm{c}\tau')k^2_2k^2_3v^4_\mathrm{c}\tau'^2\Bigg]\Bigg\}-\\ \Bigg\{\int_{-\infinity}^{\tau\rightarrow 0}\mathrm{d}\tau'\, a(\tau')\Bigg[
e^{-i(k_1+k_2+k_3)v_\mathrm{c}\tau'}(1+k_1v_\mathrm{c}\tau')k^2_2k^2_3v^4_\mathrm{c}\tau'^2\Bigg]\Bigg\}\Bigg] + \text{\textsf{sym}}(k_1,k_2,k_3)=\\
\text{{\boldmath $\vvmathbb{R}_1$}}\frac{\mathrm{H}^4}{2^4\mathrm{Q}^3v^6_\mathrm{c}} \delta(\mathbf{k_1}+\mathbf{k_2}+\mathbf{k_3})\frac{1}{(k_1k_2k_3)^3}\Bigg[\frac{k^2_2k^2_3}{(k_1+k_2+k_3)}+\frac{k_1k^2_2k^2_3}{(k_1+k_2+k_3)^2}+\text{\textsf{sym}}(k_1,k_2,k_3)\Bigg].\;\;\;\;
\end{multlined}
\end{equation}\vspace{0.15in}\\
\textbf{\tit{b}) \text{{\boldmath $\vvmathbb{R}_2$}} \tit{term}:} The \text{{\boldmath $\vvmathbb{R}_2$}} interaction part of the Hamiltonian is given by:
\begin{equation}
\tag{253}\label{eq:253}
\begin{multlined}
\mathcal{H}^{\text{{\boldmath $\vvmathbb{R}_2$}}}_\mathrm{int}(\tau')=\int\mathrm{d}^3\!x\,\Bigg[
a(\tau')\text{{\boldmath $\vvmathbb{R}_2$}}\mathcal{R}(\partial\mathcal{R})^2\Bigg]=\int\mathrm{d}^3\!x\,\Bigg[
a(\tau')\text{{\boldmath $\vvmathbb{R}_2$}}\mathcal{R}(\partial_\mu\mathcal{R})(\partial^\mu\mathcal{R})\Bigg].
\end{multlined}
\end{equation}
It follows that
\begin{align*}
\,\langle\bar\delta\mathcal{R}_\mathbf{k_1}\,\bar\delta\mathcal{R}_\mathbf{k_2}\,\bar\delta\mathcal{R}_\mathbf{k_3}\rangle_{\text{{\boldmath $\vvmathbb{R}_2$}}} = -i\text{{\boldmath $\vvmathbb{R}_2$}}\delta(\mathbf{k_1}+\mathbf{k_2}+\mathbf{k_3})\{-(\mathbf{k_1}\cdot\mathbf{k_2}+\mathbf{k_2}\cdot\mathbf{k_3}+\mathbf{k_3}\cdot\mathbf{k_1})\}\int_{-\infinity}^{\tau\rightarrow 0}\mathrm{d}\tau'\,\vspace{0.1in} \\\times\,a(\tau')[\{u_{k_1}(\tau)\,u_{k_1}^*(\tau')\}\{u_{k_2}(\tau)\,u_{k_2}^*(\tau')\}\{u_{k_3}(\tau)\,u_{k_3}^*(\tau')\}-\{u_{k_1}(\tau')\,u_{k_1}^*(\tau)\}\\\times\{u_{k_2}(\tau')\,u_{k_2}^*(\tau)\}\{u_{k_3}(\tau')\,u_{k_3}^*(\tau)\}]\xrightarrow{\tau\rightarrow 0}
-i\text{{\boldmath $\vvmathbb{R}_2$}}\frac{\mathrm{H}^6}{2^6\mathrm{Q}^3v^9_\mathrm{c}} \delta(\mathbf{k_1}+\mathbf{k_2}+\mathbf{k_3})\frac{(\mathbf{k_1}\cdot\mathbf{k_2}+\mathbf{k_2}\cdot\mathbf{k_3}+\mathbf{k_3}\cdot\mathbf{k_1})}{(k_1k_2k_3)^3}\\ \times\Bigg[\Bigg\{e^{-i(k_1+k_2+k_3)v_\mathrm{c}\tau}{{\mathlarger{\mathlarger{\mathlarger{\mathlarger{\uppi}}}}}}_{i=1}^3(1+k_iv_\mathrm{c}\tau)\vspace{0.1in}\int_{-\infinity}^{\tau\rightarrow 0}\mathrm{d}\tau'\, a(\tau')\Bigg[
e^{i(k_1+k_2+k_3)v_\mathrm{c}\tau'}{\mathlarger{\mathlarger{\mathlarger{\mathlarger{\uppi}}}}}_{i=1}^3(1-k_iv_\mathrm{c}\tau')\Bigg]\Bigg\}-\\ \Bigg\{e^{i(k_1+k_2+k_3)v_\mathrm{c}\tau}{{\mathlarger{\mathlarger{\mathlarger{\mathlarger{\uppi}}}}}}_{i=1}^3(1-k_iv_\mathrm{c}\tau)\vspace{0.1in}\int_{-\infinity}^{\tau\rightarrow 0}\mathrm{d}\tau'\, a(\tau')\Bigg[
e^{-i(k_1+k_2+k_3)v_\mathrm{c}\tau'}{\mathlarger{\mathlarger{\mathlarger{\mathlarger{\uppi}}}}}_{i=1}^3(1+k_iv_\mathrm{c}\tau')\Bigg]\Bigg\}\Bigg]=\\
-i\text{{\boldmath $\vvmathbb{R}_2$}}\frac{\mathrm{H}^6}{2^6\mathrm{Q}^3v^9_\mathrm{c}} \delta(\mathbf{k_1}+\mathbf{k_2}+\mathbf{k_3}) \frac{-(\mathbf{k_1}\cdot\mathbf{k_2}+\mathbf{k_2}\cdot\mathbf{k_3}+\mathbf{k_3}\cdot\mathbf{k_1})}{(k_1k_2k_3)^3}\Bigg[\Bigg\{\int_{-\infinity}^{\tau\rightarrow 0}\mathrm{d}\tau'\, a(\tau')\Bigg[
e^{i(k_1+k_2+k_3)v_\mathrm{c}\tau'}(1-\\i\{k_1+k_2+k_3\}v_\mathrm{c}\tau'-\{
k_1k_2+k_1k_3+k_2k_3\}v^2_\mathrm{c}\tau'^2+ik_1k_2k_3v^3_\mathrm{c}\tau'^3)\Bigg]\Bigg\}-\Bigg\{\int_{-\infinity}^{\tau\rightarrow 0}\mathrm{d}\tau'\, a(\tau')\\ \times\Bigg[
e^{-i(k_1+k_2+k_3)v_\mathrm{c}\tau'}(1+i\{k_1+k_2+k_3\}v_\mathrm{c}\tau'-\{
k_1k_2+k_1k_3+k_2k_3\}v^2_\mathrm{c}\tau'^2-ik_1k_2k_3v^3_\mathrm{c}\tau'^3\Bigg]\Bigg\}\Bigg]=\\ \text{{\boldmath $\vvmathbb{R}_2$}}\frac{\mathrm{H}^4}{2^4\mathrm{Q}^3v^8_\mathrm{c}} \delta(\mathbf{k_1}+\mathbf{k_2}+\mathbf{k_3}) \frac{(\mathbf{k_1}\cdot\mathbf{k_2}+\mathbf{k_2}\cdot\mathbf{k_3}+\mathbf{k_3}\cdot\mathbf{k_1})}{(k_1k_2k_3)^3}\Bigg[-(k_1+k_2+k_3)+\frac{k_1k_2+k_1k_3+k_2k_3}{k_1+k_2+k_3}+\\ \frac{k_1k_2k_3}{(k_1+k_2+k_3)^2}\Bigg].
\tag{254}\label{eq:254}
\end{align*}
\textbf{\tit{c}) \text{{\boldmath $\vvmathbb{R}_3$}} \tit{term}:} The \text{{\boldmath $\vvmathbb{R}_3$}} interaction part of the Hamiltonian is given by:
\begin{equation}
\tag{255}\label{eq:255}
\begin{multlined}
\mathcal{H}^{\text{{\boldmath $\vvmathbb{R}_3$}}}_\mathrm{int}(\tau')=\int\mathrm{d}^3\!x\,\Bigg[
a(\tau')^3\text{{\boldmath $\vvmathbb{R}_3$}}\dt{\mathcal{R}}^3\Bigg]=\int\mathrm{d}^3\!x\,\Bigg[\text{{\boldmath $\vvmathbb{R}_3$}}(\partial_{\tau'}\mathcal{R})^3\Bigg],
\end{multlined}
\end{equation}
It again follows that
\begin{equation}
\tag{256}\label{eq:256}
\begin{multlined}
\,\langle\bar\delta\mathcal{R}_\mathbf{k_1}\,\bar\delta\mathcal{R}_\mathbf{k_2}\,\bar\delta\mathcal{R}_\mathbf{k_3}\rangle_{\text{{\boldmath $\vvmathbb{R}_3$}}} = 
-i\text{{\boldmath $\vvmathbb{R}_3$}}\delta(\mathbf{k_1}+\mathbf{k_2}+\mathbf{k_3}){\displaystyle\int_{-\infinity}^{\tau\rightarrow 0}}\mathrm{d}\tau'\,[(\partial_{\tau'}\{u_{k_1}(\tau)\,u_{k_1}^*(\tau')\})(\partial_{\tau'}\{u_{k_2}(\tau)\,u_{k_2}^*(\tau')\})\vspace{0.1in}\\\times(\partial_{\tau'}\{u_{k_3}(\tau)\,u_{k_3}^*(\tau')\})-(\partial_{\tau'}\{u_{k_1}(\tau')\,u_{k_1}^*(\tau)\})(\partial_{\tau'}\{u_{k_2}(\tau')\,u_{k_2}^*(\tau)\})(\partial_{\tau'}\{u_{k_3}(\tau')\,u_{k_3}^*(\tau)\})] \\+ \text{\textsf{sym}}(k_1,k_2,k_3),
\end{multlined}
\end{equation}
\begin{align*}
\,\langle\bar\delta\mathcal{R}_\mathbf{k_1}\,\bar\delta\mathcal{R}_\mathbf{k_2}\,\bar\delta\mathcal{R}_\mathbf{k_3}\rangle_{\text{{\boldmath $\vvmathbb{R}_3$}}} \xrightarrow{\tau\rightarrow 0}
-i\text{{\boldmath $\vvmathbb{R}_3$}}\frac{\mathrm{H}^6}{2^6\mathrm{Q}^3v^9_\mathrm{c}} \delta(\mathbf{k_1}+\mathbf{k_2}+\mathbf{k_3})\frac{1}{(k_1k_2k_3)^3}\Bigg[\Bigg\{e^{-i(k_1+k_2+k_3)v_\mathrm{c}\tau}{{\mathlarger{\mathlarger{\mathlarger{\mathlarger{\uppi}}}}}}_{i=1}^3(1+k_iv_\mathrm{c}\tau)\times\vspace{0.1in}\\\int_{-\infinity}^{\tau\rightarrow 0}\mathrm{d}\tau'\,\Bigg[
e^{i(k_1+k_2+k_3)v_\mathrm{c}\tau'}k^2_1k^2_2k^2_3v^6_\mathrm{c}\tau'^3\Bigg]\Bigg\}-\Bigg\{e^{i(k_1+k_2+k_3)v_\mathrm{c}\tau}{{\mathlarger{\mathlarger{\mathlarger{\mathlarger{\uppi}}}}}}_{i=1}^3(1-k_iv_\mathrm{c}\tau)\vspace{0.1in}\int_{-\infinity}^{\tau\rightarrow 0}\mathrm{d}\tau'\,\Bigg[\\\times
e^{-i(k_1+k_2+k_3)v_\mathrm{c}\tau'} k^2_1k^2_2k^2_3v^6_\mathrm{c}\tau'^3\Bigg]\Bigg\}\Bigg] + \text{\textsf{sym}}(k_1,k_2,k_3)=
-i\text{{\boldmath $\vvmathbb{R}_3$}}\frac{\mathrm{H}^6}{2^6\mathrm{Q}^3v^9_\mathrm{c}} \delta(\mathbf{k_1}+\mathbf{k_2}+\mathbf{k_3})\frac{1}{(k_1k_2k_3)^3}\\\times\Bigg[\Bigg\{\int_{-\infinity}^{\tau\rightarrow 0}\mathrm{d}\tau'\,\Bigg[
e^{i(k_1+k_2+k_3)v_\mathrm{c}\tau'}k^2_1k^2_2k^2_3v^6_\mathrm{c}\tau'^3\Bigg]
\Bigg\}- \Bigg\{\int_{-\infinity}^{\tau\rightarrow 0}\mathrm{d}\tau'\, \Bigg[
e^{-i(k_1+k_2+k_3)v_\mathrm{c}\tau'}k^2_1k^2_2k^2_3v^6_\mathrm{c}\tau'^3\Bigg]\Bigg\}\Bigg] +\\ \text{\textsf{sym}}(k_1,k_2,k_3)=\text{{\boldmath $\vvmathbb{R}_3$}}\frac{\mathrm{H}^5}{2^3\mathrm{Q}^3v^6_\mathrm{c}} \delta(\mathbf{k_1}+\mathbf{k_2}+\mathbf{k_3})\frac{1}{(k_1k_2k_3)^3}\Bigg[\frac{k^2_1k^2_2k^2_3}{(k_1+k_2+k_3)^3}+ \text{\textsf{sym}}(k_1,k_2,k_3)\Bigg]
=\\
3\text{{\boldmath $\vvmathbb{R}_3$}}\frac{\mathrm{H}^5}{2^3\mathrm{Q}^3v^6_\mathrm{c}} \delta(\mathbf{k_1}+\mathbf{k_2}+\mathbf{k_3})\frac{1}{(k_1k_2k_3)^3}\Bigg[\frac{k^2_1k^2_2k^2_3}{(k_1+k_2+k_3)^3}\Bigg].\;\;\;\;\;\;\;\;\;\;\;\;\;\;\;\;\;\;\;\;\;\;\;\;\;\;\;\;\tag{257}\label{eq:257}
\end{align*}
\textbf{\tit{d}) \text{{\boldmath $\vvmathbb{R}_4$}} \tit{term}:} The \text{{\boldmath $\vvmathbb{R}_4$}} interaction part of the Hamiltonian is written using the alternative relation for $\upchi$ with $\mathcal{R}$ given in \eqref{eq:224} and \eqref{eq:229} as
\begin{equation}
\tag{258}\label{eq:258}
\begin{multlined}
\mathcal{H}^{\text{{\boldmath $\vvmathbb{R}_4$}}}_\mathrm{int}(\tau')=\int\mathrm{d}^3\!x\,\Bigg[a(t)^3\text{{\boldmath $\vvmathbb{R}_4$}}\dt{\mathcal{R}}(\partial_\mu\upchi)(\partial_\mu\mathcal{R})\Bigg]=\int\mathrm{d}^3\!x\,\Bigg[
a(\tau')^2\text{{\boldmath $\vvmathbb{R}_4$}}(\partial_{\tau'}\mathcal{R})(\partial_\mu\mathcal{R})(\partial_\mu\upchi)\Bigg].
\end{multlined}
\end{equation}\\
\begin{flushleft}\myrule[line width = 0.5mm]{fast cap reversed}{fast cap reversed}\end{flushleft}
\tit{\textbf{Sidenote}}:\\
We summarize the use of the relations \eqref{eq:224} and \eqref{eq:229} in order to arrive at the contribution of \text{{\boldmath $\vvmathbb{R}_4$}} term to the bi-spectrum. Let us rewrite
\begin{equation*}
\begin{multlined}
\partial^2\upchi=\partial^\mu\partial_\mu\upchi=\mathrm{Q}\dt{\mathcal{R}}.
\end{multlined}
\end{equation*}
In Fourier space, we can write it as:
\begin{equation}
\tag{259}\label{eq:259}
\begin{multlined}
\mathcal{F}[\partial^2\upchi]=\mathcal{F}[\partial^\mu\partial_\mu\upchi]=ik^\mu\mathcal{F}[\partial_\mu\upchi]=-k^\mu k_\mu\mathcal{F}[\upchi]\equiv -(\mathbf{k\cdot k})\mathcal{F}[\upchi] =\mathrm{Q}\,\mathcal{F}[\dt{\mathcal{R}}]=\mathrm{Q}\,\partial_{t}(\mathcal{F}[{\mathcal{R}}]).
\end{multlined}
\end{equation}
The same idea can now be applied to the Wick's theorem (in Fourier space) for the contribution of the \text{{\boldmath $\vvmathbb{R}_4$}} term the bi-spectrum.
\begin{flushright}\myrule[line width = 0.5mm]{fast cap reversed}{fast cap reversed}\end{flushright}$\;$\\
It follows that\footnote{The $\upchi$ field induces an internal degeneracy with respect to $\mathbf{k_2}$ and $\mathbf{k_3}$. The overall contribution from the \text{{\boldmath $\vvmathbb{R}_4$}} term then becomes a sum over this internal degeneracies, i.e. {$\text{\textlangle} \bar\delta\mathcal{R}_\mathbf{k_1}\,\bar\delta\mathcal{R}_\mathbf{k_2}\,\bar\delta\mathcal{R}_\mathbf{k_3}\text{\textrangle}_{\text{{\boldmath $\vvmathbb{R}_4$}}} = \text{\textlangle}\bar\delta\mathcal{R}_\mathbf{k_1}\,\bar\delta\mathcal{R}_\mathbf{k_2}\,\bar\delta{\upchi}_\mathbf{k_3}\text{\textrangle}_{\text{{\boldmath $\vvmathbb{R}_4$}}}+ \,\text{\textlangle}\bar\delta\mathcal{R}_\mathbf{k_1}\,\bar\delta{\upchi}_\mathbf{k_2}\,\bar\delta\mathcal{R}_\mathbf{k_3}\text{\textrangle}_{\text{{\boldmath $\vvmathbb{R}_4$}}}$.}},
\begin{equation*}
\,\langle\bar\delta\mathcal{R}_\mathbf{k_1}\,\bar\delta\mathcal{R}_\mathbf{k_2}\,\bar\delta\mathcal{R}_\mathbf{k_3}\rangle_{\text{{\boldmath $\vvmathbb{R}_4$}}} = \,\langle\bar\delta\mathcal{R}_\mathbf{k_1}\,\bar\delta\mathcal{R}_\mathbf{k_2}\,\bar\delta{\upchi}_\mathbf{k_3}\rangle_{\text{{\boldmath $\vvmathbb{R}_4$}}}+ \,\langle\bar\delta\mathcal{R}_\mathbf{k_1}\,\bar\delta{\upchi}_\mathbf{k_2}\,\bar\delta\mathcal{R}_\mathbf{k_3}\rangle_{\text{{\boldmath $\vvmathbb{R}_4$}}},
\end{equation*}
such that,
\begin{equation}
\tag{260}\label{eq:260}
\begin{multlined}
\langle\bar\delta\mathcal{R}_\mathbf{k_1}\,\bar\delta{\upchi}_\mathbf{k_2}\,\bar\delta\mathcal{R}_\mathbf{k_3}\rangle_{\text{{\boldmath $\vvmathbb{R}_4$}}}=
-i\text{{\boldmath $\vvmathbb{R}_4$}}\delta(\mathbf{k_1}+\mathbf{k_2}+\mathbf{k_3}){\displaystyle\int_{-\infinity}^{\tau\rightarrow 0}}\mathrm{d}\tau'\,a(\tau')^2[(\partial_{\tau'}\{u_{k_1}(\tau)\,u_{k_1}^*(\tau')\})\{u_{k_2}(\tau)\,u_{k_2}^*(\tau')\}\vspace{0.1in}\\\times\frac{-\{\mathbf{k_2\cdot k_3}\}}{-k^2_3}(\partial_{\tau'}\{u_{k_3}(\tau)\,u_{k_3}^*(\tau')\})-a(\tau')^2(\partial_{\tau'}\{u_{k_1}(\tau')\,u_{k_1}^*(\tau)\})\{u_{k_2}(\tau')\,u_{k_2}^*(\tau)\}\frac{-\{\mathbf{k_2\cdot k_3}\}}{-k^2_3}\\\times(\partial_{\tau'}\{u_{k_3}(\tau')\,u_{k_3}^*(\tau)\})]+ \text{\textsf{sym}}(k_1,k_2,k_3)\xrightarrow{\tau\rightarrow 0}
-i\text{{\boldmath $\vvmathbb{R}_4$}}\frac{\mathrm{H}^6}{2^6\mathrm{Q}^3v^9_\mathrm{c}} \delta(\mathbf{k_1}+\mathbf{k_2}+\mathbf{k_3})\frac{\mathrm{Q}(-\{\mathbf{k_2\cdot k_3})}{(k_1k_2k_3)^3}\\ \times\Bigg[\Bigg\{e^{-i(k_1+k_2+k_3)v_\mathrm{c}\tau}{{\mathlarger{\mathlarger{\mathlarger{\mathlarger{\uppi}}}}}}_{i=1}^3(1+k_iv_\mathrm{c}
\tau)\vspace{0.1in}\int_{-\infinity}^{\tau\rightarrow 0}\mathrm{d}\tau'\,a(\tau')^2\Bigg[
e^{i(k_1+k_2+k_3)v_\mathrm{c}\tau'}\{1-k_2v_\mathrm{c}
\tau'\}\{k^2_1v^2_\mathrm{c}\tau'\}\Bigg\{-\frac{1}{k^2_3}\Bigg\}\\\times\Bigg\{k^2_3v^2_\mathrm{c}\tau'\}\Bigg]
\Bigg\}-\Bigg\{e^{i(k_1+k_2+k_3)v_\mathrm{c}\tau}{{\mathlarger{\mathlarger{\mathlarger{\mathlarger{\uppi}}}}}}_{i=1}^3(1-k_iv_\mathrm{c}\tau)\vspace{0.1in}\int_{-\infinity}^{\tau\rightarrow 0}\mathrm{d}\tau'\,a(\tau')^2\Bigg[
e^{-i(k_1+k_2+k_3)v_\mathrm{c}\tau'}\{1+k_2v_\mathrm{c}
\tau'\}\{k^2_1v^2_\mathrm{c}\tau'\}\\\times\Bigg\{-\frac{1}{k^2_3}\Bigg\}\Bigg\{k^2_3v^2_\mathrm{c}\tau'\}\Bigg]\Bigg\}\Bigg] + \text{\textsf{sym}}(k_1,k_2,k_3)=\text{{\boldmath $\vvmathbb{R}_4$}}\frac{\mathrm{H}^4}{2^5\mathrm{Q}^2v^6_\mathrm{c}} \delta(\mathbf{k_1}+\mathbf{k_2}+\mathbf{k_3})\frac{1}{(k_1k_2k_3)^3}\frac{(\mathbf{-k_2\cdot k_3})k^2_1}{(k_1+k_2+k_3)}\\\times\Bigg[1+\frac{k_2}{(k_1+k_2+k_3)}\Bigg]+ \text{\textsf{sym}}(k_1,k_2,k_3),\;\;\;\;\;\;\;\;
\end{multlined}
\end{equation}
and the second degenerate term contributes,
\begin{align*}
\,\langle\bar\delta\mathcal{R}_\mathbf{k_1}\,\bar\delta\mathcal{R}_\mathbf{k_2}\,\bar\delta{\upchi}_\mathbf{k_3}\rangle_{\text{{\boldmath $\vvmathbb{R}_4$}}}=\text{{\boldmath $\vvmathbb{R}_4$}}\frac{\mathrm{H}^4}{2^5\mathrm{Q}^2v^6_\mathrm{c}} \delta(\mathbf{k_1}+\mathbf{k_2}+\mathbf{k_3})\frac{1}{(k_1k_2k_3)^3}\frac{(\mathbf{-k_2\cdot k_3})k^2_1}{(k_1+k_2+k_3)}\times\Bigg[1+\frac{k_3}{(k_1+k_2+k_3)}\Bigg].\tag{261}\label{eq:261}
\end{align*}\\
The overall contribution is then given by:
\begin{align*}
\,\langle\bar\delta\mathcal{R}_\mathbf{k_1}\,\bar\delta\mathcal{R}_\mathbf{k_2}\,\bar\delta\mathcal{R}_\mathbf{k_3}\rangle_{\text{{\boldmath $\vvmathbb{R}_4$}}}=\text{{\boldmath $\vvmathbb{R}_4$}}\frac{\mathrm{H}^4}{2^5\mathrm{Q}^2v^6_\mathrm{c}} \delta(\mathbf{k_1}+\mathbf{k_2}+\mathbf{k_3})\frac{1}{(k_1k_2k_3)^3}\frac{(\mathbf{-k_2\cdot k_3})k^2_1}{(k_1+k_2+k_3)}\\\times\Bigg[2+\frac{k_2+k_3}{(k_1+k_2+k_3)}\Bigg]+ \text{\textsf{sym}}(k_1,k_2,k_3).\tag{262}\label{eq:262}
\end{align*}
\textbf{\tit{e}) \text{{\boldmath $\vvmathbb{R}_5$}} \tit{term}:} The \text{{\boldmath $\vvmathbb{R}_5$}} interaction part of the Hamiltonian is given by:
\begin{equation}
\tag{263}\label{eq:263}
\begin{multlined}
\mathcal{H}^{\text{{\boldmath $\vvmathbb{R}_5$}}}_\mathrm{int}(\tau')=\int\mathrm{d}^3\!x\,\Bigg[a(t)^3\text{{\boldmath $\vvmathbb{R}_5$}}(\partial^2{\mathcal{R}})(\partial\upchi)^2\Bigg]=\int\mathrm{d}^3\!x\,\Bigg[a(t)^3\text{{\boldmath $\vvmathbb{R}_5$}}(\partial^2{\mathcal{R}})(\partial_\mu\upchi)(\partial^\mu\upchi)\Bigg],
\end{multlined}
\end{equation}
such that,
\begin{align*}
\,\langle\bar\delta\mathcal{R}_\mathbf{k_1}\,\bar\delta\mathcal{R}_\mathbf{k_2}\,\bar\delta\mathcal{R}_\mathbf{k_3}\rangle_{\text{{\boldmath $\vvmathbb{R}_5$}}} = 
-i\text{{\boldmath $\vvmathbb{R}_5$}}\delta(\mathbf{k_1}+\mathbf{k_2}+\mathbf{k_3}){\displaystyle\int_{-\infinity}^{\tau\rightarrow 0}}\mathrm{d}\tau'\,a(\tau')^3\Bigg[\frac{(\mathbf{-k_1\cdot k_2})}{k^2_1k^2_2}(\partial_{\tau'}\{u_{k_1}(\tau)\,u_{k_1}^*(\tau')\})\\\times(\partial_{\tau'}\{u_{k_2}(\tau)\,u_{k_2}^*(\tau')\})\vspace{0.1in}({-k^2_3})\{u_{k_3}(\tau)\,u_{k_3}^*(\tau')\}-a(\tau')^3\frac{(\mathbf{-k_1\cdot k_2})}{k^2_1k^2_2}(\partial_{\tau'}\{u_{k_1}(\tau')\,u_{k_1}^*(\tau)\})(\partial_{\tau'}\{u_{k_2}(\tau')\,u_{k_2}^*(\tau)\})\\\times({-k^2_3})\{u_{k_3}(\tau')\,u_{k_3}^*(\tau)\}\Bigg]+ \text{\textsf{sym}}(k_1,k_2,k_3)\xrightarrow{\tau\rightarrow 0}
-i\text{{\boldmath $\vvmathbb{R}_5$}}\frac{\mathrm{H}^6}{2^6\mathrm{Q}^3v^9_\mathrm{c}} \delta(\mathbf{k_1}+\mathbf{k_2}+\mathbf{k_3})\frac{\mathrm{Q^2}\{\mathbf{-k_1\cdot k_2\}}}{(k_1k_2k_3)^3}\\ \times\Bigg[\Bigg\{e^{-i(k_1+k_2+k_3)v_\mathrm{c}\tau}{{\mathlarger{\mathlarger{\mathlarger{\mathlarger{\uppi}}}}}}_{i=1}^3(1+k_iv_\mathrm{c}
\tau')\vspace{0.1in}\int_{-\infinity}^{\tau\rightarrow 0}\mathrm{d}\tau'\,a(\tau')^3\Bigg[
e^{i(k_1+k_2+k_3)v_\mathrm{c}\tau'}\{1-k_3v_\mathrm{c}
\tau\}\{-k^2_3v^4_\mathrm{c}\tau'^2\}\\\times\Bigg]
\Bigg\}-\Bigg\{e^{i(k_1+k_2+k_3)v_\mathrm{c}\tau}{{\mathlarger{\mathlarger{\mathlarger{\mathlarger{\uppi}}}}}}_{i=1}^3(1-k_iv_\mathrm{c}\tau)\vspace{0.1in}\int_{-\infinity}^{\tau\rightarrow 0}\mathrm{d}\tau'\,a(\tau')^3\Bigg[
e^{-i(k_1+k_2+k_3)v_\mathrm{c}\tau'}\{1+k_3v_\mathrm{c}
\tau'\}\{-k^2_3v^4_\mathrm{c}\tau'^2\}\Bigg]\Bigg\}\Bigg] \\+ \text{\textsf{sym}}(k_1,k_2,k_3)=\text{{\boldmath $\vvmathbb{R}_5$}}\frac{\mathrm{H}^4}{2^4\mathrm{Q}v^6_\mathrm{c}} \delta(\mathbf{k_1}+\mathbf{k_2}+\mathbf{k_3})\frac{1}{(k_1k_2k_3)^3}\frac{(\mathbf{-k_1\cdot k_2})k^2_3}{(k_1+k_2+k_3)}\Bigg[1+\frac{k_3}{(k_1+k_2+k_3)}\Bigg]+\\ \text{\textsf{sym}}(k_1,k_2,k_3).
\tag{264}\label{eq:264}
\end{align*}
\textbf{\tit{f}) \text{{\boldmath $\vvmathbb{R}_6$}} \tit{term}:} The \text{{\boldmath $\vvmathbb{R}_6$}} interaction part of the Hamiltonian is given by:
\begin{equation}
\tag{265}\label{eq:265}
\begin{multlined}
\mathcal{H}^{\text{{\boldmath $\vvmathbb{R}_6$}}}_\mathrm{int}(\tau')=\int\mathrm{d}^3\!x\,\Bigg[a(t)\text{{\boldmath $\vvmathbb{R}_6$}}(\partial^2{\mathcal{R}})(\dt{\mathcal{R}})^2\Bigg]=\int\mathrm{d}^3\!x\,\Bigg[a(t)^{-1}\text{{\boldmath $\vvmathbb{R}_6$}}(\partial^\mu\partial_\mu{\mathcal{R}})(\partial_{\tau'}\mathcal{R})^2\Bigg].
\end{multlined}
\end{equation}
It follows that,
\begin{equation}
\tag{266}\label{eq:266}
\begin{multlined}
\,\langle\bar\delta\mathcal{R}_\mathbf{k_1}\,\bar\delta\mathcal{R}_\mathbf{k_2}\,\bar\delta\mathcal{R}_\mathbf{k_3}\rangle_{\text{{\boldmath $\vvmathbb{R}_6$}}} = 
-i\text{{\boldmath $\vvmathbb{R}_6$}}\delta(\mathbf{k_1}+\mathbf{k_2}+\mathbf{k_3}){\displaystyle\int_{-\infinity}^{\tau\rightarrow 0}}\mathrm{d}\tau'\,a(\tau')^{-1}[(\partial_{\tau'}\{u_{k_1}(\tau)\,u_{k_1}^*(\tau')\})(\partial_{\tau'}\{u_{k_2}(\tau)\,u_{k_2}^*(\tau')\})\\\times\vspace{0.1in}({-k^2_3})\{u_{k_3}(\tau)\,u_{k_3}^*(\tau')\}-a(\tau')^{-1}(\partial_{\tau'}\{u_{k_1}(\tau')\,u_{k_1}^*(\tau)\})(\partial_{\tau'}\{u_{k_2}(\tau')\,u_{k_2}^*(\tau)\})({-k^2_3})\{u_{k_3}(\tau')\,u_{k_3}^*(\tau)\}]+\\ \text{\textsf{sym}}(k_1,k_2,k_3)\xrightarrow{\tau\rightarrow 0}
-i\text{{\boldmath $\vvmathbb{R}_6$}}\frac{\mathrm{H}^6}{2^6\mathrm{Q}^3v^9_\mathrm{c}} \delta(\mathbf{k_1}+\mathbf{k_2}+\mathbf{k_3})\frac{1}{(k_1k_2k_3)^3} \Bigg[\Bigg\{e^{-i(k_1+k_2+k_3)v_\mathrm{c}\tau}{{\mathlarger{\mathlarger{\mathlarger{\mathlarger{\uppi}}}}}}_{i=1}^3(1+k_iv_\mathrm{c}
\tau)\times\vspace{0.1in}\\\int_{-\infinity}^{\tau\rightarrow 0}\mathrm{d}\tau'\,a(\tau')^{-1}\Bigg[
e^{i(k_1+k_2+k_3)v_\mathrm{c}\tau'}\{1-k_3v_\mathrm{c}
\tau'\}\{-k^2_1k^2_2k^2_3v^4_\mathrm{c}\tau'^2\}\Bigg]
\Bigg\}-\Bigg\{e^{i(k_1+k_2+k_3)v_\mathrm{c}\tau}{{\mathlarger{\mathlarger{\mathlarger{\mathlarger{\uppi}}}}}}_{i=1}^3(1-k_iv_\mathrm{c}\tau)\vspace{0.1in}\\\times\int_{-\infinity}^{\tau\rightarrow 0}\mathrm{d}\tau'\,a(\tau')^{-1}\Bigg[
e^{-i(k_1+k_2+k_3)v_\mathrm{c}\tau'}\{1+k_3v_\mathrm{c}
\tau'\}\{-k^2_1k^2_2k^2_3v^4_\mathrm{c}\tau'^2\}\Bigg]\Bigg\}\Bigg] + \text{\textsf{sym}}(k_1,k_2,k_3)=\\\text{{\boldmath $\vvmathbb{R}_6$}}\frac{\mathrm{H}^6}{2^2\mathrm{Q^3}v^8_\mathrm{c}} \delta(\mathbf{k_1}+\mathbf{k_2}+\mathbf{k_3})\frac{1}{(k_1k_2k_3)}\Bigg[\frac{3k_3}{(k_1+k_2+k_3)^4}\Bigg]+ \text{\textsf{sym}}(k_1,k_2,k_3)=\\\text{{\boldmath $\vvmathbb{R}_6$}}\frac{3\mathrm{H}^6}{2^2\mathrm{Q^3}v^8_\mathrm{c}} \delta(\mathbf{k_1}+\mathbf{k_2}+\mathbf{k_3})\frac{1}{(k_1k_2k_3)}\Bigg[\frac{1}{(k_1+k_2+k_3)^3}\Bigg].\;\;\;\;\;\;\;\;\;\;\;\;
\end{multlined}
\end{equation}
\textbf{\tit{g}) \text{{\boldmath $\vvmathbb{R}_7$}} \tit{term}:} The \text{{\boldmath $\vvmathbb{R}_7$}} interaction part of the Hamiltonian is given by:
\begin{equation}
\tag{267}\label{eq:267}
\begin{multlined}
\mathcal{H}^{\text{{\boldmath $\vvmathbb{R}_7$}}}_\mathrm{int}(\tau')=\int\mathrm{d}^3\!x\,a(t)^{-1}\text{{\boldmath $\vvmathbb{R}_7$}}\Bigg[(\partial\mathcal{R})^2(\partial^2\mathcal{R})-\mathcal{R}
\partial_\mu\partial_\nu\{\partial_\mu\mathcal{R}\}(\partial_\nu\mathcal{R})\Bigg],
\end{multlined}
\end{equation}
such that,
\begin{equation}
\tag{268}\label{eq:268}
\begin{multlined}
\,\langle\bar\delta\mathcal{R}_\mathbf{k_1}\,\bar\delta\mathcal{R}_\mathbf{k_2}\,\bar\delta\mathcal{R}_\mathbf{k_3}\rangle_{\text{{\boldmath $\vvmathbb{R}_7$}}} = 
-i\text{{\boldmath $\vvmathbb{R}_7$}}\delta(\mathbf{k_1}+\mathbf{k_2}+\mathbf{k_3})\Bigg[{\displaystyle\int_{-\infinity}^{\tau\rightarrow 0}}\mathrm{d}\tau'\,a(\tau')^{-1}\Bigg\{[(\{-\mathbf{k_1\cdot k_2}\}\{u_{k_1}(\tau)\,u_{k_1}^*(\tau')\})\;\;\;\;\;\;\;\;\\\times\{u_{k_2}(\tau)\,u_{k_2}^*(\tau')\}\vspace{0.1in}({-k^2_3})\{u_{k_3}(\tau)\,u_{k_3}^*(\tau')\}-(\{-\mathbf{k_1\cdot k_2}\}\{u_{k_1}(\tau')\,u_{k_1}^*(\tau)\})\\\times\{u_{k_2}(\tau')\,u_{k_2}^*(\tau)\}({-k^2_3})\{u_{k_3}(\tau')\,u_{k_3}^*(\tau)\}]+ \text{\textsf{sym}}(k_1,k_2,k_3)\Bigg\}-a(\tau')^{-1}
\Bigg\{[\{u_{k_1}(\tau)\,u_{k_1}^*(\tau')\}\{u_{k_2}(\tau)\,u_{k_2}^*(\tau')\}\vspace{0.1in}\\\times(\mathbf{-k_1\cdot k_2})(\mathbf{-k_1\cdot k_3})\{u_{k_3}(\tau)\,u_{k_3}^*(\tau')\}-\{u_{k_1}(\tau')\,u_{k_1}^*(\tau)\}\{u_{k_2}(\tau')\,u_{k_2}^*(\tau)\}(\mathbf{-k_1\cdot k_3})(\mathbf{-k_2\cdot k_3})\times\\\{u_{k_3}(\tau')\,u_{k_3}^*(\tau)\}]+\text{\textsf{sym}}(k_1,k_2,k_3)\Bigg\}\Bigg]=
-i\text{{\boldmath $\vvmathbb{R}_7$}}\delta(\mathbf{k_1}+\mathbf{k_2}+\mathbf{k_3})\{(-k^2_3)[(\mathbf{-k_1\cdot k_2})-(\mathbf{-k_2\cdot k_3})(\mathbf{-k_1\cdot k_3})]\}\\\times{\displaystyle\int_{-\infinity}^{\tau\rightarrow 0}}\mathrm{d}\tau'\,a(\tau')^{-1}\Bigg\{[\{u_{k_1}(\tau)\,u_{k_1}^*(\tau')\}\{u_{k_2}(\tau)\,u_{k_2}^*(\tau')\}\vspace{0.1in}\{u_{k_3}(\tau)\,u_{k_3}^*(\tau')\}-\{u_{k_1}(\tau')\,u_{k_1}^*(\tau)\}\\\times\{u_{k_2}(\tau')\,u_{k_2}^*(\tau)\}\{u_{k_3}(\tau')\,u_{k_3}^*(\tau)\}]+ \text{\textsf{sym}}(k_1,k_2,k_3)\Bigg\}
\xrightarrow{\tau\rightarrow 0}
-i\text{{\boldmath $\vvmathbb{R}_7$}}\\\times\frac{\mathrm{H}^6}{2^6\mathrm{Q}^3v^9_\mathrm{c}} \delta(\mathbf{k_1}+\mathbf{k_2}+\mathbf{k_3})\frac{1}{(k_1k_2k_3)^3}\{(k^2_3)[(\mathbf{k_1\cdot k_2})-(\mathbf{k_2\cdot k_3})(\mathbf{k_1\cdot k_3})]\} \Bigg[\Bigg\{e^{-i(k_1+k_2+k_3)v_\mathrm{c}\tau}\\\times{{\mathlarger{\mathlarger{\mathlarger{\mathlarger{\uppi}}}}}}_{i=1}^3(1+k_iv_\mathrm{c}\tau)\vspace{0.1in}\int_{-\infinity}^{\tau\rightarrow 0}\mathrm{d}\tau'\, a(\tau')^{-1}\Bigg[
e^{i(k_1+k_2+k_3)v_\mathrm{c}\tau'}{\mathlarger{\mathlarger{\mathlarger{\mathlarger{\uppi}}}}}_{i=1}^3(1-k_iv_\mathrm{c}\tau')\Bigg]\Bigg\}-\\ \Bigg\{e^{i(k_1+k_2+k_3)v_\mathrm{c}\tau}{{\mathlarger{\mathlarger{\mathlarger{\mathlarger{\uppi}}}}}}_{i=1}^3(1-k_iv_\mathrm{c}\tau)\vspace{0.1in}\int_{-\infinity}^{\tau\rightarrow 0}\mathrm{d}\tau'\, a(\tau')^{-1}\Bigg[
e^{-i(k_1+k_2+k_3)v_\mathrm{c}\tau'}{\mathlarger{\mathlarger{\mathlarger{\mathlarger{\uppi}}}}}_{i=1}^3
(1+k_iv_\mathrm{c}\tau')\Bigg]\Bigg\}\Bigg]=\\
-i\text{{\boldmath $\vvmathbb{R}_7$}}\frac{\mathrm{H}^6}{2^6\mathrm{Q}^3v^9_\mathrm{c}} \delta(\mathbf{k_1}+\mathbf{k_2}+\mathbf{k_3})\frac{1}{(k_1k_2k_3)^3}\{(k^2_3)[(\mathbf{k_1\cdot k_2})-(\mathbf{k_2\cdot k_3})(\mathbf{k_1\cdot k_3})]\}\Bigg[\Bigg\{\int_{-\infinity}^{\tau\rightarrow 0}\mathrm{d}\tau'\, a(\tau')^{-1}\\\times\Bigg[
e^{i(k_1+k_2+k_3)v_\mathrm{c}\tau'}(1-i\{k_1+k_2+k_3\}v_\mathrm{c}\tau'-\{
k_1k_2+k_1k_3+k_2k_3\}v^2_\mathrm{c}\tau'^2+ik_1k_2k_3v^3_\mathrm{c}\tau'^3)
\Bigg]\Bigg\}-\\\Bigg\{\int_{-\infinity}^{\tau\rightarrow 0}\mathrm{d}\tau'\, a(\tau')^{-1}\Bigg[
e^{-i(k_1+k_2+k_3)v_\mathrm{c}\tau'}(1+i\{k_1+k_2+k_3\}v_\mathrm{c}\tau'-\{
k_1k_2+k_1k_3+k_2k_3\}v^2_\mathrm{c}\tau'^2-\\ik_1k_2k_3v^3_\mathrm{c}
\tau'^3\Bigg]\Bigg\}\Bigg]=\text{{\boldmath $\vvmathbb{R}_7$}}\times\frac{\mathrm{H}^6}{2^3\mathrm{Q}^3v^{10}_\mathrm{c}} \delta(\mathbf{k_1}+\mathbf{k_2}+\mathbf{k_3})\frac{1}{(k_1k_2k_3)^3}\{(k^2_3)[(\mathbf{k_1\cdot k_2})-(\mathbf{k_2\cdot k_3})(\mathbf{k_1\cdot k_3})]\}\\\times\Bigg[\frac{1}{(k_1+k_2+k_3)}+\frac{k_1k_2+k_1k_3+k_2k_3}{(k_1+k_2+k_3)^2}+ 3\frac{k_1k_2k_3}{(k_1+k_2+k_3)^3}\Bigg]+\text{\textsf{sym}}(k_1,k_2,k_3).\;\;\;\;\;
\end{multlined}
\end{equation}
\textbf{\tit{h}) \text{{\boldmath $\vvmathbb{R}_8$}} \tit{term}:} The \text{{\boldmath $\vvmathbb{R}_8$}} interaction part of the Hamiltonian is given by:
\begin{equation}
\tag{269}\label{eq:269}
\begin{multlined}
\mathcal{H}^{\text{{\boldmath $\vvmathbb{R}_8$}}}_\mathrm{int}(\tau')=\int\mathrm{d}^3\!x\,a(t)\text{{\boldmath $\vvmathbb{R}_8$}}\Bigg[(\partial_\mu\mathcal{R})(\partial_\mu\upchi)(\partial^2\mathcal{R})-\mathcal{R}
\partial_\mu\partial_\nu\{\partial_\mu\mathcal{R}\}(\partial_\nu{\upchi})\Bigg].
\end{multlined}
\end{equation}
The appearance of $\upchi$ field leads to similar degeneracy as in the case of \text{{\boldmath $\vvmathbb{R}_6$}} interaction term. The two degenerate contributions are given by
\begin{align*}
\langle\bar\delta\mathcal{R}_\mathbf{k_1}\,\bar\delta{\upchi}_\mathbf{k_2}\,\bar\delta\mathcal{R}_\mathbf{k_3}\rangle_{\text{{\boldmath $\vvmathbb{R}_8$}}}=
= 
-i\text{{\boldmath $\vvmathbb{R}_8$}}\,\mathrm{Q}\,\delta(\mathbf{k_1}+\mathbf{k_2}+\mathbf{k_3})\Bigg[{\displaystyle\int_{-\infinity}^{\tau\rightarrow 0}}\mathrm{d}\tau'\,a(\tau')\frac{\{-\mathbf{k_1\cdot k_2}\}}{-k^2_2}\Bigg\{[\{u_{k_1}(\tau)\,u_{k_1}^*(\tau')\})\;\;\;\;\;\;\;\;\\\times\{u_{k_2}(\tau)\,u_{k_2}^*(\tau')\}\vspace{0.1in}({-k^2_3})\{u_{k_3}(\tau)\,u_{k_3}^*(\tau')\}-\{u_{k_1}(\tau')\,u_{k_1}^*(\tau)\}\;\;\;\;\;\;\;\;\;\\\times\{u_{k_2}(\tau')\,u_{k_2}^*(\tau)\}({-k^2_3})\{u_{k_3}(\tau')\,u_{k_3}^*(\tau)\}]+ \text{\textsf{sym}}(k_1,k_2,k_3)\Bigg\}-a(\tau')\frac{1}{-k^2_2}
\Bigg\{[\{u_{k_1}(\tau)\,u_{k_1}^*(\tau')\}\{u_{k_2}(\tau)\,u_{k_2}^*(\tau')\}\vspace{0.1in}\\\times(\mathbf{-k_1\cdot k_2})(\mathbf{-k_1\cdot k_3})\{u_{k_3}(\tau)\,u_{k_3}^*(\tau')\}-\{u_{k_1}(\tau')\,u_{k_1}^*(\tau)\}\{u_{k_2}(\tau')\,u_{k_2}^*(\tau)\}(\mathbf{-k_1\cdot k_3})(\mathbf{-k_2\cdot k_3})\times\\\{u_{k_3}(\tau')\,u_{k_3}^*(\tau)\}]+\text{\textsf{sym}}(k_1,k_2,k_3)\Bigg\}\Bigg]=
-i\text{{\boldmath $\vvmathbb{R}_7$}}\delta(\mathbf{k_1}+\mathbf{k_2}+\mathbf{k_3})\{(-k^2_3)[(\mathbf{-k_1\cdot k_2})-(\mathbf{-k_2\cdot k_3})(\mathbf{-k_1\cdot k_3})]\}\\\times{\displaystyle\int_{-\infinity}^{\tau\rightarrow 0}}\mathrm{d}\tau'\,a(\tau')^{-1}\Bigg\{[\{u_{k_1}(\tau)\,u_{k_1}^*(\tau')\}\{u_{k_2}(\tau)\,u_{k_2}^*(\tau')\}\vspace{0.1in}\{u_{k_3}(\tau)\,u_{k_3}^*(\tau')\}-\{u_{k_1}(\tau')\,u_{k_1}^*(\tau)\}\;\;\;\;\;\;\;\;\;\;\;\;\;\;\;\;\;\;\\\times\{u_{k_2}(\tau')\,u_{k_2}^*(\tau)\}\{u_{k_3}(\tau')\,u_{k_3}^*(\tau)\}]+ \text{\textsf{sym}}(k_1,k_2,k_3)\Bigg\}
\xrightarrow{\tau\rightarrow 0}
-i\text{{\boldmath $\vvmathbb{R}_8$}}\;\;\;\;\;\;\;\;\;\;\;\;\\\times\frac{\mathrm{H}^6}{2^6\mathrm{Q}^2v^9_\mathrm{c}} \delta(\mathbf{k_1}+\mathbf{k_2}+\mathbf{k_3})\frac{1}{(k_1k_2k_3)^3}\{(k^2_3)[(\mathbf{k_1\cdot k_2})-(\mathbf{k_2\cdot k_3})(\mathbf{k_1\cdot k_3})]\} \Bigg[\Bigg\{e^{-i(k_1+k_2+k_3)v_\mathrm{c}\tau}\\\times{{\mathlarger{\mathlarger{\mathlarger{\mathlarger{\uppi}}}}}}_{i=1}^3(1+k_iv_\mathrm{c}\tau)\vspace{0.1in}\int_{-\infinity}^{\tau\rightarrow 0}\mathrm{d}\tau'\, a(\tau')\Bigg[
e^{i(k_1+k_2+k_3)v_\mathrm{c}\tau'}(1-k_1v_\mathrm{c}\tau')(1-k_3v_\mathrm{c}\tau')(-v^2_\mathrm{c}\tau')\Bigg]\Bigg\}-\\ \Bigg\{e^{i(k_1+k_2+k_3)v_\mathrm{c}\tau}{{\mathlarger{\mathlarger{\mathlarger{\mathlarger{\uppi}}}}}}_{i=1}^3(1-k_iv_\mathrm{c}\tau)\vspace{0.1in}\int_{-\infinity}^{\tau\rightarrow 0}\mathrm{d}\tau'\, a(\tau')\Bigg[
e^{-i(k_1+k_2+k_3)v_\mathrm{c}\tau'}{\mathlarger{\mathlarger{\mathlarger{\mathlarger{\uppi}}}}}_{i=1}^3
(1+k_1v_\mathrm{c}\tau')(1+k_3v_\mathrm{c}\tau')\\\times(-v^2_\mathrm{c}\tau')\Bigg]\Bigg\}\Bigg]=
-i\text{{\boldmath $\vvmathbb{R}_8$}}\frac{\mathrm{H}^6}{2^6\mathrm{Q}^2v^9_\mathrm{c}} \delta(\mathbf{k_1}+\mathbf{k_2}+\mathbf{k_3})\frac{1}{(k_1k_2k_3)^3}\{(k^2_3)[(\mathbf{k_1\cdot k_2})-(\mathbf{k_2\cdot k_3})(\mathbf{k_1\cdot k_3})]\}\Bigg[\Bigg\{\\\times\int_{-\infinity}^{\tau\rightarrow 0}\mathrm{d}\tau'\, a(\tau')\Bigg[
e^{i(k_1+k_2+k_3)v_\mathrm{c}\tau'}(1-i\{k_1+k_3\}v_\mathrm{c}\tau'-
k_1k_3v^2_\mathrm{c}\tau'^2)v^2_\mathrm{c}\tau'
\Bigg]\Bigg\}-\;\;\;\;\;\;\;\;\;\;\;\;\;\;\\\Bigg\{\int_{-\infinity}^{\tau\rightarrow 0}\mathrm{d}\tau'\, a(\tau')\Bigg[
e^{-i(k_1+k_2+k_3)v_\mathrm{c}\tau'}(1+i\{k_1+k_3\}v_\mathrm{c}\tau'-
k_1k_3v^2_\mathrm{c}\tau'^2)v^2_\mathrm{c}
\tau'\Bigg]\Bigg\}\Bigg]=\;\;\;\;\;\;\;\;\;\;\;\;\;\;\;\;\;\;\;\;\;\\\text{{\boldmath $\vvmathbb{R}_8$}}\times\frac{\mathrm{H}^5}{2^5\mathrm{Q}^2v^{8}_\mathrm{c}} \delta(\mathbf{k_1}+\mathbf{k_2}+\mathbf{k_3})\frac{1}{(k_1k_2k_3)^3}\{(k^2_3)[(\mathbf{k_1\cdot k_2})-(\mathbf{k_2\cdot k_3})(\mathbf{k_1\cdot k_3})]\}\\\times\Bigg[\frac{1}{(k_1+k_2+k_3)}+\frac{k_1+k_3}{(k_1+k_2+k_3)^2}+ 2\frac{k_1k_3}{(k_1+k_2+k_3)^3}\Bigg]+\text{\textsf{sym}}(k_1,k_2,k_3),\;\;\;\;\;\tag{270}\label{eq:270}
\end{align*}
and,
\begin{equation}
\tag{271}\label{eq:271}
\begin{multlined}
\langle\bar\delta\mathcal{R}_\mathbf{k_1}\,\bar\delta\mathcal{R}_\mathbf{k_2}\,\bar\delta{\upchi}_\mathbf{k_3}\rangle_{\text{{\boldmath $\vvmathbb{R}_8$}}}
= \text{{\boldmath $\vvmathbb{R}_8$}}\times\frac{\mathrm{H}^5}{2^5\mathrm{Q}^2v^{8}_\mathrm{c}} \delta(\mathbf{k_1}+\mathbf{k_2}+\mathbf{k_3})\frac{1}{(k_1k_2k_3)^3}\{(k^2_3)[(\mathbf{k_1\cdot k_2})-(\mathbf{k_2\cdot k_3})(\mathbf{k_1\cdot k_3})]\}\\\times\Bigg[\frac{1}{(k_1+k_2+k_3)}+\frac{k_3+k_2}{(k_1+k_2+k_3)^2}+ 2\frac{k_3k_2}{(k_1+k_2+k_3)^3}\Bigg]+\text{\textsf{sym}}(k_1,k_2,k_3),
\end{multlined}
\end{equation}
such that the overall contribution by the \text{{\boldmath $\vvmathbb{R}_8$}} term is given by:
\begin{equation}
\tag{272}\label{eq:272}
\begin{multlined}
\langle\bar\delta\mathcal{R}_\mathbf{k_1}\,\bar\delta\mathcal{R}_\mathbf{k_2}\,\bar\delta\mathcal{R}_\mathbf{k_3}\rangle_{\text{{\boldmath $\vvmathbb{R}_8$}}}
= \text{{\boldmath $\vvmathbb{R}_8$}}\times\frac{\mathrm{H}^5}{2^5\mathrm{Q}^2v^{8}_\mathrm{c}} \delta(\mathbf{k_1}+\mathbf{k_2}+\mathbf{k_3})\frac{1}{(k_1k_2k_3)^3}\{(k^2_3)[(\mathbf{k_1\cdot k_2})-(\mathbf{k_2\cdot k_3})(\mathbf{k_1\cdot k_3})]\}\\\times\Bigg[\frac{2}{(k_1+k_2+k_3)}+\frac{2k_3+k_2+k_1}{(k_1+k_2+k_3)^2}+ 2\frac{k_3k_2+k_3k_1}{(k_1+k_2+k_3)^3}\Bigg]+\text{\textsf{sym}}(k_1,k_2,k_3).
\end{multlined}
\end{equation}
\textbf{\tit{i}) \text{{\boldmath $\vvmathbb{R}_9$}} \tit{term}:} The \text{{\boldmath $\vvmathbb{R}_9$}} interaction part of the Hamiltonian is given by:
\begin{align*}
\mathcal{H}^{\text{{\boldmath $\vvmathbb{R}_9$}}}_\mathrm{int}(t)=\int\mathrm{d}^3\!x\,\Bigg\{-2\Bigg[\frac{1}{{\Gamma^2_1}}\Bigg\{\beth_{22}+\beth_{9}\Bigg[2\frac{{\Gamma_1}}{{\Gamma_2}}\Bigg]\Bigg\}\Bigg\{(\partial_k\mathcal{R})(\partial_k\upchi)-\frac{1}{\partial^2[\partial_\mu\partial_\nu\{(\partial_\mu\mathcal{R})(\partial_\nu\upchi\}]}\Bigg\}-2\frac{
{\Gamma_1}}{v^2_{\mathrm{c}}{\Gamma_2}}\mathcal{R}\dt{\mathcal{R}}+\\\frac{1}{{{2\Gamma_2}}a(t)^2}\Bigg[
\beth_{10}-2\Bigg\{\beth_{22}+\beth_{9}\Bigg[2\frac{{\Gamma_1}}{{\Gamma_2}}\Bigg]\Bigg\}\Bigg]\Bigg\{(\partial_k\mathcal{R})^2-\frac{1}{\partial^2[\partial_\mu\partial_\nu\{(\partial_\mu\mathcal{R})(\partial_\nu\mathcal{R}\}]}\Bigg\}\Bigg]\Bigg[a(t)^3(\dt{\mathrm{Q}_3}\dt{\mathcal{R}}
+\mathrm{Q_3}\ddt{\mathcal{R}})+\\3a(t)^2\dt{a}(t)\mathrm{Q_3}\dt{\mathcal{R}} - a(t)\mathrm{Q_3}v^2_\mathrm{c}(\partial^2\mathcal{R})\Bigg]\Bigg\}.\;\;\;\;\;
\tag{273}\label{eq:273}
\end{align*}
It can be shown that the contribution from the \text{{\boldmath $\vvmathbb{R}_9$}} term is very small compared to the other terms because the accompanying coefficients to the terms of $\mathcal{R}$, $\upchi$ and their derivatives are of the order of $\mathcal{O}(\epsilon^2_i)$ \citep{r6}. Hence, we ignore the contribution from the \text{{\boldmath $\vvmathbb{R}_9$}} term.
\subsubsection{Bi-spectrum [{\boldmath $\mathrm{B}_\mathrm{R}$}]:}
\label{br}
The definition of the bi-spectrum $\mathrm{B}_\mathrm{R}(k_1,k_2,k_3)$ is simply written as
\begin{equation}
\tag{274}\label{eq:274}
\begin{multlined}
\langle\bar\delta\mathcal{R}_\mathbf{k_1}\,\bar\delta\mathcal{R}_\mathbf{k_2}\,\bar\delta{\upchi}_\mathbf{k_3}\rangle=\delta(\mathbf{k_1}+\mathbf{k_2}+\mathbf{k_3})\mathrm{B}_\mathrm{R}(k_1,k_2,k_3).
\end{multlined}
\end{equation}
Using the previously calculated results in section \ref{intbi}, we write
\begin{equation}
\tag{275}\label{eq:275}
\begin{multlined}
\mathrm{B}_\mathrm{R}(k_1,k_2,k_3)=\frac{1}{(k_1k_2k_3)^3}\Bigg[\frac{\mathrm{H}^2}{8\uppi^2\mathrm{Q}\,v^3_\mathrm{c}}\Bigg]^2\Bigg[\frac{1}{2\mathrm{Q}(k_1+k_2+k_3)}\Bigg\{\sum_{\mu>\nu}k^2_\mu k^2_\nu-\frac{1}{2(k_1+k_2+k_3)}\sum_{\mu\neq\nu}k^2_\mu k^3_\nu\Bigg\} \text{{\boldmath $\vvmathbb{R}_1$}}+\\
 \frac{1}{4v^2_\mathrm{c}}\Bigg\{\frac{1}{2}\sum_{\mu}k^3_\mu +\frac{2}{(k_1+k_2+k_3)}\sum_{\mu>\nu}k^2_\mu k^2_\nu-\frac{1}{(k_1+k_2+k_3)^2}\sum_{\mu\neq\nu}k^2_\mu k^3_\nu\Bigg\} \text{{\boldmath $\vvmathbb{R}_2$}}+
\frac{3\mathrm{H}}{2(k_1+k_2+k_3)^3}{k^2_1k^2_2k^2_3} \text{{\boldmath $\vvmathbb{R}_3$}}+\\
\frac{\mathrm{Q}}{8}\Bigg\{\sum_{\mu}k^3_\mu +\frac{1}{2}\sum_{\mu\neq\nu}k_\mu k^2_\nu-\frac{2}{(k_1+k_2+k_3)^2}\sum_{\mu\neq\nu}k^2_\mu k^3_\nu\Bigg\} \text{{\boldmath $\vvmathbb{R}_4$}}+
\frac{\mathrm{Q}^2}{4(k_1+k_2+k_3)^2}\Bigg\{\sum_{\mu}k^5_\mu +\frac{1}{2}\sum_{\mu\neq\nu}k_\mu k^4_\nu\\-\frac{3}{2}\sum_{\mu\neq\nu}k^2_\mu k^3_\nu - k_1k_2k_3\sum_{\mu>\nu}k_\mu k_\nu\Bigg\} \text{{\boldmath $\vvmathbb{R}_5$}}+\frac{3\mathrm{H}^2}{v^2_\mathrm{c}(k_1+k_2+k_3)^3}{k^2_1k^2_2k^2_3} \text{{\boldmath $\vvmathbb{R}_6$}}+
\frac{\mathrm{H}^2}{2v^4_\mathrm{c}(k_1+k_2+k_3)}\Bigg\{1+\\ \frac{1}{(k_1+k_2+k_3)^2}\sum_{\mu>\nu}k_\mu k_\nu +\frac{3k_1k_2k_3}{(k_1+k_2+k_3)^3}\Bigg\}\Bigg\{\frac{3}{4}\sum_{\mu}k^4_\mu+\frac{3}{2}\sum_{\mu>\nu}k^2_\mu k^2_\nu \Bigg\} \text{{\boldmath $\vvmathbb{R}_7$}}+\frac{\mathrm{H}\,\mathrm{Q}}{8v^2_\mathrm{c}(k_1+k_2+k_3)^2}\\\times\Bigg\{\frac{3k_1k_2k_3}{2}\sum_{\mu}k^2_\mu -\frac{5}{2}k_1k_2k_3(k_1+k_2+k_3)^2-6\sum_{\mu\neq\nu}k^2_\mu k^3_\nu-
\sum_{\mu}k^5_\mu+\frac{7}{2}(k_1+k_2+k_3)\sum_{\mu}k^4_\mu\Bigg\} \text{{\boldmath $\vvmathbb{R}_8$}}
\Bigg],
\end{multlined}
\end{equation}
where, the non-linearity parameter $f_\mathrm{NL}$ is given by \citep{r13},
\begin{equation}
\tag{276}\label{eq:276}
\begin{multlined}
f_\mathrm{NL}=\frac{10}{3}{(k_1k_2k_3)^3}\Bigg[\frac{\mathrm{H}^2}{8\uppi^2\mathrm{Q}\,v^3_\mathrm{c}}\Bigg]^{-2}\mathrm{B}_\mathrm{R}(k_1,k_2,k_3).
\end{multlined}
\end{equation}
\appendix
\section{Appendix}
\subsection{Unitary operators}
\label{A.1}
Relevant to our discussion on Interaction picture, we give the basic identities from Linear Algebra.\\

{\noindent}For an operator $\mathrm{A}$ in some Hilbert space, the exponential $e^{\mathrm{A}}$ follows the following properties:\\ \\
1. $e^{\mathrm{A}}$ is Hermitian, if $\mathrm{A}$ is Hermitian.\\
2. $e^{\mathrm{A}}$ is Unitary, if $\mathrm{A}$ is skew-Hermitian. In other words, a Unitary operator is not necessarily Hermitian i.e. $\mathrm{U}\mathrm{U}^\dagger \neq 1$.\\
\subsection{General formulation for the Interaction Hamiltonian}
\label{A.2}
In the discussion in section \ref{quaside}, we had concluded that $\mathcal{H}_\mathrm{NI}$, $\mathcal{H}_\mathrm{int}$ and $\mathcal{H}_\mathcal{S}$ commute in our case of the perturbed Hamiltonian. However, this may not always be the case. For non-commutating set of Hamiltonians $\mathcal{H}_\mathrm{NI}$, $\mathcal{H}_\mathrm{int}$ and $\mathcal{H}_\mathcal{S}$, we can re-arrange the expressions \eqref{eq:83}, \eqref{eq:84} and \eqref{eq:85} while the identity \eqref{eq:82} doesn't hold\footnote{Remember that $\mathcal{H}_\mathrm{int} \equiv \mathcal{H}_\mathrm{IG}$.}! For some operator $\mathcal{O}$, we follow:
\begin{equation}
\tag{A.2.1}
\label{eq:A.2.1}
\mathcal{O}_{_\mathrm{\scriptstyle H}} = e^{i\mathcal{H}_\mathcal{S}t}\,e^{-i\mathcal{H}_\mathrm{NI}t}\,\mathcal{O}_{_\mathrm{\scriptstyle I}}\,e^{i\mathcal{H}_\mathrm{NI}t}\,e^{-i\mathcal{H}_\mathcal{S}t} = \hat{\mathrm{U}}^\dagger \mathcal{O}_{_\mathrm{\scriptstyle I}}, \hat{\mathrm{U}}
\end{equation}
such that the unitary propagator is given by
\begin{equation}
\tag{A.2.2}
\label{eq:A.2.2}
\hat{\mathrm{U}} = e^{i\mathcal{H}_\mathrm{NI}t}\,e^{-i\mathcal{H}_\mathcal{S}t} \neq e^{-i\mathcal{H}_\mathrm{IG}t}.
\end{equation}
Further, we can reduce the above result to a differential equation,
\begin{equation*}
i\frac{\partial\hat{\mathrm{U}}}{\partial t} = e^{i\mathcal{H}_\mathrm{NI}t}\,{(\mathcal{H}_\mathcal{S} - \mathcal{H}_\mathrm{NI})}\,e^{-i\mathcal{H}_\mathcal{S}t},
\end{equation*}
\begin{equation*}
i\frac{\partial\hat{\mathrm{U}}}{\partial t} = e^{i\mathcal{H}_\mathrm{NI}t}\,{ \mathcal{H}_\mathrm{IG}}\,e^{-i\mathcal{H}_\mathcal{S}t},
\end{equation*}
\begin{equation*}
i\frac{\partial\hat{\mathrm{U}}}{\partial t} = e^{i\mathcal{H}_\mathrm{NI}t}\,{ \mathcal{H}_\mathrm{IG}}\,e^{-i\mathcal{H}_\mathrm{NI}t}\,e^{i\mathcal{H}_\mathrm{NI}t}\,e^{-i\mathcal{H}_\mathcal{S}t},
\end{equation*}
\begin{equation*}
i\frac{\partial\hat{\mathrm{U}}}{\partial t} = e^{i\mathcal{H}_\mathrm{NI}t}\,{ \mathcal{H}_\mathrm{IG}}\,e^{-i\mathcal{H}_\mathrm{NI}t}\,\hat{\mathrm{U}},
\end{equation*}
\begin{equation}
\tag{A.2.3}
\label{eq:A.2.3}
i\frac{\partial\hat{\mathrm{U}}}{\partial t} = \mathcal{H}_\mathrm{IG}'\,\hat{\mathrm{U}},
\end{equation}
where, $\mathcal{H}_\mathrm{IG}'$ is naturally the interaction Hamiltonian in the interaction picture. The solution to this differential equation is simply given by\footnote{Peskin and Schr\"oder: \tit{An introduction to Quantum Field Theory}, Chapter \tit{4}, Sec: \tit{4.2}, Page: \tit{84--86}.}:
\begin{equation}
\tag{A.2.4}
\label{eq:A.2.4}
\hat{\mathrm{U}}(t,t_\mathrm{o}) = \mathlarger{\mathrm{T}}\,\Bigg [e^{{-i\displaystyle\int_{t_\mathrm{o}}^{t}}\mathcal{H}_\mathrm{IG}'(t')\,\mathrm{d}t'}\Bigg]\indent\text{given,}\indent t_\mathrm{o}<t,
\end{equation} 
where, $\mathlarger{\mathrm{T}}$ stands for path-ordering$^{16}$ of operators. It is easy to follow from \eqref{eq:81} that $\mathcal{H}_\mathrm{IG}' = \mathcal{H}_\mathrm{IG}$ when $[\mathcal{H}_{\mathrm{IG}},\,\mathcal{H}_{\mathrm{NI}}] = 0$.\\
\begin{flushleft}\myrule[line width = 0.5mm]{fast cap reversed}{fast cap reversed}\end{flushleft}
\tit{\textbf{Sidenote}}:\\
The idea behind path-ordering of operators is very crucial to our discussion. Let us take a closer look at this. When solved iteratively, one can arrive at the following solution for differential equation in \eqref{eq:A.2.3}:
\begin{equation}
\begin{multlined}
\tag{A.2.5}
\label{eq:A.2.5}
\hat{\mathrm{U}}(t,t_\mathrm{o}) = \mathcal{I} + (-i)\int_{t_\mathrm{o}}^{t}\mathrm{d}t_1 \mathcal{H}_\mathrm{IG}'(t_1) + (-i)^2\int_{t_\mathrm{o}}^{t}\mathrm{d}t_1 \mathcal{H}_\mathrm{IG}'(t_1)
\int_{t_\mathrm{o}}^{t_1}\mathrm{d}t_2 \mathcal{H}_\mathrm{IG}'(t_2) +\, ...\\ \,+ (-i)^\mathrm{n}\int_{t_\mathrm{o}}^{t}\mathrm{d}t_1 \mathcal{H}_\mathrm{IG}'(t_1)
\int_{t_\mathrm{o}}^{t_{\mathrm{n}-1}}\mathrm{d}t_{\mathrm{n}-1}\, ... \, \mathcal{H}_\mathrm{IG}'(t_{\mathrm{n}-1})
\int_{t_\mathrm{o}}^{t_\mathrm{n}}\mathrm{d}t_\mathrm{n} \mathcal{H}_\mathrm{IG}'(t_\mathrm{n}),
\end{multlined}
\end{equation} 
which can be further written as
\begin{equation}
\begin{multlined}
\tag{A.2.6}
\label{eq:A.2.6}
\hat{\mathrm{U}}(t,t_\mathrm{o}) = \mathcal{I} + (-i)\int_{t_\mathrm{o}}^{t}\mathrm{d}t_1 \mathcal{H}_\mathrm{IG}'(t_1) + (-i)^2\int_{t_\mathrm{o}}^{t}\mathrm{d}t_1
\int_{t_\mathrm{o}}^{t_1}\mathrm{d}t_2 \mathcal{H}_\mathrm{IG}'(t_1)
 \mathcal{H}_\mathrm{IG}'(t_2) +\, ...\\ \,+ (-i)^\mathrm{n}\int_{t_\mathrm{o}}^{t}\mathrm{d}t_1
 \, ... \,
 \int_{t_\mathrm{o}}^{t_{\mathrm{n}-1}}\mathrm{d}t_{\mathrm{n}-1}
 \int_{t_\mathrm{o}}^{t_\mathrm{n}}\mathrm{d}t_\mathrm{n} \mathcal{H}_\mathrm{IG}'(t_1)
 \, ... \,
 \mathcal{H}_\mathrm{IG}'(t_{\mathrm{n}-1})
 \mathcal{H}_\mathrm{IG}'(t_\mathrm{n}).
\end{multlined}
\end{equation}
Remember that the interval over the which the integrals are performed is the same for every integral, i.e. $(t_\mathrm{o},t),\,(t_\mathrm{o},t_1)\,... \, (t_\mathrm{o},t_\mathrm{n}) \text{ etc all span the same interval}$. Now, consider the higher order terms only, i.e. $\displaystyle{(-i)^\mathrm{n}\int_{t_\mathrm{o}}^{t}\mathrm{d}t_1
 \, ... \,
 \int_{t_\mathrm{o}}^{t_{\mathrm{n}-1}}\mathrm{d}t_{\mathrm{n}-1}
 \int_{t_\mathrm{o}}^{t_\mathrm{n}}\mathrm{d}t_\mathrm{n}\, \mathcal{H}_\mathrm{IG}'(t_1)
 \, ... \,
 \mathcal{H}_\mathrm{IG}'(t_{\mathrm{n}-1})\,
 \mathcal{H}_\mathrm{IG}'(t_\mathrm{n})}$. The order of operators appearing in the iterative expansion becomes very important, which is not the case when it comes to functions in linear algebraic calculations. This is simply because we do not intend to impose on the Hamiltonian the condition that it should commute at different times. Hence, the ordering becomes very important. In order to make the expression simpler, we can rewrite this term as a sum of all possible permutations $\sum P(\mathcal{H}_\mathrm{IG}'(t_1)
 \, ... \,
 \mathcal{H}_\mathrm{IG}'(t_{\mathrm{n}-1})\,
 \mathcal{H}_\mathrm{IG}'(t_\mathrm{n}))$ ordered in the labels of  time-arguments using a series of heaviside step functions $\uptheta(t_1 - t_2)\,\uptheta(t_2 - t_3)\,...\,\uptheta(t_{\mathrm{n}-1} - t_{\mathrm{n}})$, and then eventually using the following definition of path-ordering:
\begin{equation}
\begin{multlined}
\tag{A.2.7}
\label{eq:A.2.7}
 \mathrm{T}[\mathcal{H}_\mathrm{IG}'(t_1)
 \, ... \,
 \mathcal{H}_\mathrm{IG}'(t_{\mathrm{n}-1})
 \mathcal{H}_\mathrm{IG}'(t_\mathrm{n})] = \uptheta(t_1 - t_2)\,\uptheta(t_2 - t_3)\,...\,\uptheta(t_{\mathrm{n}-1} - t_{\mathrm{n}})\, \\ \sum P[\mathcal{H}_\mathrm{IG}'(t_1)
 \, ... \,
 \mathcal{H}_\mathrm{IG}'(t_{\mathrm{n}-1})
 \mathcal{H}_\mathrm{IG}'(t_\mathrm{n})],
\end{multlined}
\end{equation}
and the identity formally known as the Dyson expansion,
\begin{equation}
\begin{multlined}
\tag{A.2.8}
\label{eq:A.2.8}
 (-i)^\mathrm{n}\int_{t_\mathrm{o}}^{t}\mathrm{d}t_1
 \, ... \,
 \int_{t_\mathrm{o}}^{t_{\mathrm{n}-1}}\mathrm{d}t_{\mathrm{n}-1}
 \int_{t_\mathrm{o}}^{t_\mathrm{n}}\mathrm{d}t_\mathrm{n} \mathcal{H}_\mathrm{IG}'(t_1)
 \, ... \,
 \mathcal{H}_\mathrm{IG}'(t_{\mathrm{n}-1})
 \mathcal{H}_\mathrm{IG}'(t_\mathrm{n}) = \\ \frac{1}{\mathrm{n}!}(-i)^\mathrm{n}\int_{t_\mathrm{o}}^{t}\mathrm{d}t_1
 \, ... \,
 \int_{t_\mathrm{o}}^{t}\mathrm{d}t_{\mathrm{n}-1}
 \int_{t_\mathrm{o}}^{t}\mathrm{d}t_\mathrm{n} \, [\mathcal{H}_\mathrm{IG}'(t_1)
 \, ... \,
 \mathcal{H}_\mathrm{IG}'(t_{\mathrm{n}-1})
 \mathcal{H}_\mathrm{IG}'(t_\mathrm{n})],
\end{multlined}
\end{equation}
which can easily be verified for the case of $\mathrm{n}=2, 3$. The n$^\mathrm{th}$ order term can be written in the following form of the path-ordered operators:
\begin{equation}
\begin{multlined}
\tag{A.2.9}
\label{eq:A.2.9}
\hat{\mathrm{U}}^\mathrm{n}(t,t_\mathrm{o}) = \frac{1}{\mathrm{n}!}(-i)^\mathrm{n}\int_{t_\mathrm{o}}^{t}\mathrm{d}t_1
 \, ... \,
 \int_{t_\mathrm{o}}^{t}\mathrm{d}t_{\mathrm{n}-1}
 \int_{t_\mathrm{o}}^{t}\mathrm{d}t_\mathrm{n} \,\uptheta(t_1 - t_2)\,\uptheta(t_2 - t_3)\,...\,\uptheta(t_{\mathrm{n}-1} - t_{\mathrm{n}})\, \\ \sum P[\mathcal{H}_\mathrm{IG}'(t_1)
 \, ... \,
 \mathcal{H}_\mathrm{IG}'(t_{\mathrm{n}-1})
 \mathcal{H}_\mathrm{IG}'(t_\mathrm{n})],
\end{multlined}
\end{equation}
\begin{equation}
\begin{multlined}
\tag{A.2.10}
\label{eq:A.2.10}
\hat{\mathrm{U}}^\mathrm{n}(t,t_\mathrm{o}) = \frac{1}{\mathrm{n}!}(-i)^\mathrm{n}\int_{t_\mathrm{o}}^{t}\mathrm{d}t_1
 \, ... \,
 \int_{t_\mathrm{o}}^{t}\mathrm{d}t_{\mathrm{n}-1}
 \int_{t_\mathrm{o}}^{t}\mathrm{d}t_\mathrm{n} \, \mathrm{T}[\mathcal{H}_\mathrm{IG}'(t_1)
 \, ... \,
 \mathcal{H}_\mathrm{IG}'(t_{\mathrm{n}-1})
 \mathcal{H}_\mathrm{IG}'(t_\mathrm{n})].
\end{multlined}
\end{equation}
Thus, \eqref{eq:A.2.6} can now be reduced to the Taylor expansion of a path-ordered exponential functional of the Hamiltonian $\mathcal{H}_\mathrm{IG}'$ as follows:
\begin{equation}
\begin{multlined}
\tag{A.2.11}
\label{eq:A.2.11}
\hat{\mathrm{U}}(t,t_\mathrm{o}) = \mathcal{I} + (-i)\int_{t_\mathrm{o}}^{t}\mathrm{d}t_1 \mathcal{H}_\mathrm{IG}'(t_1) + \frac{1}{2!}(-i)^2\int_{t_\mathrm{o}}^{t}\mathrm{d}t_1
\int_{t_\mathrm{o}}^{t}\mathrm{d}t_2 \mathrm{T}[\mathcal{H}_\mathrm{IG}'(t_1)
 \mathcal{H}_\mathrm{IG}'(t_2)] +\, ...\\ \,+ \frac{1}{\mathrm{n}!}(-i)^\mathrm{n}\int_{t_\mathrm{o}}^{t}\mathrm{d}t_1
 \, ... \,
 \int_{t_\mathrm{o}}^{t}\mathrm{d}t_{\mathrm{n}-1}
 \int_{t_\mathrm{o}}^{t}\mathrm{d}t_\mathrm{n} \mathrm{T}[\mathcal{H}_\mathrm{IG}'(t_1)
 \, ... \,
 \mathcal{H}_\mathrm{IG}'(t_{\mathrm{n}-1})
 \mathcal{H}_\mathrm{IG}'(t_\mathrm{n})],
\end{multlined}
\end{equation}
\begin{equation}
\begin{multlined}
\tag{A.2.12}
\label{eq:A.2.12}
\hat{\mathrm{U}}(t,t_\mathrm{o}) = \mathlarger{\mathrm{T}}\,\Bigg [e^{{-i\displaystyle\int_{t_\mathrm{o}}^{t}}\mathcal{H}_\mathrm{IG}'(t')\,\mathrm{d}t'}\Bigg]\indent\text{given,}\indent t_\mathrm{o}<t.
\end{multlined}
\end{equation}
\begin{flushright}\myrule[line width = 0.5mm]{fast cap reversed}{fast cap reversed}\end{flushright}
\subsection{Relating $|\Omega\rangle$ to $|\text{\textbf{0}}\rangle$}
\label{A.3}
To begin with, let us combine the expressions \eqref{eq:97}, \eqref{eq:98} and \eqref{eq:99} such that
\begin{equation}
\tag{A.3.1}
\label{eq:A.3.1}
\begin{multlined}
\,\langle\bar\delta\hspace{-0.01in}\phi_\mathbf{k_1}\,\bar\delta\hspace{-0.01in}\phi_\mathbf{k_2}\,\bar\delta\hspace{-0.01in}\phi_\mathbf{k_3}\, ...\, \bar\delta\hspace{-0.01in}\phi_\mathbf{k_N}\rangle_{_\mathrm{\scriptstyle H}} = \,\langle \mathfrak{\Omega}|[\bar\delta\hspace{-0.01in}\phi_\mathbf{k_1}\,\bar\delta\hspace{-0.01in}\phi_\mathbf{k_2}\,\bar\delta\hspace{-0.01in}\phi_\mathbf{k_3}\, ...\, \bar\delta\hspace{-0.01in}\phi_\mathbf{k_N}]_{_\mathrm{\scriptstyle H}}|\mathfrak{\Omega}\rangle \\ = \,\langle \mathfrak{\Omega}| e^{i\mathcal{H}_\mathrm{int}t}\,[\bar\delta\hspace{-0.01in}\phi_\mathbf{k_1}\,\bar\delta\hspace{-0.01in}\phi_\mathbf{k_2}\,\bar\delta\hspace{-0.01in}\phi_\mathbf{k_3}\, ...\, \bar\delta\hspace{-0.01in}\phi_\mathbf{k_N}]_{_\mathrm{\scriptstyle I}}\,e^{-i\mathcal{H}_\mathrm{int}t}|\mathfrak{\Omega}\rangle = \,\langle 0| \hat{\mathrm{U}}^*(t,t_\mathrm{o})\,[\bar\delta\hspace{-0.01in}\phi_\mathbf{k_1}\,\bar\delta\hspace{-0.01in}\phi_\mathbf{k_2}\,\bar\delta\hspace{-0.01in}\phi_\mathbf{k_3}\, ...\, \bar\delta\hspace{-0.01in}\phi_\mathbf{k_N}]_{_\mathrm{\scriptstyle I}}\,\hat{\mathrm{U}}(t,t_\mathrm{o})| 0 \rangle.
\end{multlined}
\end{equation}
We now need to establish the relation between the true vacuum state $|\mathfrak{\Omega}\rangle$ and the free-field vacuum state $|0\rangle$. In order to derive this relation, firstly let us rewrite the Hamiltonian as follows:
\begin{equation*}
\begin{multlined}
\mathcal{H}_\mathcal{S} = \mathcal{H}_\mathrm{NI} + \mathcal{H}_\mathrm{int}e^{-(\alpha |t| + i\upbeta)} = \mathcal{H}_\mathrm{NI} + \mathcal{H'}_\mathrm{int}\indent\text{where,}\indent 0<\alpha \ll 1; \;\upbeta\in \mathcal{R},
\end{multlined}
\end{equation*}
which can be further written as a power series of the complex exponential such that
\begin{equation}
\tag{A.3.2}
\label{eq:A.3.2}
\begin{multlined}
\mathcal{H}_\mathcal{S} = \mathcal{H}_\mathrm{NI} + \mathcal{H}_\mathrm{int}\big[1 + {[-(\alpha |t|+ i\upbeta)]} + 
\frac{1}{2!}{[-(\alpha |t| + i\upbeta)]}^2\,...\,\big]\indent\text{where,}\indent 0<\alpha \ll 1; \;\upbeta\in \mathcal{R}.
\end{multlined}
\end{equation}
This is equivalent to saying that the interaction Hamiltonian vanishes at $t \rightarrow \pm\infinity$. The imaginary component $\upbeta$, as we will see shortly, is to avoid critical singularities and divergences. Secondly, we write the expression for the evolution of the ground state $|0\rangle$. It can be done as follows\footnote{Note again that $\mathcal{H}_\mathcal{S}$, $\mathcal{H}_\mathrm{NI}$ and  $\mathcal{H}_\mathrm{int}$ could be explicit/implicit functions of time themselves, which calls for the use of integrals. Therefore, we shall consider the case of time-dependent $\mathcal{H}_\mathcal{S}$, $\mathcal{H}_\mathrm{NI}$ and  $\mathcal{H}_\mathrm{int}$.}:\\  $\blacktriangleright$ Write the general time-evolution of an eigenstate $\varphi_\mathrm{k}(t)$ according to the time-dependent Schr\"odinger equation $i\hbar \dfrac{\partial\,\varphi_\mathrm{k}(t)}{\partial t} = \mathcal{H}_\mathcal{S}\,\varphi_\mathrm{k}(t)$, such that
$$\varphi_\mathrm{k}(t) = e^{\Bigg[{-i\displaystyle\int_{t_\mathrm{o}}^{t}
\scriptstyle{\mathcal{H}_\mathcal{S}\,\mathrm{d}t'}}\Bigg]}\,\varphi_\mathrm{k}(t=t_\mathrm{o}).$$
$\blacktriangleright$ Now, we know the eigenstates for the Hamiltonian (which is time-dependent!) at time $t$ from the time-independent part of the Schr\"odinger equation: $$\mathcal{H}_\mathcal{S}(t)\,\varphi_\mathrm{n}(t) = \mathcal{E}_\mathrm{n}(t)\,\varphi_\mathrm{n}(t).$$
$\blacktriangleright$ Naturally, $\varphi_\mathrm{n}(t)$ form the orthogonal basis for the Hamiltonian $\mathcal{H}_\mathcal{S}\,(t)$ specifically at time $t$. Therefore,
$$\varphi_\mathrm{k}(t) = \sum_\mathrm{n}c_\mathrm{n}(t)\,\varphi_\mathrm{n}(t)\indent\text{where,}\indent c_\mathrm{n}(t)=\,\langle\varphi_\mathrm{n}(t)\,\varphi_\mathrm{k}(t)\rangle\equiv\,\langle \mathrm{n}|\mathrm{k}\rangle,$$
$\blacktriangleright$ From the above arguments, we may conclude that,
$$e^{\Bigg[{-i\displaystyle\int_{t_\mathrm{o}}^{t}
\scriptstyle{\mathcal{H}_\mathcal{S}\,\mathrm{d}t'}}\Bigg]}\,\varphi_\mathrm{k}(t=t_\mathrm{o}) = e^{\Bigg[{-i\displaystyle\int_{t_\mathrm{o}}^{t}
\scriptstyle{\mathcal{H}_\mathcal{S}\,\mathrm{d}t'}}\Bigg]}|\mathrm{n}\rangle\,\langle \mathrm{n}|\mathrm{k}\rangle.$$
Let us now consider the case of the pure ground state as it evolves in time, i.e. $\varphi_\mathrm{k}(t=t_\mathrm{o})\equiv |\mathrm{k}\rangle=|0\rangle$, in the limit when the perturbations are switched on at very early times, i.e. $t_\mathrm{o} \rightarrow -\,\infinity$,
\begin{equation}
\tag{A.3.3}
\label{eq:A.3.3}
\begin{multlined}
\lim_{t_\mathrm{o} \to -\displaystyle\,\infinity}\;e^{\Bigg[{-i\displaystyle\int_{t_\mathrm{o}}^{t}
\scriptstyle{\mathcal{H}_\mathcal{S}\,\mathrm{d}t'}}\Bigg]}|0\rangle = \lim_{t_\mathrm{o} \to -\displaystyle\,\infinity}\; \sum_{\mathrm{n = 0}}^{\mathrm{N}} e^{\Bigg[{-i\displaystyle\int_{t_\mathrm{o}}^{t}
\scriptstyle{\mathcal{H}_\mathcal{S}\,\mathrm{d}t'}}\Bigg]}|\mathrm{n}\rangle\,\langle \mathrm{n}|0\rangle \\ =\lim_{t_\mathrm{o} \to -\displaystyle\,\infinity}\; \sum_{\mathrm{n = 0}}^{\mathrm{N}} e^{\Bigg[{-i\displaystyle\int_{t_\mathrm{o}}^{t}\mathcal{E}_\mathrm{NI}\mathrm{d}t'}\Bigg]}e^{\Bigg[{-i\displaystyle\int_{t_\mathrm{o}}^{t}\mathcal{E_\mathrm{int|n}}[1 + {[-(\alpha |t'| + i\upbeta)]} + 
\,...\,\big]\mathrm{d}t'}\Bigg]}|\mathrm{n}\rangle\,\langle \mathrm{n}|0\rangle,
\end{multlined}
\end{equation}
where, $|\mathrm{n}\rangle$ are the eigenstates of the Hamiltonian $\mathcal{H}_\mathcal{S} (= \mathcal{H}_\mathrm{NI} + \mathcal{H}_\mathrm{int})$ with eigenvalues $\mathcal{E}_\mathrm{n}$, while $\mathcal{E}_\mathrm{int|n}$ are the eigenvalues for only the interaction part $\mathcal{H}_\mathrm{int}$ of the Hamiltonian. We take the ground state out of the summation and consider only until first-order expansion of the complex exponential,
\begin{equation}
\tag{A.3.4}
\label{eq:A.3.4}
\begin{multlined}
\lim_{t_\mathrm{o} \to -\displaystyle\,\infinity}\;e^{\Bigg[{-i\displaystyle\int_{t_\mathrm{o}}^{t}
\scriptstyle{\mathcal{H}_\mathcal{S}\,\mathrm{d}t'}}\Bigg]}|0\rangle = \lim_{t_\mathrm{o} \to -\displaystyle\,\infinity}\; \sum_{\mathrm{n = 0}}^{\mathrm{N}} e^{\Bigg[{-i\displaystyle\int_{t_\mathrm{o}}^{t}
\scriptstyle{\mathcal{H}_\mathcal{S}\,\mathrm{d}t'}}\Bigg]}|\mathrm{n}\rangle\,\langle \mathrm{n}|0\rangle \\ =\lim_{t_\mathrm{o} \to -\displaystyle\,\infinity}\; \sum_{\mathrm{n = 0}}^{\mathrm{N}} e^{\Bigg[{-i\displaystyle\int_{t_\mathrm{o}}^{t}\mathcal{E}_\mathrm{n}\mathrm{d}t'}\Bigg]}e^{\Bigg[{i\alpha\displaystyle\int_{t_\mathrm{o}}^{t}\mathcal{E_\mathrm{int|n}}{|t'|\,\mathrm{d}t'}}\Bigg]}e^{\Bigg[{-\upbeta\displaystyle\int_{t_\mathrm{o}}^{t}\mathcal{E_\mathrm{int|n}}\,\mathrm{d}t'}\Bigg]}|\mathrm{n}\rangle\,\langle \mathrm{n}|0\rangle.
\end{multlined}
\end{equation}
We can now make the deduction that since the higher-order states (less stable and less localized) are affected more by the perturbations in the Hamiltonian than the lower-order states (which are more stable and strongly localized), $\mathcal{E_\mathrm{int|n}}$ increases with increasing order of the state. We give a rough intuitive treatment below in the note.\\
\begin{flushleft}\myrule[line width = 0.5mm]{fast cap reversed}{fast cap reversed}\end{flushleft}
\tit{\textbf{Sidenote}}:\\
We treat the perturbed scalar field as equivalent to a perturbed harmonic oscillator, where the perturbation is given by $\mathcal{H}_\mathrm{int}$, and it usually takes the form of power law, i.e. $\mathcal{H}_\mathrm{int} \propto (\phi)^p$ for some scalar field $\phi$ and $p > 2$. However, for the perturbation theory to be used, the magnitude of the said perturbation must be significantly smaller than the effective magnitude of the Hamiltonian itself. We quote the results for the 1$^\mathrm{st}$ and 2$^\mathrm{nd}$ order corrections to the eigenstates for a few values of $p$, such as 3 and 4. \\ \\
$\bullet$ $p = 3 \indent\Longrightarrow\indent \mathcal{E_\mathrm{int|n}}^{(1)} = 0\indent\indent\indent\;\;\; \mathcal{E_\mathrm{int|n}}^{(2)} \propto \mathrm{n}^2 + \mathrm{n} + \dfrac{11}{30}.$\\
$\bullet$ $p = 4 \indent\Longrightarrow\indent \mathcal{E_\mathrm{int|n}}^{(1)} \propto \mathrm{n}^2 + \mathrm{n} + \dfrac{1}{2}\indent \mathcal{E_\mathrm{int|n}}^{(2)} \propto O(\mathrm{n}^4).$\\ \\
It is obvious that `higher' the energy state, the less localized it is, and therefore, higher is the correction to it due to the interaction(s). Hence, we conclude that for the new ground state $|\Omega\rangle$ at any time $t$,
$$\mathcal{E}_{\mathrm{int}|\Omega} < \mathcal{E_\mathrm{int|1}} < \mathcal{E_\mathrm{int|2}} < ... < \mathcal{E}_{\mathrm{int}|\mathrm{N}}.$$
\begin{flushright}\myrule[line width = 0.5mm]{fast cap reversed}{fast cap reversed}\end{flushright}$\;$\\
Hence, we have concluded that the ground state is the most resilient state and is preserved more than the higher-order ones. Furthermore, we can also assert that the new ground state $|\Omega\rangle$ at any time $t$ evolves from the unperturbed ground state $|0\rangle$ such that $\,\langle\Omega|0\rangle \neq 0$; there must be some overlap between the two states since the interaction part of the Hamiltonian is effectively small but non-zero. This, however, may not hold true for the higher order states. We can, therefore, isolate the ground state for an arbitrarily large $\upbeta$ and neglect the higher-order contributions in \eqref{eq:A.3.4} as follows:
\begin{equation}
\tag{A.3.5}
\label{eq:A.3.5}
\begin{multlined}
\lim_{\upbeta \gg 1}\; e^{\Bigg[{-\upbeta\displaystyle\int_{t_\mathrm{o}}^{t}\mathcal{E}_{\mathrm{int}|\Omega}\,\mathrm{d}t'}\Bigg]} \gg e^{\Bigg[{-\upbeta\displaystyle\int_{t_\mathrm{o}}^{t}\mathcal{E_\mathrm{int|1}}\,\mathrm{d}t'}\Bigg]} \gg \,...\,\,e^{\Bigg[{-\upbeta\displaystyle\int_{t_\mathrm{o}}^{t}\mathcal{E_\mathrm{int|N}}\,\mathrm{d}t'}\Bigg]},
\end{multlined}
\end{equation}
which yields,
\begin{equation}
\tag{A.3.6}
\label{eq:A.3.6}
\begin{multlined}
\lim_{t_\mathrm{o} \to -\displaystyle\,\infinity}\;e^{\Bigg[{-i\displaystyle\int_{t_\mathrm{o}}^{t}
\scriptstyle{\mathcal{H}_\mathcal{S}\,\mathrm{d}t'}}\Bigg]}|0\rangle =\lim_{t_\mathrm{o} \to -\displaystyle\,\infinity}\; e^{\Bigg[{-i\displaystyle\int_{t_\mathrm{o}}^{t}\mathcal{E}_\Omega\mathrm{d}t'}\Bigg]}e^{\Bigg[{i\alpha\displaystyle\int_{t_\mathrm{o}}^{t}\mathcal{E}_{\mathrm{int}|\Omega}{|t'|\,\mathrm{d}t'}}\Bigg]}e^{\Bigg[{-\upbeta\displaystyle\int_{t_\mathrm{o}}^{t}\mathcal{E}_{\mathrm{int}|\Omega}\,\mathrm{d}t'}\Bigg]}|\mathfrak{\Omega}\rangle\,\langle \mathfrak{\Omega}|0\rangle,
\end{multlined}
\end{equation}
that can further be rewritten in terms of total energy of new ground state $\mathcal{E}_{t|\Omega}$ such that
\begin{equation}
\tag{A.3.7}
\label{eq:A.3.7}
\begin{multlined}
\lim_{t_\mathrm{o} \to -\displaystyle\,\infinity}\;e^{\Bigg[{-i\displaystyle\int_{t_\mathrm{o}}^{t}
\scriptstyle{\mathcal{H}_\mathcal{S}\,\mathrm{d}t'}}\Bigg]}|0\rangle = \lim_{t_\mathrm{o} \to -\displaystyle\,\infinity}\;e^{\Bigg[{-i\displaystyle\int_{t_\mathrm{o}}^{t}\mathcal{E}_{t|\Omega}\mathrm{d}t'}\Bigg]} |\mathfrak{\Omega}\rangle\,\langle \mathfrak{\Omega}|0\rangle.
\end{multlined}
\end{equation}
The left hand side could further be reduced to:
\begin{equation}
\tag{A.3.8}
\label{eq:A.3.8}
\begin{multlined}
\lim_{t_\mathrm{o} \to -\displaystyle\,\infinity}\;e^{\Bigg[{-i\displaystyle\int_{t_\mathrm{o}}^{t}
\scriptstyle{\mathcal{H}_\mathrm{NI}\,\mathrm{d}t'}}\Bigg]}e^{\Bigg[{-i\displaystyle\int_{t_\mathrm{o}}^{t}
\scriptstyle{\mathcal{H}_\mathrm{int}e^{\scriptstyle{-(\alpha |t| + i\upbeta)}}\,\mathrm{d}t'}}\Bigg]}|0\rangle =\lim_{t_\mathrm{o} \to -\displaystyle\,\infinity}\; \mathbbmtt{E}_{\,\Omega}(t,t_\mathrm{o})|\mathfrak{\Omega}\rangle\,\langle \mathfrak{\Omega}|0\rangle,
\end{multlined}
\end{equation}
where, we have abbreviated the exponential on the right-hand side. Recalling the identity \eqref{eq:82} for commutating operators, along with $\mathcal{H}_\mathrm{NI}|0\rangle = |0\rangle$, we get:
\begin{equation*}
\begin{multlined}
\lim_{t_\mathrm{o} \to -\displaystyle\,\infinity}\;e^{\Bigg[{-i\displaystyle\int_{t_\mathrm{o}}^{t}
\scriptstyle{\mathcal{H}_\mathrm{int}e^{\scriptstyle{-(\alpha |t| + i\upbeta)}}\,\mathrm{d}t'}}\Bigg]}|0\rangle = \lim_{t_\mathrm{o} \to -\displaystyle\,\infinity}\;\mathbbmtt{E}_{\,\Omega}(t,t_\mathrm{o})|\mathfrak{\Omega}\rangle\,\langle \mathfrak{\Omega}|0\rangle,
\end{multlined}
\end{equation*}
\begin{equation*}
\begin{multlined}
\lim_{t_\mathrm{o} \to -\displaystyle\,\infinity}\;e^{\Bigg[{-i\displaystyle\int_{t_\mathrm{o}}^{t}
\scriptstyle{\mathcal{H'}_\mathrm{int}\,\mathrm{d}t'}}\Bigg]}|0\rangle =\lim_{t_\mathrm{o} \to -\displaystyle\,\infinity}\; {\mathcal{E}^\mathrm{o}_{\Omega}}(t,t_\mathrm{o})|\mathfrak{\Omega}\rangle\,\langle \mathfrak{\Omega}|0\rangle,
\end{multlined}
\end{equation*}
\begin{equation}
\tag{A.3.9}
\label{eq:A.3.9}
\begin{multlined}
\hat{\mathrm{U}}(t,t_\mathrm{o})|0\rangle = \lim_{t_\mathrm{o} \to -\displaystyle\,\infinity}\;{\mathcal{E}^\mathrm{o}_{\Omega}}(t,t_\mathrm{o})|\mathfrak{\Omega}\rangle\,\langle \mathfrak{\Omega}|0\rangle.
\end{multlined}
\end{equation}
Finally, we have our relation between $|\mathfrak{\Omega}\rangle$ and $|0\rangle$, i.e.
\begin{equation}
\tag{A.3.10}
\label{eq:A.3.10}
\begin{multlined}
|\mathfrak{\Omega}\rangle= \lim_{t_\mathrm{o} \to -\displaystyle\,\infinity}\;\frac{\hat{\mathrm{U}}(t,t_\mathrm{o})|0\rangle}{{\mathcal{E}^\mathrm{o}_{\Omega}}(t,t_\mathrm{o})\,\langle \mathfrak{\Omega}|0\rangle}.
\end{multlined}
\end{equation}
Similarly,
\begin{equation}
\tag{A.3.11}
\label{eq:A.3.11}
\begin{multlined}
\,\langle\mathfrak{\Omega}|= \lim_{t_\mathrm{o} \to -\displaystyle\,\infinity}\;\frac{\,\langle 0|\hat{\mathrm{U}}^\dagger(t,t_\mathrm{o})}{{\mathcal{E}^\mathrm{o}_{\Omega}}^{\,*}(t,t_\mathrm{o})\,\langle 0|\mathfrak{\Omega}\rangle}.
\end{multlined}
\end{equation}
Combining \eqref{eq:A.3.10} and \eqref{eq:A.3.11},
\begin{equation}
\tag{A.3.12}
\label{eq:A.3.12}
\begin{multlined}
1 = \,\langle\mathfrak{\Omega}|\mathfrak{\Omega}\rangle= \lim_{t_\mathrm{o} \to -\displaystyle\,\infinity}\;\frac{\,\langle 0|\hat{\mathrm{U}}^\dagger(t,t_\mathrm{o})\hat{\mathrm{U}}(t,t_\mathrm{o})|0\rangle}{|{\mathcal{E}^\mathrm{o}_{\Omega}}(t,t_\mathrm{o})|^2{\,\langle 0|\mathfrak{\Omega}\rangle}{\,\langle \mathfrak{\Omega}|0\rangle}} = \lim_{t_\mathrm{o} \to -\displaystyle\,\infinity}\;\frac{\,\langle 0|\hat{\mathrm{U}}^\dagger(t,t_\mathrm{o})\hat{\mathrm{U}}(t,t_\mathrm{o})|0\rangle}{|{\mathcal{E}^\mathrm{o}_{\Omega}}(t,t_\mathrm{o})|^2{\,\langle 0|\mathfrak{\Omega}\rangle}{\,\langle \mathfrak{\Omega}|0\rangle}} \\ = \lim_{t_\mathrm{o} \to -\displaystyle\,\infinity}\;\frac{{\,\langle 0|\hat{\mathrm{U}}^\dagger(t,t_\mathrm{o})\hat{\mathrm{U}}(t,t_\mathrm{o})|0\rangle}{{\,\langle 0|\mathfrak{\Omega}\rangle}{\,\langle \mathfrak{\Omega}|0\rangle}}}{{{\,\langle 0|{\hat{\mathrm{U}}(t,t_\mathrm{o})|0\rangle}}\,{{\,\langle 0|\hat{\mathrm{U}}^\dagger(t,t_\mathrm{o})}|0\rangle}}}, 
\end{multlined}
\end{equation}
where, we can easily find from \eqref{eq:A.3.11} and \eqref{eq:A.3.12} that for the last equality,
\begin{equation*}
\begin{multlined} 
{|{\mathcal{E}^\mathrm{o}_{\Omega}}(t,t_\mathrm{o})|^2{({\,\langle 0|\mathfrak{\Omega}\rangle}{\,\langle \mathfrak{\Omega}|0\rangle}})}^2 = {{\,\langle 0|{\hat{\mathrm{U}}(t,t_\mathrm{o})|0\rangle}}\,{{\,\langle 0|\hat{\mathrm{U}}^\dagger(t,t_\mathrm{o})}|0\rangle}},
\end{multlined}
\end{equation*}
where, we have used the identity
\begin{equation}
\tag{A.3.13}
\label{eq:A.3.13}
\begin{multlined}
\hat{\mathrm{U}}(t_1,t_2)\,\hat{\mathrm{U}}(t_2,t_3)\,...\,\hat{\mathrm{U}}(t_\mathrm{n-1},t_\mathrm{n}) = \hat{\mathrm{U}}(t_1,t_{\mathrm{n}}).
\end{multlined}
\end{equation}
Note that another important identity to be remember is,
\begin{equation}
\tag{A.3.14}
\label{eq:A.3.14}
\begin{multlined}
\hat{\mathrm{U}}(t_1,t_3)\,\hat{\mathrm{U}^*}(t_2,t_3) = \hat{\mathrm{U}}(t_1,t_2).
\end{multlined}
\end{equation}
Let us now retreat to the very beginning of this section to identity \eqref{eq:A.3.1} and plug in the expressions for $|\mathfrak{\Omega}\rangle$ and $\,\langle \mathfrak{\Omega}|$ from \eqref{eq:A.3.10} and \eqref{eq:A.3.11},
\begin{equation*}
\begin{multlined}
\,\langle \mathfrak{\Omega}|[\bar\delta\hspace{-0.01in}\phi_\mathbf{k_1}\,\bar\delta\hspace{-0.01in}\phi_\mathbf{k_2}\,\bar\delta\hspace{-0.01in}\phi_\mathbf{k_3}\, ...\, \bar\delta\hspace{-0.01in}\phi_\mathbf{k_N}]_{_\mathrm{\scriptstyle H}}|\mathfrak{\Omega}\rangle = \,\langle \mathfrak{\Omega}| \hat{\mathrm{U}^*}(t,t_\mathrm{o})\,[\bar\delta\hspace{-0.01in}\phi_\mathbf{k_1}\,\bar\delta\hspace{-0.01in}\phi_\mathbf{k_2}\,\bar\delta\hspace{-0.01in}\phi_\mathbf{k_3}\, ...\, \bar\delta\hspace{-0.01in}\phi_\mathbf{k_N}]_{_\mathrm{\scriptstyle I}}\,\hat{\mathrm{U}}(t,t_\mathrm{o})| \mathfrak{\Omega} \rangle \\ = \lim_{t_\mathrm{o} \to -\displaystyle\,\infinity}\frac{\,\langle 0| \hat{\mathrm{U}}^\dagger(t,t_\mathrm{o})\hat{\mathrm{U}^*}(t,t_\mathrm{o})\,[\bar\delta\hspace{-0.01in}\phi_\mathbf{k_1}\,\bar\delta\hspace{-0.01in}\phi_\mathbf{k_2}\,\bar\delta\hspace{-0.01in}\phi_\mathbf{k_3}\, ...\, \bar\delta\hspace{-0.01in}\phi_\mathbf{k_N}]_{_\mathrm{\scriptstyle I}}\,\hat{\mathrm{U}}(t,t_\mathrm{o})\hat{\mathrm{U}}(t,t_\mathrm{o})|0 \rangle}{|{\mathcal{E}^\mathrm{o}_{\Omega}}(t,t_\mathrm{o})|^2{{\,\langle 0|\mathfrak{\Omega}\rangle}{\,\langle \mathfrak{\Omega}|0\rangle}}}.
\end{multlined}
\end{equation*}
The closing trick now is to find the factor $\hat{\mathrm{U}}(t,t_\mathrm{o})|0 \rangle$. In the limit\footnote{This is essentially the statement of the `Gell-Mann and Low theorem' upon passing the limit $0 < \alpha \ll 1$.} $0 < \alpha \ll 1$, the perturbation in the Hamiltonian becomes adiabatic\footnote{By definition of an adiabatic process, gradually changing conditions allow the system to adapt its configuration, and hence the probability density is modified by the process. If the system starts in an eigenstate of the initial Hamiltonian, it will end in the corresponding eigenstate of the final Hamiltonian. This is the so-called `Adiabatic theorem', and its formulation can be found in most standard textbooks on Quantum Mechanics or Quantum Field Theory.} in nature. Upon assuming that the ground state is non-degenerate\footnote{In case of a degenerate ground state, the perturbation in the Hamiltonian may (or, may not!) eventually switch the system from one ground state to another.}, the propagator does not evolve the ground state but only adds a phase factor ($\Theta_\mathrm{p}$) to it during its operation from $t_\mathrm{o}$ to $t$, such that
\begin{equation}
\tag{A.3.15}
\label{eq:A.3.15}
\begin{multlined}
\lim_{t_\mathrm{o} \to -\displaystyle\,\infinity}\hat{\mathrm{U}}(t,t_\mathrm{o})|0 \rangle = \Theta_\mathrm{p}(t,t_\mathrm{o})|0 \rangle\indent\text{and}\indent \lim_{t_\mathrm{o} \to -\displaystyle\,\infinity}\,\langle 0|\hat{\mathrm{U}}^\dagger(t,t_\mathrm{o}) = \Theta_\mathrm{q}^*(t,t_\mathrm{o})\,\langle 0|.
\end{multlined} 
\end{equation}
Using \eqref{eq:A.3.12} and \eqref{eq:A.3.15}, we get
\begin{equation}
\tag{A.3.16}
\label{eq:A.3.16}
\begin{multlined}
\,\langle \mathfrak{\Omega}|[\bar\delta\hspace{-0.01in}\phi_\mathbf{k_1}\,\bar\delta\hspace{-0.01in}\phi_\mathbf{k_2}\,\bar\delta\hspace{-0.01in}\phi_\mathbf{k_3}\, ...\, \bar\delta\hspace{-0.01in}\phi_\mathbf{k_N}]_{_\mathrm{\scriptstyle H}}|\mathfrak{\Omega}\rangle = \lim_{t_\mathrm{o} \to -\displaystyle\,\infinity}{\,\langle 0|\hat{\mathrm{U}}^*(t,t_\mathrm{o})\,[\bar\delta\hspace{-0.01in}\phi_\mathbf{k_1}\,\bar\delta\hspace{-0.01in}\phi_\mathbf{k_2}\,\bar\delta\hspace{-0.01in}\phi_\mathbf{k_3}\, ...\, \bar\delta\hspace{-0.01in}\phi_\mathbf{k_N}]_{_\mathrm{\scriptstyle I}}\,\hat{\mathrm{U}}(t,t_\mathrm{o})|0 \rangle}.
\end{multlined} 
\end{equation}
\subsection{Arnowitt-Deser-Misner (ADM) formalism}
\label{A.4}
We present in detail the motivation and the idea behind the ADM formalism in context of numerical relativity.
\subsubsection{The Philosophy}
\label{A.4.1}
In order to understand the philosophy behind the ADM formalism, let us write down the action for a
classical system of N discrete particles. The action in terms of the Hamiltonian $\mathcal{H}$ of such a system is given
by:
\begin{equation}
\tag{A.4.1}
\label{eq:A.4.1}
\begin{multlined}
\mathrm{I}=\int\mathrm{d}t\,\L[(p_1,p_2,p_3\,...\,p_\mathrm{N}),
(q_1,q_2,q_3\,...\,q_\mathrm{N}),t]=\int\mathrm{d}t\Bigg[
\sum_{i=1}^{\mathrm{N}}p_i\dt{q}_i-\mathcal{H}(p,q)\Bigg]\;\;\;\text{where,}\;\;\;q_i=\dt{p}_i,
\end{multlined} 
\end{equation}
where, we treat $p_i$ and $q_i$ as independent variables, and along with time $t$, they make up for a total of $\mathrm{N}+1$ independent variables. By varying the action with respect to each $p_i$ and $q_i$, one can evaluate the
equations of motion for each independent degree of freedom. This form of the action is termed as the
\tit{canonical form} and it provides complete information of the system.\\ \\
Now, let us consider the theory of General Relativity, in which all the independent degrees of
freedom are encoded within $g_{\mu\nu}$, i.e. in 10 independent terms of the symmetric metric $g_{\mu\nu}$. However, we also know that there are intrinsic gauge invariances present in the form of $g_{\mu\nu}$ via selected coordinate re-parametrizations, i.e. the the physical laws remain unchanged under a coordinate transformation of the form,
\begin{equation}
\tag{A.4.2}
\label{eq:A.4.2}
\begin{multlined}
x^\mu\rightarrow x^\mu +\xi^\mu(x^\mu),
\end{multlined} 
\end{equation}
for any differentiable set of $\xi^\mu$. The metric $g_{\mu\nu}$ transforms in this case with an additional Lie derivative term such that,
\begin{equation}
\tag{A.4.3}
\label{eq:A.4.3}
\begin{multlined}
g_{\mu\nu}\rightarrow g_{\mu\nu} +2\partial(^\mu\xi^\nu).
\end{multlined} 
\end{equation}
This implies that not all apparent degrees of freedom are physical in nature and in fact, we have some
`in-built' freedom to choose our coordinate system. Hence, upon variation of the action with respect to
all apparent degrees of freedom, we will write some equations which do not contain any evolutionary
information about the system, and these relations are in fact entirely redundant. All such non-physical degrees of
freedom are often called `gauge modes', and the corresponding system is characterized by an `unconstrained Hamiltonian'.
Our aim is to extract all true dynamical degrees of freedom in the Lagrangian/Hamiltonian formulation.
We will now show (non-exhaustively!) how such re-parametrization invariances could be devolved from
the action by means of a constraint equation accompanied by a Lagrange multiplier.\\ \\
Consider the action in \eqref{eq:A.4.1},
\begin{equation*}
\begin{multlined}
\mathrm{I}=\int\mathrm{d}t\Bigg[
\sum_{i=1}^{\mathrm{N}}p_i\dt{q}_i-\mathcal{H}(p,q)\Bigg].
\end{multlined} 
\end{equation*}
The expression above can be tweaked to be rewritten as
\begin{equation}
\tag{A.4.4}
\label{eq:A.4.4}
\begin{multlined}
\mathrm{I}=\int\mathrm{d}t\Bigg[
\sum_{i=1}^{\mathrm{N}}p_i\frac{\mathrm{d}{q}_i}{\mathrm{d}t}-\mathcal{H}(p,q)\Bigg]=\int\mathrm{d}\tau\Bigg[
\sum_{i=1}^{\mathrm{N}}p_i\frac{\mathrm{d}{q}_i}{\mathrm{d}\tau}-\mathcal{H}(p,q)\frac{\mathrm{d}t}{\mathrm{d}\tau}\Bigg]\\
=\int\mathrm{d}\tau\Bigg[
\sum_{i=1}^{\mathrm{N}}p_i{{q_i'}}-\mathcal{H}(p,q)t'\Bigg]
=\int\mathrm{d}\tau\Bigg[
\sum_{i=1}^{\mathrm{N}+1}p_i{{q_i'}}\Bigg],
\end{multlined} 
\end{equation}
where, we allow for $(\mathrm{N}+1)$th independent parameter to be such that,
\begin{equation}
\tag{A.4.5}
\label{eq:A.4.5}
\begin{multlined}
q_{\mathrm{N}+1}=t\;\;\;\;\;\;\text{and,}\;\;\;\;\;\;p_{\mathrm{N}+1}=-\mathcal{H}(p,q).
\end{multlined} 
\end{equation}
We see from \eqref{eq:A.4.4} and \eqref{eq:A.4.5} that the time re-parametrization $t\rightarrow\tau$ leads to the equation of constraint $p_{\mathrm{N}+1}=-\mathcal{H}(p,q)$. This equation of constraint can now be explicitly introduced in the action using a Lagrangian multiplier $\mathcal{N}(\tau)$ given the condition that the Lagrangian multiplier itself transforms as
\begin{equation}
\tag{A.4.6}
\label{eq:A.4.6}
\begin{multlined}
\mathcal{N}(\tau)\mathrm{d}\tau=\mathcal{N}(t)\mathrm{d}t,
\end{multlined} 
\end{equation}
yielding an effective action,
\begin{equation}
\tag{A.4.7}
\label{eq:A.4.7}
\begin{multlined}
\mathrm{I}=\int\mathrm{d}\tau\Bigg[
\sum_{i=1}^{\mathrm{N}+1}p_i{{q_i'}}-\mathcal{N}(\tau)\Big\{p_{\mathrm{N}+1}+\mathcal{H}(p,q)\Bigg\}\Bigg].
\end{multlined} 
\end{equation}
We note that by allowing a time re-parametrization, we have derived a constraint equation. In classical mechanics, the equation of constraint resulting from a time re-parametrization is called a 'Hamiltonian constraint' while the equations arising from space re-parametrization are termed as 'momentum constraints'. In nutshell, if a system of dynamics is described by a Lagrangian which has some coordinate re-parametrization built into it, this freedom of re-parametrization is encoded as a constraint equation(s). This is another way of stating the famous Noether's (first) theorem.
\subsubsection{ADM formalism}
\label{A.4.2}
The ADM formalism (also known as the 3+1 formalism) provides an intuitive way of expressing the action in the form of \eqref{eq:A.4.7}; it is especially useful in numerical relativity. ADM formalism relies on utilising the foliation of space-time into infinitesimally 'discretised' spatial hyper-surfaces evolving through time. A clearer physical interpretation of the idea of foliation of space-time is given below in figure \ref{fig:foil}.
\begin{figure}[H]
\centering
\includegraphics[width=80mm]{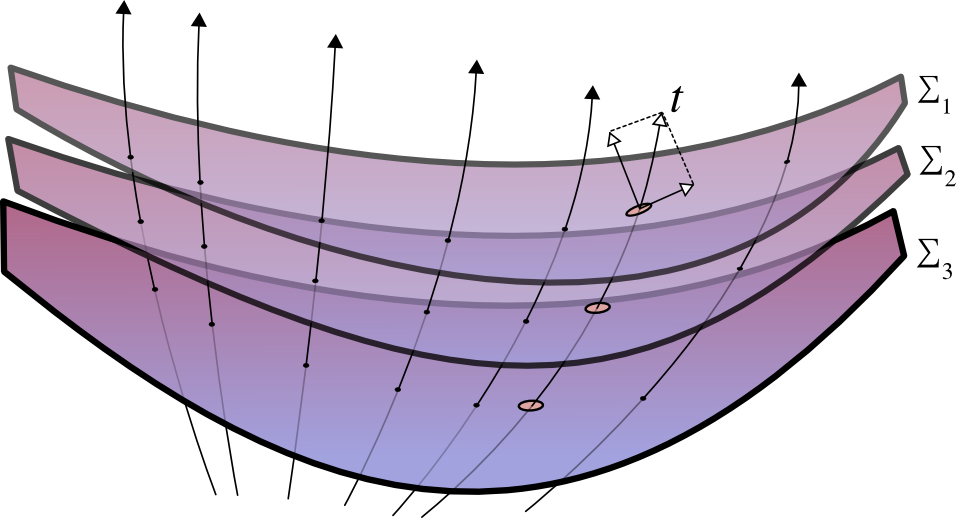}\hspace{0.3in}
\includegraphics[width=50mm]{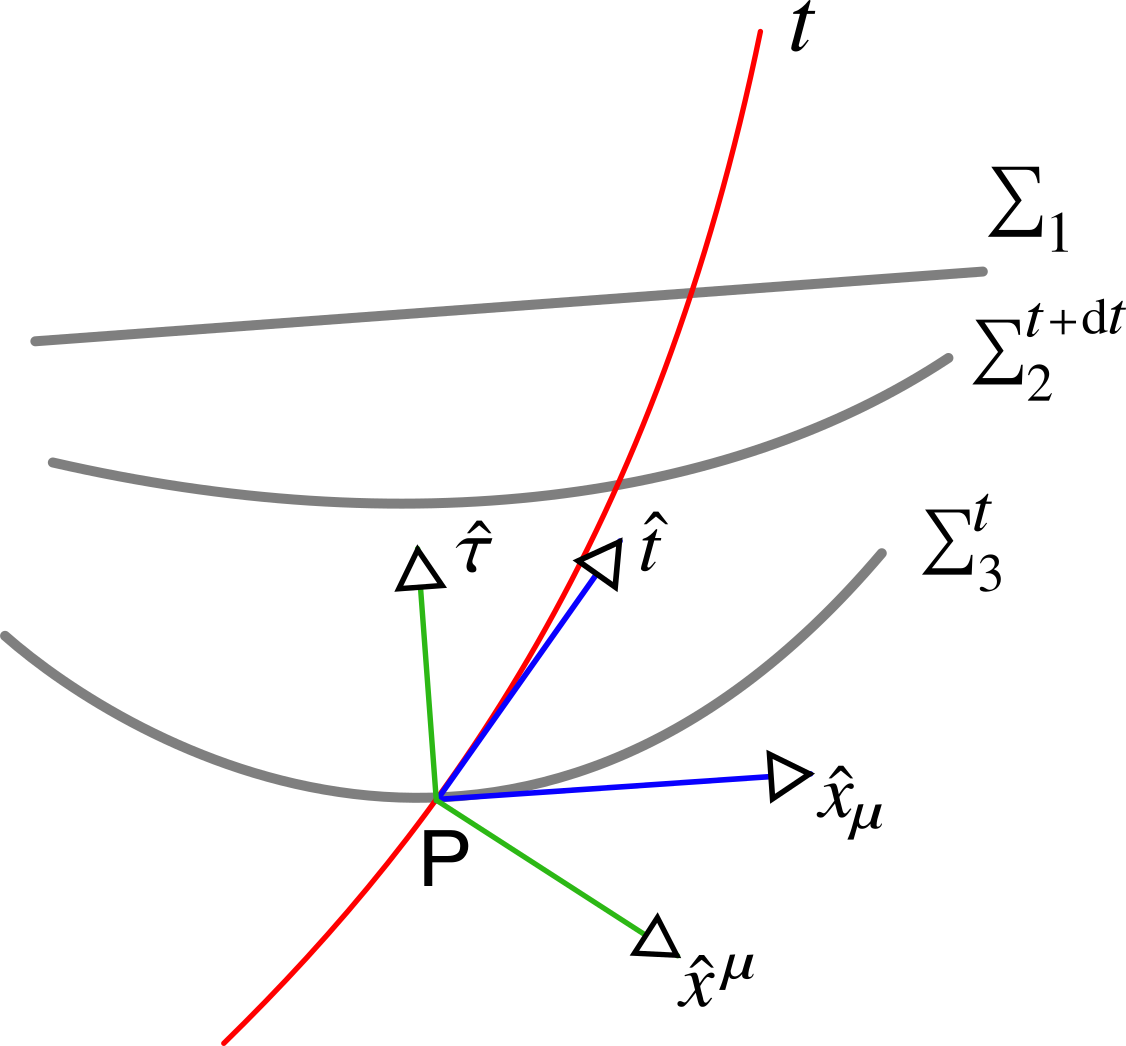}
\caption{A geometrical representation of 3+1 space-time foliation.}
\label{fig:foil}
\end{figure}$\;$\\
The process, as it appears, involves foliating space-time continuum into purely spatial hyper-surfaces ($\sum_i$) evolving in time such that the value of the scalar field $t$ is constant on a hyper-surface $\sum_i$. We can
equivalently define a vector field in the direction of evolution of $t$ by simply taking the contravariant
derivative of scalar field $t$; the components of such a vector at point P are given by $t^\mu\propto\nabla^\mu t$, where it helps to recall that $t$ represents the 'coordinate time' of the observer at point P on the hyper-surface
$\sum_i$. Moreover, at each point on the hyper-surface, we can define a time-like vector perpendicular to the hyper-surface given by the covariant derivative of the scalar component of $t$:
\begin{equation}
\tag{A.4.8}
\label{eq:A.4.8}
\begin{multlined}
\vec{\nabla}_c t\equiv\nabla_\mu t=g_{\mu\kappa}(\nabla^\kappa t)=\alpha g_{\mu\kappa}t^\kappa \;\;\;\;\;\text{such that,}\;\;\;\;\;
t^\mu=\alpha^{-1}\nabla^\mu t.
\end{multlined} 
\end{equation}
Note that in figure \ref{fig:foil}, the set of vectors $\{\hat{t},\hat{\mathbf{x}}_\mu\}$ represent the covariant basis, while the set $\{\hat{\tau},\hat{\mathbf{x}}^\mu\}$ represent the contravariant basis. By definition of covariant and contravariant dual basis set, one can deduce that $\hat{\tau}\perp\hat{\mathbf{x}}_\mu$ and $\hat{t}\perp\hat{\mathbf{x}}^\mu$, where $\tau$ is the proper time. Now, since
$\hat{t}\perp\hat{\mathbf{x}}^\mu$ and independent of $\hat{\mathbf{x}}^\mu$, the covariant derivative of $t$ will simply point in the direction of $\hat{\tau}$, and in fact, vary with only the temporal contravariant index $\tau$. Rewriting the previous expression, we get:
\begin{equation}
\tag{A.4.9}
\label{eq:A.4.9}
\begin{multlined}
\vec{\nabla}_c t\equiv\nabla_\mu t=\partial_\tau t= g_{0\gamma}(\nabla^\gamma t)=\alpha g_{0\gamma}t^\gamma\;\;\;\;\;\text{such that,}\;\;\;\;\;
\vec{\nabla}_c t\parallel\hat{\tau},
\end{multlined} 
\end{equation}
where, we recall that $\nabla^\gamma t\equiv\nabla^\gamma t=t^\gamma$. We can now construct an unit vector $\hat{\mathbf{n}}$ parallel to $\tau$ from the expression of $\vec{\nabla}_c t$:
\begin{equation}
\tag{A.4.10}
\label{eq:A.4.10}
\begin{multlined}
\hat{\mathbf{n}}^c=\frac{1}{\sqrt{-\vec{\nabla}_c t\cdot
\vec{\nabla}_c t}}\vec{\nabla}_c t=\mathrm{N}_\tau (\vec{\nabla}_c t),
\end{multlined} 
\end{equation}
such that by definition,
\begin{equation}
\tag{A.4.11}
\label{eq:A.4.11}
\begin{multlined}
\hat{\mathbf{n}}^c\cdot\hat{\mathbf{n}}^c=-1,
\end{multlined} 
\end{equation}
where, the \tit{Lapse function} $\mathrm{N}_\tau$ is given by:
\begin{equation}
\tag{A.4.12}
\label{eq:A.4.12}
\begin{multlined}
\mathrm{N}_\tau=\frac{1}{\sqrt{-\vec{\nabla}_c t\cdot
\vec{\nabla}_c t}}=\frac{1}{i(\partial_\tau t)}=-i(\partial_t \tau).
\end{multlined} 
\end{equation}
Clearly, the lapse function is nothing but a measure of the rate of change of `proper time' with
respect to the `coordinate time'. Geometrically speaking, the lapse function is a measure of the projection of $\mathrm{d}t$ onto $\hat{\tau}$ vector as shown below in figure \ref{fig:zoom}.
\begin{figure}[H]
\centering
\includegraphics[width=60mm]{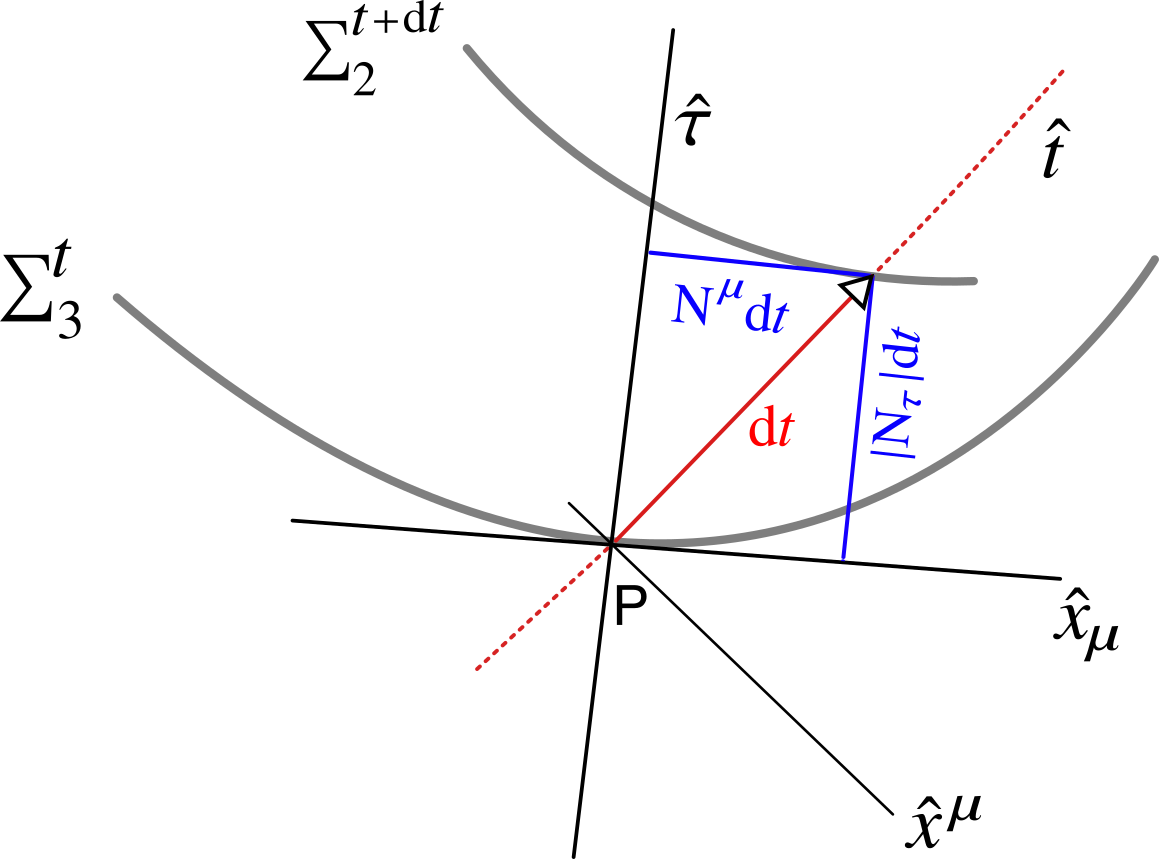}\hspace{0.3in}
\caption{A geometrical representation of the shift and the lapse function.}
\label{fig:zoom}
\end{figure}
{\noindent}In a similar fashion, one can also write the \tit{Shift function}, which is simply a measure of the movement of point P tangential to the hyper-surface $\sum_i$. In order to do so, we must first write the purely spatial metric $h_{\mu\nu}$. It can be easily proven that the metric $h_{\mu\nu}$ is given by
\begin{equation}
\tag{A.4.13}
\label{eq:A.4.13}
\begin{multlined}
h_{\mu\nu}=g_{\mu\nu}+(\hat{\mathbf{n}}_c\times
\hat{\mathbf{n}}_c),
\end{multlined} 
\end{equation}
where, $\times$ denotes the outer-product of the two vectors. In terms of the components, it is naturally written as:
\begin{equation}
\tag{A.4.14}
\label{eq:A.4.14}
\begin{multlined}
h_{\mu\nu}=g_{\mu\nu}+\mathrm{n}_\mu\mathrm{n}_\nu.
\end{multlined} 
\end{equation}\\
\begin{flushleft}\myrule[line width = 0.5mm]{fast cap reversed}{fast cap reversed}\end{flushleft}
\tit{\textbf{Sidenote}}:\\
In order to derive the purely spatial metric, let us explicitly write down the terms $\mathrm{n}_\mu\mathrm{n}_\nu$,
\begin{equation}
\tag{A.4.15}
\label{eq:A.4.15}
\begin{multlined}
\hat{\mathbf{n}}_c\times
\hat{\mathbf{n}}_c=\mathrm{n}_\mu\mathrm{n}_\nu=g_{\mu\gamma}
\mathrm{n}^\gamma \mathrm{n}_\nu.
\end{multlined} 
\end{equation}
We also know from \eqref{eq:A.4.9} that in terms of the components,
$\hat{\mathbf{n}}^c\equiv\{\mathrm{n}^\tau,\mathrm{n}^x\}\equiv\{\mathrm{n}^0,0\}$, i.e. only the
temporal contravariant component exists. Thus, \eqref{eq:A.4.15} reduces to
\begin{equation}
\tag{A.4.16}
\label{eq:A.4.16}
\begin{multlined}
\mathrm{n}_\mu\mathrm{n}_\nu=g_{\mu 0}\mathrm{n}_\nu\mathrm{n}^0.
\end{multlined} 
\end{equation}
The metric ($\hat{\mathbf{n}}_c\times\hat{\mathbf{n}}_c$) takes the form:
\begin{equation}
\tag{A.4.17}
\label{eq:A.4.17}
\begin{multlined}
(\hat{\mathbf{n}}_c\times\hat{\mathbf{n}}_c)=
\left\{ \begin{array}{cccc}
-g_{00} & -g_{01} & -g_{02} & -g_{03}\\
-g_{10} & \mathrm{n_1}\mathrm{n_1} & \mathrm{n_1}\mathrm{n_2} & \mathrm{n_1}\mathrm{n_3}\\
-g_{20} & \mathrm{n_2}\mathrm{n_1} & \mathrm{n_2}\mathrm{n_2} & \mathrm{n_2}\mathrm{n_3}\\
-g_{30} & \mathrm{n_3}\mathrm{n_1} & \mathrm{n_3}\mathrm{n_2} & \mathrm{n_3}\mathrm{n_3}\end{array} \right\},
\end{multlined} 
\end{equation}
given that $\mathrm{n}^a\mathrm{n}_a=\mathrm{n}^0\mathrm{n}_0=-1$ from \eqref{eq:A.4.10} and \eqref{eq:A.4.11}. The spatial metric is then given by:
\begin{equation}
\tag{A.4.18}
\label{eq:A.4.18}
\begin{multlined}
\mathbf{H}=\mathbf{G}+(\hat{\mathbf{n}}_c\times
\hat{\mathbf{n}}_c)=h_{\mu\nu}=
\left\{ \begin{array}{cccc}
0 & 0 & 0 & 0\\
0 & (g_{11}+\mathrm{n_1}\mathrm{n_1}) & (g_{12}+\mathrm{n_1}\mathrm{n_2}) & (g_{13}+\mathrm{n_1}\mathrm{n_3})\\
0 & (g_{21}+\mathrm{n_2}\mathrm{n_1}) & (g_{22}+\mathrm{n_2}\mathrm{n_2}) & (g_{23}+\mathrm{n_2}\mathrm{n_3})\\
0 & (g_{31}+\mathrm{n_3}\mathrm{n_1}) & (g_{32}+\mathrm{n_3}\mathrm{n_2}) & (g_{33}+\mathrm{n_3}\mathrm{n_3})\end{array} \right\}=
\left\{ \begin{array}{cc}
0 & 0 \\
0 & h_{\mu\nu}^{(3)} \\
\end{array} \right\}.
\end{multlined}
\end{equation}
The metric $h_{\mu\nu}$, geometrically speaking, is an operator for obtaining the projection of any geometrical entity on to a space-like hyper-surface. Let us take an example of a 4-vector
\begin{equation*}
\begin{multlined}
\vec{\chi}=a\,\hat{\mathbf{n}}^c+\sum_{i=1}^3 b^i\hat{\mathbf{t}}^i,
\end{multlined} 
\end{equation*}
where, $\hat{\mathbf{t}}^i$ are the spatial unit vectors for the contravariant basis. Now, the projected vector $\vec{\chi}_{_\mathrm{P}}$ is given by
\begin{equation}
\tag{A.4.19}
\label{eq:A.4.19}
\begin{multlined}
\vec{\chi}_{_\mathrm{P}}=\mathbf{H}\vec{\chi}=\mathbf{G}
\vec{\chi}+(\hat{\mathbf{n}}_c\times
\hat{\mathbf{n}}_c)\vec{\chi}.
\end{multlined} 
\end{equation}
In component form,
\begin{equation}
\tag{A.4.20}
\label{eq:A.4.20}
\begin{multlined}
\vec{\chi}_{_\mathrm{P}}=\chi_\mu=a\,g_{\mu\nu}\mathrm{n}^\nu+
\sum_{i=1}^3 b^i\mathbf{G}\,\hat{\mathbf{t}}^i+a\,\mathrm{n}_\mu
\mathrm{n}_\nu\mathrm{n}^\nu+\sum_{i=1}^3 b^i (\hat{\mathbf{n}}_c\times
\hat{\mathbf{n}}_c)\hat{\mathbf{t}}^i,
\end{multlined} 
\end{equation}
\begin{equation}
\tag{A.4.21}
\label{eq:A.4.21}
\begin{multlined}
\vec{\chi}_{_\mathrm{P}}=\chi_\mu=a\,(\mathrm{n}_\mu-\mathrm{n}_\mu)+
\sum_{i=1}^3 b^i[\mathbf{G}+(\hat{\mathbf{n}}_c\times
\hat{\mathbf{n}}_c)]\,\hat{\mathbf{t}}^i.
\end{multlined} 
\end{equation}
In detailed spatial component form,
\begin{equation}
\tag{A.4.22}
\label{eq:A.4.22}
\begin{multlined}
\vec{\chi}_{_\mathrm{P}}=\chi_\mu=
\sum_{i=1}^3 b^i(g_{\mu\nu}{\mathrm{t}_i}^\nu+\mathrm{n}_\mu\mathrm{n}
_\nu{\mathrm{t}_i}^\nu)=\sum_{i=1}^3 b^i({\mathrm{t}^i}_\mu+\mathrm{n}_\mu\mathrm{n}
_\nu{\mathrm{t}_i}^\nu)\equiv\sum_{i=1}^3 b^i\mathbf{t}_i,
\end{multlined} 
\end{equation}
where, $\hat{\mathbf{n}}_c\cdot\hat{\mathbf{t}}^i=\mathrm{n}_\nu
\,\mathrm{t}_i^\nu=0$, and $\hat{\mathbf{n}}_c\perp\hat{\mathbf{t}}^i$. Remember that the vector $\mathbf{t}_i$ is tangential to the hyper-surface,
while it is not necessarily of unit length. In fact, the magnitude
\begin{equation}
\tag{A.4.23}
\label{eq:A.4.23}
\begin{multlined}
|\vec{\chi}_{_\mathrm{P}}|=\Bigg|\sum_{i=1}^3 b^i\mathbf{t}_i\Bigg|,
\end{multlined} 
\end{equation}
is the length of the projection of the vector $\chi$ onto the spatial hyper-surface.
\begin{flushright}\myrule[line width = 0.5mm]{fast cap reversed}{fast cap reversed}\end{flushright}$\;$\\
We continue to evaluate the expression for the shift function $\mathrm{N}^\mu$ or, alternatively, the shift vector $\vec{\mathrm{N}}$. A vector field parallel to the coordinate time vector $\hat{t}$ is given by the contravariant derivative of $t$, i.e. $\nabla^\mu t$. This vector can now be projected onto the hyper-surface via the metric $h_{\mu\nu}$ in order to yield the shift vector or the shift function in covariant and contravariant components respectively:
\begin{equation}
\tag{A.4.24}
\label{eq:A.4.24}
\begin{multlined}
\mathrm{N}_\nu=h_{\mu\nu}(\nabla^\mu t)=h_{\mu\nu}(\partial^\mu t)\;\;\;\;\text{in covariant form,}
\end{multlined} 
\end{equation}
\begin{equation}
\tag{A.4.25}
\label{eq:A.4.25}
\begin{multlined}
\mathrm{N}^\nu=h_{\mu}^{\nu}(\nabla^\mu t)=h_{\mu}^{\nu}(\partial^\mu t)\;\;\;\;\text{in contravariant form.}
\end{multlined} 
\end{equation}
We have now laid all the ground work for our evaluation of the metric under the ADM formalism. Let us now write the magnitude for an abstract covariant temporal vector field, i.e. $z^\mu(\propto\nabla^\mu z)$, such that this magnitude is adjusted so as to satisfy the following relation:
\begin{equation}
\tag{A.4.26}
\label{eq:A.4.26}
\begin{multlined}
z^\mu=\nabla^\mu z=\mathrm{N}_\tau\mathrm{n}^\mu+\mathrm{N}^\mu\equiv\mathrm{N}
_\tau\hat{\mathbf{n}}^c+\vec{\mathrm{N}},
\end{multlined} 
\end{equation}
\begin{figure}[H]
\centering
\includegraphics[width=80mm]{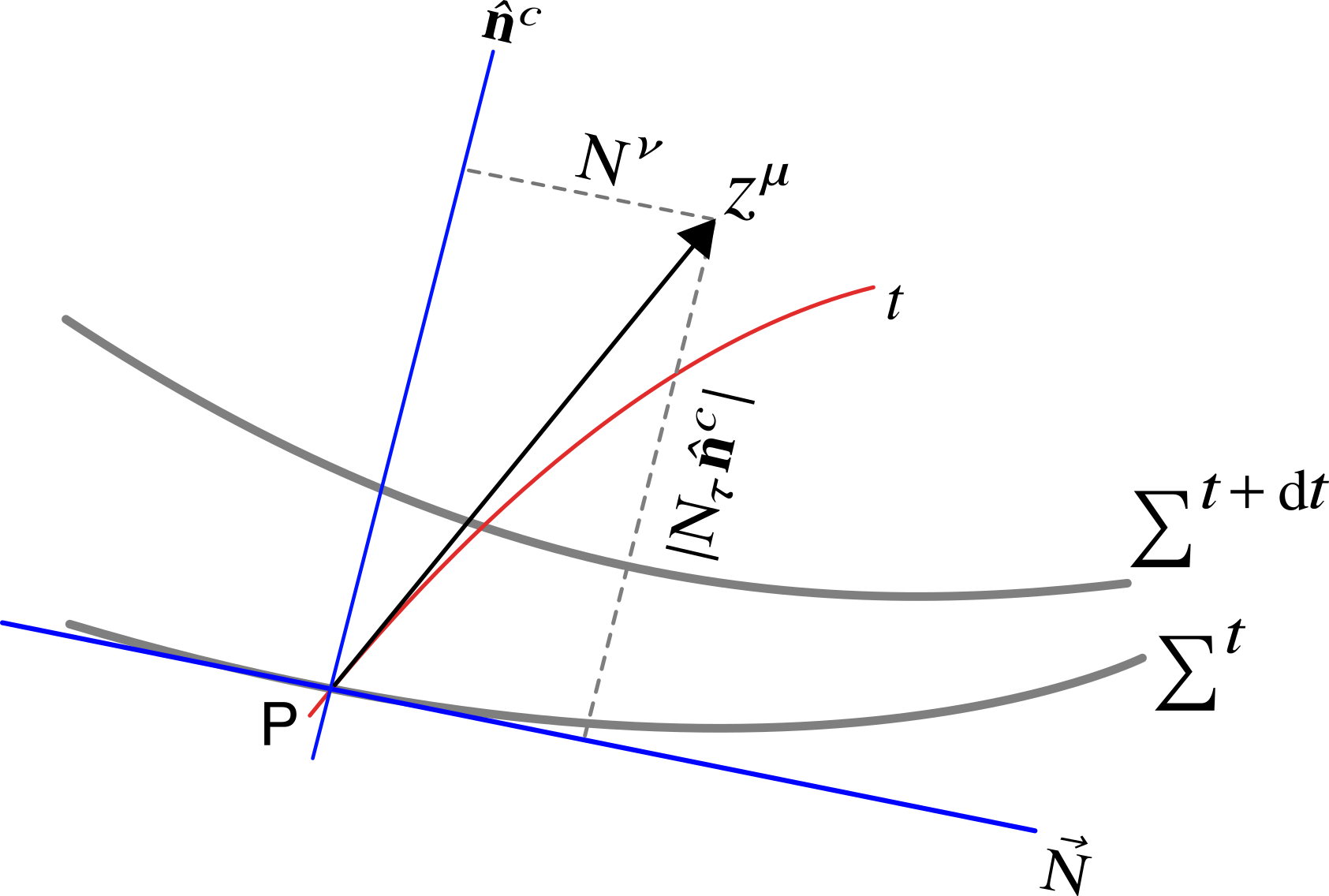}\hspace{0.3in}
\caption{An abstract vector field in ADM formalism.}
\label{fig:zoom2}
\end{figure}
{\noindent}as shown in figure \ref{fig:zoom2}\footnote{Note that the vector field $t^\mu$ is not necessarily time-like! Moreover, we consider the foliating scalar field to be the 'coordinate time' $t$ of the observer; this assumption can be given up in favor of any other general scalar field at the expense of more complicated mathematics. The conclusion however remains the same for the 3+1 decomposition.}. Recall that the directional derivatives $\vec{\partial}_i$ for $i\in\{t,x_\mu\}$ form a set of basis vectors for a vector field in the 4-dimensional space-time. Thus, the vector field $z^\mu$($\equiv \vec{\partial}_i\propto{\nabla}^\mu z$) also forms a subset of this basis vector set. By definition of the metric and using \eqref{eq:A.4.26},
\begin{equation}
\tag{A.4.27}
\label{eq:A.4.27}
\begin{multlined}
g_{00}=\vec{\partial}_t\cdot \vec{\partial}_t\equiv z^\mu z_\mu=(\mathrm{N}
_\tau\hat{\mathbf{n}}^c+\vec{\mathrm{N}})\cdot(\mathrm{N}
_\tau\hat{\mathbf{n}}^c+\vec{\mathrm{N}})=-\mathrm{N}_\tau^2+\mathrm{N}_\mu \mathrm{N}^\mu.
\end{multlined} 
\end{equation}
Similarly, 
\begin{equation}
\tag{A.4.28}
\label{eq:A.4.28}
\begin{multlined}
g_{0i}=\vec{\partial}_t\cdot \vec{\partial}_i= (\mathrm{N}
_\tau\hat{\mathbf{n}}^c+\vec{\mathrm{N}})\cdot\vec{\partial}_i=
\mathrm{N}_\mu\;\;\;\;\text{since,}\;\;\;\hat{\mathbf{n}}^c\perp\vec{\partial}_i.
\end{multlined} 
\end{equation}
We can now write the complete 4-dimensional metric as:
\begin{equation}
\tag{A.4.29}
\label{eq:A.4.29}
\begin{multlined}
g_{\mu\nu}=
\left\{ \begin{array}{cc}
g_{00} & g_{0\nu}\vspace{0.025in} \\ 
g_{\mu 0} & h_{\mu\nu}^{(3)} \\
\end{array} \right\}=
\left\{ \begin{array}{cccc}
-\mathrm{N}_\tau^2+\mathrm{N}_\mu \mathrm{N}^\mu & \mathrm{N}_1 & \mathrm{N}_2 & \mathrm{N}_3\\
\mathrm{N}_1 & (g_{11}+\mathrm{n_1}\mathrm{n_1}) & (g_{12}+\mathrm{n_1}\mathrm{n_2}) & (g_{13}+\mathrm{n_1}\mathrm{n_3})\\
\mathrm{N}_2 & (g_{21}+\mathrm{n_2}\mathrm{n_1}) & (g_{22}+\mathrm{n_2}\mathrm{n_2}) & (g_{23}+\mathrm{n_2}\mathrm{n_3})\\
\mathrm{N}_3 & (g_{31}+\mathrm{n_3}\mathrm{n_1}) & (g_{32}+\mathrm{n_3}\mathrm{n_2}) & (g_{33}+\mathrm{n_3}\mathrm{n_3})\end{array} \right\}.
\end{multlined}
\end{equation}
The line element is then given by:
\begin{equation}
\tag{A.4.30}
\label{eq:A.4.30}
\begin{multlined}
g_{\mu\nu}\mathrm{d}x^\mu\mathrm{d}x^\nu=\mathrm{N}_\tau^2
\mathrm{d}t^2+h_{\mu\nu}^{(3)}(\mathrm{d}x^\mu+\mathrm{N}^\mu\mathrm{d}t)(\mathrm{d}x^\nu+
\mathrm{N}^\nu\mathrm{d}t),
\end{multlined}
\end{equation}
while the dual metric takes the form,
\begin{equation}
\tag{A.4.31}
\label{eq:A.4.31}
\begin{multlined}
g_{\mu\nu}=
\left\{ \begin{array}{cc}
g^{00} & g^{0\nu}\vspace{0.025in} \\ 
g^{\mu 0} & h_{\mu\nu}^{(3)} \\
\end{array} \right\}=
\left\{ \begin{array}{cc}
\displaystyle{-\frac{1}{\mathrm{N}_\tau^2}} & \displaystyle{\frac{\mathrm{N}^\nu}{\mathrm{N}_\tau^2}}\vspace{0.025in} \\ 
\displaystyle{\frac{\mathrm{N}^\mu}{\mathrm{N}_\tau^2}} & \displaystyle{h^{\mu\nu}_{(3)}-
\frac{\mathrm{N}^\mu\mathrm{N}^\nu}{\mathrm{N}_\tau^2}} \\
\end{array} \right\}.
\end{multlined}
\end{equation}


\newpage

\begin{thebibliography}{16}
\providecommand{\natexlab}[1]{#1}
\providecommand{\url}[1]{\texttt{#1}}
\expandafter\ifx\csname urlstyle\endcsname\relax
\providecommand{\doi}[1]{doi: #1}\else
\providecommand{\doi}{doi: \begingroup \urlstyle{rm}\Url}\fi

\bibitem[Adshead et~al.(2009)Adshead, Easther, , and Lim]{r4}
P~Adshead, R~Easther, , and E~A Lim.
\newblock {\fontfamily{ppl}\selectfont \textit{In-In formalism and cosmological
perturbations}}.
\newblock \emph{\textup{\textbf{Phys. Rev. D}}}, 80\penalty0 (8):\penalty0
083521, 2009.

\bibitem[Arroja et~al.(2008)Arroja, Mizuno, and Koyama]{r6}
F~Arroja, S~Mizuno, and K~Koyama.
\newblock {\fontfamily{ppl}\selectfont \textit{Non-Gaussianity from the
bispectrum in general multiple field Inflation}}.
\newblock \emph{\textup{\textbf{JCAP}}}, 2008\penalty0 (08):\penalty0 015,
2008.

\bibitem[Bartolo et~al.(2004)Bartolo, Matarrese, and Riotto]{r9}
N~Bartolo, S~Matarrese, and A~Riotto.
\newblock {\fontfamily{ppl}\selectfont \textit{Non-Gaussianity in the curvaton
scenario}}.
\newblock \emph{\textup{\textbf{Phys. Rev. D}}}, 69\penalty0 (4):\penalty0
043503, 2004.

\bibitem[Baumann(2009)]{r15}
D~Baumann.
\newblock {\fontfamily{ppl}\selectfont \textit{TASI Lectures on Inflation}}.
\newblock \emph{\textup{\textbf{arXiv}}}, 0907.5424\penalty0 (--):\penalty0
159, 2009.

\bibitem[Collaboration(2006)]{r2}
The~Planck Collaboration.
\newblock {\fontfamily{ppl}\selectfont \textit{The Scientific Programme of
Planck}}.
\newblock \emph{\textup{\textbf{arXiv}}}, astro-ph/0604069\penalty0
(--):\penalty0 --, 2006.

\bibitem[De~Felice and Tsujikawa(2010)]{r7}
A~De~Felice and S~Tsujikawa.
\newblock {\fontfamily{ppl}\selectfont \textit{Generalized Brans-Dicke
theories}}.
\newblock \emph{\textup{\textbf{JCAP}}}, 2010\penalty0 (07):\penalty0 024,
2010.

\bibitem[De~Felice and Tsujikawa(2011)]{r11}
A~De~Felice and S~Tsujikawa.
\newblock {\fontfamily{ppl}\selectfont \textit{Primordial non-gaussianities in
general modified gravitational models of inflation}}.
\newblock \emph{\textup{\textbf{JCAP}}}, 2011\penalty0 (04):\penalty0 029,
2011.

\bibitem[Deffayet et~al.(2011)Deffayet, Gao, Steer, and Zahariade]{r10}
C~Deffayet, X~Gao, D~A Steer, and G~Zahariade.
\newblock {\fontfamily{ppl}\selectfont \textit{From k-essence to generalised
Galileons}}.
\newblock \emph{\textup{\textbf{Phys. Rev. D}}}, 84\penalty0 (6):\penalty0
064039, 2011.

\bibitem[Fujii and Maeda(2003)]{r5}
Y~Fujii and K~Maeda.
\newblock \emph{{\fontfamily{ppl}\selectfont \textit{The Scalar-Tensor Theory
of Gravitation}}}.
\newblock \textup{\textbf{Cambridge University Press}}, 2003.
\newblock ISBN 9780521811590.

\bibitem[Gao and Steer(2011)]{r8}
X~Gao and D~A Steer.
\newblock {\fontfamily{ppl}\selectfont \textit{Inflation and primordial
non-Gaussianities of generalized Galileons}}.
\newblock \emph{\textup{\textbf{JCAP}}}, 2011\penalty0 (12):\penalty0 019,
2011.

\bibitem[Horndeski(1974)]{r16}
G~W Horndeski.
\newblock {\fontfamily{ppl}\selectfont \textit{Second-order scalar-tensor field
equations in a four-dimensional space}}.
\newblock \emph{\textup{\textbf{IJTP}}}, 10\penalty0 (6):\penalty0 363--384,
1974.

\bibitem[Lazarides(2006)]{r1}
G~Lazarides.
\newblock {\fontfamily{ppl}\selectfont \textit{Basics of Inflationary
Cosmology}}.
\newblock \emph{\textup{\textbf{J.Phys.Conf.Ser.}}}, 53\penalty0 (1):\penalty0
528--550, 2006.

\bibitem[Maldacena(2003)]{r14}
J~M Maldacena.
\newblock {\fontfamily{ppl}\selectfont \textit{Non-Gaussian features of
primordial fluctuations in single field inflationary models}}.
\newblock \emph{\textup{\textbf{JHEP}}}, 2003\penalty0 (05):\penalty0 013,
2003.

\bibitem[Ostrogradski(1850)]{r12}
M~Ostrogradski.
\newblock {\fontfamily{ppl}\selectfont \textit{Memoires sur les equations
differentielles relatives au probleme des isoperimetres}}.
\newblock \emph{\textup{\textbf{Mem. Ac. St. Petersbourg}}}, VI\penalty0
(--):\penalty0 385, 1850.

\bibitem[Randall(1997)]{r3}
L~Randall.
\newblock {\fontfamily{ppl}\selectfont \textit{Supersymmetry and Inflation}}.
\newblock \emph{\textup{\textbf{arXiv}}}, hep-ph/9711471\penalty0
(--):\penalty0 --, 1997.

\bibitem[Takamizu et~al.(2010)Takamizu, Mukohyama, Sasaki, and Tanaka]{r13}
Y~I Takamizu, S~Mukohyama, M~Sasaki, and Y~Tanaka.
\newblock {\fontfamily{ppl}\selectfont \textit{Non-Gaussianity of superhorizon
curvature perturbations beyond $\delta$N formalism}}.
\newblock \emph{\textup{\textbf{JCAP}}}, 2010\penalty0 (06):\penalty0 019,
2010.

\end{thebibliography}
\end{document}